\documentclass[11pt,twoside,a4paper]{book}
\usepackage{aicescover}

\usepackage[english]{babel}
\usepackage{mathptmx}

\usepackage{amsmath}
\usepackage{amsfonts}
\usepackage{amssymb}
\usepackage{amsbsy}

\usepackage[pass]{geometry} 

\usepackage{algorithm}
\usepackage{algorithmic}

\usepackage{graphicx}
\usepackage[figuresright]{rotating}
\usepackage{color}
\usepackage{cite}
\usepackage{subfig}

\usepackage{fancyvrb}

\usepackage{soul}

\usepackage{framed}

\usepackage{delarray}

\usepackage[pdfauthor={Diego Fabregat Traver},
            pdftitle={Knowledge-Based Automatic Generation of Linear Algebra Algorithms and Code},
            pdfsubject={Ph.D. Dissertation},
			pdfkeywords={Automation, domain-specific languages and compilers, numerical linear algebra, FLAME, CLAK, CL1CK, generation of algorithms and code},
			colorlinks=true,
            linkcolor=blue
		]{hyperref}

\usepackage{booktabs}
\usepackage{colortbl}

\usepackage{pgf}
\usepackage{tikz}
\usetikzlibrary{shapes,arrows,shadows}

\renewcommand\chaptermark[1]{\markboth{\chaptername\ \thechapter. #1}{}}

\usepackage{float}
\floatstyle{boxed}
\newfloat{mybox}{htb}{lob}[chapter]
\floatname{mybox}{Box}
\newsubfloat{mybox}

\newcommand{\upptriRuleTwoByTwo}[6]
{
  \begin{aligned}
    \renewcommand{\arraystretch}{1.4}
      #1_{#2 \times #3} 
	  \rightarrow &
	  \left( 
		\begin{array}{c@{\;\;}|@{\;\;}c} 
		  #1_{TL} & #1_{TR} \\\hline 
		  0       & #1_{BR} 
		\end{array} 
	  \right) \\
      \small \textnormal{where } & #1_{#4} \textnormal{ is } #5 \times #6
  \end{aligned}
}

\newcommand{\lowtriRuleTwoByTwo}[6]
{
  \begin{aligned}
    \renewcommand{\arraystretch}{1.4}
      #1_{#2 \times #3} 
	  \rightarrow &
	  \left( 
		\begin{array}{c@{\;\;}|@{\;\;}c} 
		  #1_{TL} & 0 \\\hline 
		  #1_{BL} & #1_{BR} 
		\end{array} 
	  \right) \\
      \small \textnormal{where } & #1_{#4} \textnormal{ is } #5 \times #6
  \end{aligned}
}

\newcommand{\symmRuleTwoByTwo}[6]
{
  \begin{aligned}
    \renewcommand{\arraystretch}{1.4}
      #1_{#2 \times #3} 
	  \rightarrow &
	  \left( 
		\begin{array}{c@{\;\;}|@{\;\;}c} 
		  #1_{TL} & #1_{BL}^T \\\hline 
		  #1_{BL} & #1_{BR} 
		\end{array} 
	  \right) \\
      \small \textnormal{where } & #1_{#4} \textnormal{ is } #5 \times #6
  \end{aligned}
}

\newcommand{\ruleTwoByTwo}[6]
{
  \begin{aligned}
    \renewcommand{\arraystretch}{1.4}
      #1_{#2 \times #3} 
	  \rightarrow &
	  \left( 
		\begin{array}{c@{\;\;}|@{\;\;}c} 
		  #1_{TL} & #1_{TR} \\\hline 
		  #1_{BL} & #1_{BR} 
		\end{array} 
	  \right) \\
      \small \textnormal{where } & #1_{#4} \textnormal{ is } #5 \times #6
  \end{aligned}
}

\newcommand{\ruleTwoByOne}[6]
{
  \begin{aligned}
    \renewcommand{\arraystretch}{1.4}
      #1_{#2 \times #3} 
	  \rightarrow &
	  \left( 
		\begin{array}{c} 
		  #1_{T} \\\hline 
		  #1_{B}
		\end{array} 
	  \right) \\
      \small \textnormal{where } & #1_{#4} \textnormal{ is } #5 \times #6
  \end{aligned}
}

\newcommand{\ruleOneByTwo}[6]
{
  \begin{aligned}
    \renewcommand{\arraystretch}{1.4}
      #1_{#2 \times #3} 
	  \rightarrow &
	  \left( 
		\begin{array}{c@{\;\;}|@{\;\;}c} 
		  #1_{L} & #1_{R}
		\end{array} 
	  \right) \\
      \small \textnormal{where } & #1_{#4} \textnormal{ is } #5 \times #6
  \end{aligned}
}

\newcommand{\ruleOneByOne}[3]
{
  \begin{aligned}
    \renewcommand{\arraystretch}{1.4}
      #1_{#2 \times #3} 
	  \rightarrow &
	  \left( 
		\begin{array}{c} 
		  #1
		\end{array} 
	  \right) \\
      \small \textnormal{where } & #1 \textnormal{ is } #2 \times #3
  \end{aligned}
}

\usepackage{tabularx}

\newcolumntype{I}{!{\vrule width 1.3pt}}
\newlength\savedwidth
\newcommand\whline{\noalign{\global\savedwidth\arrayrulewidth
                            \global\arrayrulewidth 1.3pt}%
           \hline
           \noalign{\global\arrayrulewidth\savedwidth}}

\newcommand{\myFlaTwoByTwo}[4]{
\renewcommand{\arraystretch}{1.2}
\setlength{\arraycolsep}{0pt}
  \left(
	\begin{array}{c@{\;\;}|@{\;\;}c} 
      #1 & #2 \\ \hline
      #3 & #4 
    \end{array} 
  \right)
}

\newcommand{\myFlaTwoByOne}[2]{
\renewcommand{\arraystretch}{1.2}
  \left(
	\begin{array}{c} 
      #1 \\ \hline
      #2  
    \end{array} 
  \right)
}

\newcommand{\myFlaOneByTwo}[2]{
\renewcommand{\arraystretch}{1.2}
  \left(
	\begin{array}{c@{\;\;}|@{\;\;}c} 
      #1 & #2
    \end{array} 
  \right)
}

\newcommand{\myFlaTwoByTwoI}[4]{
\renewcommand{\arraystretch}{1.4}
\left( 
\begin{array}{c@{\;\;}I@{\;\;}c} 
#1 & #2 \\ \whline
#3 & #4 
\end{array} 
\right)
}

\newcommand{\myFlaThreeByThree}[9]{
\renewcommand{\arraystretch}{1.4}
\left( 
\begin{array}{c@{\;\;}|@{\;\;}c@{\;\;}|@{\;\;}c} 
#1 & #2 & #3 \\ \hline
#4 & #5 & #6 \\ \hline
#7 & #8 & #9
\end{array} 
\right) 
}

\newcommand{\myFlaThreeByThreeTLI}[9]{
\renewcommand{\arraystretch}{1.4}
\left( 
\begin{array}{c@{\;\;}|@{\;\;}c@{\;\;}I@{\;\;}c} 
#1 & #2 & #3 \\ \hline
#4 & #5 & #6 \\ \whline
#7 & #8 & #9
\end{array} 
\right) 
}

\newcommand{\myFlaThreeByThreeBRI}[9]{
\left( 
\begin{array}{c I c | c}
#1 & #2 & #3 \\ \whline
#4 & #5 & #6 \\ \hline
#7 & #8 & #9
\end{array} 
\right)
}

\newcommand{\myFlaThreeByThreeTRI}[9]{
\renewcommand{\arraystretch}{1.4}
\left( 
\begin{array}{c@{\;\;}I@{\;\;}c@{\;\;}|@{\;\;}c} 
#1 & #2 & #3 \\ \hline
#4 & #5 & #6 \\ \whline
#7 & #8 & #9
\end{array} 
\right) 
}

\newcommand{\myFlaThreeByThreeBLI}[9]{
\left( 
\begin{array}{c@{\;\;}|@{\;\;}c@{\;\;}I@{\;\;}c} 
#1 & #2 & #3 \\ \whline
#4 & #5 & #6 \\ \hline
#7 & #8 & #9
\end{array} 
\right)
}

\newcommand{\myFlaOneByTwoI}[2]{
\left( 
\begin{array}{@{\hspace{1pt}}c@{\hspace{2pt}}I@{\hspace{2pt}}c@{\hspace{1pt}}}
#1 & #2 
\end{array} 
\right)
}

\newcommand{\myFlaOneByThreeLI}[3]{
\left( 
\begin{array}{c@{\;\;}|@{\;\;}c@{\;\;}I@{\;\;}c} 
#1 & #2 & #3 
\end{array} 
\right)
}

\newcommand{\myFlaOneByThreeRI}[3]{
\left( 
\begin{array}{c@{\;\;}I@{\;\;}c@{\;\;}|@{\;\;}c} 
#1 & #2 & #3 
\end{array} 
\right)
}

\newcommand{\FlaTwoByTwoI}[4]{
\left( 
\begin{array}{c I c}
#1 & #2 \\ \whline
#3 & #4 
\end{array} 
\right)
}

\newcommand{\FlaThreeByThreeTLI}[9]{
\left( 
\begin{array}{c | c I c}
#1 & #2 & #3 \\ \hline
#4 & #5 & #6 \\ \whline
#7 & #8 & #9
\end{array} 
\right) 
}

\newcommand{\FlaThreeByThreeBRI}[9]{
\left( 
\begin{array}{c I c | c}
#1 & #2 & #3 \\ \whline
#4 & #5 & #6 \\ \hline
#7 & #8 & #9
\end{array} 
\right)
}

\newcommand{\PBefore}{ P_{\rm before} }
\newcommand{\PAfter} { P_{\rm after}  }
\newcommand{\PPost}  { P_{\rm post}   }
\newcommand{\PPre}   { P_{\rm pre}    }
\newcommand{\PInv}   { P_{\rm inv}    }
\newcommand{\EInv}   { E_{\rm inv}    }
\newcommand{\EBefore}{ E_{\rm before} }
\newcommand{\EAfter} { E_{\rm after}  }
\newcommand{\WSoperation}{
  $[ D, E, F, \ldots ] = {\rm op}( A, B, C, D, \ldots )$ 
}

\newcommand{\WSprecondition}{
  $\PPre$
}

\newcommand{\WSpartition}{
}

\newcommand{\WSpartitionsizes}{
  $\dots$
}

\newcommand{\WSinvariant}{
  $\PInv$
}

\newcommand{\WSguard}{ 
  $ G $
}

\newcommand{\WStopofloop}{
  $ \left\{ \left( \mbox{\WSinvariant} \right) \wedge 
    \left( \mbox{\WSguard} \right) \right\} $
}

\newcommand{\WSrepartition}{
  \begin{minipage}[t]{1in}
    \ \\
  \end{minipage}
}

\newcommand{\WSrepartitionsizes}{
  $\dots$
}

\newcommand{\WSbeforeupdate}{
  $\PBefore$ 
}

\newcommand{\WSupdate}{
  \begin{minipage}[t]{1in}
    $ S_U $
  \end{minipage}
}

\newcommand{\WSafterupdate}{
  $\PAfter$
}

\newcommand{\WSmoveboundary}{
}

\newcommand{\WSafterloop}{
  $\left\{ \left( \mbox{\WSinvariant} \right) \wedge 
   \neg \left( \mbox{\WSguard} \right) \right\} $
}

\newcommand{\WSpostcondition}{
  $\PPost$
}

\newcommand{\operation}{ [ D, E, F, \ldots ] \becomes {\rm op}( A, B, C, D, \ldots ) }
\newcommand{\routinename}{ [ D, E, F, \ldots ] \becomes {\rm op}( A, B, C, D, \ldots ) }

\newlength{\tabwidth}
\newlength{\tabmargin}
\newcommand{\resetsteps}{
\renewcommand{\WSoperation}
  { $[ D, E, F, \ldots ] = {\rm op}( A, B, C, D, \ldots )$ }
\renewcommand{\WSprecondition}{ $\PPre$ }
\renewcommand{\WSpartition}{ }
\renewcommand{\WSpartitionsizes}{}
\renewcommand{\WSinvariant}{ $\PInv$ }
\renewcommand{\WSguard}{ $ G $ }
\renewcommand{\WStopofloop}{
  $ \left\{ \left( \mbox{\WSinvariant} \right) \wedge 
    \left( \mbox{\WSguard} \right) \right\} $ }
\renewcommand{\WSrepartition}{
  \begin{minipage}[t]{1in}
    \ \\
  \end{minipage} }
\renewcommand{\WSrepartitionsizes}{}
\renewcommand{\WSbeforeupdate}{ $\PBefore$ }
\renewcommand{\WSupdate}{
  \begin{minipage}[t]{1in}
    $ S_U $
  \end{minipage} }
\renewcommand{\WSafterupdate}{ $\PAfter$ }
\renewcommand{\WSmoveboundary}{}
\renewcommand{\WSafterloop}{
  $\left\{ \left( \mbox{\WSinvariant} \right) \wedge 
   \neg \left( \mbox{\WSguard} \right) \right\} $ }
\renewcommand{\WSpostcondition}{ $\PPost$ }
\renewcommand{\operation}{ [ D, E, F, \ldots ] \becomes {\rm op}( A, B, C, D, \ldots ) }
\renewcommand{\routinename}{ [ D, E, F, \ldots ] \becomes {\rm op}( A, B, C, D, \ldots ) }
}

\newcommand{\FlaStartCompute}{
\setlength{\unitlength}{0.4in}
\begin{picture}(3,0.01)
\put(0,0){\line(1,0){4}}
\put(0,0.01){\line(1,0){4}}
\end{picture} 
}

\newcommand{\FlaEndCompute}{
\setlength{\unitlength}{0.4in}
\begin{picture}(3,0.01)
\put(0,0){\line(1,0){4}} 
\put(0,0.01){\line(1,0){4}} 
\end{picture} 
}

\newcommand{\whilecolor}{}
\newcommand{\partcolor}{}

\definecolor{redbg}{RGB}{254,158,152}

\newcommand{\FlaStartComputeLong}{
\setlength{\unitlength}{0.4in}
\begin{picture}(6,0.01)
\put(0,0){\line(1,0){11.0}}
\put(0,0.01){\line(1,0){11.0}}
\end{picture} 
}

\newcommand{\FlaEndComputeLong}{
\setlength{\unitlength}{0.4in}
\begin{picture}(6,0.01)
\put(0,0){\line(1,0){11.0}} 
\put(0,0.01){\line(1,0){11.0}} 
\end{picture} 
}

\newcommand{\myFlaAlgorithm}{
\begin{center}
\begin{tabular}{|l |} \hline\\[-1mm]
  {\bf \partcolor{Partition}} {\WSpartition}\\
	\hspace*{2.2mm} {\bf where} \WSpartitionsizes \\[2mm]

  {\bf \whilecolor{while}} \WSguard { \bf \whilecolor{do}} \\[1mm]
    \ \hspace{2.2mm} {\scriptsize \bf Repartition} \\
    \ \hspace{3.2mm} \WSrepartition \\
	\hspace{5.1mm} {\bf where} \WSrepartitionsizes \\[1mm]
    {\hspace{0.0in} \FlaStartComputeLong} \\
      {\hspace{0.0in} \WSupdate} \\
    {\hspace{0.0in} \FlaEndComputeLong} \\[1mm]
    \ \hspace{2.2mm} {\scriptsize \bf Continue with} \\
    {\ \hspace{3.2mm} \WSmoveboundary} \\
  {{\bf \whilecolor{endwhile}}} \\ \hline
\end{tabular}
\end{center}
}

\newcommand{\myFlaFullAlgorithm}{
\begin{center}
\begin{tabular}{|l |} \hline\\[-1mm]
	\raisebox{1mm}{{\bf Algorithm}: \WSoperation} \\ \whline
  {\bf \partcolor{Partition}} \\ {\tiny \WSpartition}\\
	\hspace*{2.2mm} {\bf where} \WSpartitionsizes \\[2mm]

  {\bf \whilecolor{while}} \WSguard { \bf \whilecolor{do}} \\[1mm]
    \ \hspace{2.2mm} {\scriptsize \bf Repartition} \\
    \ \hspace{3.2mm} \tiny \WSrepartition \\
	\hspace{5.1mm} {\bf where} \WSrepartitionsizes \\[1mm]
    {\hspace{0.0in} \FlaStartComputeLong} \\
      {\hspace{0.0in} \WSupdate} \\
    {\hspace{0.0in} \FlaEndComputeLong} \\[1mm]
    \ \hspace{2.2mm} {\scriptsize \bf Continue with} \\
    {\ \hspace{3.2mm} \tiny \WSmoveboundary} \\
  {{\bf \whilecolor{endwhile}}} \\ \hline
\end{tabular}
\end{center}
}

\newcommand{\FlaAlgorithmNarrowTriinv}{
\begin{center}
\begin{tabular}{|l |} \hline\\[-1mm]
	\raisebox{1mm}{{\bf Algorithm}: \WSoperation} \\\whline
	  \rowcolor[gray]{.8} \raisebox{-1mm}{$\{ L = \hat{L} \; \wedge \; LowerTriangular(L) \}$} \\[2mm]
  \raisebox{-5mm}{{\bf \partcolor{Partition}} {\WSpartition}}\\
	\hspace*{2.2mm} {\bf where} \WSpartitionsizes \\[1mm]
	  \rowcolor[gray]{.8} \raisebox{-3.3mm}{$\{ \scriptsize \myFlaTwoByTwo{L_{TL} := \hat{L}_{TL}^{-1}}{0}{\neq}{\qquad \neq \qquad \phantom{}} \}$ }\\[6mm]
  \raisebox{-1mm}{{\bf \whilecolor{While}} \WSguard { \bf \whilecolor{do}}} \\[1mm]
	  \rowcolor[gray]{.8} \raisebox{-3.3mm}{\hspace*{1.8mm} $\{ \scriptsize \myFlaTwoByTwo{L_{TL} := \hat{L}_{TL}^{-1}}{0}{\neq}{\; \neq \; \phantom{}} \wedge size( L_{TL} ) < size( L ) \}$}  \\[6mm]
    \raisebox{-1mm}{\ \hspace{0.0in} {\bf Repartition}} \\[1mm]
    \ \hspace{0.0in} \WSrepartition \\
		\hspace*{2.2mm} {\bf where} \WSrepartitionsizes \\
    {\hspace{0.0in} \FlaStartCompute} \\
    {\hspace{0.0in} \WSupdate} \\
    {\hspace{0.0in} \FlaEndCompute} \\
   \ \hspace{0.0in} {\bf Continue with} \\
    {\ \hspace{0.0in} \WSmoveboundary} \\
	  \rowcolor[gray]{.8} \raisebox{-3.3mm}{\hspace*{1.8mm} $\{ \scriptsize \myFlaTwoByTwo{L_{TL} := \hat{L}_{TL}^{-1}}{0}{\neq}{\qquad \neq \qquad \phantom{}} \}$ }\\[6mm]
   \raisebox{-1mm}{{{\bf \whilecolor{endwhile}}}} \\[1mm]
	  \rowcolor[gray]{.8} \raisebox{-3.3mm}{$\{ \scriptsize \myFlaTwoByTwo{L_{TL} := \hat{L}_{TL}^{-1}}{0}{\neq}{\; \neq \; \phantom{}} \} \wedge \neg(size( L_{TL} ) < size( L ) )$} \\[6mm]
	  \rowcolor[gray]{.8} \raisebox{-1.2mm}{$\{ L := \hat{L}^{-1} \} $} \\[2mm]\hline
\end{tabular}
\end{center}
}

\definecolor{known}{RGB}{0, 100, 0} 
\definecolor{unknown}{rgb}{1.0, 0.0, 0.0} 
\definecolor{aicesred}{RGB}{152,26,37}
\definecolor{rwthblue}{RGB}{91,162,222}
\newcommand{\known}[1]{\textcolor{known}{#1}}

\newcommand{\unknown}[1]{\textcolor{unknown}{#1}}

\newcommand{\click}{{\textsc{Cl\makebox[.58\width][c]{1}ck}}}
\newcommand{\clickplain}{\textsc{Cl\hspace{-.5mm}1\hspace{-.5mm}ck}}
\newcommand{\clak}{\textsc{Clak}}
\newcommand{\lu}{$LU$ factorization}
\newcommand{\cs}{coupled Sylvester equation}
\newcommand{\prop}[2]{{\tt{#1(}}#2{\tt{)}}}

\newcommand{\gtrsm}{g\textsc{trsm}}
\newcommand{\gchol}{g\textsc{chol}}

\newcommand{\bi}{\begin{itemize}}
\newcommand{\ei}{\end{itemize}}

\newcommand{\dv}[2]{\frac{d#1}{d#2}}
\newcommand{\dvop}[1]{{\tt dv(#1)}}

\newcolumntype{i}{!{\vrule width 1.0pt}}

\setul{}{.3ex}

\usepackage{listings}
\usepackage{caption}
\DeclareCaptionFont{black}{\color{black}}
\DeclareCaptionFormat{listing}{\colorbox[cmyk]{0.23, 0.15, 0.15, 0.01}{\parbox{\textwidth}{\hspace{15pt}#1#2#3}}}
\renewcommand{\lstlistingname}{Algorithm}

\newcommand{\LUGraphLevels}[1]
{
	\begin{tikzpicture}[scale=0.6,
					ball/.style={circle, fill=gray!70!white, draw=none, text=black, circular drop shadow},
					arrow/.style={->, shorten <= 1pt, >=stealth,thick}]
		\node[ball, fill=#1, at={( 0.0,  4.0)}] (n1) {1};
		\node[ball, fill=#1, at={(-2.0,  2.0)}] (n2) {2};
		\node[ball, fill=#1, at={( 2.0,  2.0)}] (n3) {3};
		\node[ball, fill=#1, at={( 0.0,  0.0)}] (n4) {4};
		\node[ball, fill=#1, at={( 0.0, -2.0)}] (n5) {5};
		\node[text=gray, at={(-4, 4)}]() {Level 1};
		\draw[-, thick, color=lightgray] (-4, 3) -- (4, 3);
		\node[text=gray, at={(-4, 2)}]() {Level 2};
		\draw[-, thick, color=lightgray] (-4, 1) -- (4, 1);
		\node[text=gray, at={(-4, 0)}]() {Level 3};
		\draw[-, thick, color=lightgray] (-4,-1) -- (4,-1);
		\node[text=gray, at={(-4, -2)}]() {Level 4};
		\draw[arrow] (n1.south) -- (n2.north);
		\draw[arrow] (n1.south) -- (n3.north);
		\draw[arrow] (n2.south) -- (n4.north);
		\draw[arrow] (n3.south) -- (n4.north);
		\draw[arrow] (n4.south) -- (n5.north);
	\end{tikzpicture}
}

\newcommand{\depGraphLU}[5]
{
	\begin{tikzpicture}[scale=0.6,
					ball/.style={circle, fill=gray!70!white, draw=none, text=black, circular drop shadow},
					arrow/.style={->, shorten <= 1pt, >=stealth,thick}]
		\node[ball, fill=#1, at={(-2.5,  0.9)}] (n1) {1};
		\node[ball, fill=#2, at={( 2.5,  0.9)}] (n2) {2};
		\node[ball, fill=#3, at={(-2.5, -0.9)}] (n3) {3};
		\node[ball, fill=#4, at={( 2.5, -0.9)}] (n4) {4};
		\node[ball, fill=#5, at={( 2.5, -2.6)}] (n5) {5};
		\draw[-, thick, color=lightgray] (-5,0) -- (5,0);
		\draw[-, thick, color=lightgray] (0,-2.0) -- (0,2.0);
		\draw[arrow] (n1.south) -- (n3.north);
		\draw[arrow] (n1.east)  -- (n2.west);
		\draw[arrow] (n3.east)  -- (n4.west);
		\draw[arrow] (n2.south) -- (n4.north);
		\draw[arrow] (n4.south) -- (n5.north);
	\end{tikzpicture}
}

\newcommand{\smallDepGraphLU}[5]
{
	\begin{tikzpicture}[
					ball/.style={circle, fill=gray!70!white, draw=none, text=black},
					arrow/.style={->, shorten <= 1pt, >=stealth,thick}]
		\node[ball, fill=#1, at={(-0.75,  0.32)}] (n1) {};
		\node[ball, fill=#2, at={( 0.75,  0.32)}] (n2) {};
		\node[ball, fill=#3, at={(-0.75, -0.32)}] (n3) {};
		\node[ball, fill=#4, at={( 0.75, -0.32)}] (n4) {};
		\node[ball, fill=#5, at={( 0.75, -0.88)}] (n5) {};
		\draw[-, thick, color=lightgray] (-1.5,0) -- (1.5,0);
		\draw[-, thick, color=lightgray] (0, -0.8) -- (0, 0.8);
		\draw[arrow] (n1.south) -- (n3.north);
		\draw[arrow] (n1.east)  -- (n2.west);
		\draw[arrow] (n3.east)  -- (n4.west);
		\draw[arrow] (n2.south) -- (n4.north);
		\draw[arrow] (n4.south) -- (n5.north);
	\end{tikzpicture}
}

\newcommand{\smallDepGraphgChol}[5]
{
	\begin{tikzpicture}[
					ball/.style={circle, fill=gray!70!white, draw=none, text=black},
					arrow/.style={->, shorten <= 1pt, >=stealth,thick}]
		\node[ball, fill=#1, at={(-0.75,  0.32)}] (n1) {};
		\node[ball, fill=#2, at={(-0.75, -0.32)}] (n2) {};
		\node[ball, fill=#3, at={(-0.75, -0.88)}] (n3) {};
		\node[ball, fill=#4, at={( 0.75, -0.32)}] (n4) {};
		\node[ball, fill=#5, at={( 0.75, -0.88)}] (n5) {};
		\draw[-, thick, color=lightgray] (-1.5,0) -- (1.5,0);
		\draw[-, thick, color=lightgray] (0, -0.8) -- (0, 0.8);
        \draw[arrow] (n1.south) -- (n2.north);
        \draw[arrow] (n2.south) -- (n3.north);
        \draw[arrow] (n3.east)  -- (n4.west);
        \draw[arrow] (n4.south) -- (n5.north);
	\end{tikzpicture}
}

\newcommand{\smallDepGraphCoupSylv}[5]
{
	\begin{tikzpicture}[
					ball/.style={circle, fill=gray!70!white, draw=none, text=black},
					arrow/.style={->, shorten <= 1pt, >=stealth,thick}]
		\node[ball, fill=aicesred,  at={(-0.40,  1.20)}] (n1)  {};
		\node[ball, fill=#3, at={(-0.60, -0.20)}] (n5)  {};
		\node[ball, fill=#4, at={(-0.20, -0.20)}] (n6)  {};
		\node[ball, fill=#5, at={(-0.40, -0.70)}] (n7)  {};
		\node[ball, fill=#1, at={( 0.60,  0.70)}] (n2)  {};
		\node[ball, fill=#2, at={( 1.00,  0.70)}] (n3)  {};
		\node[ball, fill=#5, at={( 0.80,  0.20)}] (n4)  {};
		\node[ball, fill=#5, at={( 0.20, -1.20)}] (n9)  {};
		\node[ball, fill=#5, at={( 0.60, -1.20)}] (n11) {};
		\node[ball, fill=#5, at={( 1.00, -1.20)}] (n8)  {};
		\node[ball, fill=#5, at={( 1.40, -1.20)}] (n10) {};
		\node[ball, fill=rwthblue, at={( 0.80, -1.70)}] (n12) {};
		\draw[-, thick, color=lightgray] (-1.0,0) -- (1.6,0);
		\draw[-, thick, color=lightgray] (0,-2.0) -- (0,1.5);
		\draw[arrow] (n1.east) -- (n2.north west);
		\draw[arrow] (n1.east) -- (n3.north west);
		\draw[arrow] (n2.south) -- (n4.north);
		\draw[arrow] (n3.south) -- (n4.north);
		\draw[arrow] (n1.south)  -- (n5.north);
		\draw[arrow] (n1.south)  -- (n6.north);
		\draw[arrow] (n4.south) -- (n8.north);
		\draw[arrow] (n4.south) -- (n10.north);
		\draw[arrow] (n5.south) -- (n7.north);
		\draw[arrow] (n6.south) -- (n7.north);
		\draw[arrow] (n7.east) -- (n9.north west);
		\draw[arrow] (n7.east) -- (n11.north west);
		\draw[arrow] (n8.south) -- (n12.north);
		\draw[arrow] (n9.south) -- (n12.north);
		\draw[arrow] (n10.south) -- (n12.north);
		\draw[arrow] (n11.south) -- (n12.north);
	\end{tikzpicture}
}

\begin{document}

\frontmatter

\aicescovertitle{
\Huge
	Knowledge-Based Automatic Generation of Linear Algebra Algorithms and Code
}
\aicescoverauthor{Diego Fabregat Traver}
\aicescoverpage{}

\newgeometry{margin=4cm}
\begin{titlepage}

\begin{center}


{\Large
	Knowledge-Based Automatic Generation \\[1mm] of Linear Algebra Algorithms and Code
}

\vspace{1cm}

{

	Von der Fakult\"at f\"ur Mathematik, Informatik und Naturwissenschaften
	der RWTH Aachen University zur Erlangung des akademischen Grades eines
	Doktors der Naturwissenschaften genehmigte Dissertation
}

\vspace{2cm}

{\large
	vorgelegt von
}

\vspace{6mm}

{\large
	Dipl.-Ing. Diego Fabregat Traver
}

\vspace{6mm}

{\large
	aus Castell\'on de la Plana, Spanien.
}

\end{center}

\vfill

\noindent
Berichter: Prof. Paolo Bientinesi, Ph.D. \\
\phantom{Berichter:} Prof. Dr. Uwe Naumann \\
\phantom{Berichter:} Prof. Dr. Christian Bischof

\vspace{5mm}
\noindent
Tag der m\"undlichen Pr\"ufung: 06.12.2013

\vspace{5mm}
\noindent
Diese Dissertation ist auf den Internetseiten der Hochschulbibliothek online
verf\"ugbar.

\vspace{2cm}

\end{titlepage}

\restoregeometry

\chapter*{Abstract}

This dissertation focuses on the design and the implementation of
domain-specific compilers for linear algebra matrix equations.
The development of efficient libraries for such equations, which lie at 
the heart of most software for scientific computing, is a 
complex process that requires expertise in a variety of areas, 
including the application domain, algorithms, numerical 
analysis and high-performance computing.
Moreover, the process involves the collaboration of several people for 
a considerable amount of time.
With our compilers, we aim to relieve the developers from both
designing algorithms and writing code, and to generate 
routines that match or even surpass the performance of those 
written by human experts.

We present two compilers, \clak{} and \click{}, that take as input the 
description of a target equation together with domain-specific
knowledge, and generate efficient customized algorithms and 
routines.
\clak{} targets high-level matrix equations, possibly 
encompassing the solution of multiple instances of 
interdependent problems. This compiler generates algorithms 
consisting in 
a sequence of library-supported building blocks.
It builds on top of a methodology that
combines a model of human expert reasoning with the 
power of computers; the search for algorithms 
makes use of the available domain knowledge to prune the 
search space and to tailor the algorithms to the application.
Along the process, \clak{} prioritizes the reduction of the 
computational cost, the elimination of redundant computations,
and the selection of the most suitable building blocks.
For one target equation, many algorithms, with different 
properties, are generated.

\click{}, instead, addresses the generation of algorithms for
specialized building blocks.
To this end, this compiler adopts the FLAME 
methodology for the derivation of formally correct loop-based
algorithms.
\click{} takes a three-stage approach:
First, the PME(s) ---a recursive definition of the target operation
in a divide and conquer fashion--- is found;
then, the PME is analyzed to identify a family of loop invariants;
finally, each loop invariant is transformed into a
corresponding loop-based algorithm.
\click{} fully automates the application
of this methodology;
in this dissertation, we dissect the mechanisms necessary to 
make it possible.

As we show, for our compilers, the exploitation of both linear
algebra and application-specific knowledge is crucial to 
generate efficient customized solvers. 
In order to facilitate the management of knowledge and to 
increase productivity, we raise the abstraction level and 
provide the users with an expressive domain-specific language 
that allows them to reason at the matrix equation level.
The {\em users} are thus able to state {\em what} needs to be 
solved, providing as much domain knowledge as possible,
and delegating the {\em compilers} to find {\em how} to 
efficiently solve it.

We illustrate the potential of our compilers by applying them
to real world problems.
For instance, for a challenging problem arising in 
computational biology, the exploitation of the available 
knowledge leads to algorithms that lower the 
complexity of existing methods by orders of magnitude. Our algorithms, at the heart of 
the publicly available library OmicABEL, have become the 
state-of-the-art.
We also carry out a thorough study of the application of our
compilers to derivative operations, quantifying how much
tools used in algorithmic differentiation can benefit from 
incorporating the techniques discussed in this dissertation.
The experiments demonstrate the compilers' potential to produce
efficient derivative versions of part of LAPACK and of the entire 
BLAS, an effort that requires thousands of routines and for which a 
manual approach is unfeasible.

This dissertation provides evidence that a linear 
algebra compiler, which increases experts' productivity and 
makes efficiency accessible to non-experts, is within reach.

\chapter*{Acknowledgements}

First and foremost, I thank Paolo Bientinesi for being
such an inspiring advisor.
He has been a model of scientific passion and honesty, 
hard work, and perseverance.
I am extremely grateful for his support and for
the respect he has always shown towards my ideas;
from the very first day, he gave me the freedom to make
my own decisions, while steering me towards meaningful goals.
We have shared an excellent relationship, and I cannot
think of a better advisor for me.

I am also grateful to Prof. Christian Bischof for his 
firm belief in automation and his direct and indirect
supervision.
I wish to thank Prof. Bischof and Prof. Uwe Naumann as well
for agreeing to review this dissertation, and 
Prof. Martin B\"ucker and Matthias Petschow for their valuable
comments and suggestions on early drafts of this manuscript.

\sloppypar
Financial support from the 
Deutsche Forschungsgemeinschaft (German Research Association) through grant GSC 111 and the
DAAD (Deutscher Akademischer Austausch Dienst) through project 50225798 PARSEMUL
is gratefully acknowledged; 
I extend the acknowledgement to 
the RWTH Computing Center (RZ) for the access to their computational resources and their support.

I wish to express my deep gratitude to everyone at AICES
who makes an effort to provide us (the students)
with the best environment and resources so that we 
can concentrate on our research.
The many organized team-building activities, talks, seminars, and courses
are also a great complement to our daily work.
A special thanks goes to the service team for their support, relieving
AICES students from many administrative tasks; particularly, thanks to
Nicole Faber, Nadine Bachem, Annette de Haes, and Joelle Janssen.

I am indebted to Prof. Enrique Quintana Ort\'i,
under whose supervision I completed my undergraduate studies.
He awakened my interest in high-performance matrix computations,
and guided my first steps into science.
I owe him having introduced me to Paolo, and
encouraging me to pursue a Ph.D.

Science is much more fun when exchanging ideas and joining efforts.
In this respect, I am thankful to 
Dr. Yurii Aulchenko 
and 
Prof. B\"ucker 
for sharing with me a number of exciting problems that opened up
multiple research directions.
Their collaboration greatly contributes to this dissertation.
I also thank the FLAME research group at Austin, Texas, and especially
Prof. Robert van de Geijn, for developing the foundations 
of an important part of my research, and for their insight and numerous suggestions.

\newpage
Of course, I thank all AICES students for the relaxing chat
over all sorts of international sweets and cakes, and
in particular, Aravind Balan for his predisposition to give me a hand whenever needed.

For the countless hours of scientific discussion and fun both 
at work and outside, I am grateful to the HPAC group:
Paolo Bientinesi, Edoardo di Napoli, Matthias Petschow, 
Roman Iakymchuk, Elmar Peise, Lucas Beyer, Daniel Tameling, 
Paul Springer and Viola Wierschem.
I am especially grateful to Matthias, Roman, Edo, and Paolo, who
accompanied me during this entire trip.
You have been like a family to me.

I cannot list everyone who joined our countless
game nights. We played thousands of games,
ate tones of pizzas, and most importantly, we had lots of fun.
I also thank Edo and Jess for organizing many dinner and
(bad) movie nights. 
Having Edo commenting on bad sci-fi movies is priceless.

I want to spend a few words to thank my Spanish friends in Aachen. 
Sports and food have been the best excuse to hang out and talk about 
the good things we have at home and how much we miss them. 
I do not forget my friends in Spain. There is no need
to name them, they know who they are.
Distance and time does not matter for them; 
every time we meet, it feels like I never left.
I know you will always be there for me.


I owe the biggest and deepest thanks to my family,
who always took care of me, supported me, and encouraged me
to pursue my dreams. You made everything at your reach to
turn my life into a wonderful adventure.

Finally, I find no words to express my infinite gratitude to Ana.
You stood by me through thick and thin during all these years,
and, despite the distance, you have been my biggest emotional support.
{\em Todo al negro.}

\tableofcontents

\mainmatter 

\chapter{Introduction}

This dissertation focuses on the design and the implementation of
domain-specific compilers for linear algebra matrix equations.
Matrix equations constitute the computational bottleneck of 
most scientific and engineering applications;
the development of efficient libraries for such equations
has proven to be a complex and time consuming task that
requires expertise in a variety of areas,
from the application domain, through numerical 
analysis, to high-performance computing.
A typical development process begins with an application expert 
  modeling a problem in terms of matrix equations,
continues with the discovery of efficient algorithms to solve them, and
completes with the writing of high-performance code;
normally, the process requires the collaboration of several people for months or even years.
With our compilers, we aim to relieve the developers from both
designing algorithms and writing code, and to generate 
routines that match or even surpass the performance of those 
written by human experts.

Given a target problem in terms of one or more matrix equations,
no further work is required
only if these equations can be solved directly by an existing
high-performance library (e.g., LAPACK~\cite{laug}).
Most often, however, libraries do not offer routines that take advantage
of domain knowledge, and their use results in suboptimal solutions; 
the burden is thus shifted to the developers who have
to modify or extend existing libraries to tailor the computation to their needs.
We present here two examples of such equations.

\paragraph{Example 1: The genome-wide association study.}
At the heart of the genome-wide association 
study (GWAS),\footnote{When carried out by means of the
variance components method based on linear mixed models~\cite{Astle-Balding-2009,Yu2006}.}
an important problem in computational biology~\cite{GWAScatalog}, lies
the generalized least-squares (GLS) problem:
$$ b := (X^T M^{-1} X)^{-1} X^T M^{-1} y, $$
where 
$M \in R^{n \times n}$, 
$X \in R^{n \times p}$, and
$y \in R^{n}$; 
$b \in R^{p}$ is the sought-after solution.
While LAPACK offers routines for closely related problems,
such as the ordinary least-squares $b := (X^T X)^{-1} X^T y$,
some effort is needed to extend it to solve a GLS.
Moreover, GWAS requires the solution of not one single
GLS, but many of them; specifically, it requires solving
the two-dimensional grid of $m \times t$ GLSs
\begin{equation}
\label{eq:GWAS}
	b_{ij} := (X_i^T M_j^{-1} X_i)^{-1} X_i^T M_j^{-1} y_j,
	\;
	\quad \text{ with } 1 \le i \le m \; \text{ and } \; 1 \le j \le t,
\end{equation}
where $M_j$ is built as a function of a matrix $\Phi$, the identity matrix $I$, and a scalar $h_j$:
$M_j := h_j \Phi + (1 - h_j) I$.
The key to an efficient solver is to exploit all the available
knowledge:
The specific structure of the matrix $M_j$, and
the interdependence among the GLS problems that allows the reuse of computation
across them.
Unfortunately,
this knowledge is more complex than what traditional libraries may take.
In fact, the only alternative offered by these libraries is the
computation of $m \times t$ such GLSs independently, in a black box fashion.
In practice, the computational cost of this approach makes it unfeasible.
The burden of designing competitive routines is put on the user.

\paragraph{Example 2: The derivative of Cholesky.}
The derivative of the Cholesky factorization (\gchol{})
represents a building block required, 
for instance, in sensitivity analyses and optimization problems~\cite{smith-1995}.
One way of computing this derivative is to solve the equation
\begin{equation}
  G L^T + L G^T = B
  \label{eq:gchol}
\end{equation}
for unknown $G$, where $L$ and $G \in R^{n\times n}$ are lower triangular,
and $B \in R^{n\times n}$ is symmetric.
Even though traditional libraries offer solvers for closely-related operations,
none of them supports the solution of Equation~\eqref{eq:gchol}.
For this type of operations, an unexperienced programmer
is likely to only find inefficient unblocked algorithms;
the discovery of efficient blocked algorithms, which enable data reuse 
to overcome the memory bandwidth bottleneck,
requires a high-performance expert.

\vspace{5mm}
To further complicate matters, 
multiple algorithms may exist to solve one single target equation. 
It is well known that the performance of an algorithm
depends on multiple factors, including the problem size and
the underlying architecture, and that in general no single algorithm
performs best in all scenarios.
Therefore, to attain high performance in a range of scenarios,
it is desirable to develop not one but a family of algorithms.
This adds extra complexity to the development task:
Multiple routines must be coded and maintained.

Since the mechanisms required for the management of domain knowledge
and the derivation of families of algorithms are beyond the
scope of traditional general-purpose compilers,
we focus on the development of domain-specific linear algebra compilers.
Our goal is to allow application experts to reason at the matrix
equation level, and to relieve them from the design and the coding of algorithms.
Following the approach of other domain-specific compilers,
we provide a high-level interface,
in the form of a domain-specific language,
that allows the expert to express the matrix equations to
be solved, together with as much knowledge as possible.
Then, our compilers take this information
and automatically produce efficient solvers.
More specifically, they take care of 
the exploitation of knowledge,
the derivation of families of algorithms,
the efficient mapping onto library-supported building blocks, and
the code generation.
In short, the {users} are able to state {\em what} needs to be 
solved, providing as much domain knowledge as possible,
and delegating the {compilers} to find {\em how} to 
efficiently solve it.

Our ultimate goal is not different from that of the
first compilers~\cite{PreliminaryFortran}. At that time,
the community was skeptical because obviously a human expert 
could generate better code than a program (compiler).
Nowadays, nobody conceives computer science, and especially
the branch of programming, without general-purpose compilers and
interpreters.
Due to the complexity of matrix equations, similar and 
even stronger objections may be raised against our work.
However, while the community concern is understandable, we argue that
a) the loss in performance will be, in general, marginal,
b) the gain in productivity is substantial, and
c) for non-trivial problems, 
   especially given that it is desirable to generate families of algorithms,
   it is very likely that our compilers find algorithms that even experts would miss.

\section{Our compilers: A short overview}

Applications require customized routines for both
high-level matrix equations (e.g., Example 1), and 
specialized building blocks (e.g., Example 2).
Since the concepts behind the algorithms for 
these two classes of operations are different,
we developed two different prototypes of domain-specific compilers:
1) \clak{}, for matrix equations, and
2) \click{}, for building blocks.

\subsection{\clak{}}
\clak{} is our compiler for matrix equations comprising, for instance,
linear systems, matrix inversions, and least-squares-like problems.
With \clak{}, we aim at modeling the reasoning of a human expert for the 
derivation of algorithms, and extending it with computers'
exploration power.
The approach
may be described as follows:
The same way that a traditional compiler breaks a program into 
assembly instructions directly supported by the processor,
attempting different types of optimizations,
\clak{} breaks a target operation down to library-supported kernels, 
tailoring the algorithm to the application.
In general, the decomposition is not unique, and the number
of possible algorithms may be large;
our compiler makes use of knowledge
to prune the space of algorithms during the search,
yielding only the most promising ones.

\sloppypar
\paragraph{Example: The genome-wide association study.}
\clak{} takes as input the description of a target operation
expressed in a high-level mathematical notation (which we
discuss in Section~\ref{sec:clak-input});
for instance, Box~\ref{box:intro-GWASOpDesc} contains the
description of the GWAS equation.
Given this input, 
\clak{} generates a family algorithms
that cast the computation in terms of kernels from the BLAS and LAPACK libraries.
Algorithms~\ref{alg:gwas-var1}~and~\ref{alg:gwas-var2} are two members
of the family;
in brackets, we specify the kernel corresponding to each statement
(for a list of acronyms and their meaning, please see Appendix~\ref{app:blas-lapack}).

\begin{mybox}[!ht]
\vspace{3mm}
\begin{verbatim}
      Equation GWAS
    
          Matrix X   <Input, FullRank, ColumnPanel>;
          Vector Y   <Input>;
          Scalar h   <Input>;
          Matrix Phi <Input, SymmetricLower>;
          Vector b   <Output>;
    
          Matrix M   <Intermediate, SPD>;
    
          b{ij} = inv( trans(X{i}) * inv(M{j}) * X{i} ) * 
                       trans(X{i}) * inv(M{j}) * y{j};
          M{j}  = h{j} * Phi + (1 - h{j}) * I;
\end{verbatim}
\caption{Description of the GWAS equation.}
\label{box:intro-GWASOpDesc}
\end{mybox}

\begin{center}
\renewcommand{\lstlistingname}{Algorithm}
\begin{minipage}{0.47\linewidth}
	\begin{lstlisting}[numbers=left,caption={GWAS variant 1.}, escapechar=!, label=alg:gwas-var1]
for $j$ in 1:t
  $M_j := h_j \Phi + (1-h_j)I$     (!\sc sc-add!)
  $L L^T = M_j$                (!\sc potrf!)
  $y_j := L^{-1} y_j$                (!\sc trsv!)
  for $i$ in 1:m
    $W := L^{-1} X_i$              (!\sc trsm!)
    $S := W^T W$              (!\sc syrk!)
    $G G^T = S$              (!\sc potrf!)
    $b_{ij} := W^T y_j$              (!\sc gemv!)
    $b_{ij} := G^{-1} b_{ij}$              (!\sc trsv!)
    $b_{ij} := G^{-T} b_{ij}$              (!\sc trsv\vspace{8.5mm}!)
	\end{lstlisting}
\end{minipage}
\hfill
\begin{minipage}{0.47\linewidth}
	\begin{lstlisting}[numbers=left,caption={GWAS variant 2.}, escapechar=!,label=alg:gwas-var2]
$Z \Lambda Z^T = \Phi$                  (!\sc syevr!)
for $i$ in 1:m
  $K_i := X_i^T Z$                (!\sc gemm!)
for $j$ in 1:t
  $D := h_j \Lambda + (1 - h_j) I$                (!\sc sc-add!)
  $y_j := Z^T y_j$                (!\sc gemv!)
  for $i$ in 1:m
    $V := K_i D^{-1}$              (!\sc scal!)
    $A := V K_i^T$              (!\sc gemm!)
    $Q R = A$              (!\sc geqrf!)
    $b_{ij} := V y_j$              (!\sc gemv!)
    $b_{ij} := Q^T b_{ij}$              (!\sc ormqr!)
    $b_{ij} := R^{-1} b_{ij}$              (!\sc trsv!)
	\end{lstlisting}
\end{minipage}
\end{center}

\begin{figure}[!h]
\centering
	\includegraphics[scale=0.75]{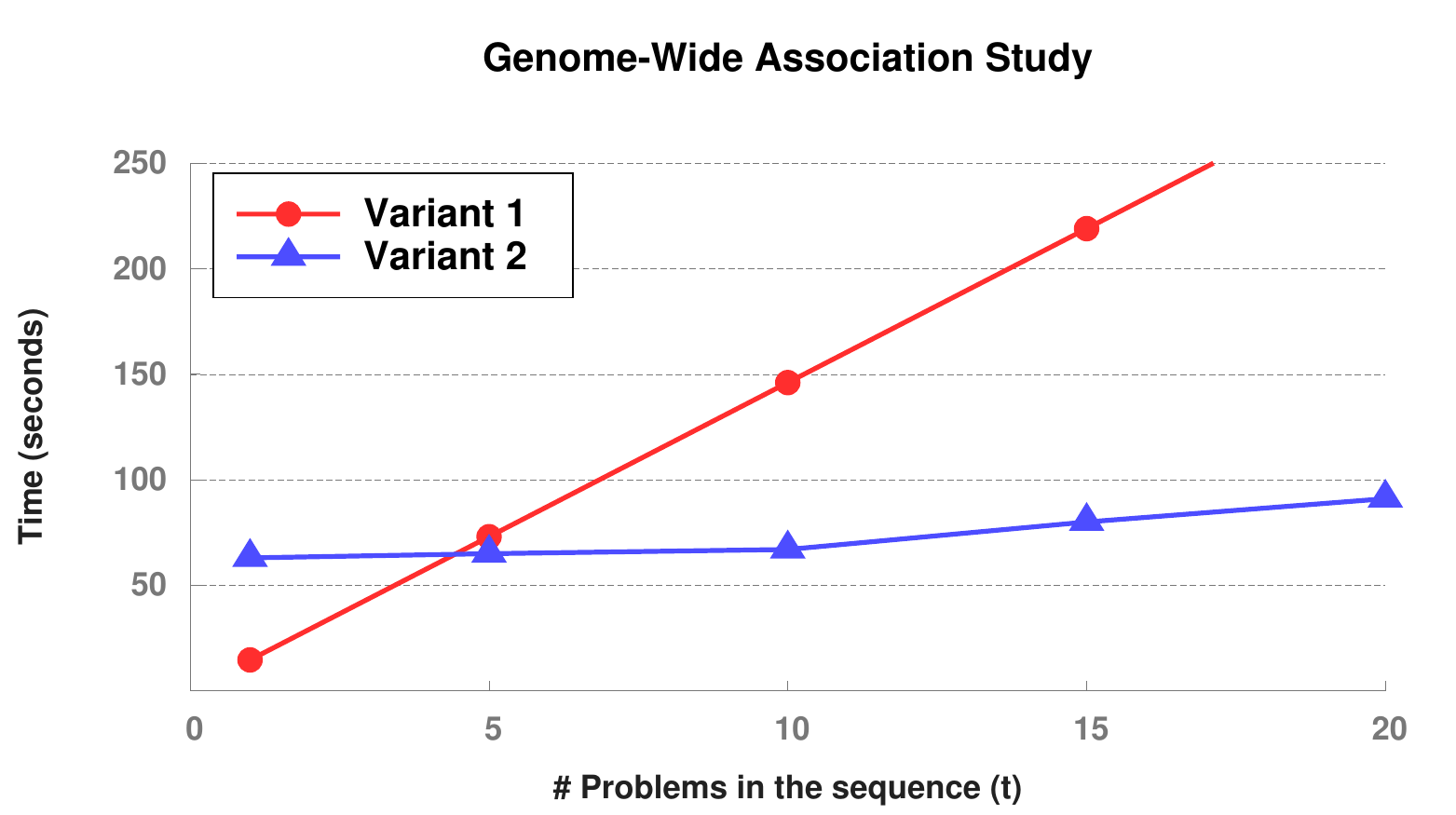}
	\caption{Performance of Algorithms~\ref{alg:gwas-var1}~and~\ref{alg:gwas-var2} for GWAS.
	         Values for $n$, $p$, and $m$ are $1{,}000$, $4$, and $1{,}000{,}000$, respectively.}
	\label{fig:gwas-intro}
\end{figure}

We choose these two example algorithms for their practical relevance.
We recall that GWAS computes the $m \times t$ grid of GLS problems displayed in Equation~\eqref{eq:GWAS}.
In a typical study, $m$ takes large values (from millions to hundreds of millions),
and $t$ is either 1 ---{\em Scenario 1}--- or 
ranges from thousands to hundreds of thousands ---{\em Scenario 2}---.
Even though the mathematical problem is the same for both scenarios, its parameters are not;
different ways of exploiting the available domain knowledge, result in radically
different algorithms that suit best each specific case.
As illustrated by Figure~\ref{fig:gwas-intro},
while Algorithm~\ref{alg:gwas-var1} is best suited for Scenario 1 (very small values of $t$), 
Algorithm~\ref{alg:gwas-var2} is to be preferred for Scenario 2 (large values of $t$).
A deeper discussion of this application is carried out in Section~\ref{sec:cost}.

\subsection{\clickplain{}}
\click{} targets the generation of algorithms and code for
building blocks such as matrix and vector products, and matrix factorizations.
\click{} builds on a methodology born in the context of 
the FLAME project~\cite{Gunnels:2001:FFL}
for the generation of loop-based blocked algorithms.
Based on formal methods, the methodology takes
a high-level description of a target operation,
and derives a family of provably correct algorithms.
The key lies in finding the operation's Partitioned
Matrix Expression (PME), a divide-and-conquer definition
of the operation, from which a pool of loop invariants are identified;
each loop invariant leads to a different algorithm,
resulting in a family of them.

\paragraph{Example: The derivative of Cholesky.}
In line with FLAME's methodology, \click{} takes two predicates
---Precondition and Postcondition--- as formalism for the input.
Box~\ref{box:Intro-gCholOpDesc} contains the input to \click{}
for the derivative of the Cholesky factorization (\gchol{});
given this description,
\click{} derives four algorithms to compute \gchol{}.
One example (Variant 1) is given in Figure~\ref{fig:gchol-var1} in FLAME notation;
the complete family is collected in Section~\ref{sec:gchol-algs}.

\begin{mybox}[!ht]
$$
\small
G = gChol(L, B) \equiv
\left\{
\begin{split}
	P_{\rm pre}: \{ & \prop{Output}{G} \; \wedge \; \prop{Input}{L}  \; \wedge \prop{Input}{B}  \;\; \wedge \\
	                & \prop{Matrix}{G} \; \wedge \; \prop{Matrix}{L} \; \wedge \prop{Matrix}{B} \;\; \wedge \\
	                & \prop{LowerTriangular}{G} \; \wedge \; \prop{LowerTriangular}{L} \;\; \wedge \\
                    & \prop{Symmetric}{B} \} \\
\\
P_{\rm post}: \{ &  G L^T + L G^T = B \}
\end{split}
\right.
$$
\caption{Description of the derivative of the Cholesky factorization.}
\label{box:Intro-gCholOpDesc}
\end{mybox}

\begin{figure}[!ht]
\centering
	\input{Chapter1_Introduction/tex/gchol}
	\caption{One of the four algorithms generated by \clickplain{} for the derivative of the Cholesky factorization.}
	\label{fig:gchol-var1}
\end{figure}

\begin{figure}[!h]
\centering
	\includegraphics[scale=0.65]{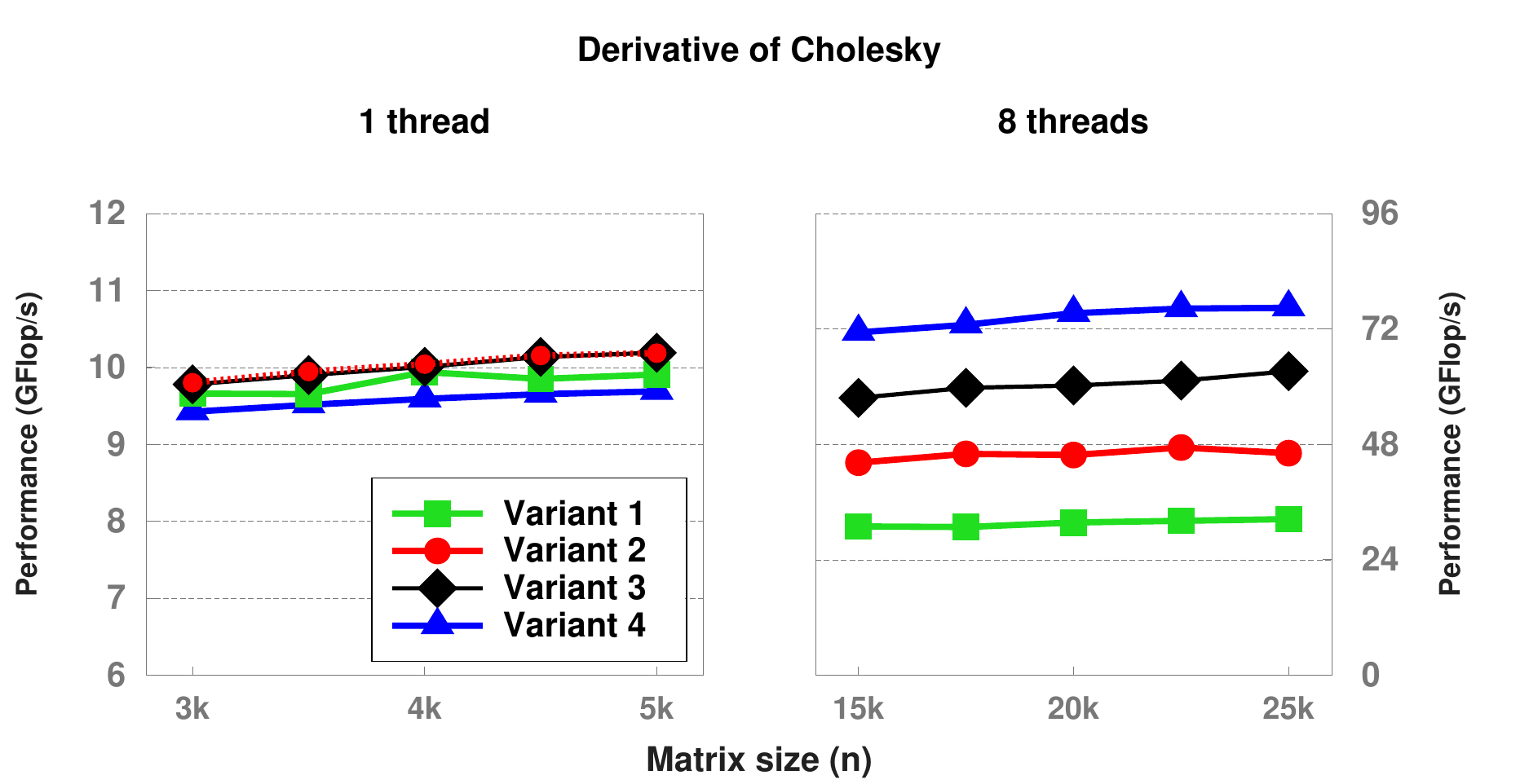}
	\caption{Performance of the four variants for the derivative of Cholesky.}
	\label{fig:intro-gchol}
\end{figure}

In Section~\ref{sec:gchol-experiments}, we carry out a
thorough study of the performance of the algorithms.
Here, we briefly discuss the need for deriving multiple variants.
Figure~\ref{fig:intro-gchol} contains performance results
using one thread (left) and eight threads (right).
We make two points:
First, despite the fact that all variants have the same computational cost,
we observe differences in performance of up to 2.4x;
second, while Variants 2 and 3 attain the best performance when one single
thread is used, a different one, Variant 4, performs best in the multi-threaded case.

\section{Contributions}

This dissertation makes the following contributions.

\begin{itemize}

	\item {\bf Methodology for the derivation of algorithms for matrix equations.}
		We propose a methodology inspired by the reasoning of a human expert.
		The methodology takes into account 
		knowledge from the application domain, numerical linear algebra, and
		high-performance libraries;
		given the mathematical specification of a target equation, 
		it yields a family of algorithms that exploit as much
		available domain knowledge as possible.
		The algorithms produced consist in efficient mappings 
		onto a sequence of optimized library-supported kernels.

	\item {\bf Automatic generation of algorithms and code for matrix equations.}
		We introduce \clak{}, a prototype of linear algebra compiler
		that automates the application of the aforementioned methodology for matrix equations. 
		For well-studied operations,
		including linear systems, matrix inversions, and least-squares problems,
		\clak{} finds already known algorithms;
		here, we concentrate on 
		1) the discovery of novel algorithms for challenging equations, and
		2) the application of \clak{} to problems that require customized
		libraries comprising a large number of routines, where a manual approach
		is unfeasible.
		As case studies, we make use of the genome-wide association study (GWAS), and 
		derivative operations arising in the context of algorithmic differentiation (AD).

	\item {\bf Complete automation of FLAME's methodology.}
		The FLAME project provides a methodology for the systematic derivation of 
		loop-based algorithms for a class of linear algebra building blocks.
		While FLAME researchers believed that the methodology could be fully
		mechanized, i.e., automatically carried out by a computer,
		little evidence existed.
		We introduce a second compiler, \click{}, which demonstrates that 
		the automatic application of the methodology is indeed feasible.
		We illustrate the application of \click{} to standard operations,
		such as the LU factorization and the triangular
		Sylvester equation, and to unsupported kernels arising
		in AD, such as the 
		derivative of the Cholesky factorization.

	\item {\bf Knowledge management.}
		We implement an engine to manipulate the available knowledge associated to
		the target problem, and
		to dynamically deduce structure and properties of expressions from that
		of individual operands.
		This engine encodes linear algebra rules and theorems related to
		properties such as matrix positive-definiteness, orthogonality and rank.
		Such a knowledge management is often sought after when
		performing symbolic computations; however, even the most
		advanced computer-algebra systems, such as Mathematica~\cite{Mathematica} and Maple~\cite{Maple},
		lack powerful deduction modules.
		Thanks to the knowledge management engine, our compilers can produce highly customized
		routines.

\end{itemize}

Two more contributions are made to specific fields of computational science and engineering.

\begin{itemize}

	\item {\bf Algorithmic Differentiation.} 
		We perform a thorough performance comparison of the routines
		generated by our compilers with those of ADIFOR (a characteristic AD tool)
		for the derivative of a number of BLAS and LAPACK operations.
		The results quantify how much AD tools can benefit from the application
		of the techniques described in this dissertation when targeting 
		linear algebra problems.
		Furthermore, the use of characteristic operations, such as linear systems,
		matrix factorizations, and matrix products, provide evidence that
		our compilers are capable of generating a differentiated version
		of a subset of LAPACK and of the entire BLAS.

	\item {\bf Genome-wide association studies.} 
		The application of \clak{} to Equation~\eqref{eq:GWAS} yielded
		novel algorithms for the computation of linear mixed-models in the context
		of genome-wide association studies. 
		These algorithms are now included in the OmicABEL 
		package\footnote{\texttt{http://www.genabel.org/packages/OmicABEL}}
		as part of the widely-used R library GenABEL~\cite{genabel}, and
		are considered the state-of-the-art.

\end{itemize}

\section{Outline of the thesis}

The organization of this dissertation follows.
Chapters~\ref{ch:compiler}~and~\ref{ch:compiler-ad} are devoted to \clak{}.
Chapter~\ref{ch:compiler}, introduces the compiler's engine.
The input language and the class of accepted matrix equations
are presented in Section~\ref{sec:clak-input}.
Section~\ref{sec:heuristics}, discusses the heuristic-based model of the human expert reasoning,
while Sections~\ref{sec:engine} and~\ref{sec:multiple-instances} uncover the core modules of the compiler's engine
that support the application of these heuristics.
Cost analysis and code generation are discussed in Sections~\ref{sec:cost}~and~\ref{sec:clak-codegen},
respectively.

Chapter~\ref{ch:compiler-ad} is dedicated to showing the broad applicability and extensibility
of the compiler; to this end, we target the generation of libraries for the
derivative of BLAS and LAPACK operations, as they arise in 
the field of algorithmic differentiation.
The challenge behind the development of such libraries is discussed in Section~\ref{sec:ad-challenge}.
The extension to \clak{} for the support of derivative operations is 
presented in Section~\ref{sec:ad-ext}.
Section~\ref{sec:ad-clak-experiments} carries out a deep
comparison of \clak{}'s routines with those produced by ADIFOR
to evaluate \clak{}'s potential.

Chapters~\ref{ch:click}~and~\ref{ch:flame-ad} concentrate on \click{}.
Chapter~\ref{ch:click} details thoroughly the steps behind the complete
automation of the FLAME methodology.
An overview of the approach is given in Section~\ref{sec:flame}, while 
Sections~\ref{sec:GenPME}~through~\ref{sec:AlgConstruction} are dedicated to each of 
the three major steps in the process: 
1) The generation of PMEs, 2) the identification of loop invariants, and
3) the construction of the algorithms.
LAPACK and RECSY\footnote{RECSY is a specialized library for control theory equations.} 
operations are used as examples to illustrate the application of \click{}.

Chapter 5 focuses on showing the application of \click{}
to non-standard operations.
Section~\ref{sec:gchol}, gives a complete example of \click{}'s application to
the derivative of the Cholesky factorization, 
while Section~\ref{sec:gtrsm}
provides a more general example by means of a kernel representative 
of many other specialized kernels.
In both cases, performance results evidence that \click{} not only
increases productivity, but it also delivers efficient customized routines.

Chapter~\ref{ch:relwork} reviews related work,
and Chapter~\ref{ch:conclusions} concludes with a summary of the results of this thesis and
a discussion of open research directions.

\chapter{\clak{}}
\label{ch:compiler}

We introduce \clak{}, a compiler
for linear algebra matrix equations.
\clak{} takes as input the mathematical 
description of a target equation together with domain knowledge, and 
returns a family of algorithms and routines that solve the equation.
The goal is to replicate the exploitation of
knowledge and the optimizations carried out by
a human expert,
and combine them with the computational power
of a computer to address from simple to challenging problems.

\sloppypar
Simple examples of matrix operations are 
$x := Q^T L \, y$,
$b := (X^T X)^{-1} v$, and
$B_i := A^T_i M^{-1} A_i$; 
in all cases,
the quantities on the right-hand side of the assignment are known (matrices in capital
letters and vectors in lower case), and the left-hand side has to be
computed. Despite their mathematical simplicity, these operations pose
challenges so significant that even the best tools for linear algebra
produce suboptimal results. 
For instance, Matlab\footnote{When comparing to Matlab, we 
refer to version 7.11.0.584 (R2010b).}
uses a cubic---instead of quadratic---algorithm in the first
equation,
incurs possibly critical numerical errors in the second 
one, and
fails to reuse intermediate results---and thus save computation---in the last one. 

Let us take a closer look at $x := Q^T L \, y$, with 
$Q, L \in R^{n \times n}$, and $x, y \in R^n$: 
Algorithms~\ref{alg:matlab-qly} and~\ref{alg:compiler-qly}
display two alternative ways of computing $x$.
In the algorithm on the left, the one used by Matlab,\footnote{
With the extra help of parentheses, Matlab can be forced to use 
Algorithm~\ref{alg:compiler-qly}. We
highlight the fact that in absence of user's intervention, 
Matlab falls back on a suboptimal algorithm.
}
the input equation is decomposed into a {\sc gemm} 
(matrix-matrix multiplication), followed by a 
{\sc gemv} (matrix-vector multiplication), for a total of $O(n^3)$ 
floating point operations (flops); 
the algorithm on the right, generated by \clak{}, instead
maps the equation onto two {\sc gemv}s,
for a cost of $O(n^2)$ flops.
The difference lays in
how the input operation is decomposed and mapped
onto available kernels. 
In more complex matrix equations, it is not uncommon to face 
dozens and dozens of alternative decompositions, all corresponding 
to viable, but not equally effective, algorithms. 
We will illustrate how unfruitful branches can be avoided 
by propagating knowledge, as the algorithm unfolds,  
from the input operands to intermediate results.

\begin{center}
\renewcommand{\lstlistingname}{Algorithm}
\begin{minipage}{0.40\linewidth}
	\begin{lstlisting}[caption={Matlab's algorithm for $x := Q^T L \, y$}, escapechar=!, label=alg:matlab-qly]
	$T := Q^T L$            !({\sc gemm})!
	$x := T y$            !({\sc gemv})!
\end{lstlisting}
\end{minipage}
\hspace*{2cm}
\begin{minipage}{0.40\linewidth}
	\begin{lstlisting}[caption={\clak{}'s alternative algorithm for $x := Q^T L \, y$}, escapechar=!, label=alg:compiler-qly]
	$t := L \, y$            !({\sc gemv})!
	$x := Q^T t$            !({\sc gemv})!
\end{lstlisting}
\end{minipage}
\end{center}

Challenging matrix equations appear in 
applications as diverse as 
machine learning, sensitivity analysis, and computational biology.
In most cases, one
has to solve not one instance of the problem, but thousands or even
billions of them.
A characteristic example is given by the aforementioned computation of mixed models in
the context of the genome-wide association study,
which requires the solution of up to $10^{12}$ (trillions) instances of 
the generalized least-squares problem
\[
 b := (X^T M^{-1} X)^{-1} X^T M^{-1} y,
 \quad {\rm where} \quad
 M = h^2 \Phi + (1 - h^2) I.
\]
These instances are related to one another
suggesting that intermediate results could be saved and
reused; unfortunately, none of the currently available libraries allows this.

We developed \clak{} with the objective of 
overcoming the deficiencies discussed so far.
Very much like a standard compiler takes a computer program and maps
it onto the instruction set provided by the processor, our approach is
to decompose the input equations into kernels provided by linear algebra libraries
such as BLAS and LAPACK.
The mapping is, in general, not unique, and the number of
alternatives may be very large. For this reason, our compiler carries
out a search within the space of possible algorithms, and yields the
most promising ones.  The search is guided by a number of heuristics
which, in conjunction with a mechanism for inferring properties, aim
at simulating the thought-process of an expert in the
field.  
Moreover, by means of dependency analyses, \clak{} actively
seeks to avoid redundant computation, both within a single equation
and across multiple instances of them.  

The application of these techniques heavily relies on pattern matching 
and symbolic computations; due to its
powerful engine for pattern matching and expression rewriting,
we chose Mathematica~\cite{Mathematica} to implement the compiler's engine.
In this chapter, we discuss the mechanisms incorporated into \clak{}
to automate the process of generation of algorithms and code:
From the input to the compiler, through the mapping onto building blocks
and knowledge management, to the code generation.

\section{Defining input equations}
\label{sec:clak-input}

We consider equations that involve scalar, 
vector and matrix operands, combined 
with the binary operators ``$+$'' (addition) and ``$*$'' 
(multiplication, used both for scaling and matrix products),
and the unary operators
``$-$'' (negation),
``{\small $T$}'' (transposition), and ``{\small ${-1}$}'' 
(inversion, for scalars and square matrices). 
Equations come with what we refer to as {\em knowledge}: 
Each operand is annotated with a list of zero or 
more properties such as ``square'', ``orthogonal'', ``full rank'',
``symmetric'', ``symmetric positive definite'', ``diagonal'', and so on. 
Additionally, 
we allow operands to be subscripted, 
indicating that the problem has to be solved multiple times.
As an example, Box~\ref{box:input-example} illustrates the description of 
the solution of multiple linear systems 
that share the same symmetric coefficient matrix: $x_i := A^{-1} b_i$.
\begin{mybox}[!h]
\vspace{3mm}
\begin{verbatim}
            Equation MultSymmSolve
                Matrix A <Input, SymmetricLower>;
                Vector b <Input>;
                Vector x <Output>;
                        
                x{i} = inv(A) * b{i};
\end{verbatim}
\caption{Description of the solution of multiple linear systems $x_i := A^{-1} b_i$,
with a common symmetric coefficient matrix $A$. 
The description includes 
the definition of the operands together with their properties,
and the target equation where the operands are labeled with the corresponding subindex.}
\label{box:input-example}
\end{mybox}

\indent
The notation is straightforward:
First, the operands are declared specifying their type
and a number of properties;
then, the equation to be solved is stated in a
high-level notation similar to that of Matlab.
Essentially, the valid equations are formed by a left-hand side
consisting of an output operand, the unknown,
and a right-hand side consisting of input operands combined
using the aforementioned operators.
In the remainder of this section, we briefly formalize the syntax of
the language accepted by our compiler, and thus the class
of accepted input equations.
These details do not affect the description of the mechanisms behind \clak{};
readers mainly interested in the generation of algorithms may skip the rest of this section.

As in a natural language, \clak{} defines a collection of admissible 
words, the {\em tokens}. These tokens are grouped in lexical
categories, as shown in Table~\ref{tab:lexic};
for each category (indicated in the left column), we
provide the regular expression that defines it (middle column)
and its meaning (right column).
We use the traditional notation for the specification of regular expressions:
\bi
   \item {\tt *} matches zero or more occurrences of the expression on the left,
   \item {\tt +} matches one or more occurrences of the expression on the left,
   \item {\tt ?} matches zero or one occurrences of the expression on the left,
   \item $|$ matches one of the many options,
   \item {\tt []} indicates a class of characters, e.g., {\tt[a-z]} matches any lower case letter, and
   \item {\tt "[*"} indicates the literals within the quotes, in this case an opening square bracket followed by a star.
\ei
\begin{table}[!ht]
	\centering
	\small
	\renewcommand{\arraystretch}{1.4}
	\begin{tabular}{l l l}\toprule
	{\bf Category} & {\bf Regular Expression} & {\bf Meaning} \\\midrule
		id         & {\tt [a-zA-Z][a-zA-Z0-9\_]*}   & Identifier \\
		optype    & {\tt Scalar $|$ Vector $|$ Matrix} & Type of operand \\
		iotype     & {\tt Input $|$ Output $|$} & I/O type of the operand \\
		           & {\tt InOut $|$ Intermediate} & \\
		property   & {\tt Square $|$} {\tt ColumnPanel $|$} & Properties of the operands \\
				   & {\tt RowPanel $|$ Diagonal $|$} & \\
		           & {\tt LowerTriangular $|$} & \\
		           & {\tt UpperTriangular $|$} & \\
		           & {\tt Symmetric $|$} & \\
		           & {\tt SymmetricLower $|$} & \\
		           & {\tt SymmetricUpper $|$} & \\
		           & {\tt SPD $|$} {\tt SPDLower $|$} & \\
		           & {\tt SPDUpper $|$} & \\
		           & {\tt Orthogonal $|$} {\tt FullRank} & \\
		subscript  & {\tt \{[a-z](,[a-z])*\}} & Operand subscripts \\
		number     & {\tt [0-9]+($\backslash$.[0-9]+)? }              & Numeric constant \\
						 & {\tt \hspace{4mm} ([Ee][+-]?[0-9]+)? }              &                  \\
		opeq       & {\tt =}             & Equality operator \\
		opadd      & {\tt [+-]}          & Addition operator \\
		opmul      & {\tt [*]}          & Multiplication operator \\
		unary      & {\tt trans $|$ inv} & Transpose and inverse operators \\
		init       & {\tt init} & Initial contents of an InOut operand \\\bottomrule
	\end{tabular}
	\caption{Lexical categories of \clak{}'s input language.}
	\label{tab:lexic}
\end{table}

\vspace{5mm}
\noindent
We clarify the meaning of a few keywords included in the {\em iotype} and the {\em property} lexical categories:
\paragraph{iotype.}
The input/output (I/O) type of the operands may be not only {\tt Input} or {\tt Output},
but also {\tt InOut} and {\tt Intermediate}.
{\tt InOut} is used to specify operands that are overwritten; this type
is used in conjunction with the macro {\tt init}, which specifies
the initial contents of the operand. For instance, the equation
\begin{center}
	{\tt A = 3 * init(A)},
\end{center}
scales {\tt InOut} matrix {\tt A} and overwrites it with the result.

The {\tt Intermediate} type is used for temporary operands,
and it allows the specification of properties of expressions that are empirical, 
i.e., properties that cannot be deduced analytically.
An example of {\tt Intermediate} operand will appear in Section~\ref{sec:detailed-example}.

\paragraph{property.}
Typically, symmetric matrices are only partially stored in
either their upper or lower triangle;
the suffixes {\tt Lower} and {\tt Upper} for symmetric and SPD matrices 
indicate in which part of the matrix is data actually stored,
and are essential for the generation of code.
The {\tt ColumnPanel} and {\tt RowPanel} keywords are
used for rectangular matrices:
A matrix $A \in R^{m \times n}$ is a column panel if $m > n$, i.e.,
it has more rows than columns, 
while it is a row panel if it has more columns than rows.

\vspace{2em}
Tokens are combined to construct program sentences or {\em statements}. 
The structure of a correct \clak{} program is specified
by means of a grammar $G$, defined as the quadruple $G = (N, \Sigma, P, \text{\tt<}S\text{\tt>})$, where
\bi
	\item $N$ is the set of {\em non-terminal} symbols ({\tt<}Model{\tt>}, {\tt<}OpDecl{\tt>}, {\tt<}Equation{\tt>}, {\tt<}Expression{\tt>}, {\tt<}Term{\tt>}, {\tt<}Factor{\tt>}).
	\item $\Sigma$ is the set of {\em terminal} symbols (id, optype, iotype, property, subscript, number, opeq, opadd, opmul, unary, init).
	\item $P \subseteq N \times (N \cup \Sigma)^*$ is the set of {\em production rules} (provided in Box~\ref{box:grammar}).
	\item $\text{\tt<}S\text{\tt>} \in N$ is the {\em starting symbol}. In this case, the starting symbol is {\tt<}Model{\tt>}.
\ei

\begin{mybox}
	\centering
	\renewcommand{\arraystretch}{1.4}
	\begin{tabular}{ r c l }
		{\tt<}Model{\tt>}      & $\rightarrow$ & {\tt "Equation"} {\bfseries id} \\
														  &               & \hspace{5mm} ({\tt<}OpDecl{\tt>} {\tt ";"}$)^+$ \\
							   &               & \hspace{5mm} ({\tt<}Equation{\tt>} {\tt ";"}$)^+$ \\
		{\tt<}OpDecl{\tt>}     & $\rightarrow$ & {\bfseries optype} {\bfseries id} {\tt "<"} {\bfseries iotype} ({\tt "},{\tt "} {\bfseries property})$^*${\tt ">"} \\
		{\tt<}Equation{\tt>}   & $\rightarrow$ & {\bfseries id} {\bfseries opeq} {\tt<}Expression{\tt>} \\
		{\tt<}Expression{\tt>} & $\rightarrow$ & {\tt<}Term{\tt>} ({\bfseries opadd} {\tt<}Term{\tt>}$)^*$ \\
		{\tt<}Term{\tt>}       & $\rightarrow$ & {\tt<}Factor{\tt>} ({\bfseries opmul} {\tt<}Factor{\tt>}$)^*$ \\
		{\tt<}Factor{\tt>}     & $\rightarrow$ & {\bfseries id} ({\bfseries subscript}$)^?$ $|$ \\
							   & & {\bfseries number} $|$  \\
							   & & {\tt"("} {\tt<}Expression{\tt>} {\tt")"} $|$ \\
							   & & {\bfseries unary} {\tt"("} {\tt<}Expression{\tt>} {\tt")"} $|$ \\
							   & & {\bfseries init} {\tt"("} {\bfseries id} {\tt")"} $|$ \\
							   & & ({\bfseries opadd}$)^+$ {\tt<}Factor{\tt>}
	\end{tabular}
\caption{Production rules of the grammar defining \clak{}'s language.
String literals within double quotes must be found as-is in the input program.}
\label{box:grammar}
\end{mybox}

\noindent
As is the case for most programming languages, this is a context-free grammar~\cite{LLR}.
More specifically, it is an LLR(1) grammar:
{\bf L}eft-to-right parsing,
{\bf L}eftmost derivation,
with {\bf R}egular expressions in the production's right-hand sides,
and {\bf 1} lookahead token.

Any program that can be produced by means of the production rules,
starting from the non-terminal symbol {\tt<}Model{\tt>}, and
only containing terminal symbols, is grammatically correct.
However, similarly to natural languages, not every
grammatically-correct statement is meaningful. 
Therefore, a number of semantic rules are needed.
We will not get into the details of the semantic analysis,
but we emphasize that these are the rules in charge to
assert, for instance, that 
the left-hand side operand is an output operand, and
the expression in the right-hand side only includes input operands.

Finally, we recall \clak{} is written in Mathematica; therefore, 
we also developed a parser 
that reads a file containing the description of a target operation in \clak{}'s
language and translates it into a Mathematica representation accepted by \clak{}'s 
core.

\section{Heuristics for the generation of algorithms}
\label{sec:heuristics}

\begin{figure}
\centering
	\centering
	\includegraphics[scale=0.6]{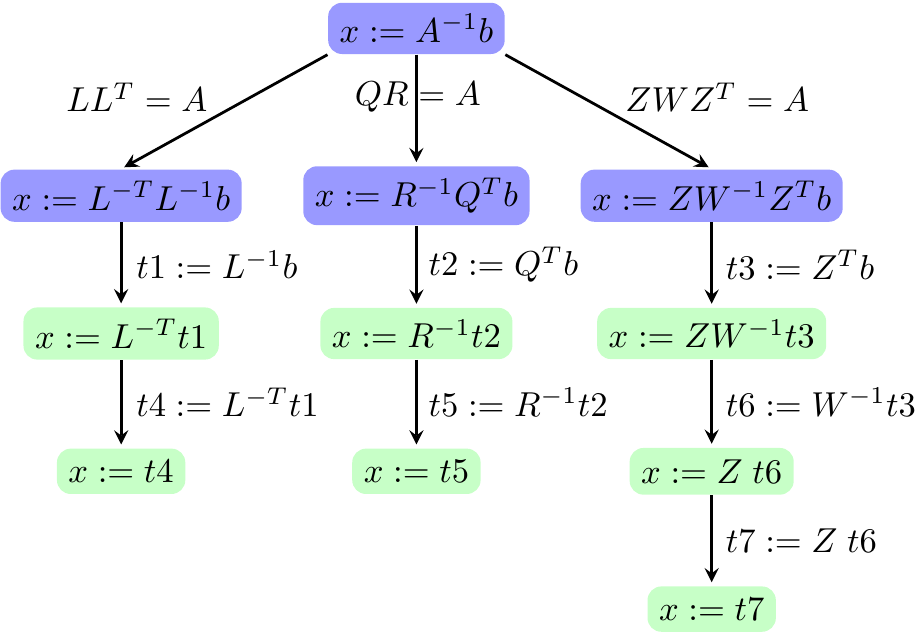}
	\caption{Full tree spawned by the compiler when processing the solution of a linear system of
		equations $x := A^{-1} b$, with an SPD coefficient matrix $A$, and a single right-hand
		side $b$.}
	\label{fig:linsys-full}
\end{figure}

Starting from a target equation, 
\clak{} explores a subset of the space of possible algorithms, 
dynamically generating a ``tree of decompositions''.
For instance, Figure~\ref{fig:linsys-full} contains the complete tree 
generated for the solution of a linear system, 
when the coefficient matrix is symmetric positive definite (SPD).
The root node corresponds to the input equation $x := A^{-1} b$, 
and every edge represents the mapping onto a building block;
in the example, 
the three branches are originated by three different factorizations of the matrix $A$:
a Cholesky factorization, a QR factorization, and an eigendecomposition.
Once the process is over, the operations along the edges from the root to each leaf  
constitute a valid algorithm.
In practice, the tree is built in two phases, corresponding to 
the blue (dark) and green (light) nodes, respectively. 
In the first phase, \clak{} deals with the inverse operator
via matrix factorizations; in the second phase, the decomposition
completes with the mapping of expressions onto kernels.
In order to limit the size of the tree,
the compiler uses the heuristics described hereafter. 

\subsection{Dealing with the inverse operator}
\label{subsec:heuristics-one}

The inversion of matrices is a delicate operation.
There are only rare occurrences of problems in which one 
is interested in the actual matrix inverse;
most often, the operation appears in the context of linear systems, least squares problems,
or more complex expressions; in the majority of cases, the inversion 
can---and should---be avoided altogether. 
Because of this, \clak{} splits the generation of algorithms 
in two phases, 
the first of which is solely devoted to the treatment of inverses;
the objective is to reduce the input equations to an expression in which 
the inverse is only applied to matrices in factored form,
i.e., triangular or diagonal (see blue subtree in Figure~\ref{fig:linsys-full}).
In the second phase, the resulting expression is 
mapped onto computational kernels (see green branches in Figure~\ref{fig:linsys-full}).

This first phase takes as input the target equation, and
generates the subtree characterized by leaf nodes that require 
no further treatment of the inverses.
This is an iterative process in which the tree is constructed 
in a breadth-first fashion;
at each iteration, the current expression is inspected for inverse operators,
the innermost of which is 
handled.
The inversion is applied to either a full matrix, such as $A^{-1}$, 
or to a non-simplifiable expression, e.g., $(A^T A)^{-1}$ with $A$ rectangular.

\begin{figure}
\centering
\includegraphics[scale=0.88]{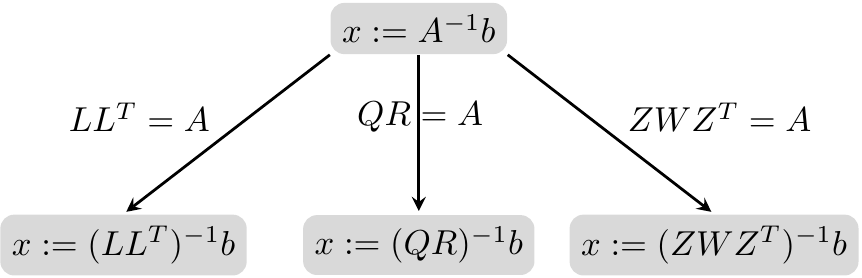}
\caption{Solution of an SPD linear system. 
In the first iteration, 
\clak{} considers three possible factorizations 
for the coefficient matrix $A$; 
three branches are originated, corresponding to 
a Cholesky factorization (left), 
a QR factorization (middle), 
and an eigendecomposition (right).}
\label{fig:linsys-tree}
\end{figure}

\paragraph{Inversion of a full matrix.}
In the first case, the matrix is factored by means of one or more
of the many matrix decompositions provided by LAPACK, 
but instead of exhaustively trying all possibilities,
the factorizations are chosen according 
to the properties of the matrix.
For instance, if $A$ is 
a symmetric positive definite  matrix, 
we limit the viable options to
the Cholesky factorization ($L L^T = A$), 
the QR factorization ($QR = A$), 
and the eigendecomposition ($Z W Z^T$); 
vice versa, the LU ($LU = A$), and LDL ($LDL^T = A$) 
factorizations are not considered.
As depicted in Figure~\ref{fig:linsys-tree}, 
the compiler constructs as many branches as factorizations,
while altering the initial expression. 
All the branches are subsequently  explored.

Limiting the search to a subset of all possible factorizations 
has two advantages:
On the one hand, non-promising algorithms are discarded
and the search space is pruned early on;
on the other hand, the algorithms are 
tailored to the specific properties of the application.
Table~\ref{tab:factorizations} contains 
the set of factorizations currently used by \clak{},
together with the matrix properties that enable them.

\begin{table}
\centering
\setlength\extrarowheight{2pt}
\renewcommand{\arraystretch}{1.0}
\begin{tabular}{l@{\hspace*{8mm}} l} \toprule
	{\bf Matrix Property} & {\bf Factorizations} \\\midrule 
	Symmetric & LDL, QR, Eigendecomposition \\
	SPD       & Cholesky, QR, Eigendecomposition \\
	Column Panel (FullRank) & QR \\
	Column Panel (RankDef) & SVD \\
	Row Panel (FullRank) & LQ \\
	Row Panel (RankDef)  & SVD \\
	General & LU, SVD \\\bottomrule
\end{tabular}
\caption{Factorizations currently used by \clak{}, 
  and matrix properties that enable them.}
\label{tab:factorizations}
\end{table}

\begin{figure}
\centering
	{
		\centering
		\hspace{2mm}
		\includegraphics[scale=0.74]{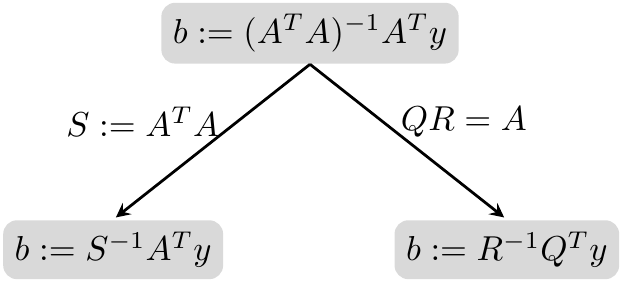}
		\hspace{2mm}
		\label{fig:ols_tree_fr}
	}
	\caption{Snippet of the tree spawned by \clak{} when processing the ordinary least-squares equation
	$b := (A^T A)^{-1} A^T y$, where $A \in R^{m \times n} (m > n)$ is full rank.
        }
	\label{fig:ols} 
\end{figure}

\paragraph{Inversion of an expression.}
We concentrate now on the case of  
an inverse operator applied to a non-simplifiable expression.
A characteristic example
comes from the normal equations, arising for instance as
part of the ordinary least-squares problem
\begin{equation}
	b := (A^T A)^{-1} A^T y,
  \label{eq:ols}
\end{equation}
where $A \in R^{m \times n}$ (with $m > n$) is full rank.
In this scenario, as depicted in Figure~\ref{fig:ols}, 
our compiler explores two alternative routes:
1) the multiplication of the expression $A^T A$, thus reducing it
to the inverse of a single SPD operand $S$; and
2) the decomposition of one of the matrices in the expression, 
in this case $A$, thus spawning a branch per 
suitable factorization.
As dictated by Table~\ref{tab:factorizations}, 
in Equation~\eqref{eq:ols} $A$ is decomposed by means of a QR factorization.

The treatment of inverses continues
until the inverse operator is only applied to triangular or diagonal matrices.
For the example in Figure~\ref{fig:ols}, 
the left branch would be further processed 
by factoring the matrix $S$,
yielding three more nodes;
the right branch instead, 
since $R$ is a triangular matrix, 
is complete.

\subsection{Mapping onto kernels}
\label{subsec:mapping}

The goal of this second phase is 
to find efficient mappings from expressions
to kernels provided by numerical libraries,
e.g., BLAS and LAPACK.
The number of possible mappings grows exponentially with the number
of operators in the expression, therefore 
heuristics are necessary to constrain 
the amount of explored alternatives.
We discuss two examples of such heuristics.

\paragraph{Common segments.}
The objective is 
to reduce the complexity of the algorithms
by avoiding redundant computations;
common segments of the expression are identified,
thus allowing the reuse of intermediate results.
We emphasize that this is by no means a trivial optimization. 
In fact, 
even for the simplest cases,
sophisticated tools such as Matlab do not adopt it.
For instance, when evaluating the operation
$$ \alpha := x^T y x^Ty, $$
where $x$ and $y$ are vectors of size $n$, 
Matlab makes use of Algorithm~\ref{alg:matlab-comseg},
which computes the products from left to right; 
\clak{} instead recognizes that the expression $x^T y$ appears twice,
and generates Algorithm~\ref{alg:compiler-comseg}, 
reducing the number of flops from $5n$ to $2n$.

\begin{center}
\renewcommand{\lstlistingname}{Algorithm}
\begin{minipage}{0.40\linewidth}
	\begin{lstlisting}[caption={Matlab's computation for $\alpha := x^T y x^T y $}, escapechar=!, label=alg:matlab-comseg]
  t1 := x' * y
  t2 := t1 * x'
  alpha := t2 * y
\end{lstlisting}
\end{minipage}
\hspace*{2cm}
\begin{minipage}{0.40\linewidth}
	\begin{lstlisting}[caption={\clak{}'s algorithm for $\alpha := x^T y x^T y $}, escapechar=!, label=alg:compiler-comseg]
  t1 := x' * y
  alpha := t1 * t1 !\vspace{4mm}!
\end{lstlisting}
\end{minipage}
\end{center}

More challenging is the case where one of the occurrences of
the common segment appears in transposed or inverted form.
As an example, let us consider the expression
$$ v := X^T L^{-1} L^{-T} X,$$
where both operands $X$ and $L$ are matrices, and $L$ is triangular.
In order to recognize that $L^{-T} X$ is the transpose of
$X^T L^{-1}$, and in general, to recognize that two segments are
the negation, inverse, or transpose of one another,
our compiler incorporates a large set of ground linear algebra
knowledge. This is covered in Section~\ref{sec:engine}.

\paragraph{Prioritization.}
In an attempt to minimize the cost of the generated algorithms, 
the kernels available to the compiler are classified according to
a precedence system. 
In Table~\ref{tab:precedence}, we give an example of a subset
of these kernels, sorted from high to low priority.
The precedences are driven by the dimensionality of the operands
in the kernels: The idea is to reduce the number of required flops
by keeping the dimensionality of the resulting operands as low as possible.~%
\footnote{The problem at hand---minimizing the cost of a sequence of
	(symbolic) matrix products---resembles the matrix chain multiplication 
	problem. However, \clak{} targets general size computations, and 
	operates with symbolic sizes. Hence, it does not 
	necessarily have enough information to determine whether the cost of a specific 
	parenthesization has a lower cost than another.}
For the example in Table~\ref{tab:precedence},
the first two kernels reduce the dimensionality of
the output operand with respect to that of the input, while
the third kernel maintains it, and the fourth increases it. 
Finally, the inversion of a triangular matrix is given the lowest precedence: 
A matrix will only be inverted if no other option is available.

\begin{table}
	\centering
	\begin{tabular}{c @{\hspace{3mm}} l @{\hspace{3mm}} l @{\hspace{3mm}} c @{\hspace{3mm}} c @{\hspace{3mm}} c} \toprule
		{\bf \#} & {\bf Kernels} & {\bf Example} & {\bf Dim(in)} & {\bf Dim(out)} \\\midrule
	1 & inner product & $\alpha := x^T y$ & 1/1 & 0 \\
	2 & matrix-vector operations & $y := A x$, $b := L^{-1} x$ & 2/1 & 1 \\
	3 & matrix-matrix operations & $C := A B$, $B := L^{-1} A$ & 2/2 & 2 \\
	4 & outer product & $A := x y^T$ & 1/1 & 2 \\
	5 & inversion of a triangular matrix & $C := L^{-1}$ & -- & -- \\\bottomrule
	\end{tabular}
\caption{Example of the classification of kernels based on a system of precedences.
The kernels that reduce the dimensionality of the output operands with respect to the input ones are
given higher precedence. The inversion is only selected when no other option exists.}
\label{tab:precedence}
\end{table}

The benefits of the prioritization were already outlined in this chapter's introduction
(Algorithms~\ref{alg:matlab-qly}~and~\ref{alg:compiler-qly}):
There, by favoring the matrix-vector over 
the matrix-matrix product, the complexity
was lowered by an order of magnitude. 
Here, we provide more examples.
Consider the operation
$$ \alpha := x^T z x^T y,$$
where $x$, $y$, and $z$ are vectors, and $\alpha$ is a scalar.
When inspecting the expression for kernels, \clak{} finds 
two inner products ($x^T z$, and $x^T y$), and one outer product ($z x^T$).
While all three options lead to valid algorithms,
the inner products are favored,
producing, for instance, Algorithm~\ref{alg:inner}; 
the cost of this algorithm is $O(n)$ flops, instead of 
a cost of $O(n^2)$, had the compiler favored 
the outer product (Algorithm~\ref{alg:outer}).

\begin{center}
\renewcommand{\lstlistingname}{Algorithm}
\begin{minipage}{0.43\linewidth}
	\begin{lstlisting}[caption={Computation of $ \alpha~:=~x^T z x^T y$ favoring inner products}, escapechar=!, label=alg:inner]
  t1 := x' * z
  t2 := x' * y
  alpha := t1 * t2
\end{lstlisting}
\end{minipage}
\hfill
\begin{minipage}{0.43\linewidth}
	\begin{lstlisting}[caption={Computation of $ \alpha~:=~x^T z x^T y$ favoring outer products}, escapechar=!, label=alg:outer]
  T1 := z * x'
  t2 := x' * T1
  alpha := t2 * y
\end{lstlisting}
\end{minipage}
\end{center}

A third example is given by
\begin{equation}
	\beta := v^T L^{-1} L^{-T} u,
	\nonumber
\end{equation}
where $L$ is a square lower triangular matrix, and $v$ and $u$ are vectors.
The inspection for kernels yields the following matches: $v^T L^{-1}$, $L^{-1}$,
and $L^{-T} u$. Again, the inversion of $L$ is avoided, unless no alternatives exist;
this is captured by the precedences listed in Table~\ref{tab:precedence}, which give priority to the solution
of linear systems over the inversion of matrices. Therefore, the second
option ($L^{-1}$) is dismissed, and the compiler only explores the branches
spawned by the first and third kernels. While the inversion of $L$ would lead to
a cubic algorithm (Algorithm~\ref{alg:matlab-inv}), the ones generated 
(e.g., Algorithm~\ref{alg:compiler-trsv}) have a quadratic cost.

\begin{center}
\renewcommand{\lstlistingname}{Algorithm}
\begin{minipage}{0.45\linewidth}
	\begin{lstlisting}[caption={Computation of $\beta~:=~v^T L^{-1} L^{-T} u$
	favoring the inversion of matrices}, escapechar=!, label=alg:matlab-inv]
	T1 := inv(L)
	t2 := v' * T1
	t3 := t2 * T1'
	beta := t3 * u
\end{lstlisting}
\end{minipage}
\hfill
\begin{minipage}{0.45\linewidth}
	\begin{lstlisting}[caption={Computation of $\beta~:=~v^T L^{-1} L^{-T} u$
	favoring the solution of triangular systems}, escapechar=!, label=alg:compiler-trsv]
	t1 := v' / L
	t2 := L' \ u
	beta := t1 * t2 !\vspace{3.8mm}!
\end{lstlisting}
\end{minipage}
\end{center}

Notice that if implemented naively, the rules discussed so far 
may lead to an infinite process: 
For instance, a matrix could be factored and built again, 
as in 
$(A^T A)^{-1} \xrightarrow{QR = A} ((QR)^T QR)^{-1} \xrightarrow{A := QR} (A^T A)^{-1}$; 
also, a matrix could be factored indefinitely, as in 
$A \xrightarrow{Q_1 R_1 = A} Q_1 R_1
\xrightarrow{Q_2 R_2 = Q_1} Q_2 R_2 R_1 
\xrightarrow{\dots} \dots 
\xrightarrow{Q_i R_i = Q_{i-1}} Q_i R_i \dots R_2 R_1 $.
To avoid such situations, our compiler incorporates a mechanism to measure and guarantee progress.

\section{Compiler's engine}
\label{sec:engine}

The availability of knowledge 
is crucial for a successful application of the heuristics.
Equally important is the capability of algebraically manipulating
expressions with the objective of simplifying them 
or finding common segments.
Here, we detail the different modules that constitute
\clak{}'s engine, and how these modules enable:
1) the algebraic manipulation of expressions,
2) the mapping onto building blocks, and
3) the management of both input and inferred knowledge.

\subsection{Matrix algebra}
\label{subsec:algebra}

The {\em Matrix algebra} module deals with the algebraic
manipulation of expressions.
It incorporates a considerable amount of knowledge regarding
properties of the operators, such commutativity 
and distributivity, and linear algebra equalities,
such as
``the inverse of an orthogonal matrix equals its transpose''.
This knowledge is encoded as 
an extensive list of {\em rewrite rules} that allow the compiler to
rearrange expressions, simplify them, and find subexpressions that are the 
inverse, transpose, etc, of one another. 

A rewrite rule consists of a left-hand and a right-hand side.
The left-hand side contains a pattern, possibly restricted via constraints to be satisfied by the operands;
the right-hand side specifies how the pattern, if matched, should be replaced.
For instance, the rule
$$
	Q^{-1} \wedge \text{Orthogonal($Q$)} \rightarrow Q^T \\
$$
reads as follows: The inverse of a matrix $Q$, provided that $Q$ is orthogonal,
may be replaced with the transpose of $Q$.
Box~\ref{box:simplify-rules} includes examples of rewrite rules
relative to transposition, product, and inversion of matrices.
\begin{mybox}
	\vspace{1em}
	\centering
	\begin{minipage}{0.90\textwidth}
		$\begin{aligned}
		(A \times B)^T & \;\, \rightarrow \;\, B^T \times A^T \\
		Q^T \times Q \, \wedge \text{Orthogonal($Q$)} & \;\, \rightarrow \;\, I \\
		A \times I \, \wedge \neg \text{Scalar($A$)} & \;\, \rightarrow A \\
		(A \times B)^{-1} \wedge \text{Square($A$)} \wedge \text{Square($B$)} & \;\, \rightarrow \;\, B^{-1} \times A^{-1} \\
		A^{-1} \times A & \;\, \rightarrow I \\
		\end{aligned}$
  \end{minipage}
  \caption{Examples of rewrite rules included in the Matrix Algebra module. $I$ is the identity matrix.}
  \label{box:simplify-rules}
\end{mybox}

The example in Box~\ref{box:simplification} gives an idea of
how the compiler is capable of 
eliminating unnecessary calculations
by means of algebraic transformations.
The initial expression is $(X^T X)^{-1} X^T L^{-1} y$;
we assume $X$ is a full rank column panel, and 
$L$ a lower triangular matrix.
As dictated by the heuristics presented in Section~\ref{sec:heuristics},
one alternative in the processing of the expression
is through a QR factorization of the matrix $X$:
The symbol $X$ is replaced by $Q R$---line 2---(where 
$Q$ and $R$ are orthogonal and upper triangular, respectively),
and a series of transformations are triggered. 
First, the transposition is distributed over the product---line 3---;
next, due to the orthogonality of $Q$, the product 
$Q^T Q$ is removed as it equals the identity---line 4---.
Since $R$ is square, the inverse may be distributed 
over the product $R^T R$ resulting in $R^{-1} R^{-T}$---line 5---. 
Another simplification rule establishes that the product 
of a square matrix with its inverse equals the identity; because of this, 
the $R^{-T} R^T$ is removed---line 6---. 
After all these algebraic steps, 
the expression $((Q R)^T Q R)^{-1} (Q R)^T L^{-1} y$
simplifies to $R^{-1} Q^T L^{-1} y$. 
Box~\ref{box:simplify-rules} contains the necessary 
set of rewrite rules for this manipulation.

\begin{mybox}
\begin{align}
	1) \quad b &:= (X^T X)^{-1} X^T L^{-1} y; \nonumber \\
	2) \quad b &:= ((Q R)^T Q R)^{-1} (Q R)^T L^{-1} y; \nonumber \\
	3) \quad b &:= (R^T Q^T Q R)^{-1} R^T Q^T L^{-1} y; \nonumber \\
	4) \quad b &:= (R^T R)^{-1} R^T Q^T L^{-1} y; \nonumber \\
	5) \quad b &:= R^{-1} R^{-T} R^T Q^T L^{-1} y; \nonumber \\
	6) \quad b &:= R^{-1} Q^T L^{-1} y. \nonumber
\end{align}
\caption{Example of expression simplification carried out by \clak{}.
$X$ is a full rank column panel, and $L$ a lower triangular matrix.}
\label{box:simplification}
\end{mybox}

Rewrite rules are algebraic identities, i.e., 
they may be applied in both directions. 
For instance, the expression $(A B)^T$ may be rewritten as $B^T A^T$, and vice versa,
leading to multiple equivalent representations for the same expression. 
Since this fact complicates the manipulation and identification 
of building blocks, one may be tempted to use rules as
``always distributing the transpose over the product''
for reducing every expression to a canonical form.
Unfortunately, there exists no ``best'' representation for matrix expressions. 
Indeed, imposing a canonical form would lower the effectiveness of the compiler.

A typical example is given by the distribution of the product
over the addition: 
$(A + B)C$ may be transformed into $A C + B C$ and vice versa, 
but neither representation is superior in all scenarios.
Consider, for instance, the expression 
$ \alpha x x^T + \beta y x^T + \beta x y^T$,
where $\alpha$ and $\beta$ are scalars, and $x$ and $y$ are vectors.
In this format, 
it is straightforward to realize that the expression is 
symmetric---the first term is symmetric, and the second
and third are one the transpose of the other---;
if instead $x^T$ is factored out as in 
$ (\alpha x + \beta y) x^T + \beta x y^T $,
the symmetry is not visible, 
and redundant computation would be performed.
This is an example in which the distribution of the product over the addition 
seems to be the choice to favor.

On the contrary, let us consider the expression in Box~\ref{box:factor-out}:
$(Z W Z^T + Z Z^T)^{-1},$ where $Z$ is square and orthogonal, and $W$ is diagonal. 
Factoring $Z$ and $Z^T$ out---$(Z (W + I) Z^T)^{-1}$, where $I$ is the identity matrix---is 
an indispensable first step towards the simplification of the expression.
Next, since all matrices are square, 
the inverse may be distributed over the product, 
and the orthogonality of $Z$ allows the rewriting of its inverse as its transpose, 
resulting in $Z (W + I)^{-1} Z^T$. 
This transformation---absolutely crucial in practical cases---is only possible 
thanks to the initial factoring;
hence, this is a contrasting example 
in which the distribution of the product is not the best option.
In light of this dichotomy, 
\clak{} always operates with multiple alternative representations.

\begin{mybox}
\begin{align}
M &:= (Z W Z^T + Z Z^T)^{-1}; \nonumber \\
M &:= (Z (W + I) Z^T)^{-1}; \nonumber \\
M &:= Z^{-T} (W + I)^{-1} Z^{-1}; \nonumber \\
M &:= Z (W + I)^{-1} Z^T; \nonumber
\end{align}
\caption{Another example of expression manipulation carried out by the compiler.
$Z$ is square and orthogonal; $W$ is diagonal.}
\label{box:factor-out}
\end{mybox}

\vspace*{-3mm}

\subsection{Interface to building blocks}

We have claimed repeatedly that the goal of \clak{} is to decompose 
the target equation in terms of building blocks that 
can be directly mapped to library invocations;
it remains to be discussed what are the available building blocks. 
The exact list is configurable,
and is provided to the compiler via the {\em Interface to building blocks} module.
This module contains 
a list of patterns associated to the corresponding computational kernels.
As of now, this list includes a subset of the operations provided by 
BLAS and LAPACK, e.g.,
matrix products and the solution of linear systems;
a sample is given in Box~\ref{box:bblocks-sample}.

\begin{mybox}
	\centering
	\begin{minipage}{\textwidth}
%
  {\sc Matrix products:} \\[1mm]
  {\tt 
    \hspace*{2mm} plus[times[alpha\_, A\_, B\_], times[beta\_, C\_]] \\
    \hspace*{2mm} plus[times[alpha\_, trans[A\_], B\_], times[beta\_, C\_]] \\
    \hspace*{2mm} plus[times[alpha\_, trans[A\_], A\_], times[beta\_, C\_]] \\
    \hspace*{2mm} times[A\_, trans[A\_]] /; isTriangularQ[A] \\[2mm]
  }
  {\sc Linear systems:} \\[1mm]
  {\tt
    \hspace*{2mm} plus[times[inv[A\_], B\_]] /; isTriangularQ[A] \&\& isMatrixQ[B] \\
    \hspace*{2mm} plus[times[inv[A\_], b\_]] /; isTriangularQ[A] \&\& isVectorQ[b] 
  }
  \end{minipage}
  \caption{A snippet of the interface to available building blocks.
  }
  \label{box:bblocks-sample}
\end{mybox}

\clak{} is by no means limited to this set of operations. 
Should an additional or a different set of building blocks be available, 
say RECSY~\cite{RECSY1,RECSY2}
or an extension of the BLAS library~\cite{2002:USB:567806.567807}, 
this can be made accessible to the compiler with only minimal effort,
by including in this module the corresponding patterns. 
For instance, in order to add support for the operation $w := \alpha x + \beta y$, 
as proposed in the extension to the BLAS library, 
we only need to incorporate the pattern
\begin{verbatim}
     plus[ times[ alpha_, x_ ], times[ beta_, y_ ] ] /;
                  isVectorQ[x,y] && isScalarQ[alpha, beta];
\end{verbatim}
the compiler is then ready to make use of this building block in the generation of algorithms.

\subsection{Inference of properties}
\label{sec:inf-props}

Properties play a central role in the search for efficient algorithms;
the more knowledge is available,
the more opportunities arise for optimizations.
A distinguishing feature of \clak{} is the propagation
of properties: We developed an engine for
inferring properties of expressions from those of the individual operands.
Thanks to this engine, the initial knowledge (from the input equation) is augmented dynamically.

This mechanism is activated every time a mapping takes place:
1) when mapping onto factorizations, properties
are propagated from the input matrix to its factors;
2) when mapping onto other kernels, properties are propagated
from the segment to the output quantity.
The gained knowledge on the intermediate operands is then used
by the compiler for further tailoring the algorithms.
Boxes~\ref{box:inf-fact}~and~\ref{box:inf-kernels} provide
examples of inference of knowledge in factorizations and products, 
respectively.

\begin{mybox}
	\centering
	\begin{minipage}{.75\textwidth}
		{\sc eigendecomposition} ($Z W Z^T = A$): \\
		\begin{tabular}{@{\hspace{5mm}}l @{\hspace{5mm}}l}
			Input  & $A$: matrix, square, symmetric \\
			Output & $Z$: matrix, square, orthogonal \\
				   & $W$: matrix, square, diagonal \\
		\end{tabular}

		\vspace{3mm}

		{\sc qr} ($Q R = A$): \\
		\begin{tabular}{@{\hspace{5mm}}l @{\hspace{5mm}}l }
			Input  & $A$: matrix, column-panel, full rank \\
			Output & $Q$: matrix, orthogonal, column-panel, full rank \\
				   & $R$: matrix, square, upper triangular, full rank
		\end{tabular}
	\end{minipage}
	\caption{Inference of properties for two representative factorizations. }
	\label{box:inf-fact}
\end{mybox}

\begin{mybox}
	\centering
	\begin{minipage}{.6\textwidth}
		$W := L^{-1} X$: \\
		\begin{tabular}{@{\hspace{5mm}}l @{\hspace{5mm}}l }
			Input  & $L$: matrix, square, full rank \\
				   & $X$: matrix, column-panel, full rank \\
			Output & $W$: matrix, column-panel, full rank \\
		\end{tabular}

		\vspace{3mm}

		$S := W^T W$: \\
		\begin{tabular}{@{\hspace{5mm}}l @{\hspace{5mm}}l }
			Input  & $W$: matrix, column-panel, full rank \\
			Output & $S$: matrix, square, SPD \\
		\end{tabular}
	\end{minipage}
	\caption{Inference of properties for two mappings onto kernels. }
	\label{box:inf-kernels}
\end{mybox}

It is important to notice that the inference of rules and the mapping onto kernels are
completely independent actions. 
For instance, in the absence of the second rule in Box~\ref{box:inf-kernels},
\clak{} would still be able to match a product of the form $A^T A$ (provided the pattern is
included in the {\em Interface to building blocks} module);
however, if $A$ is a full rank, column panel matrix, the compiler would not be able to
infer, and then exploit, the positive definiteness of $S$.

We regard the inference engine as a growing database of linear algebra knowledge.
In its current form, the database is populated with a sample of rules and theorems,
but the flexible design of the module allows it to be easily extended
with new inference rules. 

\section{A detailed example: GWAS (Part I)}
\label{sec:detailed-example}

We use the computationally challenging 
genome-wide association study (GWAS) problem to
illustrate the potential of \clak{}'s engine and heuristics.
We recall that, as part of GWAS, one has to solve the equation
\begin{equation}
\label{eq:probDef}
\left\{ 
{\begin{aligned}
	b_{ij} & := (X_i^T M_j^{-1} X_i)^{-1} X_i^T M_j^{-1} y_j \\
	M_j    & := h_j \Phi + (1 - h_j) I
\end{aligned}}
\right.
\;
\quad \text{ with }
{\begin{aligned}
 & 1 \le i \le m \\
 & 1 \le j \le t,
\end{aligned}}
\end{equation}
where $X_i$, $y_j$, $h_j$, and $\Phi$ are known quantities, and $b_{ij}$ is sought after.
The size and properties of the operands are as follows:
$b_{ij} \in {R}^{p}$, 
$X_i \in {R}^{n \times p}$ is a full rank column panel ($n > p$), 
$M_j \in {R}^{n \times n}$ is symmetric positive definite,
$y_j \in {R}^{n}$, 
$h_j \in {R}$,
$\Phi \in {R}^{n \times n}$ is symmetric, and
$I$ is the identity matrix.
Box~\ref{box:input} contains the representation of Equation~\eqref{eq:probDef} in \clak{}'s language. 

\begin{mybox}[!h]
\vspace{3mm}
\begin{verbatim}
      Equation GWAS
    
          Matrix X   <Input, FullRank, ColumnPanel>;
          Vector Y   <Input>;
          Scalar h   <Input>;
          Matrix Phi <Input, SymmetricLower>;
          Vector b   <Output>;
    
          Matrix M   <Intermediate, SPD>;
    
          b{ij} = inv( trans(X{i}) * inv(M{j}) * X{i} ) * 
                       trans(X{i}) * inv(M{j}) * y{j};
          M{j}  = h{j} * Phi + (1 - h{j}) * I;
\end{verbatim}
\caption{\clak{}'s representation of the GWAS equation.}
\label{box:input}
\end{mybox}

\begin{figure}
\centering
\includegraphics[scale=0.88]{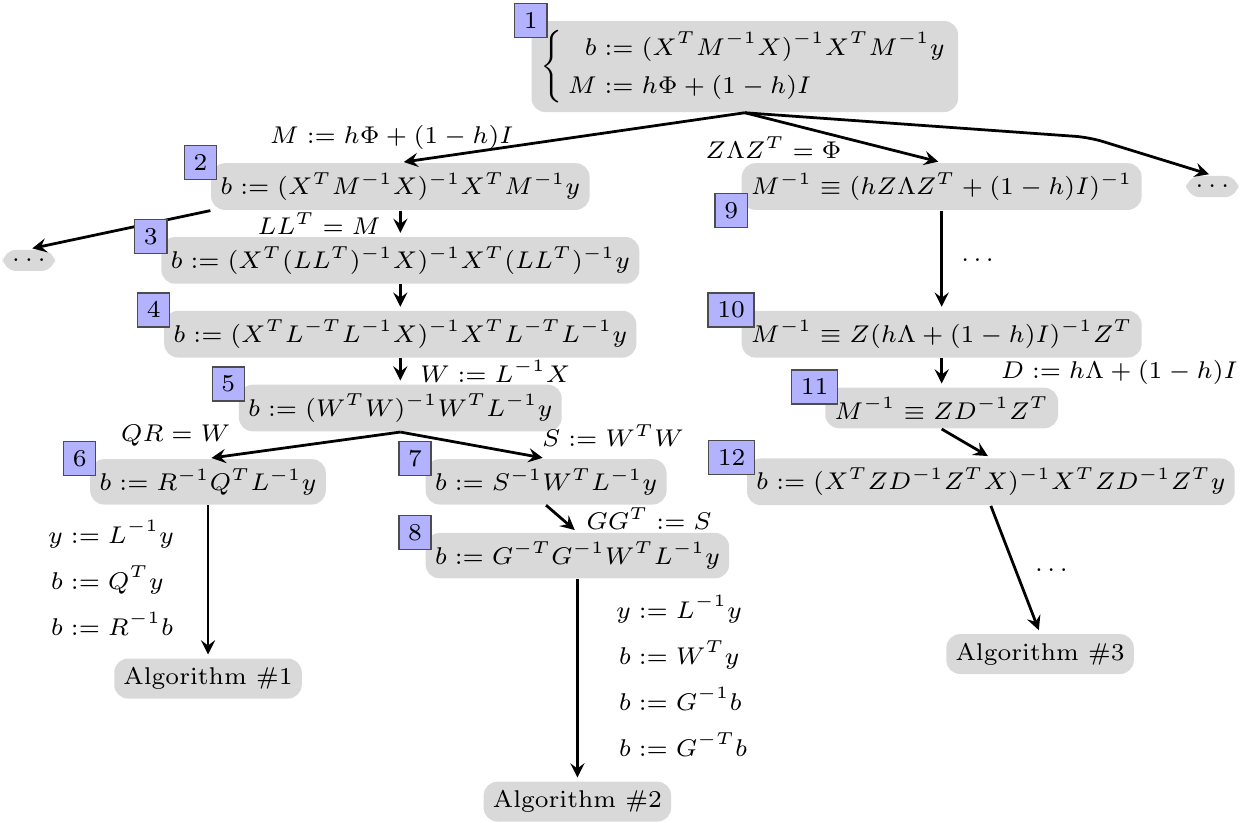}
\caption{Snippet of the tree spawned by \clak{} while constructing algorithms
for the computation of GWAS.}
\label{fig:gwas-tree}
\end{figure}

Due to the complexity of GWAS, a large number of alternatives are generated. 
For the sake of this discussion, we focus now on the solution of a single instance 
of Equation~\eqref{eq:probDef}, as if both $m$ and $t$ were 1, and
defer the case of multiple instances to next section.
In Figure~\ref{fig:gwas-tree}, we provide a snippet of the tree spawned by \clak{}
while constructing algorithms. Among the dozens of different branches,
we describe three representative ones. 

At the root node,
 the compiler starts by dealing with the innermost inverse, 
$M^{-1}$, and equivalently, $ (h_j \Phi + (1 - h_j) I)^{-1} $. 
As explained in Section~\ref{sec:heuristics}, 
the options are either to reduce the expression 
to a single operand ($M$, which is known to be SPD),
or to factor one of the matrices in the expression, in this case $\Phi$. 
The former choice leads directly to Node 2 
(modulo the order in which addition and scaling are performed), 
while the latter opens up a number of branches, corresponding to 
all the admissible factorizations of $\Phi$;
the middle branch in Figure~\ref{fig:gwas-tree}
follows the eigendecomposition of $\Phi$.
One might argue that
based on the available knowledge ($M$ is SPD),
the compiler should decide against the eigendecomposition,
since a Cholesky factorization is about ten times as fast.
In actuality, although the eigendecomposition is suboptimal for the solution of one single instance,
in the general case (Equation~\eqref{eq:probDef})
it leads to the fastest algorithms of all~\cite{OmicABEL}.%
\footnote{
This is because the eigendecomposition can be reused across
the entire two-dimensional sequence, 
while the Cholesky factorization cannot.
}

Let us now concentrate on the subtree rooted at Node 2. 
The input equation was reduced to $b := (X^T M^{-1} X)^{-1} X^T M^{-1} y$; 
again, \clak{} looks for the innermost inverse, $M^{-1}$, and
spawns a branch per factorization allowed for SPD matrices: 
QR, Cholesky, and eigendecomposition (Table~\ref{tab:factorizations});
here, we only describe the Cholesky factorization ($L L^T = M$),
which generates Node 3:
The equation becomes $b := (X^T (L L^T)^{-1} X)^{-1} X^T (L L^T)^{-1} y$, and  
the inference engine asserts a number of properties for $L$: 
square, lower triangular, and full rank.
The innermost inverse now is $(L L^T)^{-1}$;
since $L$ is square, rewrite rules allow the distribution of the 
inverse over the product $L L^T$, resulting in Node 4.

Once more, the compiler looks at the innermost inverse operators: 
In this case, they all are applied to triangular matrices and,
according to our guidelines, they do not require further treatment.
Therefore the focus shifts on the expression $(X^T L^{-T} L^{-1} X)^{-1}$;
$L$ is already in factored form, 
while according to a heuristic,
the factorizations of $X$ would not be useful;
hence the compiler resorts to mappings onto kernels.
Matching the expression against the list of available kernels yields two
segments: $L^{-1}$ and $L^{-1} X$. 
The latter has higher priority,
so it is exposed ($W := L^{-1} X$), 
every occurrence is replaced with $W$ (generating Node 5),
and it is established that $W$ is a full rank, column panel (Box~\ref{box:inf-kernels}).

Similar to the example depicted in Figure~\ref{fig:ols}, 
the inspection of Node 5
causes two branches to be constructed:
In the right one, \clak{} multiplies out $S := W^T W$,
producing the $SPD$ matrix $S$ (Node 7).
In the left one, in accordance to the properties of $W$, 
the matrix is factored via a QR factorization;
after replacing $W$ with the product $Q R$, the simplifications
exposed in Box~\ref{box:simplification} are carried out, 
resulting in Node 6.
At this point, all inverses are processed, as 
the remaining ones are only applied to triangular matrices. 
For this node, the first phase (as described in Section~\ref{subsec:heuristics-one}) is completed, thus
the remaining expression is now to be mapped onto available kernels. 
The compiler identifies the following building blocks: 
$R^{-1}$, $R^{-1} Q^T$, $Q^T L^{-1}$, $L^{-1}$, and $L^{-1} y$.
The first four are either matrix inversions or matrix-matrix operations, 
while the last one corresponds to a matrix-vector operation. 
Based on the list of priorities (Table~\ref{tab:precedence}), 
the matrix-vector operation $L^{-1} y$ is chosen. 
The same reasoning is applied subsequently, 
leading to the sequence of operations 
$y' := L^{-1} y$,
$b := Q^T y'$, 
and $b := R^{-1} b$. 
A similar discussion leads from Node 7 to Algorithm \#2.

Finally, we focus on the subtree rooted at Node 9. After the eigendecomposition
of $\Phi$, the innermost inverse is given by $M^{-1} \equiv (h Z \Lambda Z^T + (1-h)I)^{-1}$.
Analogous to the reasoning previously illustrated in Box~\ref{box:factor-out}, 
\clak{} carries out a number of algebraic transformations that lead to the simplified
expression $M^{-1} \equiv Z (h \Lambda + (1-h)I)^{-1} Z^T$ (Node 10). 
Here, the innermost inverse is applied to a diagonal object ($\Lambda$ is diagonal and $h$ a scalar);
no more factorizations are needed, and $D := h \Lambda + (1-h)I$ is exposed (Node 11). 
The inverse of $M$ is then replaced in $b := (X^T M^{-1} X)^{-1} X^T M^{-1} y$, resulting in Node 12.
The subsequent steps develop similarly to the case of Node 4, generating Algorithm \#3.

Once the search completes, the algorithms
are built by assembling the operations that label each edge along the path
from the root node to each of the leafs. 
The three algorithms are provided in 
Algorithms~\ref{alg:alg-qr}~(\#1:~{\sc qr-gwas}),
\ref{alg:alg-chol}~(\#2:~{\sc chol-gwas}),
and~\ref{alg:alg-eigen}~(\#3:~{\sc eig-gwas}),
together with the corresponding screenshot of Mathematica's output.
In brackets, we provide the names of the matching building blocks.
Next, we discuss how the produced algorithms are tailored
for the computation of multiple instances of problems.

\begin{center}
\renewcommand{\lstlistingname}{Algorithm}
\begin{minipage}{0.45\linewidth}
\begin{lstlisting}[caption=\normalsize \sc qr-gwas, escapechar=!, label=alg:alg-qr]
$M := h\Phi + (1-h)I$       (!\sc scal-add!)
$L L^T = M$              (!{\sc potrf}!)
$W := L^{-1} X$              (!{\sc trsm}!)
$Q R = W$              (!{\sc geqrf}!)
$y := L^{-1} y$              (!{\sc trsv}!)
$b := Q^T y$              (!{\sc gemv}!)
$b := R^{-1} b$              (!{\sc trsv}!)
\end{lstlisting}
\end{minipage}
\hspace*{2cm}
\begin{minipage}{0.30\linewidth}
\centering
\fbox{ \includegraphics[]{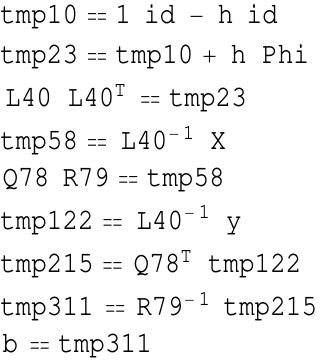} }
\end{minipage}
\end{center}

\begin{center}
\renewcommand{\lstlistingname}{Algorithm}
\begin{minipage}{0.45\linewidth}
\begin{lstlisting}[caption={\normalsize \sc chol-gwas}, escapechar=!, label=alg:alg-chol]
$M := h\Phi + (1-h)I$     (!\sc scal-add!)
$L L^T = M$              (!\sc potrf!)
$W := L^{-1} X$              (!\sc trsm!)
$S := W^T W$              (!\sc syrk!)
$G G^T = S$              (!\sc potrf!)
$y := L^{-1} y$              (!\sc trsv!)
$b := W^T y$              (!\sc gemv!)
$b := G^{-1} b$              (!\sc trsv!)
$b := G^{-T} b$              (!\sc trsv!)
\end{lstlisting}
\end{minipage}
\hspace*{2cm}
\begin{minipage}{0.30\linewidth}
\fbox{ \includegraphics[]{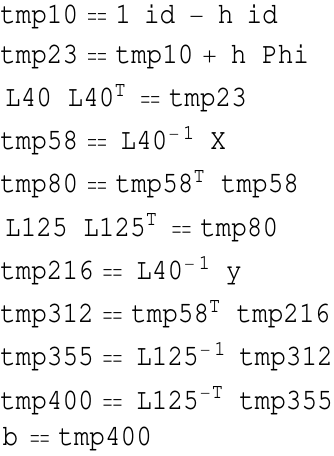} }
\end{minipage}
\end{center}

\begin{center}
\renewcommand{\lstlistingname}{Algorithm}
\begin{minipage}{0.45\linewidth}
\begin{lstlisting}[caption={\normalsize \sc eig-gwas}, escapechar=!,label=alg:alg-eigen]
$Z \Lambda Z^T$ = $\Phi$           (!{\sc syevr}!)
$D := h \Lambda + (1 - h) I$              (!{\sc scal-add}!)
$K := X^T Z$              (!{\sc gemm}!)
$V := K D^{-1}$              (!{\sc scal}!)
$A := V K^T$              (!{\sc gemm}!)
$Q R = A$              (!{\sc geqrf}!)
$y := Z^T y$              (!{\sc gemv}!)
$b := V y$              (!{\sc gemv}!)
$b := Q^T b$              (!{\sc gemv}!)
$b := R^{-1} b$              (!{\sc trsv}!)
\end{lstlisting}
\end{minipage}
\hspace*{2cm}
\begin{minipage}{0.30\linewidth}
\fbox{ \includegraphics[]{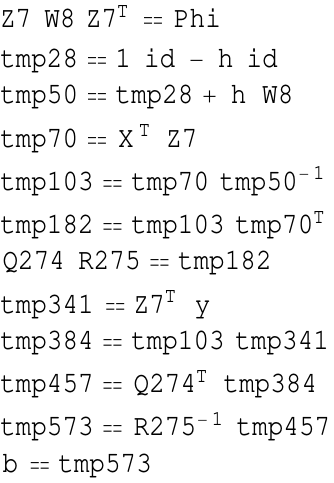} }
\end{minipage}
\end{center}

\section{Multiple instances of problems: GWAS (Part II)}
\label{sec:multiple-instances}

It is not uncommon that scientific and engineering applications require
the solution of not a single instance of a problem, but an $n$-dimensional grid of them.
Typically, libraries and languages for scientific computing follow a black-box approach, 
i.e., they provide a routine to solve one instance, 
which is then used repeatedly for the entire grid. 
While this approach is effective for problems that are completely 
independent from one another, 
its rigidity leads to a suboptimal strategy when the problems are related, 
and intermediate results may be reused.
To overcome this limitation, \clak{} breaks the black-box approach by
1) exposing the computation within the single-instance algorithm, 
2) performing an analysis of data dependencies, and 
3) rearranging the operations so that redundant computations are avoided. 

The generation of algorithms for grids of problems is divided in two steps:
First, the compiler creates a family of algorithms for a single instance, 
following the techniques described in Sections~\ref{sec:heuristics}~and~\ref{sec:engine};
then, each of the algorithms is customized for the solution of the entire grid.
We proceed by describing the latter step, using as a case study the GWAS example; 
specifically, we focus on Algorithm~\ref{alg:alg-eigen}.

The single-instance algorithm is wrapped with as many loops as different dimensions;
in this case, with a double loop along the $m$ and $t$ dimensions.
Both loop transpositions ---{\tt for i, for j}~~and~~{\tt for j, for i}---
are generated and analyzed; we concentrate on the latter. 
Next, \clak{} identifies operations that are loop-invariant, i.e., 
operations that do not change across iterations of the loop;
and applies the so-called {\em code motion} optimization,
which consists in moving loop-invariant operations to the preheader of the 
loop, i.e., the region right before of the loop.

In details, invariant operations are identified by means of an 
analysis of the
dependencies between operands and loop indices: 
The compiler labels each input operand
according to the input equation description (Algorithm~\ref{alg:loop-init-label}).
The subscripts are then propagated with
a single pass, from top to bottom, through the algorithm:
For each operation, 
the union of the indices appearing in the right-hand side
is attached to the operand(s) on the left-hand side, 
and to all their occurrences thereafter (Algorithm~\ref{alg:loop-labeled}).
\begin{center}
\begin{minipage}{0.4\linewidth}
\begin{lstlisting}[numbers=left,caption={\normalsize \sc eig-gwas}. Initial dependencies., escapechar=!,label=alg:loop-init-label]
for $j$ in 1:t
  for $i$ in 1:m
    $Z \Lambda Z^T = \Phi$
    $D := h_j \Lambda + (1 - h_j) I$
    $K := X_i^T Z$
    $V := K D^{-1}$
    $A := V K^T$
    $Q R = A$
    $y := Z^T y_j$
    $b := V y_j$
    $b := Q^T b$
    $b := R^{-1} b$
\end{lstlisting}
\end{minipage}
\hfill
\begin{minipage}{0.4\linewidth}
\begin{lstlisting}[numbers=left,caption={\normalsize \sc eig-gwas}. Propagated dependencies., escapechar=!,label=alg:loop-labeled]
for $j$ in 1:t
  for $i$ in 1:m
    $Z \Lambda Z^T = \Phi$
    $D_j := h_j \Lambda + (1 - h_j) I$
    $K_i := X_i^T Z$
    $V_{ij} := K_i D_j^{-1}$
    $A_{ij} := V_{ij} K_i^T$
    $Q_{ij} R_{ij} = A_{ij}$
    $y_j := Z^T y_j$
    $b_{ij} := V_{ij} y_j$
    $b_{ij} := Q_{ij}^T b_{ij}$
    $b_{ij} := R_{ij}^{-1} b_{ij}$
\end{lstlisting}
\end{minipage}
\end{center}
At this point, 
the subindices in the left-hand side operands indicate 
each operation's dependencies. Operations not labeled with the $i$
subindex may now be moved to the preheader of the innermost loop;
these operations are computed once per iteration over $j$, and
reused across the loop over $i$.
Additionally, the operation in line 3 is invariant with respect 
to both loops; thus it is moved outside the loops.
The rearranged algorithm is provided in Algorithm~\ref{alg:loop-rearranged}.

We note that Algorithm~\ref{alg:loop-rearranged} still performs some redundant
computation: line 6 depends on the iterator $i$ but not on $j$, thus
being computed redundantly for each iteration over $j$. Similarly to
the application of code motion, \clak{} identifies this situation
and drags the operation outside the loops, 
precomputes each of the $i$ products,
and reuses the results in the loops.\footnote{
	This optimization reduces the computational cost but
	increases the temporary storage requirements.
	Currently, \clak{} focuses on flop-efficient algorithms;
	in future versions of the compiler, the application of this and 
	similar optimizations should be configurable.
}
The final algorithm for the two-dimensional grid of problems is
provided in Algorithm~\ref{alg:loop-final}.
In the following section, we discuss the impact of the analysis of dependencies.

\noindent
\hspace*{2mm}
\begin{minipage}{0.4\linewidth}
\begin{lstlisting}[numbers=left,caption={\normalsize \sc eig-gwas}. After applying code motion., escapechar=!,label=alg:loop-rearranged]
$Z \Lambda Z^T = \Phi$
for $j$ in 1:t
  $D_j := h_j \Lambda + (1 - h_j) I$
  $y_j := Z^T y_j$
  for $i$ in 1:m
    $K_i := X_i^T Z$
    $V_{ij} := K_i D_j^{-1}$
    $A_{ij} := V_{ij} K_i^T$
    $Q_{ij} R_{ij} = A_{ij}$
    $b_{ij} := V_{ij} y_j$
    $b_{ij} := Q_{ij}^T b_{ij}$
    $b_{ij} := R_{ij}^{-1} b_{ij}$ !\vspace{4.3mm}! 
\end{lstlisting}
\end{minipage}
\hfill
\begin{minipage}{0.4\linewidth}
\begin{lstlisting}[numbers=left,caption={\normalsize \sc eig-gwas}. Final algorithm., escapechar=!,label=alg:loop-final]
$Z \Lambda Z^T = \Phi$
for $i$ in 1:m
  $K_i := X_i^T Z$
for $j$ in 1:t
  $D_j := h_j \Lambda + (1 - h_j) I$
  $y_j := Z^T y_j$
  for $i$ in 1:m
    $V_{ij} := K_i D_j^{-1}$
    $A_{ij} := V_{ij} K_i^T$
    $Q_{ij} R_{ij} = A_{ij}$
    $b_{ij} := V_{ij} y_j$
    $b_{ij} := Q_{ij}^T b_{ij}$
    $b_{ij} := R_{ij}^{-1} b_{ij}$
\end{lstlisting}
\end{minipage}
\hspace*{4mm}

\section{Cost analysis} \label{sec:cost}

As detailed throughout this chapter, \clak{} produces a family
of algorithmic variants to solve one target equation.
Since the amount of generated algorithms might be (fairly) large, it is
convenient to provide the user with a metric to select the variant
that best suits his needs.
As part of the ``Interface to building blocks'' module,
each of the building blocks onto which \clak{} may map the algorithms
is labeled with its asymptotic cost.
By combining the individual costs, the compiler documents
each of the output algorithms with their computational cost.

As an example, we provide the cost of the three selected algorithms for
GWAS: {\sc qr-gwas}, {\sc chol-gwas}, and {\sc eig-gwas}.
Table~\ref{tab:cost} includes the cost of the three
algorithms after the tailoring for one instance of GLS problem, 
as well as for the two most common 
grids of GLS problems in GWAS (Equation~\ref{eq:probDef}): 
a one-dimensional grid, where $t=1$, and 
a two-dimensional grid, where $t \approx 10^5$.

\begin{table}[!h]
\centering
\renewcommand{\arraystretch}{1.4}
  \begin{tabular}{l@{\hspace*{5mm}} c@{\hspace*{5mm}} c@{\hspace*{5mm}} c} \toprule
    {Scenario} & 
	{\bf {\phantom{y}{\sc qr-gwas}\phantom{y}} } &
	{\bf {\phantom{y}{\sc chol-gwas}\phantom{y}} } &
    {\bf {\phantom{y}{\sc eig-gwas}\phantom{y}} } \\ \midrule
	{One instance} & $O(n^3)$               & $O(n^3)$               & $O(n^3)$ \\[2mm]
	{1D grid}  & $O(n^3 + m p n^2)$     & $O(n^3 + m p n^2)$     & $O(n^3 + m p n^2 +  m p^2 n)$ \\[2mm]
	{2D grid}  & $O(t n^3 + m t p n^2)$ & $O(t n^3 + m t p n^2)$ & $O(n^3 + m p n^2 +  m t p^2 n)$ \\[2mm]
	\bottomrule
  \end{tabular}
  \caption{Computational cost for the three selected algorithms generated by \clak{}.}
\label{tab:cost}
\end{table}

{\sc qr-gwas} and {\sc chol-gwas} share the same computational cost 
for both types of grids, suggesting a very similar behavior in practice.
When compared to {\sc eig-gwas}, these two algorithms present a lower
cost for the one-dimensional case;
in contrast, for the two-dimensional case,
the cost of {\sc eig-gwas} is considerably lower.
This analysis suggests that {\sc qr-gwas} and {\sc chol-gwas} are better suited for the one-dimensional
grid, while {\sc eig-gwas} is better suited for the two-dimensional one. 
Experimental results in~\cite{SingleGWAS,MultiGWAS} confirm these predictions.

These numbers illustrate two of the problems that motivate this research,
and emphasize the benefits of our approach.
First, a generic solver for a GLS, e.g., the Matlab routine {\tt lscov}
(LAPACK does not provide an expert routine for this operation), has a computational cost of $O(n^3)$, as
the algorithms generated by \clak{}. However, Matlab provides no means to
exploit domain-knowledge such as the structure of the matrix $M$, or the
linkage among problems; thus, unless the user develops his own algorithms,
the only approach supported by Matlab is to use {\tt lscov} for each individual
problem in the grid, for a total cost of $O(m t n^3)$
versus a cost of $O(n^3 + mpn^2 + mtp^2n)$ for
\clak{}'s best algorithm.
Second, the availability of multiple variants allow the user to choose the
one that fits best his needs. If the solution of a one-dimensional grid ($t=1$)
is sought after, {\sc chol-gwas} or {\sc qr-gwas} should be used;
if the target is the two-dimensional grid, {\sc eig-gwas} is to be 
preferred.

Besides their theoretical interest, these three algorithms are also
of practical relevance. 
State-of-the-art tools offer routines only for the 1D case;
the alternative for the 2D case is to repeatedly use the
algorithms for the 1D scenario $t$ times in a black-box fashion.
In both scenarios, our algorithms improve the state-of-the-art ones:
For the 1D scenario, {\sc qr-gwas} and {\sc chol-gwas} perform half the computation of
the best existing algorithms,
and for the 2D case {\sc eig-gwas} reduces their computational cost by $O(n)$~\cite{OmicABEL,SingleGWAS,MultiGWAS}.

\section{Code generation: Matlab and Fortran}
\label{sec:clak-codegen}

Algorithms have been generated; we now turn the attention
towards their translation into code.
We incorporated into \clak{} two prototypes of code generators,
which produce Matlab and Fortran routines.

As an example, we provide the routines generated
for the {\sc eig-gwas} algorithm (Algorithm~\ref{alg:loop-final}):
Routine~\ref{fig:eig-gwas-matlab} and Routine~\ref{fig:eig-gwas-fortran} 
contain, respectively, the Matlab and Fortran implementations.
The name of the routines, {\tt GWAS\_26\_2}, stands for 
the operation (GWAS), 
the algorithm number (26 out of the 99 generated), and
the permutation number (second generated permutation of the iterators $i$ and $j$, as 
described in Section~\ref{sec:multiple-instances}).
As a direct consequence of our approach,
each of the operations in Algorithm~\ref{alg:loop-final} has
a unique mapping onto available building blocks.
In the case of Matlab, its high level notation for matrix operations 
such as product, addition, inverse and transposition
---which are internally mapped onto the corresponding kernels provided by high-performance libraries---,
relieves the user (or the code generator) from tedious and error-prone low-level details
and enables a one to one translation of the algorithm statements into code.

In contrast, the generation of Fortran code is much more complicated. 
The mapping onto library routines is now explicit, and 
the user is exposed to the libraries' internals.
In the case of BLAS and LAPACK, special attention must be paid to both
the overwriting of input operands and
the implicit storage of the operands with special structure.
The former enforces further data dependency analysis and expression rewriting, while 
the latter requires carefully tracking the storage representation of the operands of each kernel.
We illustrate these issues via the QR factorization in line 10
of Algorithm~\ref{alg:loop-final}, and the subsequent use of the 
$Q$ and $R$ operands in lines 12 and 13.
The factorization is performed by the call to LAPACK's {\tt dgeqrf} routine 
in line 52 of the Fortran code. 
The routine factors the operand {\tt temp9};
however, instead of creating two new operands for the corresponding $Q$ and $R$ matrices, 
the contents of the computed $R$ are stored in the elements on and above the diagonal of {\tt temp9},
and $Q$ is implicitly stored in the combination of the elements below the diagonal
and the extra buffer {\tt tau10}. 
As a consequence, the compiler has to ensure
1) that if {\tt temp9} is needed in subsequent calls, a copy is kept, and 
2) that subsequent uses of $Q$ and $R$ refer now to parts of {\tt temp9}.
Further, the compiler must detect that $Q$ is not stored explicitly 
as an orthogonal matrix but implicitly as the product of a number of Householder reflectors~\cite{Golub:1996:MC:248979},
and consequently map the statement representing the matrix product $Q^T_{ij} b_{ij}$
to the special routine {\tt dormqr}, instead of the more general {\tt dgemv}.

This operation clearly exemplifies the challenge behind a Fortran (and similarly a C)
code generator; and also how \clak{} alleviates the developer's burden not
only in generating specialized algorithms but also in the tedious and error
prone translation into code.

\renewcommand{\lstlistingname}{Routine}
\newcounter{listingtmp}
\noindent
\begin{minipage}{\linewidth}
	\begin{lstlisting}[caption={Matlab code for {\sc eig-gwas} (Algorithm~\ref{alg:loop-final}) generated by \clak{}.},
				       label=fig:eig-gwas-matlab,
					   numbers=left,
				       basicstyle={\tt},
				       fancyvrb=true,language=Fortran,columns=fixed,basewidth=.5em,frame=b,
framexleftmargin=-0pt,
framexrightmargin=0pt,
xleftmargin=10pt
				   ]
function [b] = GWAS_26_2(X, y, h, Phi, sm, sn, nXs, nys)
   b = zeros(sm, nXs * nys);
   T3 = zeros(sm, sn * nXs);
   [Z1, W1] = eig( Phi );

   for i = 1:nXs
      T3(:, sn*(i-1)+1:sn*i) = X(:, sm*(i-1)+1:sm*i)' * Z1;
   end
   for j = 1:nys
      T1 = 1 * eye(sn) + - h(j) * eye(sn);
      T2 =  T1 + h(j) * W1;
      T6 =  Z1' * y(:, j);
      for i = 1:nXs
         T4 =  T3(:, sn*(i-1)+1 : sn*i) / T2;
         T5 =  T4 * T3(:, sn*(i-1)+1 : sn*i)';
         [Q1, R1] = qr( T5, 0 );
         T7 =  T4 * T6;
         T8 =  Q1' * T7;
         T9 = R1 \ T8;
         b(:, i + (j-1)*nXs) = T9;
      end
   end
end
	\end{lstlisting}
\end{minipage}

\noindent
\begin{minipage}{\linewidth}
	\setcounter{listingtmp}{\value{lstlisting}}
	\begin{lstlisting}[caption={Fortran code for {\sc eig-gwas} (Algorithm~\ref{alg:loop-final}) 
	                           generated by \clak{} (I).}, 
				       label=fig:eig-gwas-fortran,
					   numbers=left,
				       basicstyle={\tt},
				       fancyvrb=true,language=Fortran,columns=fixed,basewidth=.5em,frame=b,
framexleftmargin=-0pt,
framexrightmargin=0pt,
xleftmargin=10pt,
name=EIG
				   ]
SUBROUTINE GWAS_26_2( X, csX, dsX, y, csy, h, Phi, csPhi, 
                      b, csb, sn, sm, nXs, nys )
   INTEGER sn, sm, nXs, nys, csX, dsX, csy, csPhi, csb
   DOUBLE PRECISION X(csX, dsX, nXs), y(csy, nys), h(nys), 
                    Phi(csPhi, sn), b(csb, nXs, nys)

   DOUBLE PRECISION ZERO
   PARAMETER (ZERO=0.0D+0)
   DOUBLE PRECISION ONE
   PARAMETER (ONE=1.0D+0)
    
   DOUBLE PRECISION tmp50(sn), tmp182(sm, sm), temp13(sm), 
                    temp8(sm, sn), tmp28(sn), 
                    tmp70(sm, sn, nXs), tmp384(sm), [...]
   [...]
   EXTERNAL dtrsv, dgemm, dgeqrf, dgemv, dsyevr, dscal, 
            dormqr, dcopy
   
   call dsyevr( 'V', 'A', 'L', sn, Phi( 1, 1 ), sn, ddummy, 
                ddummy, idummy, idummy, ddummy, nCompPairs5, 
                W8( 1 ), Z7( 1, 1 ), sn, isuppz4, ... )
   DO i = 1, nXs
      call dgemm( 'T', 'N', sm, sn, sn, ONE, X( 1, 1, i ), 
                  sn, Z7( 1, 1 ), sn, ZERO, 
                  tmp70( 1, 1, i ), sm ) 
   END DO
   DO j = 1, nys
      DO iter1 = 1, sn
         tmp28( iter1 ) = 1 + ( - h(j))
      END DO
      DO iter1 = 1, sn
         tmp50( iter1 ) = tmp28(iter1) + h(j) * W8(iter1)
      END DO
      call dgemv( 'T', sn, sn, ONE, Z7( 1, 1 ), sn, 
                  y( 1, j ), 1, ZERO, tmp341( 1 ), 1 )
	\end{lstlisting}
\end{minipage}

\noindent
\begin{minipage}{\linewidth}
	\setcounter{lstlisting}{\value{listingtmp}}
	\renewcommand{\thelstlisting}{\thechapter.\arabic{lstlisting}b}
	\begin{lstlisting}[caption={Fortran code for {\sc eig-gwas} (Algorithm~\ref{alg:loop-final}) 
	                           generated by \clak{} (II).}, 
				       label=fig:eig-gwas-fortran-cont,
					   escapechar=!,
					   numbers=left,
				       basicstyle={\tt},
				       fancyvrb=true,language=Fortran,columns=fixed,basewidth=.5em,frame=b,
framexleftmargin=-0pt,
xrightmargin=0em,
xleftmargin=10pt,
name=EIG
				   ]
      DO i = 1, nXs
         DO iter1 = 1, sn
            call dcopy( sm, tmp70( 1, iter1, i ), 1,
                        temp8( 1, iter1 ), 1 )
         END DO
         DO iter1 = 1, sn
            call dscal( sm, 1/tmp50(iter1), 
                        temp8( 1, iter1 ), 1 )
         END DO
         call dgemm( 'N', 'T', sm, sm, sn, ONE, 
                     temp8( 1, 1 ), sm, tmp70( 1, 1, i ), 
                     sm, ZERO, tmp182( 1, 1 ), sm )
         DO iter1 = 1, sm
           call dcopy( sm, tmp182( 1, iter1 ), 1, 
                       temp9( 1, iter1 ), 1 )
         END DO
         call dgeqrf( sm, sm, temp9( 1, 1 ), sm, !\label{line:qr}!
                      tau10( 1 ), work12, sm*192, info )
         call dgemv( 'N', sm, sn, ONE, temp8( 1, 1 ), sm, 
                     tmp341( 1 ), 1, ZERO, tmp384( 1 ), 1 )
         call dcopy( sm, tmp384( 1 ), 1, temp13( 1 ), 1 )
         call dormqr( 'L', 'T', sm, 1, sm, temp9( 1, 1 ), 
                      sm, tau10( 1 ), temp13( 1 ), sm, 
                      work15( 1 ), sm*192, info )
         call dcopy( sm, temp13( 1 ), 1, temp16( 1 ), 1 )
         call dtrsv( 'U', 'N', 'N', sm, temp9( 1, 1 ), sm, 
                     temp16( 1 ), 1 )
         DO iter1 = 1, sm
            b( iter1, i, j ) = temp16( iter1 )
         END DO
      END DO
   END DO
   RETURN
END
	\end{lstlisting}
\end{minipage}

\newpage

\section{Scope and limitations}
\label{sec:clak-scope}

As input, \clak{} accepts target operations of the form
\begin{verbatim}
                   op = <Expression>,
\end{verbatim}
where {\tt op} is a single output operand, and
{\tt <Expression>} is a combination of input 
operands and the operators 
$+$, 
$*$, 
$-$, 
$T$, 
$-1$. 
In addition, as we illustrate in the next chapter, 
operations easily translated into such a form,
e.g., linear systems like $A X = B$, where $X$ is the unknown,
are also allowed (with a slight modification of the grammar
presented in Section~\ref{sec:clak-input}).
For a given operation in this class, 
\clak{} returns a family of algorithmic variants, 
documented with their computational cost, and
translated into the corresponding Matlab and Fortran routines.

While powerful, \clak{} presents a
number of limitations that should be addressed in the future:

\paragraph{Accepted knowledge and equations.}
We successfully showcased how to handle and exploit 
different pieces of domain knowledge such
as operands' properties and the relation among instances of problems.
Yet, the power of the compiler would be increased with the
support for an extended set of properties, e.g., 
banded matrices.
Also, while broad, the range of supported equations is still limited.
We plan an extension to deal with more complex operations, ranging from
explicit equations (as opposed to only assignments) 
to determinants, logarithms, and matrix functions in general.

\paragraph{Choosing an algorithm.}
Each of the generated algorithms is currently documented with a 
rather simple performance analysis based on the operation count (flops).
This is, in general, not a reliable metric, and we aim at 
incorporating more advanced techniques for performance prediction
which account for the underlying architecture and libraries.
A promising research project relies on a sample-based approach:
The idea is to create performance models not for the competing algorithms, 
but only for those routines that are used as building blocks. 
By combining the models, it is then possible to make  
accurate performance predictions~\cite{Peise2012:50}.

\paragraph{Code generation.}
\clak{}'s code generator currently produces sequential code,
which may take advantage of multi-threaded implementations
of the BLAS library to exploit shared-memory parallelism. 
Nevertheless, the variety of available computing platforms 
(e.g, multi- and many-core processors,
clusters, and co-processors such as GPGPUs)
demands the generation of algorithms that are tailored not
only to the application but also to the architecture.
To this end, we envision the development of a number of modules
responsible for the tailoring to each specific architecture and type of parallelism.

\paragraph{Stability analysis.}
The main goal of \clak{} (beyond increasing development productivity)
is the generation of efficient algorithms.
However, while mathematically correct, the produced algorithms may be
numerically unstable when executed in finite precision arithmetic.
For instance, for the ordinary least-squares (OLS) problem
$x := (A^T A)^{-1} A b$, \clak{} generates, among other stable ones,
Algorithm~\ref{alg:ols-chol};
depending on the condition number of the matrix $A$,
the algorithm may yield highly inaccurate results~\cite{Golub:1996:MC:248979}.

\noindent
\begin{center}
\renewcommand{\lstlistingname}{Algorithm}
\begin{minipage}{0.87\linewidth}
\begin{lstlisting}[caption={Potentially unstable algorithm to solve the OLS problem.}, escapechar=!,label=alg:ols-chol]
		$S := A^T A$              (!\sc syrk!)
		$L L^T = S$              (!\sc potrf!)
		$x := A^T b$              (!\sc gemv!)
		$b := L^{-T} x$              (!\sc trsv!)
		$b := L^{-1} x$              (!\sc trsv!)
\end{lstlisting}
\end{minipage}
\end{center}

Unfortunately, a completely automatic analysis is extremely challenging,
and, to the best of our knowledge, no methodology exists that addresses this issue.
Thus, currently, the generated algorithms need to be manually validated or tested for
numerical robustness.

\section{Summary}

We introduced \clak{}, a linear algebra compiler
for the generation of application-tailored algorithms and routines.
\clak{} takes as input the mathematical description of a target
operation together with domain-specific knowledge;
in a process that closely replicates the reasoning of a human expert,
the target equation is mapped onto a sequence of calls to high-performance
library-supported kernels.
Along the process, the compiler applies a number of optimizations and
exploits the available knowledge to produce specialized algorithms.
The contents of this chapter extend our work published 
in~\cite{CLAK-VECPAR12,CLAK-IJHPCA}.

This chapter makes the following contributions:

\bi
    \item A model of expert reasoning.
		We observed the reasoning carried out
		by linear algebra experts in the derivation of algorithms, and 
        set up a heuristic-based model of this reasoning, making it systematic.
		A number of optimizations, from simple to advanced ones 
		but all overlooked by sophisticated environments like Matlab, 
		are also discussed and incorporated into our compiler.

    \item Advanced management of domain knowledge.
		We exposed the importance of exploiting domain-specific knowledge, and 
		introduced an engine that enables the inference of operands' properties 
		to dynamically expand the available knowledge.
		This allows for a more precise tailoring of the algorithms and routines.
		Such an inference engine is not provided~\footnote{At the time of this dissertation writing.}
		by any (symbolic) linear algebra package.

    \item High-performance algorithms for GWAS.
		Part of the examples in this chapter are taken from an operation arising in the context
		of genome-wide association studies, a popular tool in computational biology.
		The generated algorithms led to high-performance out-of-core routines that largely outperform
		state-of-the-art libraries~\cite{SingleGWAS,MultiGWAS}. These routines have been
		incorporated into a widely-used R package for statistical genomics, GenABEL~\cite{genabel},
		as part of the OmicABEL library~\cite{OmicABEL}.~\footnote{Available at \texttt{http://www.genabel.org/packages/OmicABEL}}
\ei

\chapter{\clak{}: High-Performance BLAS and LAPACK Derivatives}
\chaptermark{HP BLAS and LAPACK Derivatives}
\label{ch:compiler-ad}

\sloppypar
In Chapter~\ref{ch:compiler}, we detailed the mechanisms behind \clak{},
and demonstrated its potential by applying it to the challenging
GWAS problem.
The purpose of this chapter is to provide evidence of 
the broad applicability and extensibility of our approach.
Moreover, we show how our automated system comes in handy
where a manual approach is not viable.
We concentrate on the field of Algorithmic Differentiation (AD)
---often referred to as Automatic Differentiation---,
and illustrate the application of \clak{} to generate efficient algorithms
and code for computing the derivative of BLAS and LAPACK operations.

Derivatives are needed in a wide range of fields: 
from cost-optimization in finance, through sensitivity analysis of 
simulation models, to parameter estimation and design optimization.
A popular technique for the computation of derivatives 
is that of AD~\cite{0021725,Naumann2012TAo}.
AD tools take a function given as a computer program, and
change the semantics of the program to compute both the function and its derivative
with respect to a set of selected input parameters.
The conceptual idea behind AD is to 
first decompose the input program into elementary operations 
(addition, multiplication, division,~...) and functions (sin, cos, log, exp,~...),
and then differentiate them. 
The derivatives are accumulated according to the chain rule,
resulting in the computation of the derivative of the overall program.

Depending on how the chain rule is applied, AD distinguishes between
two basic modes: {forward mode} and {reverse mode}. 
We focus on the forward mode, which follows the control flow of the
program, accumulating the derivative of intermediate variables with respect
to the input variables.
For each of these modes, there exist two approaches to semantic transformation:
Source transformation and operator overloading (see~\cite{Bischof2000CDo} for a short review).
In this chapter, we consider the source transformation approach,
which consists in rewriting the input program $f$ to 
produce an {\em extended program} that computes both $f$ and its derivative $f'$.
For instance, the statement $a = b * c$ is rewritten into two
statements: $a = b * c$ and $a' = b' * c + b * c'$.
This scheme is semantic-oblivious, e.g., whenever a routine call is encountered, 
no knowledge of the operation is used; the scheme is blindly applied recursively
to the routine's code.
Although this is an effective and scalable approach,
when the input program relies on highly-optimized BLAS and LAPACK kernels,
the extended program suffers from a significant loss in performance.

Due to the widespread usage of BLAS and LAPACK in scientific software,
the availability of an optimized differentiated version of these libraries 
is relevant to the AD community.
In this chapter, we extend \clak{} to enable the generation of algorithms
and routines for the derivative of BLAS and LAPACK operations.

\paragraph{Preliminaries.}
In the remainder of this chapter, we consider exclusively 
AD based on source transformation and the forward mode. 
We also restrict the discussion to AD tools for imperative programming 
languages with statements that return scalars;\footnote{As opposed to vector-valued statements
in languages like Matlab.}
this is an important class of languages in scientific computing, which includes C and Fortran 77.
For any comparison with \clak{}, we will use ADIFOR~\cite{Bischof1996AAD} 
as representative of such tools. \\[3mm]
Below, we briefly introduce the key concepts for the discussion in this chapter:\\[1mm]
{\em 1. Dependent, independent, and active variables.}\\
In AD, an input variable is called {\em independent}
if derivatives with respect to that variable are desired. 
An output variable is called {\em dependent} if its derivatives 
with respect to the independent variables are to be computed. 
{\em Active} variables depend on one or more of the independent variables, and 
also contribute to the computation of one or more dependent variables.

Let us consider, as an example, the following pseudocode

\floatname{algorithm}{Pseudocode}
\begin{algorithm}
	{\bf Input:} $\nu, y$ \\
	{\bf Output:} $w, z$
    
    \vspace{1mm}

  \algsetup{indent=2em}
  \begin{algorithmic}[1]
	\STATE $\alpha := f(\nu)$\\
	\STATE $x := g(\nu)$\\
	\STATE $w := h(x)$\\
	\STATE $z := \alpha x + y$\\
  \end{algorithmic}
  \caption{{\bf:} Dependent, independent, and active variables.}
  \label{ps:dep}
\end{algorithm}

\noindent
and let us assume that the derivative of $z$ with respect to $\nu$ is desired.
In this case, $z$ and $\nu$ are the dependent and independent variables, respectively.
Also, since both $\alpha$ and $x$ depend on $\nu$ (lines 1 and 2) and
contribute to the computation of $z$ (line 4), they are active.
Since $y$ does not depend on $\nu$ and $w$ does not contribute to the 
computation of $z$, they are inactive.\\

\noindent
{\em 2. Activity pattern.}\\
The {\em activity pattern} indicates which operands in a statement 
are active and which ones are not, making possible computation savings~\cite{%
Hascoet:2005:RAR:1149000.1708228}.
Based on an activity analysis, 
statements that do not contribute to the dependent variable are not
differentiated, e.g., line 3 in Pseudocode~\ref{ps:dep} is not differentiated.
The only inactive variables left are those that do not depend on the 
independent variable, e.g., $y$, whose derivative is thus $0$.
Accordingly, in the derivative of the statement in line 4
$$\dv{z}{\nu} := \dv{\alpha}{\nu} * x + \alpha * \dv{x}{\nu} + \dv{y}{\nu},$$
the term $\dv{y}{\nu}$ equals $0$, and the expression simplifies to
$$\dv{z}{\nu} := \dv{\alpha}{\nu} * x + \alpha * \dv{x}{\nu}.$$
Assuming $z$, $x$, and $y$ are vectors of size $n$, and $\alpha$ 
is a scalar, exploiting the activity pattern saves $n$ out of $4n$ flops.
The resulting extended program is given in Pseudocode~\ref{ps:gdep};
functions $g\_f$ and $g\_g$ calculate the derivative of $f$ and $g$, respectively.

\floatname{algorithm}{Pseudocode}
\begin{algorithm}
	{\bf Input:} $\nu, \nu', y$ \\
	{\bf Output:} $w, z, z'$
    
    \vspace{1mm}

  \algsetup{indent=2em}
  \begin{algorithmic}[1]
	\STATE $\alpha  := f(\nu)$\\
	\STATE $\alpha' := g\_f(\nu, \nu')$\\
	\STATE $x  := g(\nu)$\\
	\STATE $x' := g\_g(\nu, \nu')$\\
	\STATE $w  := h(x)$\\
	\STATE $z  := \alpha x + y$\\
	\STATE $z' := \alpha' x + \alpha x'$\\
  \end{algorithmic}
  \caption{{\bf:} Differentiated Pseudocode~\ref{ps:dep} for the computation of $\dv{z}{\nu} v'$.}
  \label{ps:gdep}
\end{algorithm}

\noindent
{\em 3. Multiple derivatives.}\\
The independent variable $\nu$ is not necessarily a scalar,
it may also be a vector-valued variable. 
When $\nu \in R^p$, derivatives with respect to each $\nu_i$ are desired,
and each derivative formula is to be computed $p$ times:
$$\dv{z}{\nu_i} := \dv{\alpha}{\nu_i} * x + \alpha * \dv{x}{\nu_i}, \quad \quad 1 \le i \le p.$$
Conceptually, the derivative operands gain an extra dimension:
scalars become vectors, vectors become matrices, and matrices become
three-dimensional arrays.

\paragraph{Disadvantages of AD's approach.}
AD tools based on source transformation present two significant 
disadvantages when differentiating BLAS (and LAPACK) operations.
The main drawback lies in the low performance of the generated code.
The difficulty resides in the automatic generation of 
high-performance code for BLAS-like operations. 
Manually optimized BLAS implementations, such as 
the Intel's Math Kernel Library (MKL)~\cite{MKL}, attain
from 10\% (BLAS 1) to more than 90\% (BLAS 3) of the 
architectures' peak performance.
Even the most prominent projects focused on automatically producing optimized
BLAS libraries, such as ATLAS~\cite{atlas-sc98,Whaley:PhD}, require hand-tuned 
microkernels to achieve a relatively high percent of such performance.
AD-generated code, instead, achieves only about 5\% of the peak 
(see performance results in Section~\ref{sec:ad-clak-experiments});
for BLAS 3 operations, this means code that is almost 20 times slower.

The second disadvantage is related to the treatment of different
activity patterns.
It may occur that a same routine, e.g., vec\_mul in Pseudocode~\ref{rou:vec-mul}, 
is encountered multiple times in the original program, 
each time with a different activity pattern. 
\floatname{algorithm}{Pseudocode}
\begin{algorithm}
  \vspace{1mm}
  \algsetup{indent=2em}
  \begin{algorithmic}[1]
	\STATE {\bf function} v := vec\_mul(x, y, z)
	\FOR{i := 1 \TO length(x)}
	\STATE v(i) := x(i) * y(i) * z(i)
	\ENDFOR
  \end{algorithmic}
  \caption{{\bf:} vec\_mul. Element-wise vector product.}
  \label{rou:vec-mul}
\end{algorithm}

\noindent
Suppose that in every occurrence $x$ is active and $y$ is inactive;
then, there are two possible cases: $z$ is active and $z$ is inactive.
It is not uncommon that in such a situation source transformation tools 
generate one single differentiated version of the routine that covers 
for both scenarios.
In the example above, such a tool generates g\_vec\_mul (Pseudocode~\ref{rou:gvec-mul});
when the routine is called in the extended program with $z$ inactive, 
a zero vector is passed for $z$'s derivative g\_z. 
Notice the unnecessary computation in line 4.

\floatname{algorithm}{Pseudocode}
\begin{algorithm}
  \vspace{1mm}
  \algsetup{indent=2em}
  \begin{algorithmic}[1]
 	\STATE {\bf function} (v, g\_v) := g\_vec\_mul(x, g\_x, y, z, g\_z)
	\FOR{i := 1 \TO length(x)}
	\STATE v(i) :=  x(i) * y(i) * z(i)
	\STATE g\_v(i) :=  g\_x(i) * y(i) * z(i) + x(i) * y(i) * g\_z(i)
	\ENDFOR
  \end{algorithmic}
  \caption{{\bf:} Derivative of vec\_mul. Argument $y$ is inactive.}
  \label{rou:gvec-mul}
\end{algorithm}

\noindent
While the approach may beneficial in that it simplifies the
code generation process and reduces the, possibly very large, size of the extended program,
it may also result in considerable redundant computation.

\section{The challenge}
\label{sec:ad-challenge}

The need for an efficient differentiated version of BLAS was brought up
by the AD community already in 2000 at the {\em Automatic Differentiation Conference}
(AD 2000). 
Given the simplicity, from a mathematical perspective, of BLAS operations, 
analytic formulas for their derivatives can be derived easily. 
For instance,
the formula for the derivative of {\sc axpy} 
($y := \alpha x + y$, $\alpha \in R$, $x,y \in R^n$)
with respect to an
independent variable $\nu$ is
\begin{equation}
    \dv{y}{\nu} := \frac{d\alpha}{d\nu} x + \alpha \frac{dx}{d\nu} + \frac{dy}{d\nu}.
    \label{eq:daxpy}
\end{equation}

M.~H.~Bucker and P.~Hovland approached the aforementioned efficiency problem
by deriving and coding such derivatives manually.
Although successful in providing efficient routines
for a small subset of BLAS 1 and 2 operations,
the authors realized that in practice, due to the vast number of operations and variants to support,
the problem becomes unmanageable.
On the one hand, BLAS already contains a fairly large number of operations (about 40); 
each of them accepting several options to indicate multiple flavors of the operation,
and supporting multiple datatypes.
For instance, the routine for a general matrix-matrix product ({\sc gemm}:
$C := \alpha A^{(T)} B^{(T)} + \beta C$), accepts two options to indicate
whether $A$ and $B$ are to be transposed or not (for a total of four combinations),
and provides support for four datatypes (single and double precision, real and complex data).
On the other hand, the activity pattern adds another multiplicative factor
to the number of flavors: 
Each of the $n$ input operands may be active or inactive,
resulting in $2^n - 1$ different formulas.
For instance, Equation~\eqref{eq:daxpy} yields seven different equations:
\begin{center}
\begin{minipage}{\textwidth}
	\centering
	\begin{minipage}[t]{.4\textwidth}
		\begin{enumerate}
			\item $\dv{y}{\nu} := \dv{\alpha}{\nu} x + \alpha \dv{x}{\nu} + \dv{y}{\nu}$
			\item $\dv{y}{\nu} := \dv{\alpha}{\nu} x + \alpha \dv{x}{\nu} \phantom{+ \dv{y}{\nu}}$
			\item $\dv{y}{\nu} := \dv{\alpha}{\nu} x \phantom{+ \alpha \dv{x}{\nu}} + \dv{y}{\nu}$
			\item $\dv{y}{\nu} := \dv{\alpha}{\nu} x \phantom{+ \alpha \dv{x}{\nu} + \dv{y}{\nu}}$
		\end{enumerate}
	\end{minipage}
	\hspace{5mm}
	\begin{minipage}[t]{.4\textwidth}
		\begin{enumerate}
				\setcounter{enumi}{4}
			\item $\dv{y}{\nu} := \phantom{\dv{\alpha}{\nu} x +} \alpha \dv{x}{\nu} + \dv{y}{\nu}$
			\item $\dv{y}{\nu} := \phantom{\dv{\alpha}{\nu} x +} \alpha \dv{x}{\nu} \phantom{+ \dv{y}{\nu}}$
			\item $\dv{y}{\nu} := \phantom{\dv{\alpha}{\nu} x + \alpha \dv{x}{\nu} +} \dv{y}{\nu}$
		\end{enumerate}
	\end{minipage}
\end{minipage}
\end{center}

To give a taste of the magnitude of the challenge, in Table~\ref{tab:BLAS-variants}
we provide the number of variants required for a subset of BLAS 3 operations.
Let $m$ be the number of possible options (which typically take one of two values)
and $n$ the number of input operands (which may be active or inactive),
a differentiated BLAS must support, for each of the operations,
$2^m \times (2^n -1)$ variants;
further, the support of four different datatypes multiplies the above
quantity by four.
For the operations in the table, this totals 2624 variants.
The differentiation of the entire library would require more than 8000 variants.

\begin{table}
    \centering
	\footnotesize
	\begin{tabular}{ l@{\hspace{1.5mm}}l c  c  c } \toprule
		\multicolumn{2}{l}{\bf Operation} & {\bf \# Options} & {\bf \# Operands} & {\bf \# Variants}\\\toprule
		{\sc gemm}  & ($C := \alpha A B + \beta C$)                  & 2 & 5 & $2^2 \times (2^5 -1) \times 4 = 496$ \\
		{\sc symm}  & ($C := \alpha A B + \beta C$)                  & 2 & 5 & $496$ \\
		{\sc syrk}  & ($C := \alpha A A^T + \beta C$)                & 2 & 4 & $240$ \\
		{\sc syr2k} & ($C := \alpha A B^T + \alpha B A^T + \beta C$) & 2 & 5 & $496$ \\
		{\sc trmm}  & ($C := \alpha A B$)                            & 4 & 3 & $448$ \\
		{\sc trsm}  & ($C := \alpha A^{-1} B$)                       & 4 & 3 & $448$ \\\midrule
        \multicolumn{4}{r}{\bf Total:} & $2624$ \\\bottomrule
    \end{tabular}
	\caption{Number of variants required to fully support a differentiated version of a subset of BLAS 3 operations.}
    \label{tab:BLAS-variants}
\end{table}

Clearly, the manual development and maintenance of such a library is unfeasible.
The most common approach these days is to either sacrifice efficiency or 
to manually code only the specific variants needed in each case and plug them in the extended program.
Still, the need remains: The issue was brought up again by P.~Hovland in
the {\em Seventh European Workshop on Algorithmic Differentiation}
in 2008, where
he insisted on the benefits that efficient differentiated versions of BLAS and LAPACK would bring
to the community.

\section{\clak{} for high-performance derivatives}
\label{sec:ad-ext}

We propose a third alternative that has the potential to automatically
generate derivative code for BLAS and part of LAPACK 
while attaining high performance.
Similarly to Hovland and B\"ucker, the idea is
to raise the abstraction level from scalar operations
to matrix equations (the analytic formulas)
and exploit \clak{}'s capabilities to find efficient mappings 
onto BLAS and LAPACK kernels.

To enable \clak{} to find algorithms for derivative operations,
the engine presented in the previous chapter is augmented
with a number of features.
We discuss the inclusion of support for the derivative operator and 
the extension of the {\em Inference of properties} module.

\subsection{The derivative operator}

The support for the derivative operator implied two modifications
to \clak{}'s engine.
First, we modified the input language so to accept a new unary operator: 
\dvop{$\cdot$}.
The new grammar allows, on the one hand, the declaration of derivative operands, e.g.,
\begin{center}
	{\tt Vector dv(x)<Input>},
\end{center}
and, on the other hand, the specification of derivative equations 
by means of the following extra production rule:
\begin{center}
	{\tt<}Factor{\tt>} $\rightarrow$ {\tt"dv("} {\tt<}Expression{\tt>} {\tt")"}.
\end{center}
An example of valid expression is \dvop{A * B + C}.

Second, we incorporated into \clak{} the rewrite rules necessary to
encode the so-called chain rule, so that the system can differentiate a given expression.
These rules are displayed in Box~\ref{box:dv-rules}. A collection of results for
the derivative of the matrix operations in Box~\ref{box:dv-rules} and more complex 
ones can be found in~\cite{Giles2008CMD}.
\begin{mybox}
$$
	\begin{aligned}
		1. \;\; & {\tt dv}(A+B)    & \longrightarrow & \quad {\tt dv}(A) + {\tt dv}(B) \\
		2. \;\; & {\tt dv}(A-B)    & \longrightarrow & \quad {\tt dv}(A) - {\tt dv}(B) \\
		3. \;\; & {\tt dv}(A \times B)    & \longrightarrow & \quad {\tt dv}(A) \times B + A \times {\tt dv}(B) \\
		4. \;\; & {\tt dv}(A^T)    & \longrightarrow & \quad {\tt dv}(A)^T \\
		5. \;\; & {\tt dv}(A^{-1}) & \longrightarrow & \quad - A^{-1} \times {\tt dv}(A) \times A^{-1}  \\
	\end{aligned}
$$
\caption{Rewrite rules encoding the chain rule.}
\label{box:dv-rules}
\end{mybox}

After these modifications have been incorporated into the 
engine, the derivative of {\sc axpy} (Box~\ref{box:f}) may
be expressed as illustrated by Box~\ref{box:gf}.
We recall that the macro {\tt init} is used
when an operand is overwritten to refer to
its initial contents.
\begin{mybox}[!h]
\vspace{3mm}
\begin{verbatim}
                Equation Axpy
                    Scalar alpha<Input>;
                    Vector x<Input>;
                    Vector y<InOut>;
                  
                    y = alpha * x + init(y);
\end{verbatim}
\caption{Description of {\sc axpy} in \clak{}.}
\label{box:f}
\end{mybox}

\begin{mybox}[!h]
\vspace{3mm}
\begin{verbatim}
    Equation gAxpy
        Scalar    alpha <Input>;
        Scalar dv(alpha)<Input>;
        Vector    x <Input>;
        Vector dv(x)<Input>;
        Vector dv(y)<InOut>;
    
        dv(y) = dv(alpha) * x + alpha * dv(x) + init(dv(y));
\end{verbatim}
\caption{Description of the derivative of {\sc axpy} in \clak{}.
All of $\alpha$, $x$, and $y$ are active.}
\label{box:gf}
\end{mybox}

\subsection{The AD mode: Inference of properties and activity patterns}
\label{par:dv-inference}

The derivative of {\sc axpy} in Box~\ref{box:gf} corresponds to the case where
all input operands are active. However, manually writing such descriptions
for all possible patterns is time consuming and error-prone; for instance, {\sc gemm} presents
31 different patterns (see Table~\ref{tab:BLAS-variants}).
Furthermore, it is the user who has to deduce the properties of the derivative
operands. To simplify this task, we include an ``AD mode'' in \clak{}
to carry out this process automatically.

In the AD mode, \clak{} takes the description of the original
equation (Box~\ref{box:f}) as input and,
by means of the chain rule (Box~\ref{box:dv-rules}), it generates
the derivative expression
\begin{center}
	{\tt dv(y) = dv(alpha) * x + alpha * dv(x) + init(dv(y))}.
\end{center}
Then, it deduces a number of properties for the operands, and generates
the input corresponding to each activity pattern.

\paragraph{Inference.}
We extended the Inference of properties module, described in Section~\ref{sec:inf-props},
to infer properties for the operands of derivative formulas.
First, the module deduces the type of operand for the derivative operands.
As we mentioned earlier, when differentiating with respect to a vector-valued variable ($\nu \in R^p$),
the derivative operands gain an extra dimension. 
Alternatively, following \clak{}'s design, we regard the problem as multiple instances
of the scalar-valued case, where the derivative operands vary along the $p$ dimension.
Thus, {\tt dv(x)} and {\tt dv(y)} are assigned the type {\tt Vector}, {\tt dv(alpha)}
is a {\tt Scalar}, and the three of them are attached a subindex:
\begin{verbatim}
	dv(y{i}) = dv(alpha{i}) * x + alpha * dv(x{i}) + init(dv(y{i})).
\end{verbatim}

Next, based on the description of the original function $f$, the module
determines which operands are input and which ones 
are output to the derivative $f'$.
In the forward mode, the following rules apply:
\begin{enumerate}
	\item The (nonlinear) input operands to $f$ are also input operands to $f'$.
	\item The output of $f$, if required, becomes an input to $f'$.
	\item The derivative of active inputs to $f$ are also inputs to $f'$.
	\item The derivative of the output operand is sought after, and therefore it is the output in $f'$.
\end{enumerate}
We illustrate the application of these rules by means of {\sc axpy}:
The input operands in {\sc axpy} ---{\tt alpha}, {\tt x} and {\tt init(y)}---
and their derivative counterparts ---{\tt dv(alpha)}, {\tt dv(x)} and {\tt dv(init(y))}---, are input to its derivative;
the final contents of {\tt y}, if they appeared in the derivative formula,
would also be an input;
the final contents of {\tt dv(y)} are sought after.

Finally, further knowledge of the properties of the derivative operands 
may also be inferred from the properties of the original operands.
As an example, the derivative
of matrix operands presenting some type of zero-pattern,
maintain such pattern; for instance, the derivative of
diagonal and triangular matrices are diagonal and
triangular, respectively.
Similarly, the derivative of constant scalars and
matrices, e.g, the identity matrix, is zero
(either the scalar 0 or the zero matrix).
Other properties instead are inferred
only partially. This is the case of the positive
definiteness of an SPD operand {\tt A}:
While the derivative inherits the symmetry 
(if the (i,j) and (j,i) entries of {\tt A} are equal,
so are in the derivative {\tt dv(A)}),
the positive-definiteness of {\tt dv(A)} is not guaranteed.

\paragraph{Activity patterns.}
Once properties are deduced, the input corresponding to each
possible activity pattern is generated. 
This is accomplished by first replacing
the derivative of the inactive variables with 0, and then simplifying
the resulting expression.
Box~\ref{box:one-gaxpy} reproduces the
input generated for the case where $x$ and $y$ are active,
and $\alpha$ is inactive.

\begin{mybox}[!h]
\vspace{3mm}
\begin{verbatim}
    Equation gAxpy
        Scalar alpha<Input>;
        Vector dv(x)<Input>;
        Vector dv(y)<InOut>;
    
        dv(y{i}) = alpha * dv(x{i}) + init(dv(y{i}));
\end{verbatim}
\caption{Description of the derivative of {\sc axpy} in \clak{}.
$x$ and $y$ are active, $\alpha$ is inactive.}
\label{box:one-gaxpy}
\end{mybox}

\noindent
For each such description, \clak{} produces algorithms and code
as illustrated in Chapter~\ref{ch:compiler}.

\section{An example: Differentiating $\; A X = B$}
\label{sec:application}

We provide now a brief example of the process
carried out by \clak{} to generate
algorithms for derivative operations.
We use as example the derivative of the linear system $AX=B$,
where the coefficient matrix $A \in R^{n\times n}$ is SPD, and
$X$ and $B \in R^{n\times m}$.
The description of the equation in \clak{}'s language
is provided in Box~\ref{box:SPDSystem};
the extension for the generation of derivative code 
is activated via the ``{\tt --AD-mode}'' argument:
\begin{center}
	\begin{minipage}{.7\textwidth}
	\begin{verbatim}
	      ./clak  --AD-mode  SPDSolve.ck
	\end{verbatim}
	\end{minipage}
\end{center}

\begin{mybox}[!h]
\vspace{3mm}
\begin{verbatim}
                   Equation SPDSolve
                       Matrix A<Input, SPD>;
                       Matrix B<Input>;
                       Matrix X<Output>;
              
                       A * X = B;
\end{verbatim}
\caption{SPDSolve.ck. Description of the SPD system $AX=B$ in \clak{}.}
\label{box:SPDSystem}
\end{mybox}

First, \clak{} takes the description
and, by means of the rules in Box~\ref{box:dv-rules} (the chain rule),
it produces the most general derivative formula
(the one where all input operands are active):
\begin{equation}
	\text{{\tt dv(A) * X + A * dv(X) = dv(B)}}
	\label{eq:gSPDSystem}
\end{equation}

Next, properties for the operands are inferred:
Since $A$, $B$, and $X$ are matrices, so are {\tt dv(A)}, {\tt dv(B)}, and {\tt dv(X)};
each derivative operand is attached the index $i$.
Also, the inference rules for the input and output determine that
$A$, $B$, $X$, {\tt dv(A)}, and {\tt dv(B)}, are input to Equation~\eqref{eq:gSPDSystem},
and {\tt dv(X)} is the output.
In terms of structure, {\tt dv(A)} inherits the symmetry of $A$,
and neither $B$ and $X$, nor {\tt dv(B)} and {\tt dv(X)}, present any structure.

Then, formulas for each of the possible activity patterns are produced:

\begin{center}
\begin{minipage}{.8\textwidth}
	\begin{enumerate}
		\item {\tt dv(A\{i\}) * X + A * dv(X\{i\}) = dv(B\{i\})}
		\item {\tt dv(A\{i\}) * X + A * dv(X\{i\}) = 0}
		\item \hspace{2.93cm} {\tt A * dv(X\{i\}) = dv(B\{i\})}
	\end{enumerate}
\end{minipage}
\end{center}

\noindent
and, for each of them, \clak{} input is generated 
(Boxes~\ref{box:gSPDSystem-a}~to~\ref{box:gSPDSystem-c}).

\begin{mybox}
\vspace{3mm}
\begin{verbatim}
          Equation gSPDSolve
              Matrix    A <Input, SPD>;
              Matrix dv(A)<Input, Symmetric>;
              Matrix dv(B)<Input>;
              Matrix    X <Input>;
              Matrix dv(X)<Output>;
            
              dv(A{i}) * X + A * dv(X{i}) = dv(B{i});
\end{verbatim}
\caption{Description of the derivative of the SPD linear system $AX=B$ in \clak{}.
        $A$ and $B$ are active.}
\label{box:gSPDSystem-a}
\end{mybox}

\begin{mybox}
\vspace{3mm}
\begin{verbatim}
          Equation gSPDSolve
              Matrix    A <Input, SPD>;
              Matrix dv(A)<Input, Symmetric>;
              Matrix    X <Input>;
              Matrix dv(X)<Output>;
            
              dv(A{i}) * X + A * dv(X{i}) = 0;
\end{verbatim}
\caption{Description of the derivative of the SPD linear system $AX=B$ in \clak{}.
        $A$ is active, $B$ is inactive.}
\label{box:gSPDSystem-b}
\end{mybox}

\begin{mybox}
\vspace{3mm}
\begin{verbatim}
          Equation gSPDSolve
              Matrix    A <Input, SPD>;
              Matrix dv(B)<Input>;
              Matrix dv(X)<Output>;
            
              A * dv(X{i}) = dv(B{i});
\end{verbatim}
\caption{Description of the derivative of the SPD linear system $AX=B$ in \clak{}.
        $A$ is inactive, $B$ is active.}
\label{box:gSPDSystem-c}
\end{mybox}

Finally, for each of the many activity patterns, algorithms are generated.
We outline the process for the general derivative (Box~\ref{box:gSPDSystem-a})
$$A'X + A X' = B'.$$
First, a step of algebraic manipulation rewrites the equation so that the input
operands lie on the right-hand side, and the output on the left-hand side:
\begin{equation}
	X' = A^{-1}(B' -A'X).
	\label{eq:io-rearranged}
\end{equation}
Then, \clak{} applies the heuristics discussed in Chapter~\ref{ch:compiler}.
Since the inverse operator is applied to $A$, a full SPD matrix, the matrix
is factored using multiple factorizations (Cholesky, QR, and eigendecomposition);
in the case of a Cholesky factorization, $L L^T = A$, the inference engine
deduces properties for $L$ (lower triangular, square, full rank), and Equation~\eqref{eq:io-rearranged}
is rewritten as 
$$X' = (L L^T)^{-1}(B' -A'X).$$
Now, since $L$ is square, the inverse may be distributed over the product, resulting in
\begin{equation}
	X' = L^{-T} L^{-1}(B' -A'X).
	\label{eq:inv-done}
\end{equation}
In Equation~\ref{eq:inv-done}, the inverse is applied only to triangular operands, and no further processing
of inverses is required; \clak{} proceeds with the mapping of the equation onto
kernels.
The compiler matches multiple kernels: $L^{-1}$, $L^{-1}B'$, $L^{-1}A'$, and $B' - A'X$.
The inversion has the lowest priority and is therefore discarded; since all operands in
the remaining three kernels are matrices, the kernels have the same priority and
the three paths are considered; for the sake of brevity, we only describe the latter ($S := B' - A'X$).
The decomposition of the remaining expression,
$$X' = L^{-T} L^{-1} S,$$
completes with the identification of two {\sc trsm}s:
$T := L^{-1} S$, and $X' := L^{-T} T$.
The resulting algorithm is assembled in Algorithm~\ref{alg:alg-gspd-one}.

The process completes with the tailoring of the algorithm
to the computation of multiple instances $\nu \in R^p$
(the case of scalar-valued $\nu$ is captured by $p=1$).
The analysis of dependencies determines that, since $A$
does not vary along the derivative direction, the Cholesky factorization 
in line 1 may be performed once and reused; the remaining
operations depend on the iterator $i$, thus must be kept within the loop.
The final algorithm for the computation of multiple derivatives
is provided in Algorithm~\ref{alg:alg-gspd-many}.

\begin{center}
\renewcommand{\lstlistingname}{Algorithm}
\begin{minipage}{0.46\linewidth}
\begin{lstlisting}[caption=Single-instance gSPD, escapechar=!, label=alg:alg-gspd-one]
$L L^T = A$              (!{\sc potrf}!)
$S := B' - A' X$              (!{\sc gemm}!)
$T := L^{-1} S$              (!{\sc trsm}!)
$X' := L^{-T} T$              (!{\sc trsm}!)!\vspace{4mm}! 
\end{lstlisting}
\end{minipage}
\hfill
\begin{minipage}{0.48\linewidth}
\begin{lstlisting}[caption=Multiple-instance gSPD, escapechar=!, label=alg:alg-gspd-many]
$L L^T = A$                (!{\sc potrf}!)
for i in 1:p
  $S_i := B_i' - A_i' X$              (!{\sc gemm}!)
  $T_i := L^{-1} S_i$              (!{\sc trsm}!)
  $X_i' := L^{-T} T_i$              (!{\sc trsm}!)
\end{lstlisting}
\end{minipage}
\end{center}

AD tools based on a black box approach, e.g. ADIFOR, generate code that
computes every statement $p$ times regardless of whether the computation 
is redundant.
In contrast, not only does Algorithm~\ref{alg:alg-gspd-many} benefit from
a mapping onto optimized BLAS and LAPACK kernels, it also
reduces the computational cost with respect to that of ADIFOR's equivalent
routine. While the cost of ADIFOR to compute the derivate is
$\frac{2}{3}pn^3 + 4p n^2m$, \clak{} reduces the cost to
$\frac{1}{3} n^3 + 4p n^2m$.

\section{Experimental results}
\label{sec:ad-clak-experiments}

We present now performance results for the derivative of
two example operations: 
the previously discussed solution of a linear system 
with an SPD coefficient matrix (LAPACK's {\sc posv} routine),
and the so-called symmetric rank-update ({\sc syrk}) BLAS operation.
In both cases, we compare the performance of the routines
generated by \clak{} with those generated by the AD tool ADIFOR (version 2.0).

We recall that the code generated by ADIFOR mixes the computation
of both the function and its derivative.
For a fair comparison, the timings corresponding to the routines
generated by \clak{} also include the computation of both operation
and derivative.
Samples of the routines generated by \clak{} are provided in Appendix~\ref{app:code}.

The experiments were performed on an SMP system consisting of two Intel Xeon E5450
multi-core processors. Each processor comprises four cores operating at 3 GHz.
The system is equipped with 16 GB of RAM.
The routines were compiled using the GNU C (version 4.4.5) and Fortran (version 4.4.6) 
compilers, and linked to the Intel MKL library (version 12.1).
The compiler flags ``-O2 -mcmodel=medium'' were used.
Computations were performed in double precision.

\subsection{Example 1: Solution of a SPD linear system}

We commence by presenting experimental results for the 
solution of the SPD system discussed in the previous section:
$A X = B$, where $A \in R^{n \times n}$, and
$B$ and $X \in R^{n \times m}$. 
In the experiments, we concentrate on the activity pattern
where all operands are active, i.e.,
$$A' X + A X' = B'.$$

\begin{figure}
    \centering
    \includegraphics[scale=0.8]{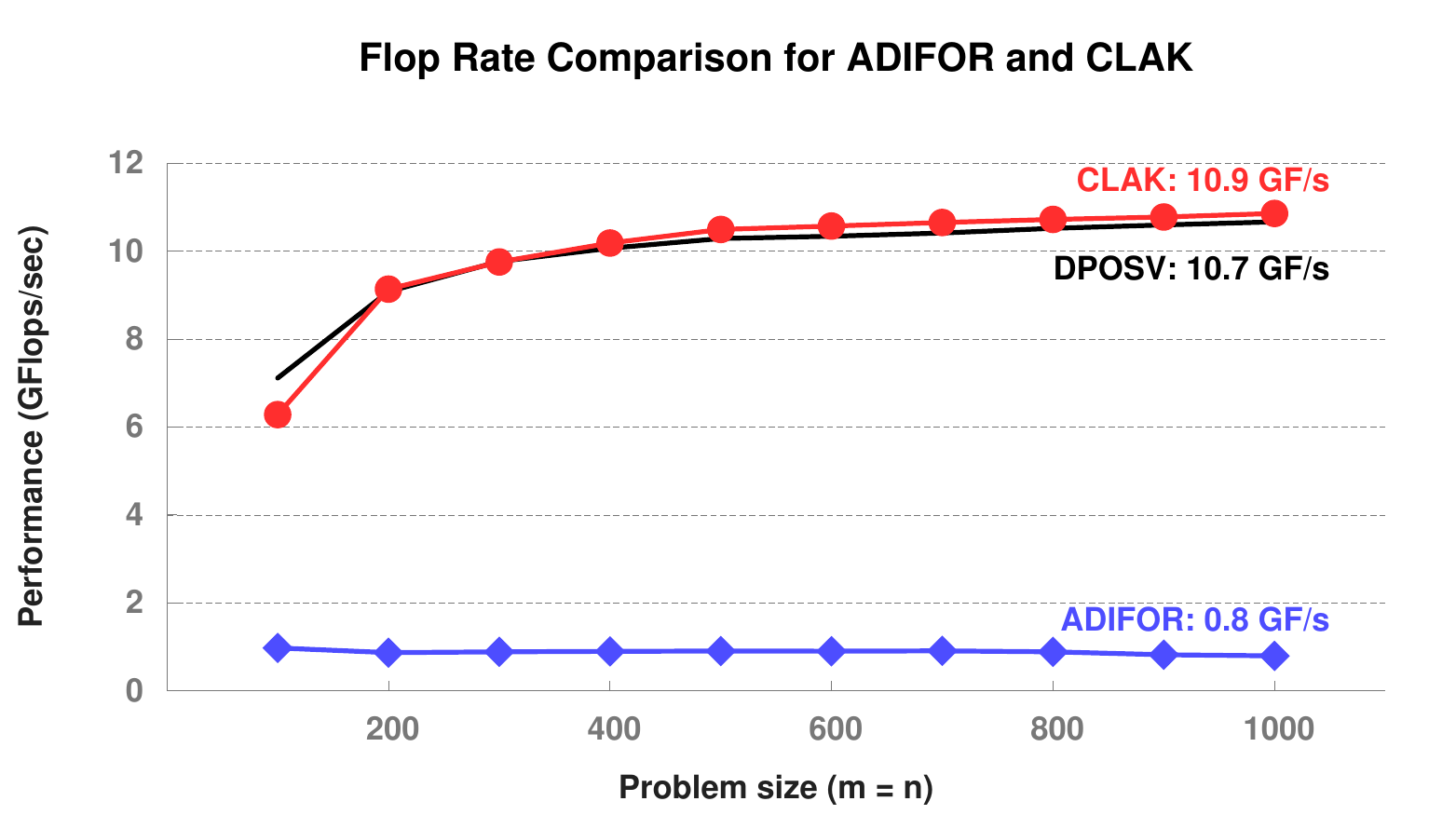}
    \caption{Performance comparison between the routines generated by ADIFOR
	and \clak{} for the solution of an SPD system and its derivative.
	The flop rate attained by MKL for the SPD system ({\sc dposv})
	is given as a reference.
    Results obtained for a single derivative ($p=1$), and a single core.}
    \label{fig:dposv-perf}
\end{figure}

First, we give a sense of the performance differences between the code
generated by ADIFOR and that generated by \clak{}. 
In Figure~\ref{fig:dposv-perf}, we report on the flop rate attained 
for a single derivative ($p = 1$) and an increasing size of the matrices ($m = n$).
As a reference, the top of the figure represents the theoretical
peak performance of the architecture for a single core (12 GFlops/sec),
and the line labeled ``DPOSV'' indicates the flop rate attained by MKL's
routine for the solution of the original SPD system.
While ADIFOR's code delivers rather poor performance (below 1 GFlop/sec),
\clak{}'s code attains a performance of 11 GFlops/sec, comparable to that of LAPACK,
and close to the peak performance.
The message is that
the usage of \clak{}'s routine in the extended program sustains LAPACK's
performance levels, and prevents a loss in performance.

\begin{figure}
    \centering
    \includegraphics[scale=0.8]{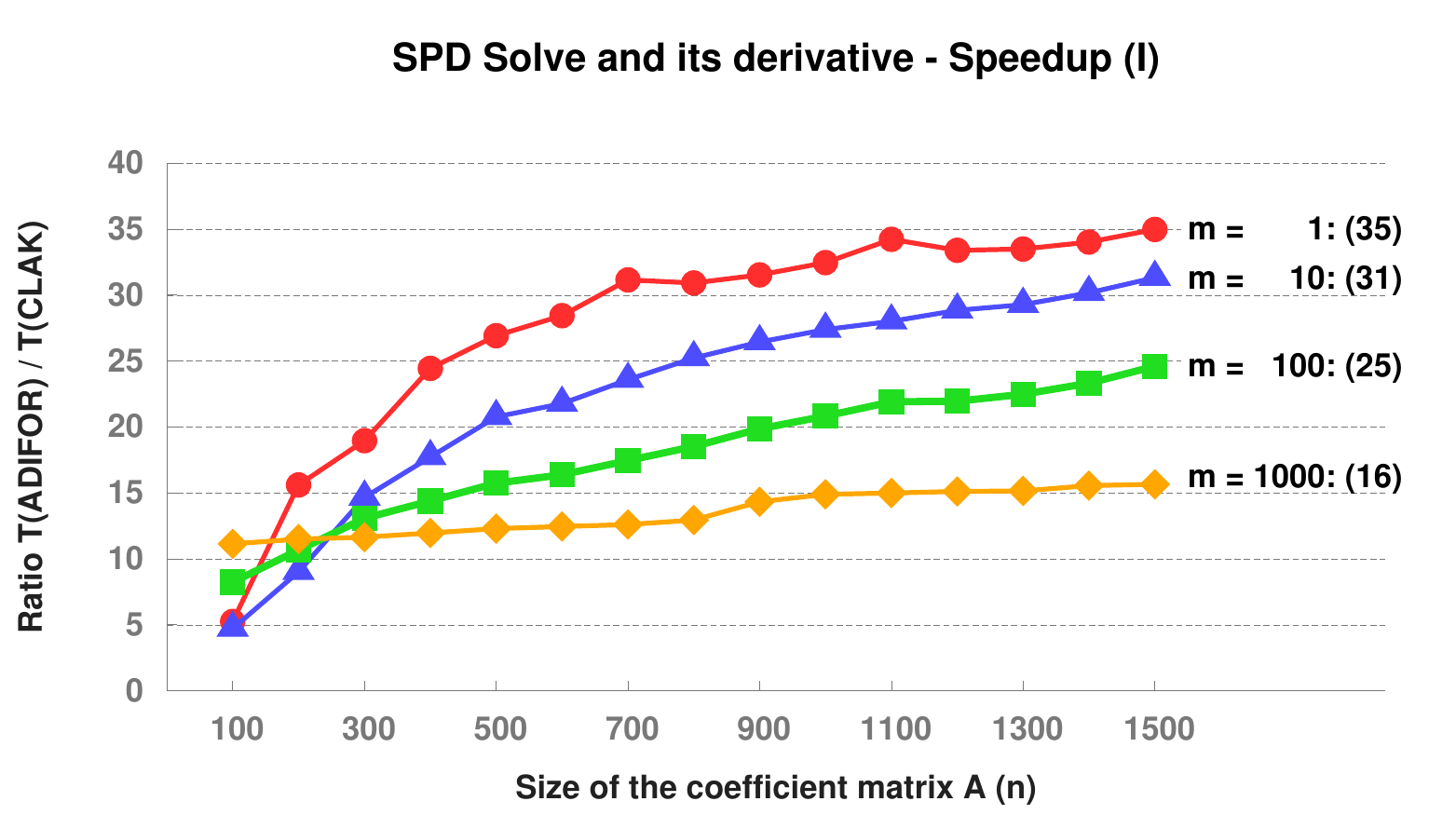}
	\caption{Speedup of \clak{}'s routine over ADIFOR's for a variety
		     of coefficient matrices and number of right-hand sides.
			Results obtained for a single derivative ($p=1$), and a single core.
		    In brackets, the attained speedup.}
    \label{fig:dposv-incr-n}
\end{figure}

Next, in Figure~\ref{fig:dposv-incr-n}, we show how this gap in performance
translates into large speedups.
The speedup (ratio of execution time for ADIFOR over execution time for \clak{})
ranges from 5x for a small coefficient matrix ($100 \times 100$)
and one right-hand side, up to 35x for large problems.

\begin{figure}
    \centering
    \includegraphics[scale=0.8]{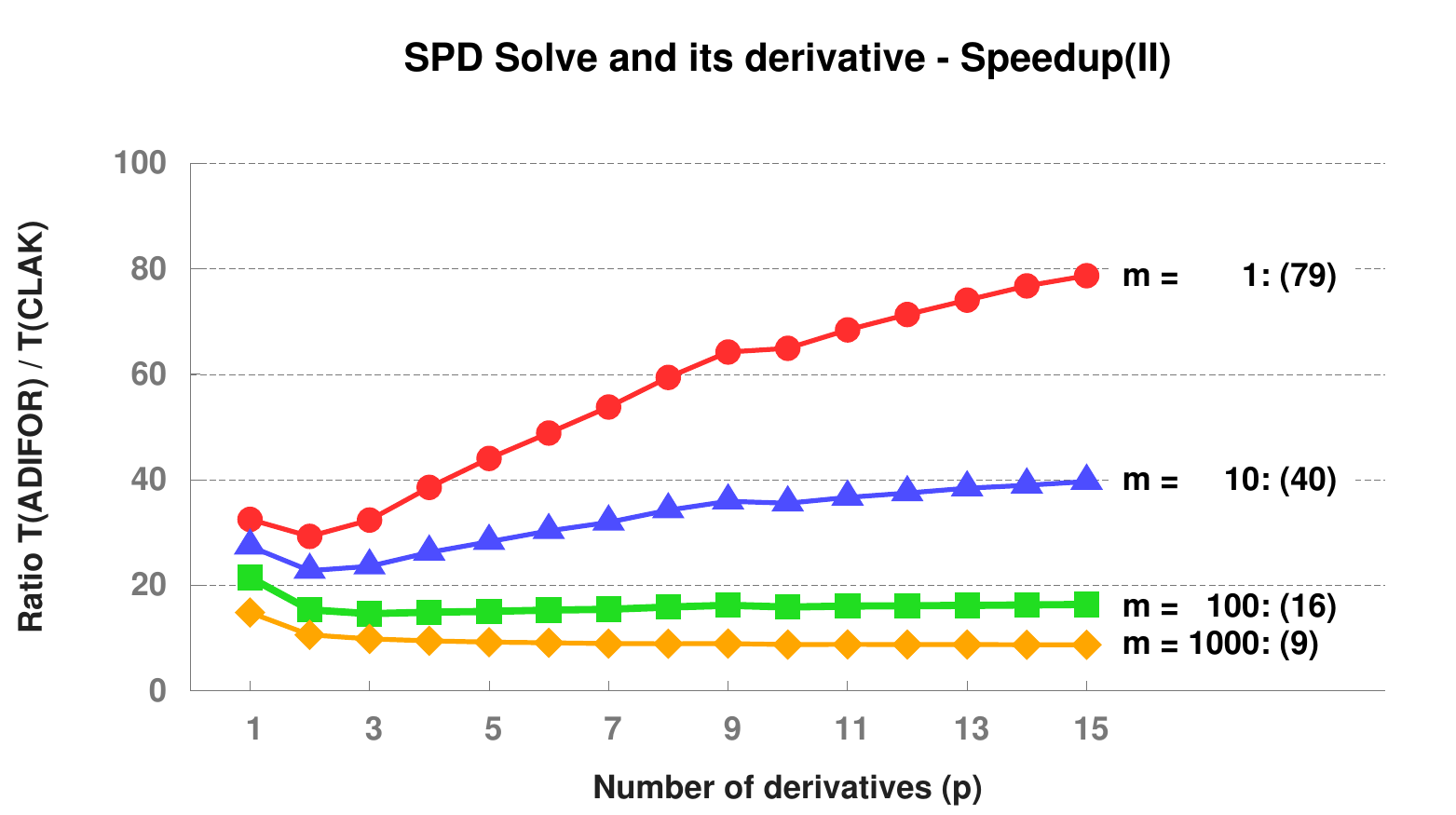}
	\caption{Speedup of \clak{} over ADIFOR when differentiating
			with respect to a vector-valued variable ($p > 1$).
			Results obtained for $n=1{,}000$, and a single core.
		    In brackets, the attained speedup.}
    \label{fig:dposv-incr-p}
\end{figure}

In Figure~\ref{fig:dposv-incr-p}, we provide
further experiments where multiple derivatives
($p > 1$) are computed. The goal is to emphasize the even larger speedups
achieved thanks to the analysis of dependencies carried out by \clak{}.
As anticipated by the computational cost formulas, 
the largest ratio is attained when computing multiple derivatives with a single 
right-hand side, for a speedup of about 80x.

\subsection{Example 2: Symmetric rank update}

We now concentrate on the second example, the computation of the BLAS {\sc syrk} operation:
$C := \alpha A A^T + \beta C$,
where $\alpha$ and $\beta \in R$, $A \in R^{n \times k}$, and $C \in R^{n \times n}$ is symmetric.
This operation involves four input operands ($\alpha$, $A$, $\beta$, $C$),
and thus admits 15 different activity patterns; a subset of them is
collected in Table~\ref{tab:gdsyrk-cost} together with their computational cost.

\begin{table}
	\centering
	\renewcommand{\arraystretch}{1.4}
	\begin{tabular}{l c l} \toprule
		\multicolumn{1}{c}{\bf Activity Pattern} & {\bf Cost for ADIFOR} & {\bf Cost for \clak{}} \\ \midrule
		\footnotesize $ C' := \alpha' A A^T + \alpha A' A^T + \alpha A A'^T + \beta' C + \beta C'$ & 
		$(2p+1)n^2k$ & 
		$(3p+1)n^2k$ \\
		\footnotesize $C' := \alpha' A A^T \phantom{ + \alpha A'^T A + \alpha A^T A'} + \beta' C + \beta C'$ & 
		$(\phantom{3}p+1)n^2k$ &
		$(\phantom{3}p+1)n^2k$ \\
		\footnotesize $C' := \phantom{\alpha' A^T A + } \alpha A' A^T + \alpha A A'^T \phantom{+ \beta' C} + \beta C'$ & 
		$(2p+1)n^2k$ &
		$(2p+1)n^2k$ \\
		\footnotesize $C' := \phantom{\alpha' A^T A + \alpha A'^T A + \alpha A^T A' + } \beta' C + \beta C'$ & 
		$(\frac{3}{2}p+k)n^2$ &
		$(\frac{3}{2}p+k)n^2$ \\
		\footnotesize $C' := \phantom{\alpha' A^T A + \alpha A'^T A + \alpha A^T A' + \beta' C +} \beta C'$ & 
		$(\frac{1}{2}p+k)n^2$ &
		$(\frac{1}{2}p+k)n^2$ \\ \bottomrule
	\end{tabular}
	\caption{Derivative of {\tt dsyrk}.
		    Cost of the routines generated by \clak{} and ADIFOR for a variety of
		activity patterns. The cost of computing {\tt dsyrk} itself ($n^2 k$) is included.}
	\label{tab:gdsyrk-cost}
\end{table}

\begin{figure}
    \centering
    \includegraphics[scale=0.8]{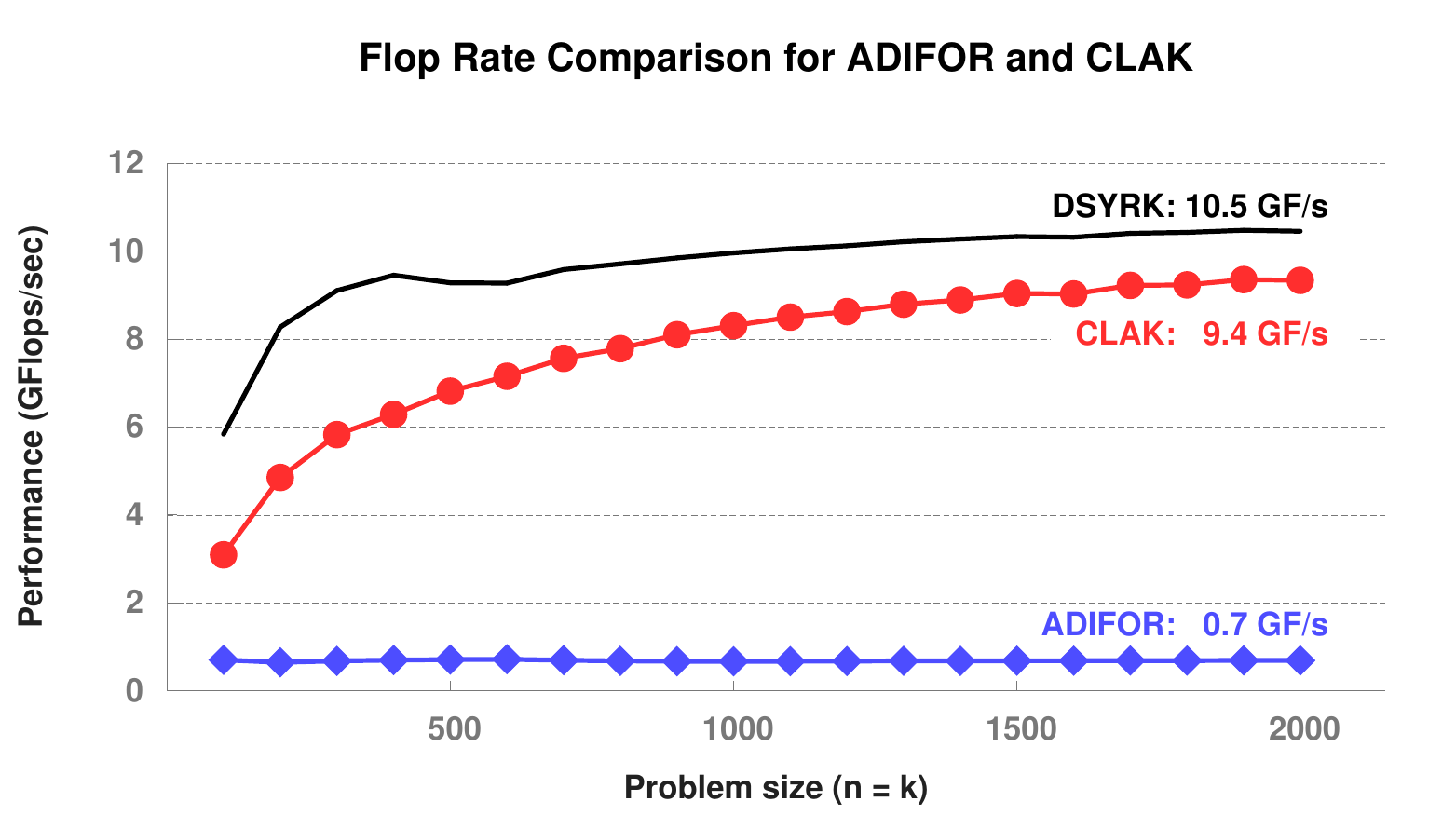}
    \caption{Performance comparison between the code generated by ADIFOR
		and \clak{} for the computation of {\tt dsyrk} and its derivative.
		The flop rate attained by MKL for {\sc dsyrk} is given as a reference.
    	Results obtained for a single derivative ($p=1$), and a single core.}
    \label{fig:gdsyrk-perf}
\end{figure}

Figure~\ref{fig:gdsyrk-perf} provides further evidence of the
large gap between the performance attained by \clak{}'s and ADIFOR's code;
the experiments were run for
the most general derivative (row 1 in Table~\ref{tab:gdsyrk-cost}), and $p=1$.
Again, the top of the graph represents the peak performance of the architecture,
and the line labeled with ``DSYRK'' shows the performance attained by MKL for 
the computation of {\sc dsyrk} only.
As it was the case for the SPD system, the performance of ADIFOR's code is about 0.7 GFlops/sec, 
while \clak{}'s routine attains a performance of almost 10 GFlops/sec, similar to that of 
BLAS for {\sc dsyrk} (10.5 GFlops/sec), and close to the peak.

\begin{figure}
    \centering
    \includegraphics[scale=0.8]{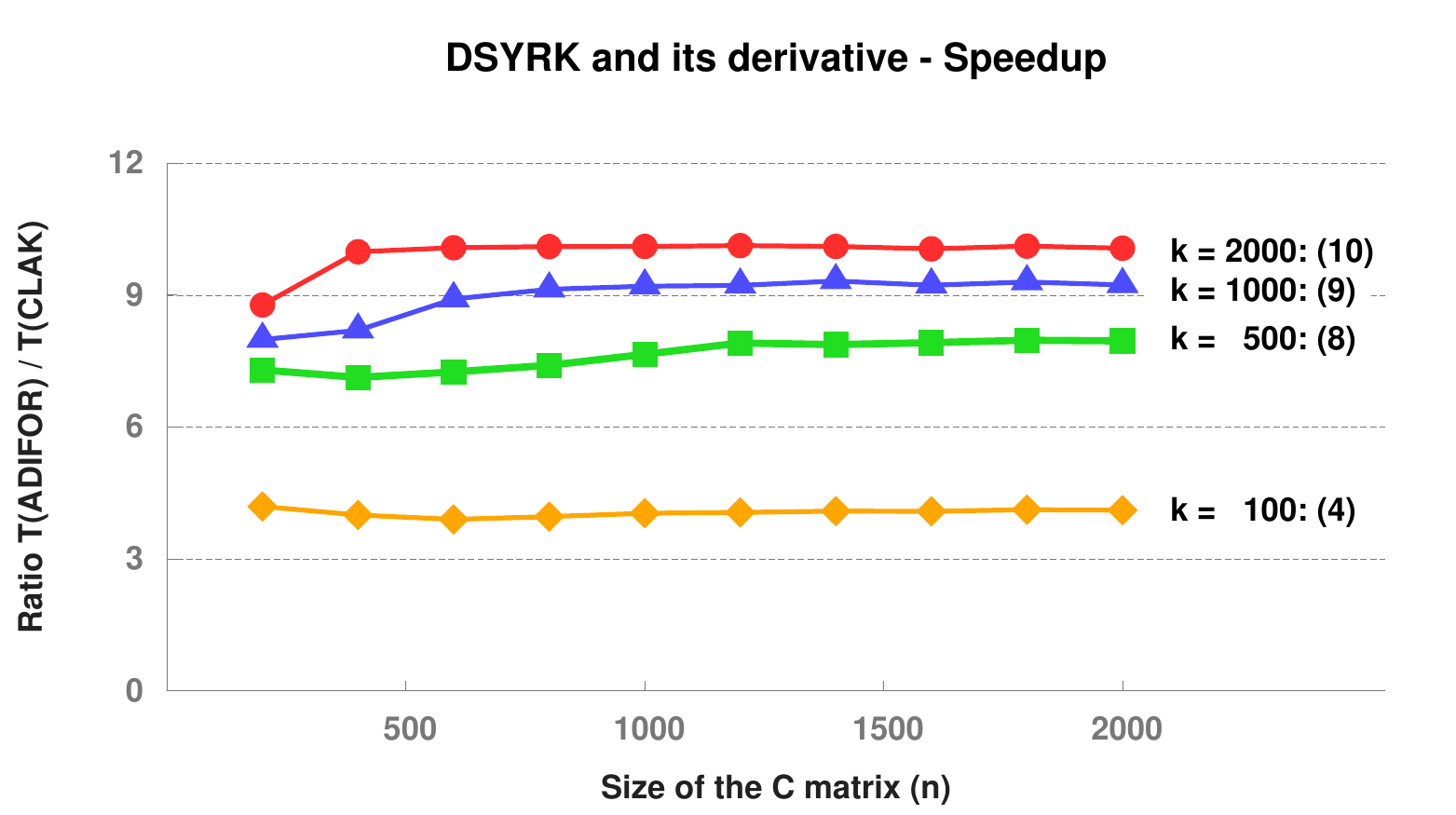}
	\caption{Speedup of \clak{}'s code with respect to ADIFOR's for a variety
		     of matrix sizes.
			Results obtained for a single derivative ($p=1$), and a single core.
		    In brackets, the attained speedup.}
    \label{fig:dsyrk-p1}
\end{figure}

The differences in performance translate into large speedups across the spectrum of problem sizes.
In Figure~\ref{fig:dsyrk-p1}, we report on the speedup of \clak{} over ADIFOR for
the most general derivative, and $p=1$.
The speedup ranges from 4x to 10x, the larger the size of the
matrix $A$ the higher the efficiency attained by BLAS, and thus the larger the speedup.

\begin{figure}
    \centering
    \includegraphics[scale=0.8]{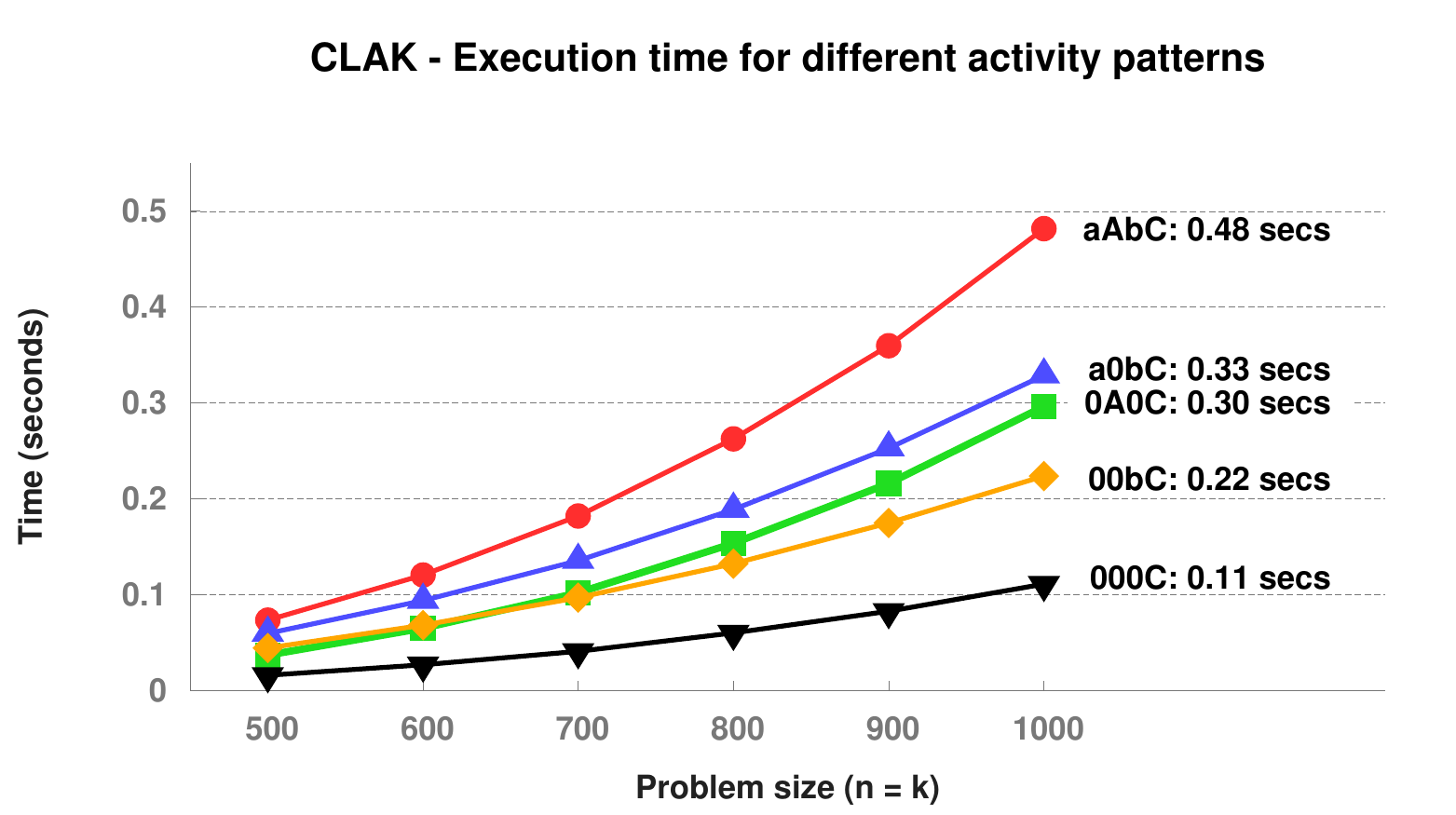}
    \caption{Timings for different activity patterns arising in the differentiation
		of the {\tt dsyrk} BLAS operation.
		Timings for a single derivative ($p=1$), and a single core.}
    \label{fig:act-patt-timings}
\end{figure}

To conclude the study, we illustrate the benefit of having routines available for all activity
patterns. 
In Figure~\ref{fig:act-patt-timings}, we provide timings for the routines generated
by \clak{} for the computation of the different activity patterns in Table~\ref{tab:gdsyrk-cost}.
Each pattern is labeled as follows: 
the label ``aAbC'' (for $\alpha$, $A$, $\beta$, and $C$) means all four variables are active;
whenever one of the characters is set to 0, it means that the corresponding variable is inactive.
For instance, ``a0bC'' means $\alpha$, $\beta$, and $C$ are active, while $A$ is inactive.
As the figure shows, time to solution may be further reduced by using the
most specific routine. 
For instance, using the routine for the case ``0A0C'',
instead of the most general ``aAbC'',
results in an extra 40\% speedup.

\section{Summary}

In this chapter, we focused on illustrating the broad applicability and 
the extensibility of our compiler.
To this end, we chose the application of \clak{} to matrix operations arising in 
the field of algorithmic differentiation;
as examples, we used the computation of the derivative of BLAS and LAPACK operations.

The core of the compiler presented in the 
previous chapter was extended by
adding support for the derivative operator (with the encoding of the chain rule to generate the analytical derivative formulas), and
by augmenting the engine for the inference of properties to determine properties of the derivative operands.
The mapping of the derivative formulas onto BLAS and LAPACK kernels
resulted in efficient algorithms and routines that,
compared to the code generated by ADIFOR, attained considerable speedups.

Beyond further showcasing the properties of our approach and compiler,
this chapter also makes a contribution to the AD community in terms
of the study of the potential benefits, should high-performance 
differentiated versions of BLAS and LAPACK be available.
First, the experimental results provide evidence that large speedups 
should be expected for a broad range of BLAS and
LAPACK operations, as illustrated by the characteristic examples used
in the experiments (a linear system and a matrix product).
Second, the analysis of data dependencies at a higher level of abstraction (the analytic formulas),
may lead to reductions in the computational cost with respect to traditional approaches,
resulting in further speedups.
And third, we show the convenience of providing specific routines for every activity pattern,
especially in the context of expensive matrix computations where the computational savings
may be substantial.

While it is not uncommon that these benefits can be achieved
by manually coding the kernels that represent the bulk of the computation,
having a differentiated version of the libraries available, and
enabling AD tools to make use of them, would have an impact in productivity.
The work presented in this chapter represents a step forward towards this goal,
demonstrating that efficient differentiated versions of BLAS and LAPACK are within reach.

\chapter{\clickplain{}}
\label{ch:click}

In this second part of the dissertation, we focus on the generation of 
algorithms for computational building blocks,
such as matrix products and factorizations.
For this class of operations, instead of the decomposition of a
target operation into a sequence of building blocks, 
we seek the derivation of loop-based blocked algorithms.
The design of such algorithms is in general a complex task,
which has long been considered a fine art.
Fortunately, in the last decade, in the frame of the
Formal Linear Algebra Methods Environment (FLAME) project~\cite{Gunnels:2001:FFL},
a methodology has been developed for the systematic
derivation of provably correct 
blocked algorithms~\cite{PaulDj:PhD,Bientinesi:2005:SDD}.
We adopt the FLAME methodology, and develop \click{},
a compiler that demonstrates 
how the methodology can be applied fully automatically, i.e., 
 without any human intervention.

The FLAME methodology enables the derivation of {multiple algorithmic variants}
for one same target operation.
In fact, for many operations, such as the Cholesky and LU factorizations, 
all the previously known algorithms are systematically discovered
and unified under a common root~\cite{TSoPMC}.
For more involved operations, such as the triangular continuous-time Sylvester equation and
the reduction of a generalized eigenproblem to standard form, the
generated family of algorithms include new and {better performing}
ones~\cite{Quintana-Orti:2003:FDA,FLAWN56}.
A quick review of FLAME-related literature~\cite{TSoPMC,Quintana-Orti:2003:FDA,FLAWN56,%
flame-lyapunov,Bientinesi:2005:SDD,1377606} reveals that the methodology 
has been consistently tested against well-known matrix operations 
available from numerical linear algebra libraries such as BLAS, LAPACK,
and RECSY.
Indeed, the project also provides libFLAME~\cite{libflame}, a library regarded
as a modern rewrite of LAPACK, that codes
hundreds of algorithms derived via this methodology. 

We emphasize that the methodology is not restricted to these
example problems, and is of far more {general applicability}.
We are especially interested in specialized operations not supported by 
high-performance libraries, ranging from 
slight variations of available kernels, such as the matrix product $C = AB$ with $A$ and $B$ triangular,
to the derivative of matrix factorizations.
While in this chapter we make use of classical examples to illustrate
our work,
in Chapter~\ref{ch:flame-ad} we provide an example
where the methodology yields high-performance algorithms for two
operations not directly supported by any numerical library.

Even though systematic, the methodology heavily relies on
pattern matching and symbolic manipulation of algebraic expressions,
hence becoming a tedious and error-prone process.
In fact, in~\cite{Quintana-Orti:2003:FDA}, a mistake in the
derivation led to an incorrect algorithm.
As the complexity of the target equation increases,
the methodology requires longer and more involved algebraic manipulation,
quickly surpassing what is manageable by hand.
The situation is aggravated by the fact that not one but multiple algorithmic variants
are desired.

We developed \click{} with the objective of relieving the developer from this burden and 
enabling the automatic generation of entire libraries.
In this chapter, we describe how \click{} is capable of generating 
a family of loop-based blocked algorithms to compute a
target operation from its sole mathematical description.

\section{Automating FLAME: A three-stage approach}
\label{sec:flame}

\begin{figure}[h]
\centering
    \includegraphics{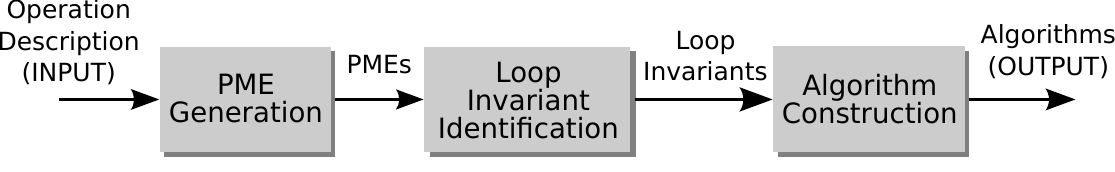}
    \caption{The three stages in the algorithm generation process.} 
	\label{fig:steps}
\end{figure}

The FLAME methodology enables the systematic 
derivation of formally correct loop-based
linear algebra algorithms.
The main idea is that the correctness is not proved 
a posteriori, once the algorithm is built;
instead, the proof of correctness and the algorithm grow
hand in hand.
To this end, loop invariants are identified first, and
then, for each of them, a skeleton of proof is created 
and the algorithm is built so that the proof is satisfied.
To automate the application of this methodology,
we first characterize the minimal input information about the
target operation necessary to automate the entire process,
and then we follow the constructive three-stage approach 
illustrated by Figure~\ref{fig:steps}.
Here, we outline these stages;
we devote the next sections to discuss each of them in detail.

\paragraph{Input.}
We define a target operation by means of two predicates:
the Precondition ($\PPre$) and the Postcondition ($\PPost$).
The postcondition states the equation to be solved, while the
precondition enumerates the properties of the operands.
Box~\ref{box:OpDescEx} contains the definition of the
inversion of a triangular matrix $L$; $L$ is overwritten with
its inverse; the notation $\hat{L}$ indicates the initial contents of $L$.
This is the only information about the operation required by
\click{} to automate the generation of algorithms.

\begin{mybox}
\vspace{3mm}
$$
\left\{
	\text{\hspace{-2mm}
	\begin{tabular}{l@{\hspace{1mm}}l}
		$\PPre:$  & $\{ \prop{Overwritten}{L} \; \wedge \; \prop{Matrix}{L} \; \wedge \; \prop{LowerTriangular}{L} \}$ \\[2mm]
		$\PPost:$ & $\{ L = \hat{L}^{-1} \}$
	\end{tabular}}
\right.
$$
\caption{Formal description for the inversion of a lower triangular matrix.}
\label{box:OpDescEx}
\end{mybox}

\paragraph{PME generation.}
The first stage of the process takes the description of the input operation, and 
yields its {\em Partitioned Matrix Expression}.
The PME is a decomposition of the target problem into simpler sub-problems
in a divide-and-conquer fashion; it exposes how each part of the output 
matrices is computed from parts of the input matrices.
Equation~\eqref{eq:PMEEx} represents the PME for the triangular inverse,
which states that the inverse may be decomposed as a two-sided triangular system
and two smaller inverses.
\begin{equation}
	\myFlaTwoByTwo{L_{TL} := \hat{L}_{TL}^{-1}}
				  {0}
				  {L_{BL} := -\hat{L}_{BR}^{-1} \hat{L}_{BL} \hat{L}_{TL}^{-1}}
				  {L_{BR} := \hat{L}_{BR}^{-1}}
	\label{eq:PMEEx}
\end{equation}

\paragraph{Loop invariant identification.}
The second stage of the process deals with the identification of
loop invariants. 
A loop invariant is a boolean predicate that encodes the state of the computation
at specific points of a loop:
It must be satisfied before the loop is entered and 
at the top and the bottom of each iteration~\cite{GrSc:92}. 
Loop invariants can be extracted as subsets of the computation encapsulated in the PME.
Equation~\eqref{eq:LoopInvEx} contains one loop invariant (out of eight)
for the triangular inverse; 
it indicates that the inverse of the top-left part of $\hat{L}$ has been computed,
while the other parts of $\hat{L}$ remain to be computed.
\begin{equation}
  \small
	\myFlaTwoByTwo{L_{TL} := \hat{L}_{TL}^{-1}}
				  {0}
				  {\neq}
				  {\qquad \neq \qquad \phantom{}} 
   \label{eq:LoopInvEx}
\end{equation}

\begin{figure}
\centering
	\subfloat[FLAME template for a formal proof of correctness for algorithms
		      consisting of an initialization step followed by a loop.]{
		\fbox{
		\begin{minipage}{0.40\textwidth}
			\centering
			\vspace{5mm}
			\renewcommand{\PBefore}{ P_{\rm before} }
\renewcommand{\PAfter} { P_{\rm after}  }
\renewcommand{\PPost}  { P_{\rm post}   }
\renewcommand{\PPre}   { P_{\rm pre}    }
\renewcommand{\PInv}   { P_{\rm inv}    }
\renewcommand{\EInv}   { E_{\rm inv}    }
\renewcommand{\EBefore}{ E_{\rm before} }
\renewcommand{\EAfter} { E_{\rm after}  }

\renewcommand{\WSupdate} { \rm Loop Body }

\newcommand{\skeleton}[0]
{
		\begin{tabular}{ l }
		  \\[-2mm]
		  \rowcolor[gray]{.8} \raisebox{-.8mm}{$\{ \PPre \}$} \\[1.5mm]
		  \raisebox{-.8mm}{{\bf Partition}} \\[1.5mm]
		  \rowcolor[gray]{.8} \raisebox{-.8mm}{$\{ \PInv \}$} \\[1.5mm]
		  \raisebox{-.8mm}{{\bf While} $G$ {\bf do}}\\[1.5mm]
		  \rowcolor[gray]{.8} \raisebox{-.8mm}{\hspace*{7.5mm}$\{ \PInv \wedge G\}$} \\[1.5mm]
		  \raisebox{-.8mm}{\hspace*{7.5mm}{\bf Repartition}} \\[1.5mm]
		  \raisebox{-.8mm}{\hspace*{7.5mm}$\WSupdate$} \\[1.5mm]
		  \raisebox{-.8mm}{\hspace*{7.5mm}{\bf Continue With}} \\[1.5mm]
		  \rowcolor[gray]{.8} \raisebox{-.8mm}{\hspace*{7.5mm}$\{ \PInv \}$} \\[1.5mm]
		  \raisebox{-.8mm}{{\bf end}} \\[1.5mm]
		  \rowcolor[gray]{.8} \raisebox{-.8mm}{$\{ \PInv \wedge \neg G \}$} \\[2mm]
		  \rowcolor[gray]{.8} \raisebox{-.8mm}{$\{ \PPost \}$} \\[1.5mm]
		\end{tabular}
}

\newcommand{\skeletonzoom}[0]
{
    \begin{tabular}{ l }
      \\[-2mm]
      \hspace*{7.5mm}$\{ \PInv \wedge G\}$ \\[1.5mm]
      \hspace*{7.5mm}{\bf Repartition} \\[1.5mm]
	  \hspace*{7.5mm}$\{ \PBefore \equiv \PInv|_{\text{\rm Repartitioning}} \}$ \\[1.5mm]
      \hspace*{7.5mm}${ \rm Algoritm Updates }$ \\[1.5mm]
	  \hspace*{7.5mm}$\{ \PAfter \equiv \PInv|_{\text{\rm Continue with}^{-1}} \}$ \\[1.5mm]
      \hspace*{7.5mm}{\bf Continue With} \\[1.5mm]
      \hspace*{7.5mm}$\{ \PInv \}$ \\[1.5mm]
    \end{tabular}
}

			\skeleton{}
			\vspace{5mm}
		\end{minipage}
	    }
		\label{fig:skeleton}
    }
    \qquad
	\subfloat[
		      Algorithm to compute the inverse of a lower triangular matrix
		  derived from loop invariant~\eqref{eq:LoopInvEx}.]{
		\begin{minipage}{0.45\textwidth}
			\scriptsize
			\input{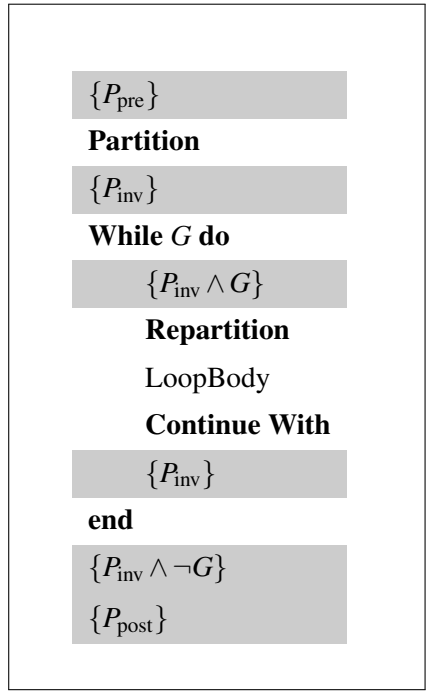}
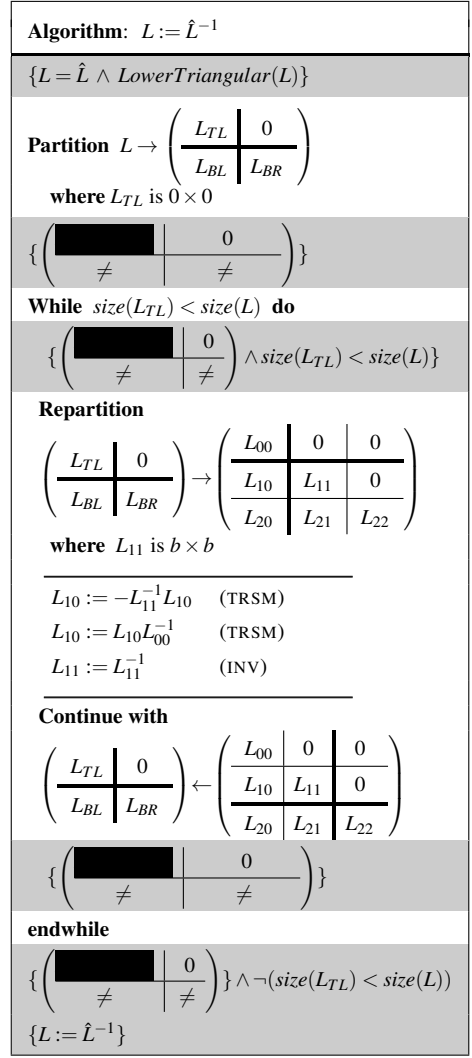
			\FlaAlgorithmNarrowTriinv
		\end{minipage}
        \label{fig:AlgEx}
    }
	\caption{FLAME template for a proof of correctness, and
	         example of FLAME algorithm. The shaded predicates
		     are part of the proof of correctness.}
	\label{fig:SkelAlgEx}
\end{figure}

\paragraph{Algorithm construction.}
In the third and last stage, each loop invariant is
transformed into its corresponding loop-based algorithm.
To this end, FLAME's methodology provides a template 
for a proof of correctness (Figure~\ref{fig:skeleton});
the predicates $\PPre$, $\PPost$, and $\PInv$ are replaced, respectively, with
the precondition, postcondition, and the loop invariant at hand.
Then, the algorithm statements (boldface labels) are filled in
so that the proof is satisfied (Figure~\ref{fig:AlgEx}).
The details in Figure~\ref{fig:SkelAlgEx} are not important now; they will
become clear by the end of the chapter.

\vspace{3mm}
By automating each of these three stages, we achieve,
for the first time, the complete automation
of the FLAME methodology:
From the mathematical description of the target operation, 
a family of algorithms that compute it are generated.
In the following, we describe in detail how \click{}
carries out each of the three stages.

\section{Input to \clickplain{}} 
\label{sec:input}
To unequivocally describe a target operation, 
we choose the language traditionally used to reason about program correctness:
Equations shall be specified by means of the predicates 
Precondition ($\PPre$) and Postcondition ($\PPost$)~\cite{GrSc:92}. 
The precondition enumerates the operands that appear in the equation and 
describes their properties, while
the postcondition specifies the equation to be solved.

We commence the discussion using the Cholesky factorization as an example:
Given a symmetric positive definite (SPD) matrix $A$, the goal is to find a
lower triangular matrix $L$ such that $L L^T = A$.
In Box~\ref{box:CholOpDesc}, we provide the description of such a factorization; 
the notation $L := \Gamma(A)$ indicates that $L$ is the Cholesky factor of $A$.
\begin{mybox}
$$
\small
L := \Gamma(A) \equiv
\left\{
\begin{split}
	P_{\rm pre}: \;\, \{ &\prop{Output}{L} \; \wedge \; \prop{Matrix}{L} \; \wedge \; \prop{LowerTriangular}{L} \;\; \wedge \\
					     &\prop{Input}{A}  \; \wedge \; \prop{Matrix}{A} \; \wedge \; \prop{SPD}{A} \} \\[2mm]
	P_{\rm post}:     \{ &L L^T = A \}
\end{split}
\right.
$$
\caption{Formal description for the Cholesky factorization.}
\label{box:CholOpDesc}
\end{mybox}

The definition is unambiguous, and it includes all the
information specific to the operation needed by \click{} to 
fully automate the derivation process.  

\subsubsection{Pattern Learning}\label{subsec:pattLearn}

\click{} takes the pair of predicates in Box~\ref{box:CholOpDesc} 
and creates the pattern in Box~\ref{box:chol-patt} that identifies 
the Cholesky factorization.  
The pattern establishes that 
matrices $L$ and $A$ are one the Cholesky factor of the other, 
provided that 
the constraints in the precondition are satisfied, and
that $L$ and $A$ are related as dictated by the postcondition 
($L L^T = A$). 

\begin{mybox}
	\small
\begin{verbatim}


            equal[ times[ L_, trans[L_] ], A_ ] /;
                    isOutputQ[L] && isInputQ[A] &&
                    isMatrixQ[L] && isMatrixQ[A] &&
                    isLowerTriangularQ[L] && isSPDQ[A] 
\end{verbatim}
\caption{Mathematica pattern representing the Cholesky factorization.}
\label{box:chol-patt}
\end{mybox}

\noindent
For instance, in the expression
$$X X^T = A - B C,$$ 
in order to determine whether $X = \Gamma(A - B C)$, 
the following facts need to be asserted:
i) $X$ is an unknown (output) lower triangular matrix; 
ii) the expression $A - B C$ is a known (input) quantity ($A, B$ and $C$ are known); 
iii) the matrix $A - B C$ is symmetric positive definite.

The strategy for decomposing an equation in terms of simpler problems
greatly relies on pattern matching. 
Initially, \click{} only knows the patterns for a basic set
of operations: addition, multiplication, inversion, and transposition
of matrices, vectors and scalars. This information is built-in in the compiler.
More complex patterns are instead dynamically learned during 
the process of algorithm derivation. 
As \click{}'s pattern knowledge
increases, also does its capability of tackling complex operations.

\section{PME generation}
\label{sec:GenPME}
\begin{figure}[!h]
  \centering
  \includegraphics[scale=1.00]{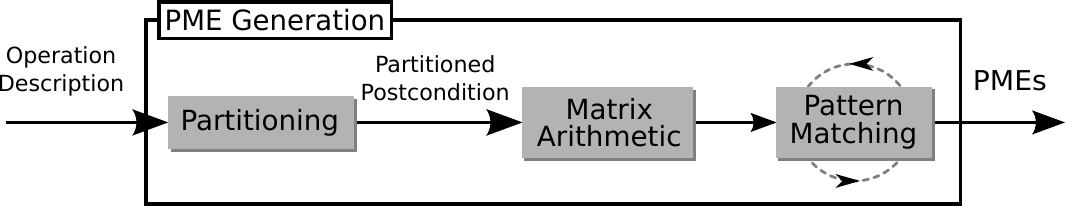}
  \caption{Steps for the automatic generation of PMEs.} \label{fig:stepsPME}
\end{figure}

This section centers around the first stage of the derivation process, the generation of PMEs.
As Figure~\ref{fig:stepsPME} shows, such process involves three steps: 
1) the partitioning of the operands in the equation, 
2) matrix arithmetic involving the partitioned operands, and
3) a sequence of iterations, each consisting of algebraic manipulation and
pattern matching, that yield the sought-after PMEs.

\subsection{Operands partitioning} \label{sec:descToPMEs}

We illustrate all the steps performed by
\click{} to transform the description of the input equation into one
or more PMEs. The idea is to first rewrite the postcondition in terms
of partitioned matrices and then apply pattern matching to identify
known operations. To this end, we introduce a set of rules to
partition and combine operands and to assert properties of expressions
involving sub-operands. The application of these rules to the
postcondition yields one or more predicates called {\em partitioned
  postcondition}.  In next section, an iterative process consisting of
algebraic manipulation and pattern matching will take us to the PMEs.

\subsubsection{Operands partitioning and direct inheritance}

The discussion commences with a set of rules for partitioning matrices
and vectors and for transferring properties to sub-matrices and
sub-vectors. These rules are part of the basic engine of
\click{}. Depending on constraints imposed by both the structure of the input
operands and the postcondition, only few partitioning rules will be
meaningful.

As shown in Box~\ref{box:part}, a generic matrix $A$ can be
partitioned in four different ways. The $1 \times 1$ rule
(Box~\ref{box:part}\subref{sbox:part1x1}) is special, as it does not affect the operand;
we refer to it as the {\em identity}.
For a vector, only the $2 \times 1$ and $1 \times 1$ rules apply,
while for scalars only the identity is admissible.
When referring to any of the parts resulting from a non-identity rule, we use the
terms {\em sub-matrix} or {\em sub-operand}, and for $2 \times 2$ partitionings,
we also use the term {\em quadrant}.
 
\begin{mybox}
\vspace{2mm}
\begin{center}
    \subfloat[$2 \times 2$ rule]{
        \label{sbox:part2x2}
        \begin{minipage}{3.6cm}
		  \centering
		  $\ruleTwoByTwo{A}{m}{n}{TL}{k_1}{k_2}$
        \end{minipage}
    }
    \qquad
    \subfloat[$2 \times 1$ rule]
    {\label{sbox:part2x1}
        \begin{minipage}{3.6cm}
		  \centering
		  $\ruleTwoByOne{A}{m}{n}{T}{k_1}{n}$
        \end{minipage}
    }
    \\
    \subfloat[$1 \times 2$ rule]
    {\label{sbox:part1x2}
        \begin{minipage}{3.6cm}
		  \centering
		  $\ruleOneByTwo{A}{m}{n}{L}{m}{k_2}$
        \end{minipage}
    }
    \qquad
    \subfloat[$1 \times 1$ (identity) rule]
    {\label{sbox:part1x1}
        \begin{minipage}{3.6cm}
		  \centering
		  $\ruleOneByOne{A}{m}{n}$
        \end{minipage}
    }

\end{center}
 
\caption{
  Rules for  partitioning  a generic matrix operand A.
  We use the subscript letters $T$, $B$, $L$, and $R$ for $T$op, $B$ottom,
  $L$eft, and $R$ight, respectively.
}\label{box:part}
\end{mybox}

The inheritance of
properties plays an important role in subsequent stages of the
algorithm derivation.
Thus, when the operands have a special structure, it is beneficial to choose
partitioning rules that respect it.  
For a symmetric matrix, for instance,
it is convenient to create sub-matrices that exhibit the same property.
The $1 \times 2$ and $2 \times 1$ rules break the structure of a
symmetric matrix, as neither of the two sub-matrices inherit the
symmetry. Therefore, we only allow $1 \times 1$ or $2 \times 2$
partitionings, with the extra constraint that the $TL$ quadrant has to
be square.

Box~\ref{box:partLM} illustrates
the admissible partitionings for lower triangular ($L$) and symmetric ($M$) matrices.
On the left, the identity rule is applied and the operands remain unchanged.
On the right instead, a constrained $2 \times 2$ rule is applied, 
so that some of the resulting quadrants inherit properties.
For a lower triangular matrix $L$,
both $L_{TL}$ and $L_{BR}$ are square and lower triangular, $L_{TR}$ is zero, and $L_{BL}$ is
a generic matrix. For a symmetric matrix $M$,
both $M_{TL}$ and $M_{BR}$ are square and symmetric,
and $M_{BL} = M_{TR}^T$ (or vice versa $M_{TR} = M_{BL}^T$).

\begin{mybox}
\vspace{2mm}
\begin{center}
\subfloat[Viable partitionings for a lower triangular matrix.]
{\label{sbox:partL}
         \begin{tabular}{lcl}
        \begin{minipage}{3.3cm}
		  \centering
		  $\ruleOneByOne{L}{m}{m}$
        \end{minipage}
         \quad or \qquad &
        \begin{minipage}{3.6cm}
		  \centering
		  $\lowtriRuleTwoByTwo{L}{m}{m}{TL}{k}{k}$
        \end{minipage}
         \end{tabular}
}
\\
\subfloat[Viable partitionings for a symmetric matrix.]
{\label{sbox:partM}
         \begin{tabular}{lcl}
        \begin{minipage}{3.3cm}
		  \centering
		  $\ruleOneByOne{M}{m}{m}$
        \end{minipage}
         \quad or \qquad &
        \begin{minipage}{3.6cm}
		  \centering
		  $\symmRuleTwoByTwo{M}{m}{m}{TL}{k}{k}$
        \end{minipage}
         \end{tabular}
}
\caption{
  Partitioning rules for structured matrices.}\label{box:partLM}
\end{center}
\end{mybox}

\paragraph{Theorem-aware inheritance}

Although frequent, direct inheritance of properties is only the simplest form of inheritance.
Here we expose a more complex situation.
Let A be an SPD matrix. Because of symmetry, the only allowed partitioning rules are
the ones listed in Box~\ref{box:partLM}\subref{sbox:partM}; applying the
$2 \times 2$ rule, we obtain
\begin{equation}
\label{eqn:SPDPart}
    \symmRuleTwoByTwo{A}{m}{m}{TL}{k}{k},
\end{equation}
and both $A_{TL}$ and $A_{BR}$ are symmetric. More properties about
the quadrants of $A$ can be stated. For example, it is well known that
{\it if $A$ is SPD, then all its principal sub-matrices are SPD}~\cite{Golub:1996:MC:248979}.  
As a consequence, the quadrants $A_{TL}$ and $A_{BR}$ inherit
such a property.  Moreover, it can be proved that given a $2 \times
2$ partitioning of an SPD matrix as in (\ref{eqn:SPDPart}), the
matrices $A_{TL} - A_{BL}^{T} A_{BR}^{-1} A_{BL}$ and
$A_{BR} - A_{BL} A_{TL}^{-1} A_{BL}^{T}$ (known as Schur complements) are also symmetric
positive definite.
The knowledge emerging from this theorem is included in \click{}'s engine.
In Section~\ref{sec:PattMatch} it will become apparent how this
information is essential for the generation of PMEs.

\subsubsection{Combining the partitionings}

The partitioning rules are now applied to rewrite the
postcondition equation.  Since in general each operand can be decomposed in
multiple ways, not one, but many partitioned postconditions are
created. As an example, in the Cholesky factorization
(Box~\ref{box:CholOpDesc}) both the $1\times 1$ and $2\times 2$ rules
are viable for both $L$ and $A$, leading to four different sets of partitionings:
\begin{itemize}
\item Both $L$ and $A$ are partitioned in $1 \times 1$.
\item $L$ and $A$ are partitioned in $1 \times 1$ and $2 \times 2$, respectively.
\item $L$ and $A$ are partitioned in $2 \times 2$ and $1 \times 1$, respectively.
\item Both $L$ and $A$ are partitioned in $2 \times 2$.
\end{itemize}

\begin{table}
\centering
\renewcommand{\arraystretch}{1.4}
\scriptsize
\begin{tabular}{c i c | c | c } \toprule
	{\bf \#} & {\bf L} & {\bf A} & {\bf Partitioned Postcondition} \\ 
	\addlinespace[0.5em]
	\toprule 
\rule[-0.3cm]{0cm}{0.7cm} 1 &
$
L \rightarrow \left( L \right)$ & $A \rightarrow \left( A \right)$ &
\renewcommand{\arraystretch}{1.4}
$
\left( L \right)
\left( L \right)^T
=
\left( A \right)
$ \\ \midrule
\rule[-0.4cm]{0cm}{1cm} 2 &
$ L \rightarrow \left( L \right)$ &
\renewcommand{\arraystretch}{1.4}
$
         A \rightarrow
         \left( 
	    	\begin{array}{@{}c@{\,}|@{\,}c@{}} 
			  A_{TL} & A_{BL}^T \\\hline 
			  A_{BL} & A_{BR} 
			\end{array} 
		\right)
$ &
\renewcommand{\arraystretch}{1.4}
$
\left( L \right)
\left( L \right)^T
=
\left( 
  \begin{array}{@{}c@{\,}|@{\,}c@{}} 
    A_{TL} & A_{BL}^T \\\hline 
	A_{BL} & A_{BR} 
  \end{array} 
\right)
$ \\ \midrule
\rule[-0.4cm]{0cm}{1cm} 3 &
\renewcommand{\arraystretch}{1.4}
$
         L \rightarrow
		 \myFlaTwoByTwo{L_{TL}}{0}{L_{BL}}{L_{BR}}
$ & $A \rightarrow \left( A \right)$ &
\renewcommand{\arraystretch}{1.4}
$
	 \myFlaTwoByTwo{L_{TL}}{0}{L_{BL}}{L_{BR}}
	 \myFlaTwoByTwo{L_{TL}^T}{L_{BL}^T}{0}{L_{BR}^T}
	=
	\left( A \right)
$ \\ \midrule
\rule[-0.4cm]{0cm}{1cm} 4 &
\renewcommand{\arraystretch}{1.4}
$
         L \rightarrow
         \left( \begin{array}{@{}c@{\,}|@{\,}c@{}} L_{TL} & 0 \\\hline L_{BL} & L_{BR} \end{array} \right)
$ &
\renewcommand{\arraystretch}{1.4}
$
         A \rightarrow
         \left( \begin{array}{@{}c@{\,}|@{\,}c@{}} A_{TL} & A_{BL}^T \\\hline A_{BL} & A_{BR} \end{array} \right)
$ &
\renewcommand{\arraystretch}{1.4}
$
\left( \begin{array}{@{}c@{\,}|@{\,}c@{}} L_{TL} & 0 \\\hline L_{BL} & L_{BR} \end{array} \right)
\left( \begin{array}{@{}c@{\,}|@{\,}c@{}} L_{TL}^{T} & L_{BL}^{T} \\\hline 0 & L_{BR}^{T} \end{array} \right)
=
\left( \begin{array}{@{}c@{\,}|@{\,}c@{}} A_{TL} & A_{BL}^T \\\hline A_{BL} & A_{BR} \end{array} \right) $ \\
\bottomrule
\end{tabular}
\caption{Application of the different combinations of partitioning rules to the postcondition.} 
\label{tab:partPostcond}
\end{table}

Table~\ref{tab:partPostcond} contains the resulting four partitioned
postconditions.  It is apparent that some of the expressions in the
fourth column are not algebraically well-defined.  Consequently, in
addition to constraints on each individual oper\-and, the 
partitioning rules need to
be such that the partitioned operands can be combined together
according to standard matrix arithmetic.  For instance, in the
expression $X+Y$, if the $2\times 1$ rule is applied to matrix $X$,
the $+$ operator imposes that the same rule is applied to $Y$ too.

With reference to Table~\ref{tab:partPostcond}, the rules in the
second row lead to an expression whose left-hand and right-hand sides 
are a $1 \times 1$ and a $2 \times 2$ object, respectively.
The reverse is true in the third row.
Since, both lead to ill-defined partitioned postconditions,
they are discarded.
Despite leading to a well defined expression, the first row of the table
is discarded too, as it leads to an expression 
in which none of the operands has been partitioned,
while the goal is to obtain a {\it Partitioned} Matrix Expression.
In light of these additional restrictions, the only viable set
of rules for the Cholesky factorization is the one given in the last row of 
Table~\ref{tab:partPostcond}, 
with the additional constraint that the $A_{TL}$ and $L_{TL}$ quadrants are square.

In summary, partitioning rules must satisfy
both the constraints due to the nature of the individual operands, 
and those due to the operators appearing in the postcondition. 
Next, we detail the algorithm used by \click{} to
generate only the viable sets of partitioning rules.

\subsubsection{Automation}
\label{subsec:automation}

We show how \click{} performs the partitioning process automatically.
A naive approach would exhaustively search among all the
rules applied to all the operands, leading to a search space of
exponential size in the number of operands. 
Instead, \click{} utilizes an algorithm that traverses
the postcondition (represented as a tree) 
just once, and yields only
the viable sets of partitioning rules. 

\begin{figure}
\begin{center}
\includegraphics[scale=0.55]{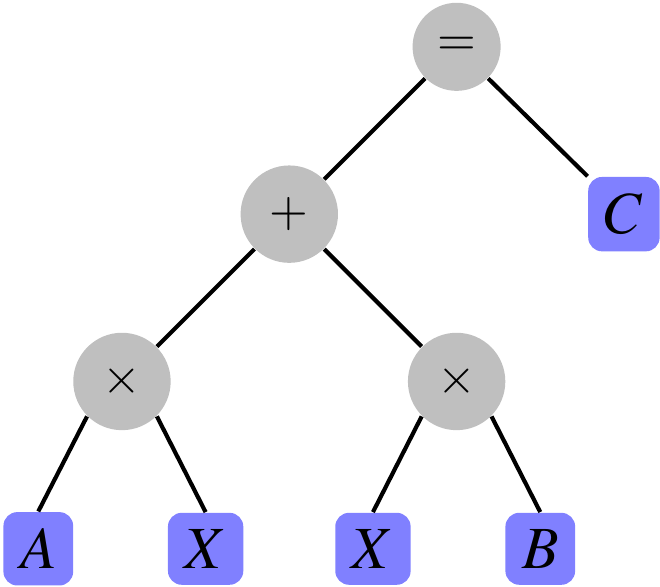}
\caption{Tree representation for the Sylvester equation $A X + X B = C$.} \label{fig:tree}
\end{center}
\end{figure}

The algorithm builds around two main ideas:
1) the properties of an operand impose restrictions on the viable rules;  
2) the operators in the postcondition also constrain the partitionings of their operands.
The input to the algorithm is a target operation, in the form of the 
predicates $P_{\rm pre}$ and $P_{\rm{post}}$.
As an example, we look at the triangular Sylvester equation
$$A X + X B = C,$$
defined formally in Box~\ref{box:sylvdesc}.
Henceforth, we will also refer to the equation as the {\em Sylvester equation}.
\begin{mybox}[!t]
{\small
\begin{equation} \nonumber
X := \Omega(A, B, C) \equiv
\left\{
\begin{split}
P_{\rm pre}: \{ & \prop{Input}{A} \wedge \prop{Matrix}{A} \wedge \prop{UpperTriangular}{A} \, \wedge \\
                & \prop{Input}{B} \wedge \prop{Matrix}{B} \wedge \prop{UpperTriangular}{B} \, \wedge \\
                & \prop{Input}{C} \wedge \prop{Matrix}{C} \wedge \prop{Output}{X} \wedge \prop{Matrix}{X} \} \\
\\[-2mm]
P_{\rm post}: \{ &  A X + X B = C \}.
\end{split}
\right.
\end{equation}
}
\caption{Formal description for the triangular Sylvester equation.}
\label{box:sylvdesc}
\end{mybox}
First, the algorithm transforms the postcondition to prefix notation
(Figure~\ref{fig:tree}) and collects name and dimensionality of each operand. 
A list of disjoint sets, one per dimension of the operands is then
created.
For the Sylvester equation, this initial list is
$$\left[ \; \{ A_{r}\}, \{ A_{c}\}, \{ B_{r}\}, \{ B_{c}\}, \{ C_{r}\}, \{ C_{c}\}, \{ X_{r}\}, \{ X_{c}\} \; \right],$$
where $r$ and $c$ stand for {\it rows} and {\it columns} respectively.
The algorithm traverses the tree, in a post-order fashion, to
determine if and which dimensions are bound together. Two dimensions
are bound to one another if the partitioning of one implies the
partitioning of the other. 
If two dimensions are found to be bound, then their corresponding
sets are merged. As the algorithm moves from the leaves to
the root of the tree, it keeps track of the dimensions of the operands'
subtrees.

The algorithm starts by visiting the node corresponding to the upper triangular operand $A$.
There it establishes
that the identity and the $2 \times 2$ partitioning rules are the only admissible
ones. Thus, the rows and the columns of $A$ are bound together, and the list becomes
$$
\left[ \; \{ A_{r}, A_{c}\}, \{ B_{r}\}, \{ B_{c}\}, \{ C_{r}\}, \{ C_{c}\}, \{ X_{r}\}, \{ X_{c}\} \; \right].
$$
\noindent
The next node to be visited is that of the operand $X$.
Since $X$ has no specific structure, its analysis causes no bindings.
Then, the node corresponding to the $\times$ operator is analyzed.
The dimensions of $A$ and $X$ have to agree according to the matrix product, therefore,
a binding between $A_c$ and $X_r$ is imposed:
$$\left[ \; \{ A_{r}, A_{c}, X_{r}\}, \{ B_{r}\}, \{ B_{c}\}, \{ C_{r}\}, \{ C_{c}\}, \{ X_{c}\} \; \right].$$
\noindent
At this stage, the dimensions of the product $A X$ are also determined to be
$A_{r} \times X_{c}$.

The procedure continues by analyzing the subtree corresponding to the product $X B$.
Similarly to the product $A X$, the lack of structure in $X$ does not cause any binding,
while the triangularity of $B$ imposes a binding between
$B_{r}$ and $B_{c}$ leading to 
$$\left[ \; \{ A_{r}, A_{c}, X_{r}\}, \{ B_{r}, B_{c}\}, \{ C_{r}\}, \{ C_{c}\}, \{ X_{c}\} \; \right].$$
Then, the node for the $\times$ operator is analyzed, and a binding between $X_c$ and $B_r$ is found:
$$\left[ \; \{ A_{r}, A_{c}, X_{r}\}, \{ B_{r}, B_{c}, X_{c}\}, \{ C_{r}\}, \{ C_{c}\} \; \right].$$
The dimensions of the product $X B$ are determined to be $X_{r} \times B_{c}$. 

The next node to be considered is the corresponding to the $+$ operator. It imposes a binding between
the rows and the columns of the products $A X$ and $X B$, i.e., between $A_{r}$ and $X_{r}$,
and between $X_{c}$ and $B_{c}$. Since each of these pairs of dimensions already belong to the same
set, no modifications are made to the list. The algorithm establishes that the dimensions of 
the $+$ node are $A_r \times B_c$. Next, the node associated to the operand $C$
is analyzed. Since $C$ has no particular structure, its analysis does not cause any
modification. The last node to be processed is the equality operator
$=$. This node binds the rows of $C$ to those of $A$ ($C_r$, $A_r$)
and the columns of $C$ to those of $B$ ($C_c$, $B_c$). The final list
consists of two separate groups of dimensions:
$$\left[ \; \{ A_{r}, A_{c}, X_{r}, C_{r}\}, \{ B_{r}, B_{c}, X_{c}, C_{c}\} \; \right].$$

Having created $g$ groups of bound dimensions, 
the algorithm generates $2^g$ combinations of rules
(the dimensions within each group being either partitioned or not), 
resulting in a family of partitioned postconditions, one per combination.
In practice, since the combination including solely identity rules
does not lead to a PME, only $2^g-1$ combinations are acceptable.
In our example, the algorithm found two groups of bound dimensions;
therefore three possible combinations of rules are generated:
1) only the dimensions in the second group are partitioned, 
2) only the dimensions in the first group are partitioned, or
3) all dimensions are partitioned.
The resulting partitionings are listed in Table~\ref{tab:sylvPart}.

\begin{table}
\centering
\scriptsize
\renewcommand{\arraystretch}{1.4}
\begin{tabular}{c i c | c | c | c} \toprule
{\bf \#} & {\bf A}  & {\bf B} & {\bf C} & {\bf X} \\
\addlinespace[0.5em]
	\toprule
\scriptsize 1 \rule[-0.35cm]{0cm}{0.9cm} &
$(A)$ & 
$\myFlaTwoByTwo{B_{TL}}{B_{TR}}{0}{B_{BR}}$ &
$\left( \begin{array}{c|c} C_{L} & C_{R} \end{array} \right)$ &
$\left( \begin{array}{c|c} X_{L} & X_{R} \end{array} \right)$ \\\midrule 
\scriptsize 2 \rule[-0.35cm]{0cm}{0.9cm} &
$\myFlaTwoByTwo{A_{TL}}{A_{TR}}{0}{A_{BR}}$ &
$(B)$ &
$\myFlaTwoByOne{C_{T}}{C_{B}}$ &
$\myFlaTwoByOne{X_{T}}{X_{B}}$ \\\midrule 
\scriptsize 3 \rule[-0.35cm]{0cm}{0.9cm} &
$\myFlaTwoByTwo{A_{TL}}{A_{TR}}{0}{A_{BR}}$ &
$\myFlaTwoByTwo{B_{TL}}{B_{TR}}{0}{B_{BR}}$ &
$\myFlaTwoByTwo{C_{TL}}{C_{TR}}{C_{BL}}{C_{BR}}$ &
$\myFlaTwoByTwo{X_{TL}}{X_{TR}}{X_{BL}}{X_{BR}}$ \\\bottomrule
\end{tabular}
\caption{Viable combinations of partitioning rules for the Sylvester equation.} \label{tab:sylvPart}
\end{table}

By means of this algorithm,  \click{} efficiently generates, for every
target operation, 
only the acceptable sets of partitioning rules.

\subsection{Matrix arithmetic and pattern matching} \label{sec:PattMatch}

This section covers the second and third steps in the stage of PME generation (Figure~\ref{fig:stepsPME}).
Given a partitioned postcondition,
within the {\it Matrix Arithmetic} step, 
symbolic arithmetic is performed and the = operator is distributed over the partitions, 
originating multiple equations.
In Equation~\eqref{eqn:matArit1}, we display the result of these actions for the Cholesky factorization,
where the symbol $\star$ means that the equation in the top-right quadrant
is the transpose of the one in the bottom-left quadrant.

{\small
\vspace{-2mm}
\setlength\extrarowheight{2pt}
\begin{equation}
  \renewcommand{\arraystretch}{1.0}
  \left( \begin{array}{@{}c@{\,}|@{\,}c@{}} L_{TL} & 0 \\\hline L_{BL} & L_{BR} \end{array} \right)
  \left( \begin{array}{@{}c@{\,}|@{\,}c@{}} L_{TL}^{T} & L_{BL}^{T} \\\hline 0 & L_{BR}^{T} \end{array} \right)
  =
  \left( \begin{array}{@{}c@{\,}|@{\,}c@{}} A_{TL} & A_{BL}^T \\\hline A_{BL} & A_{BR} \end{array} \right)
  \; \Rightarrow \;
  \renewcommand{\arraystretch}{1.0}
  \left( \begin{array}{@{}c@{\,}|@{\,}c@{}} L_{TL} L_{TL}^T = A_{TL} &
      \star \\\hline
      L_{BL} L_{TL}^T = A_{BL} &
      L_{BL} L_{BL}^T + L_{BR} L_{BR}^T = A_{BR}
          \end{array} \right) .  
  \label{eqn:matArit1}
\end{equation}
} 

The last step, {\em Pattern Matching}, carries out an iterative process during
which \click{} finds the solution for each of the equations resulting from the previous step.
When such a solution is found, it is expressed as an assignment to the unknown(s) of 
an implicit or explicit function of known quantities.
The unknown(s) or output(s) are then labeled as computable 
(the system knows now the formula to compute them)
and become known quantities to the remaining equations.
Upon completion, the process delivers the sought-after PME.

Success of this step is dependent
on the ability to identify expressions with known structure and properties. 
In order to facilitate pattern matching, we force equations to be in
their {\em canonical form}. We state that an equation is in canonical
form if 
a) its left-hand side only consists of those terms that contain at least one unknown object,
and 
b) its right-hand side only consists of those terms that solely contain known objects. 

The iterative process comprises three separate actions:
1) algebraic manipulation:
The equations are rearranged in canonical form;
2) structural pattern matching:
Equations are matched against known patterns;
3) exposing new available operands:
Once a known pattern is matched,
the equation becomes an assignment, and
the unknown operands are flagged as known.

We clarify the iterative process by illustrating, action by action, how
\click{} works through the Cholesky factorization.
During the discussion, \known{green} and \unknown{red} are
used to highlight the \known{known} and \unknown{unknown} operands, respectively.
The {\bf first iteration} is depicted in
Box~\ref{box:it1}, in which the top-left formula (\ref{box:it1}\subref{sbox:it1a})
displays the initial state. Initially, parts of \known{$A$} are known, and parts of \unknown{$L$} are
unknown.

\begin{description}
\item[Algebraic manipulation] \hfill \\
All the equations in Box~\ref{box:it1}\subref{sbox:it1a} are in canonical form,
and no manipulation is required.

\item[Structural pattern matching] \hfill \\
The three equations are tested against known patterns.
\click{} recognizes the top-left quadrant,
which matches the pattern for the Cholesky factorization in Box~\ref{box:chol-patt}.
The system rewrites the equation as an assignment (Box~\ref{box:it1}\subref{sbox:it1b}),
and labels the output quantity, $L_{TL}$, as computable.

\item[Exposing new available operands] \hfill \\
Having matched the equation in the top-left quadrant, 
\click{} turns the unknown quantity \unknown{$L_{TL}$}
into \known{$L_{TL}$}, and propagates the information to all the other quadrants:
Box~\ref{box:it1}\subref{sbox:it1c}.
The first iteration ends.
\end{description}

\begin{mybox} \centering
\footnotesize
\setlength\extrarowheight{2pt}
\setlength\arraycolsep{0.5pt}
\subfloat[Initial state. No manipulation is required.]
{ \label{sbox:it1a}
	\begin{minipage}[t]{.45\textwidth}
        $\left( \begin{array}{c|c}
                         \unknown{L_{TL}} \unknown{L_{TL}^T} = \known{A_{TL}} &
                            \star \\\hline
                            \unknown{L_{BL}} \unknown{L_{TL}^T} = \known{A_{BL}} &
                            \unknown{L_{BL}} \unknown{L_{BL}^T} + \unknown{L_{BR}} \unknown{L_{BR}^{T}} = \known{A_{BR}}
         \end{array} \right) $
		\end{minipage}
}
\hspace{2em}
\subfloat[The top-left equation is identified as a Cholesky sub-problem.]
{ \label{sbox:it1b}
		\begin{minipage}[t]{.45\textwidth}
         $\left( \begin{array}{c|c}
                         \unknown{\boxed{L_{TL}}} := \Gamma(\known{A_{TL}}) &
                            \star \\\hline
                            \unknown{L_{BL}} \unknown{L_{TL}^T} = \known{A_{BL}} &
                            \unknown{L_{BL}} \unknown{L_{BL}^T} + \unknown{L_{BR}} \unknown{L_{BR}^{T}} = \known{A_{BR}}
         \end{array} \right) $
		\end{minipage}
}
\\
\subfloat[The operand $L_{TL}$ becomes known for the rest of equations. 
         ]
{ \label{sbox:it1c}
         $\left( \begin{array}{c|c}
                         \known{\boxed{L_{TL}}} := \Gamma(\known{A_{TL}}) &
                            \star \\\hline
                            \unknown{L_{BL}} \known{\boxed{L_{TL}^T}} = \known{A_{BL}} &
                            \unknown{L_{BL}} \unknown{L_{BL}^T} + \unknown{L_{BR}} \unknown{L_{BR}^{T}} = \known{A_{BR}}
         \end{array} \right) $
}
\caption{First iteration towards the PME generation.} \label{box:it1}
\end{mybox}

In this first iteration, one unknown operand, $L_{TL}$, has become known, 
and one equation has turned into an assignment.
The {\bf second iteration} is shown in Box~\ref{box:it2}.

\begin{description}
\item[Algebraic manipulation] \hfill \\
Box~\ref{box:it2}\subref{sbox:it2a}
reproduces the final state from the previous iteration.
The remaining equations are still in canonical form,
thus no operation takes place.

\item[Structural pattern matching] \hfill \\
Among the two outstanding equations,
the one in the bottom-left quadrant is identified (Box~\ref{box:it2}\subref{sbox:it2b}),
as it matches the pattern 
of a triangular system of equations with multiple right-hand sides ({\sc trsm}).
The pattern for a {\sc trsm} is
\begin{mybox}
	\small
\begin{verbatim}

          equal[ times[ X_, trans[L_] ], B_ ] /;
                isOutputQ[X] && isInputQ[L] && isInputQ[B] && 
                isLowerTriangularQ[L].
\end{verbatim}
\vspace{-3mm}
\end{mybox}

\noindent
For the sake of brevity,
we assume that \click{}
learned such pattern from a previous derivation; 
in practice, 
a nested task of PME generation could be initiated, 
yielding the required pattern.

\item[Exposing new available operands] \hfill \\
Once the {\sc trsm} is identified, the output operand $L_{BL}$ becomes
available and turns to green in the bottom-right quadrant
(Box~\ref{box:it2}\subref{sbox:it2c}).
\end{description}

\begin{mybox} \centering
\footnotesize
\setlength\extrarowheight{2pt}
\setlength\arraycolsep{0.5pt}
\subfloat[Initial state. No manipulation is required.]
{ \label{sbox:it2a}
	\begin{minipage}[t]{.5\textwidth}\centering
         $\left( \begin{array}{c|c}
                         \known{L_{TL}} := \Gamma(\known{A_{TL}}) &
                            \star \\\hline
                            \unknown{L_{BL}} \known{L_{TL}^T} = \known{A_{BL}} &
                            \unknown{L_{BL}} \unknown{L_{BL}^T} + \unknown{L_{BR}} \unknown{L_{BR}^{T}} = \known{A_{BR}}
         \end{array} \right) $
		 \end{minipage}
}
\subfloat[The equation in the bottom-left quadrant is identified as a triangular system of equations.]
{ \label{sbox:it2b}
	\begin{minipage}[t]{.45\textwidth}\centering
         $\left( \begin{array}{c|c}
                         \known{L_{TL}} := \Gamma(\known{A_{TL}}) &
                            \star \\\hline
                            \unknown{\boxed{L_{BL}}} := \known{A_{BL}} \known{L_{TL}^{-T}} &
                            \unknown{L_{BL}} \unknown{L_{BL}^T} + \unknown{L_{BR}} \unknown{L_{BR}^{T}} = \known{A_{BR}}
         \end{array} \right) $
		 \end{minipage}
}
\\
\subfloat[$L_{BL}$ becomes a known operand.] 
{ \label{sbox:it2c}
	\begin{minipage}[t]{.52\textwidth}\centering
         $\left( \begin{array}{@{}c@{\,}|@{\,}c@{}}
                         \known{L_{TL}} := \Gamma(\known{A_{TL}}) &
                            \star \\\hline
                            \known{\boxed{L_{BL}}} := \known{A_{BL}} \known{L_{TL}^{-T}} &
                            \known{\boxed{L_{BL}^{\phantom{T}}}} \known{\boxed{L_{BL}^T}} + \unknown{L_{BR}} \unknown{L_{BR}^{T}} = \known{A_{BR}}
         \end{array} \right) $
		 \end{minipage}
}
\caption{Second iteration towards the PME generation.} \label{box:it2}
\end{mybox}

The process continues until all the equations are turned into assignments. 
The third and {\bf final iteration} for the Cholesky factorization
is shown in Box~\ref{box:it3}.

\begin{description}
\item[Algebraic manipulation] \hfill \\
The bottom-right equation is not in canonical form anymore:
The product $L_{BL} L_{BL}^T$, now a known quantity,
does not lay in the right-hand side. A simple manipulation
brings the equation back to canonical form (Box~\ref{box:it3}\subref{sbox:it3a}).

\item[Structural pattern matching] \hfill \\
Only the equation in the bottom-right quadrant remains unprocessed.  
At a first glance, one might recognize a Cholesky factorization, but the corresponding
pattern in Box~\ref{box:CholOpDesc} requires $A$ to be SPD. The question is whether the expression $A_{BR} -
L_{BL} L_{BL}^T$ represents an SPD matrix. In
order to answer the question, \click{} applies rewrite rules and
symbolic simplifications.

In Section~\ref{subsec:automation}, we explained
that the following facts regarding the quadrants of $A$ are known:

\begin{minipage}{0.45\textwidth}
\begin{itemize}
\label{spdlist}
\item SPD( $A_{TL}$ )
\item SPD( $A_{BR}$ )
\end{itemize}
\end{minipage}
\begin{minipage}{0.45\textwidth}
\vspace{.5mm}
\begin{itemize}
\item SPD( $A_{TL} - A_{BL}^{T} A_{BR}^{-1} A_{BL}$ )
\item SPD( $A_{BR} - A_{BL} A_{TL}^{-1} A_{BL}^{T}$ )
\end{itemize}
\end{minipage}

\noindent
In order to determine whether $A_{BR} - L_{BL} L_{BL}^T$
is equivalent to any of the expressions listed above,
\click{} makes use of the knowledge acquired throughout the previous iterations.
Specifically, in the first two iterations it was discovered that
\begin{itemize}
\item $L_{TL} L_{TL}^T = A_{TL}$, and
\item $L_{BL} = A_{BL} L_{TL}^{-T}$.
\end{itemize}
Using these identities as rewrite rules, the expression $A_{BR} -
L_{BL} L_{BL}^T$ is manipulated: First, $L_{BL} =
A_{BL} L_{TL}^{-T}$ is used to replace the instances of $L_{BL}$,
yielding $A_{BR} - A_{BL} L_{TL}^{-T} L_{TL}^{-1} A_{BL}^T$, and
equivalently, $A_{BR} - A_{BL} (L_{TL} L_{TL}^{T})^{-1}
A_{BL}^T$; then, by virtue of the identity $L_{TL} L_{TL}^T = A_{TL}$,
$L_{TL} L_{TL}^T$ is replaced by $A_{TL}$, yielding $A_{BR} - A_{BL}
A_{TL}^{-1} A_{BL}^{T}$, which is known to be SPD.
Now that \click{} can assert the SPDness of $A_{BR} - L_{BL} L_{BL}^T$,
it
successfully associates the
equation in the bottom-right quadrant with the pattern for a Cholesky factorization,
and $L_{BR}$ is labeled as computable.

\item[Exposing new available operands] \hfill \\
Once the expression in the bottom-right quadrant is identified, 
the system exposes the quantity $L_{BR}$ as known.
Since no equation is left, the process completes and the PME---formed by the three
assignments---is returned as output. 
\end{description}

\begin{mybox} \centering
\footnotesize
\setlength\extrarowheight{2pt}
\setlength\arraycolsep{0.5pt}
\subfloat[Simple algebraic manipulation takes the bottom-right equation back to canonical form.]
{ \label{sbox:it3a}
	\centering
	\begin{minipage}[t]{.45\textwidth}\centering
         $\left( \begin{array}{c|c}
                         \known{L_{TL}} := \Gamma(\known{A_{TL}}) &
                            \star \\\hline
                            \known{L_{BL}} := \known{A_{BL}} \known{L_{TL}^{-T}} &
                            \unknown{L_{BR}} \unknown{L_{BR}^{T}} = \known{A_{BR}} - \known{L_{BL}} \known{L_{BL}^T}
         \end{array} \right) $
		 \end{minipage}
}
\hspace{2em}
\subfloat[The bottom-right equation is identified as a Cholesky factorization.]
{ \label{sbox:it3b}
	\centering
	\begin{minipage}[t]{.45\textwidth}\centering
         $\left( \begin{array}{c|c}
                         \known{L_{TL}} := \Gamma(\known{A_{TL}}) &
                            \star \\\hline
                            \known{L_{BL}} := \known{A_{BL}} \known{L_{TL}^{-T}} &
                            \unknown{\boxed{L_{BR}}} := \Gamma(\known{A_{BR}} - \known{L_{BL}} \known{L_{BL}^T})
         \end{array} \right) $
		 \end{minipage}
}
\\
\subfloat[The operand $L_{BR}$ becomes known.]
{ \label{sbox:it3c}
	\centering
	\begin{minipage}[t]{.45\textwidth}\centering
         $\left( \begin{array}{c|c}
                         \known{L_{TL}} := \Gamma(\known{A_{TL}}) &
                            \star \\\hline
                            \known{L_{BL}} := \known{A_{BL}} \known{L_{TL}^{-T}} &
                            \known{\boxed{L_{BR}}} := \Gamma(\known{A_{BR}} - \known{L_{BL}} \known{L_{BL}^T})
         \end{array} \right) $
		 \end{minipage}
}
\caption{Final iteration towards the PME generation.} \label{box:it3}
\end{mybox}

By means of the described process, PMEs for a target equation are automatically
generated. The PME for the Cholesky factorization is given in Box~\ref{box:PMEChol}.
We point out that the decomposition encoded by the PME is correct
independently of the size of the quadrants 
(as long as $A_{TL}$ and $L_{TL}$ are square, as specified by the initial partitioning of the matrices $L$ and $A$).

\begin{mybox} \centering
	\renewcommand{\arraystretch}{1.4}
	$$\left( {\begin{array}{@{\,}c@{\,}|@{\,}c@{\,}} 
			L_{TL} := \Gamma(A_{TL}) &
			\star \\\hline
			L_{BL} := A_{BL} L_{TL}^{-T} &
			L_{BR} := \Gamma(A_{BR} - L_{BL} L_{BL}^T)
	 \end{array}} \right)$$ \\
	\caption{Partitioned Matrix Expression for the Cholesky factorization.} \label{box:PMEChol}
\end{mybox}

Before proceeding with the second stage ({\em Loop Invariant Identification}),
we briefly discuss 
the existence of multiple PMEs for a single operation, and
the relation of the object PME with recursive divide-and-conquer algorithms.

\subsection{Non-uniqueness of the PME}
\label{subsec:NonUniquePME}

For the Cholesky factorization, \click{} identifies that only one set
of partitioning rules is feasible (Table~\ref{tab:partPostcond}),
which corresponds to one way of decomposing the problem and 
to the generation of one PME. 
In general, the PME {\it is not unique} since, for one target operation,
multiple sets of viable rules may be found,
each of them leading to a different problem decomposition and a different PME. 
To illustrate such a situation, we look once more at the triangular Sylvester equation
(Box~\ref{box:sylvdesc}).

The procedure described in Section~\ref{subsec:automation} is used to
obtain the sets of admissible partitioning rules, listed in
Table~\ref{tab:sylvPart}.
Each of them is then applied to the postcondition equation,
obtaining three different partitioned postconditions, as shown
in Table~\ref{tab:SylvPMEs} (left).
By applying the iterative process
described in Section~\ref{sec:PattMatch}, 
three PMEs are generated: Table~\ref{tab:SylvPMEs}
(right). In Box~\ref{box:sylvSteps}, we illustrate the steps performed by \click{} to
transform the second partitioned postcondition into a PME.

\begin{sidewaystable}
\centering
\scriptsize
\begin{tabular}{c | c | c} \toprule
\renewcommand{\arraystretch}{1.4}
{\bf \#} & {\bf \small Partitioned Postcondition} & {\bf \small Partitioned Matrix Expression} \\ \midrule
1 &
$
\renewcommand{\arraystretch}{1.2}
		\left( A \right) 
		\left( \begin{array}{@{}c@\;|@\;c@{}} X_{L} & X_{R} \end{array} \right) 
		+
		\left( \begin{array}{@{}c@\;|@\;c@{}} X_{L} & X_{R} \end{array} \right) 
		\left( \begin{array}{@{}c@\;|@\;c@{}} B_{TL} & B_{TR} \\\hline 0 & B_{BR} \end{array} \right) 
		=
		\left( \begin{array}{@{}c@\;|@\;c@{}} C_{L} & C_{R} \end{array} \right)
$ & 
$
\renewcommand{\arraystretch}{1.2}
		\left( {\begin{array}{@{}c@\;|@\;c@{}}
			X_{L} := \Omega(A, B_{TL}, C_{L}) &
			X_{R} := \Omega(A, B_{BR}, C_{R} - X_{L} B_{TR})
		\end{array}} \right)
$ \\[8mm]
2 &
$
\renewcommand{\arraystretch}{1.2}
		\myFlaTwoByTwo{A_{TL}}{A_{TR}}{0}{A_{BR}}
		\left( \begin{array}{@{}c@{}} X_{T} \\\hline X_{B} \end{array} \right) 
		+
		\left( \begin{array}{@{}c@{}} X_{T} \\\hline X_{B} \end{array} \right) 
		\left( B \right) 
		=
		\left( \begin{array}{@{}c@{}} C_{T} \\\hline C_{B} \end{array} \right) 
$ & 
$
\renewcommand{\arraystretch}{1.2}
		\left( {\begin{array}{@{}c@{}} 
			X_{T} := \Omega(A_{TL}, B, C_{T} - A_{TR} X_{B}) \\\hline
			X_{B} := \Omega(A_{BR}, B, C_{B})
		\end{array}} \right)
		$ \\[8mm]
3 &
$
\renewcommand{\arraystretch}{1.2}
		\myFlaTwoByTwo{A_{TL}}{A_{TR}}{0}{A_{BR}}
		\myFlaTwoByTwo{X_{TL}}{X_{TR}}{X_{BL}}{X_{BR}}
		+
		\myFlaTwoByTwo{X_{TL}}{X_{TR}}{X_{BL}}{X_{BR}}
		\myFlaTwoByTwo{B_{TL}}{B_{TR}}{0}{B_{BR}}
		=
		\left( \begin{array}{@{}c@\;|@\;c@{}} C_{TL} & C_{TR} \\\hline C_{BL} & C_{BR} \end{array} \right)
$ & 
$
\renewcommand{\arraystretch}{1.2}
	\left( {\begin{array}{@{}c@\;|@\;c@{}} 
		X_{TL} := \Omega(A_{TL}, B_{TL}, C_{TL} - A_{TR} X_{BL}) & 
		\begin{aligned}
		X_{TR} := \Omega(& A_{TL}, B_{BR}, \\
			   & C_{TR} - A_{TR} X_{BR} - X_{TL} B_{TR}) 
		\end{aligned} \\\hline
		X_{BL} := \Omega(A_{BR}, B_{TL}, C_{BL}) &
		X_{BR} := \Omega(A_{BR}, B_{BR}, C_{BR} - X_{BL} B_{TR})
	\end{array}} \right)
$ \\[4mm]
\bottomrule
\end{tabular}
\caption{Partitioned postconditions and Partitioned Matrix Expressions for the triangular Sylvester equation.} \label{tab:SylvPMEs}
\end{sidewaystable}

\begin{mybox*} 
\centering
	\small
	\renewcommand{\arraystretch}{1.4}
	\subfloat[Initial state.]
	{ \label{sbox:sylvStep1}
		\begin{minipage}{5.0cm}\centering
		$\left( \begin{array}{c}
				\known{A_{TL}} \unknown{X_{T}} + \known{A_{TR}} \unknown{X_{B}} + \unknown{X_{T}} \known{B} = \known{C_{T}} \\\hline
				\known{A_{BR}} \unknown{X_{B}} + \unknown{X_{B}} \known{B} = \known{C_{B}}
		\end{array} \right) $
		\end{minipage}
	}
	\hspace{2em}
	\subfloat[The bottom equation is identified as a Sylvester equation, where all the input operands are known.]
	{ 
	    \label{sbox:sylvStep2}
		\begin{minipage}{5.2cm}\centering
		$\left( \begin{array}{c}
				\known{A_{TL}} \unknown{X_{T}} + \known{A_{TR}} \unknown{X_{B}} + \unknown{X_{T}} \known{B} = \known{C_{T}} \\\hline
				\unknown{\boxed{X_{B}}} := \Omega(\known{A_{BR}}, \known{B}, \known{C_{B}})
		\end{array} \right) $
		\end{minipage}
	}
	\\\vspace{1em}
	\subfloat[The operand $X_{T}$ becomes available for the equation in the top quadrant.]
	{ \label{sbox:sylvStep3}
		\begin{minipage}{5.5cm}\centering
		$\left( \begin{array}{c}
				\known{A_{TL}} \unknown{X_{T}} + \known{A_{TR}} \known{\boxed{X_{B}}} + \unknown{X_{T}} \known{B} = \known{C_{T}} \\\hline
				\known{\boxed{X_{B}}} := \Omega(\known{A_{BR}}, \known{B}, \known{C_{B}})
		\end{array} \right) $
		\end{minipage}
	}
	\hspace{2em}
	\subfloat[State after algebraic manipulation.]
	{ \label{sbox:sylvStep4}
		\begin{minipage}{5.0cm}\centering
		$\left( \begin{array}{c}
				\known{A_{TL}} \unknown{X_{T}} + \unknown{X_{T}} \known{B} = \known{C_{T}} - \known{A_{TR}} \known{X_{B}} \\\hline
				\known{X_{B}} := \Omega(\known{A_{BR}}, \known{B}, \known{C_{B}})
		\end{array} \right) $
		\end{minipage}
	}
	\\\vspace{1em}
	\subfloat[The equation in the top quadrant is also identified as a Sylvester equation.]
	{ \label{sbox:sylvStep5}
		\begin{minipage}{5.2cm}\centering
		$\left( \begin{array}{c}
				\unknown{\boxed{X_{T}}} := \Omega(\known{A_{TL}}, \known{B}, \known{C_{T}} - \known{A_{TR}} \known{X_{B}}) \\\hline
				\known{X_{B}} := \Omega(\known{A_{BR}}, \known{B}, \known{C_{B}})
		\end{array} \right) $
		\end{minipage}
	}
	\hspace{2em}
	\subfloat[Resulting PME.]
	{ \label{sbox:sylvStep6}
		\begin{minipage}{5.5cm}\centering
		$\left( \begin{array}{c}
				X_{T} := \Omega(A_{TL}, B, C_{T} - A_{TR} X_{B}) \\\hline
				X_{B} := \Omega(A_{BR}, B, C_{B})
		\end{array} \right) $
		\end{minipage}
	}
	\caption{Generation of the PME corresponding to the second partitioned postcondition (Table~\ref{tab:SylvPMEs})
	         of the Sylvester equation.} \label{box:sylvSteps}
\end{mybox*}

\subsubsection{Learning the PMEs}

As we discuss in the {\em Loop Invariant Identification} stage,
the generated PMEs represent a pool of loop invariants. 
However, this is not the only place
where the information encoded in the PMEs is used.
In the last stage of the algorithm generation process,
the {\em Algorithm Construction}, \click{} makes use
of the PMEs 
to rewrite expressions involving partitioned operands
into multiple expressions. For instance, the triangular Sylvester
equation in its implicit form $X = \Omega(A, B, C)$, with the following combination
of partitioned operands:
$$
\myFlaTwoByOne{X_T}{X_B} := \Omega \left(
	   \myFlaTwoByTwo{A_{TL}}{A_{TR}}{0}{A_{BR}},
	   (B),
	   \myFlaTwoByOne{C_T}{C_B}
\right),
$$
needs to be rewritten as
$$
\left( {\begin{array}{@{}c@{}} 
	X_{T} := \Omega(A_{TL}, B, C_{T} - A_{TR} X_{B}) \\\hline
	X_{B} := \Omega(A_{BR}, B, C_{B})
\end{array}} \right).
$$

To enable such transformations, 
\click{} produces one rewrite rule per PME,
and incorporates them into its knowledge-base: 
For the example of the Sylvester equation, the system translates the PMEs
in Table~\ref{tab:SylvPMEs}, into the rewrite rules in Box~\ref{box:sylv-PME-rules}.

\begin{mybox}
\scriptsize
\setlength{\arraycolsep}{0pt}
\hspace*{-6mm}
\begin{minipage}{\textwidth}
\begin{enumerate}
	\item $ \myFlaOneByTwo{X_L}{X_R} := \Omega \left( \hspace{-0.7mm}
	                                       (A),
										   \myFlaTwoByTwo{B_{TL}}{B_{TR}}{0}{B_{BR}},
										   \myFlaOneByTwo{C_L}{C_R} \hspace{-0.7mm}
	                                    \right)
			\longrightarrow
		\left( {\begin{array}{@{}c@\;|@\;c@{}}
			X_{L} := \Omega(A, B_{TL}, C_{L}) &
			X_{R} := \Omega(A, B_{BR}, C_{R} - X_{L} B_{TR})
		\end{array}} \right)
		$\\[2mm]

	\item $ \myFlaTwoByOne{X_T}{X_B} =: \Omega \left(
										   \myFlaTwoByTwo{A_{TL}}{A_{TR}}{0}{A_{BR}},
	                                       (B),
										   \myFlaTwoByOne{C_T}{C_B}
	                                    \right)
			\longrightarrow 
		\left( {\begin{array}{@{}c@{}} 
			X_{T} := \Omega(A_{TL}, B, C_{T} - A_{TR} X_{B}) \\\hline
			X_{B} := \Omega(A_{BR}, B, C_{B})
		\end{array}} \right)
		$\\[2mm]

	\item $ \myFlaTwoByTwo{X_{TL}}{X_{TR}}{X_{BL}}{X_{BR}} := \Omega \left(
															   \myFlaTwoByTwo{A_{TL}}{A_{TR}}{0}{A_{BR}},
															   \myFlaTwoByTwo{B_{TL}}{B_{TR}}{0}{B_{BR}},
															   \myFlaTwoByTwo{C_{TL}}{C_{TR}}{C_{BL}}{C_{BR}}
															\right)
			\longrightarrow \\[3mm]
		\hspace*{1cm} 
	\left( {\begin{array}{@{}c@\;|@\;c@{}} 
		X_{TL} := \Omega(A_{TL}, B_{TL}, C_{TL} - A_{TR} X_{BL}) & 
		X_{TR} := \Omega(A_{TL}, B_{BR}, C_{TR} - A_{TR} X_{BR} - X_{TL} B_{TR}) \\\hline
		X_{BL} := \Omega(A_{BR}, B_{TL}, C_{BL}) &
		X_{BR} := \Omega(A_{BR}, B_{BR}, C_{BR} - X_{BL} B_{TR})
	\end{array}} \right) \\[2mm]
		  $
\end{enumerate}
\end{minipage}
\caption{Rewrite rules representing the knowledge acquired by {\sc Cl1ck} after
generating the three PMEs for the Sylvester equation.}
\label{box:sylv-PME-rules}
\end{mybox}

\subsection{Recursive algorithms}

The concept of PME leads naturally to recursive divide-and-conquer algorithms.
Such algorithms consist of three main parts:
The decomposition of the problem into smaller sub-problems,
the computation of these sub-problems, and
the composition of the solution from the partial results.
In this scheme, the PME acts as the decomposition operator, 
showing how to decompose the problem into sub-problems;
then, each partial result is obtained, and
the output matrix is composed by assembling its subparts.

In the case of Cholesky, the PME determines that the operation may be computed as 
1) a recursive call involving smaller matrices, followed by 
2) the solution of a triangular system, and 
3) one more recursive call applied to a matrix product.
The base case of the recursion involves the corresponding scalars from
$L$ and $A$, $\lambda := \Gamma(\alpha)$.
This base case is equivalent to the equation 
$\lambda \times \lambda = \alpha$ for unknown $\lambda$,
whose solution is $\lambda := \sqrt{\alpha}$.
Algorithm~\ref{alg:rec-chol} presents such a recursive algorithm
for the computation of the Cholesky factorization.

\begin{center}
\renewcommand{\lstlistingname}{Algorithm}
\begin{minipage}{0.90\linewidth}
	\begin{lstlisting}[caption={Recursive algorithm to compute the Cholesky factorization.}, escapechar=!, label=alg:rec-chol]
	L := Cholesky(A)
	  if size(A) is !$1\times1$!
	  then
	    L := !$\sqrt{\text A}$!
	  else
	    (n,n) := size(A)
	    !$L \rightarrow \myFlaTwoByTwo{L_{TL}}{0}{L_{BL}}{L_{BR}}$!, with !$L_{TL} \in R^{\frac{n}{2} \times \frac{n}{2}}$!
	    !$A \rightarrow \myFlaTwoByTwo{A_{TL}}{A_{TR}}{A_{BL}}{A_{BR}}$!, with !$A_{TL} \in R^{\frac{n}{2} \times \frac{n}{2}}$!

	    !$L_{TL}$! := Cholesky(!$A_{TL}$!)
	    !$L_{BL}$! := !$A_{BL} L_{TL}^{-T}$!
	    !$L_{BR}$! := Cholesky(!$A_{BR} - L_{BL} L_{BL}^T$!)

	    !$L \leftarrow \myFlaTwoByTwo{L_{TL}}{0}{L_{BL}}{L_{BR}}$!
	  end
	\end{lstlisting}
\end{minipage}
\end{center}

While it is worth noticing that once the PME is generated,
recursive algorithms can already be derived, 
for performance reasons we focus on the derivation of families of loop-based algorithms.

\section{Loop invariant identification}
We focus now on the second stage of the algorithm generation process, the
Loop Invariant Identification.
We recall that a loop invariant expresses the contents
of the output matrices (the state of the computation)
at different points of a loop.
Inherently, a loop invariant describes an intermediate
result towards the complete computation of the target operation.
Thus, loop invariants can be identified by selecting different
subsets of the operations in the PME that satisfy
the data dependencies.

\begin{figure}
  \centering
  \includegraphics[scale=1]{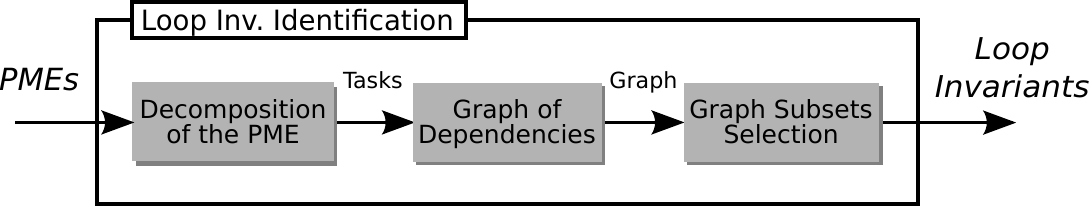}
  \caption{Steps for the identification of loop invariants from a PME.} \label{fig:stepsLInv}
\end{figure}

In this second stage, \click{} takes the PME(s) of the target equation, 
and produces a family of loop invariants.
The identification of loop invariants consists of three steps (Figure~\ref{fig:stepsLInv}):
1) each of the assignments in the PME is decomposed into its
building blocks, the {\em tasks};
2) an analysis of dependencies among tasks is carried out
to build a dependency graph;
3) the graph is traversed, selecting all possible subgraphs that
satisfy the dependencies.
The subgraphs correspond to predicates that are candidates to
becoming loop invariants. \click{} checks the feasibility of such predicates,
discarding the non-feasible ones and promoting the remaining ones to loop invariants.

\subsection{Decomposition of the PME}

\click{} commences by analyzing the assignments in the PME.
All assignments share the same structure: the left-hand
side includes one or more output quantities, which are computed
according to the expression on the right-hand side.
Similarly to the approach discussed in Chapter~\ref{ch:compiler},
\click{} decomposes the right-hand sides into one or more
building blocks. 

Here, additionally to
kernels such as matrix products and additions, the expressions
to decompose may also include implicit functions. 
Functions will be considered building blocks; 
when one or more of the arguments is an expression, 
the expression is also decomposed.
We illustrate the application of the decomposition rules 
by example. The following patterns will arise in the discussion:

\begin{enumerate}
	\item {\tt times[inv[A\_], B\_] /; isTriangularQ[A]}
	\item {\tt times[B\_, inv[A\_]] /; isTriangularQ[A]}
	\item {\tt plus[times[A\_, B\_], C\_]}
	\item {\tt f\_[$x_1$, ..., $x_n$] /; $\forall_i$ isOperandQ[$x_i$]}
	\item {\tt f\_[$x_1$, ..., $x_n$] /; $\exists_i$ isExpressionQ[$x_i$]}.
\end{enumerate}
The first two patterns represent triangular systems ({\sc trsm}),
the third one corresponds to a matrix product $A B + C$ ({\sc gemm}),
the fourth matches functions where all arguments are simple operands, and
the fifth matches functions where at least one argument is an expression.

\vspace{1mm}
As an example, we choose the \lu{}, which is defined as
{\footnotesize
$$
\{L, U\} := LU(A) \equiv
\left\{
\begin{aligned}
\PPre: \{ & \prop{Output}{L} \, \wedge \, \prop{Matrix}{L} \, \wedge \, \prop{LowerTriangular}{L} \, \wedge \\ 
	      & \prop{UnitDiagonal}{L} \, \wedge \, \prop{Output}{U} \, \wedge \, \prop{Matrix}{U} \, \wedge \\
          & \prop{UpperTriangular}{U} \, \wedge \, \prop{Input}{A}  \, \wedge \, \prop{Matrix}{A} \, \wedge \, \prop{\exists \; LU}{A} \} \\
\\
\PPost: \{ &  L U = A \}
\end{aligned}
\right. .
$$}

\noindent
The corresponding PME comprises four assignments:

	$$\myFlaTwoByTwo{\{L_{TL}, U_{TL}\} := LU(A_{TL})}
	              {U_{TR} := L_{TL}^{-1} A_{TR}}
				  {L_{BL} := A_{BL} U_{TL}^{-1}}
				  {\{L_{BR}, U_{BR}\} := LU(A_{BR} - L_{BL} U_{TR})}.$$

The decomposition of the assignments can be performed
independently from one another; \click{} arbitrarily
commences from the top-left quadrant:
{\small $\{L_{TL}, U_{TL}\} := LU(A_{TL}).$}
Since the right-hand side matches the pattern 
associated to a function with simple operands,
no decomposition is necessary; the system returns one task:
{\small $\{L_{TL}, U_{TL}\} := LU(A_{TL}).$}

The analysis proceeds with the top-right quadrant:
{\small $U_{TR} := L_{TL}^{-1} A_{TR}.$}
The expression is identified as a {\sc trsm} operation;\footnote{%
The triangularity of $L_{TL}$ is inferred during
the partitioning of $L$.} \click{} recognizes it as a basic task and returns it.
Similarly, in the bottom-left quadrant a third task is matched and yielded.

Only one assignment remains to be studied:
{\small $\{L_{BR}, U_{BR}\} := LU(A_{BR} - L_{BL} U_{TR})$},
whose right-hand side corresponds to a function with an expression
as input argument.
In such a case, the system first decomposes the expression,
yielding a number of tasks, and then returns the function itself as a task.
The expression {\small $A_{BR} - L_{BL} U_{TR}$} matches the pattern for 
a matrix product ({\sc gemm}), corresponding
to a basic task. Accordingly, \click{} returns two tasks:
{\small $A_{BR} := A_{BR} - L_{BL} U_{TR}$, and
$\{L_{BR}, U_{BR}\} := LU(A_{BR})$}.
In total, the following five tasks are produced:

\begin{enumerate}
\small
\item $\{L_{TL}, U_{TL}\} := LU(A_{TL});$ \\[-5mm]
\item $U_{TR} := L_{TL}^{-1} A_{TR};$ \\[-5mm]
\item $L_{BL} := A_{BL} U_{TL}^{-1};$ \\[-5mm]
\item $A_{BR} := A_{BR} - L_{BL} U_{TR};$ \\[-5mm]
\item $\{L_{BR}, U_{BR}\} := LU(A_{BR}).$ \\[-5mm]
\end{enumerate}

\subsection{Graph of dependencies}

Once the decomposition into tasks is available, \click{} proceeds with the study of
the dependencies among them. Three different kinds of dependencies may occur.

\begin{itemize}
\item {\bf True dependency.} One of the input arguments of a task is 
also the result of a previous task:
$$
  \begin{array}{@{}c@{\;}c@{\;}c@{}}
    A & := & B + C \\
    X & := & A + D
  \end{array}
$$
The order of the assignments cannot be reversed because the second
one requires the value of $A$ computed in the first one.

\item {\bf Anti dependency.} One of the input arguments of a task is 
also the result of a subsequent task:
$$
  \begin{array}{@{}c@{\;}c@{\;}c@{}}
    X := A + D \\
    A := B + C 
  \end{array}
$$
The order of the statements cannot be reversed because the
first one needs the value of $A$ before the second one overwrites it.

\item {\bf Output dependency.} The result of a task is 
also the result of a different task:
$$
  \begin{array}{@{}c@{\;}c@{\;}c@{}}
    A := B + C \\
    A := D + E
  \end{array}
$$
The second assignment cannot be performed until the first is computed to
ensure the correct final value of $A$.
\end{itemize}

Since, in general, there is no explicit ordering among the produced tasks,
the distinction between true and anti dependencies
is not straightforward.
However, since assignments from different quadrants
compute different parts of the output matrices,
any time the output of a statement is
found as an input argument of another one, it
implies a true dependency: first the quantity is computed,
then it is used elsewhere.
The only exception is the case when the occurrence of the
operand as input is labeled with a hat, refering to
the initial contents of an overwritable operand.
For instance, given the following pair of tasks:
$$
  \begin{array}{l}
	X_{BL} = \hat{X}_{BL} - A_{BL} B_{TL} \\[2mm]
	X_{BR} = \hat{X}_{BR} - \hat{X}_{BL} B_{TR},
  \end{array}
$$
the second needs the initial contents of $X_{BL}$
before it is overwritten by the first,
thus imposing an anti dependency.

Similarly, it is not easy to
distinguish the direction of an output dependency.
Since output dependencies only occur among
tasks belonging to the same quadrant (each
quadrant writes to a different part of the output
matrices), the order is determined because one of 
the involved tasks comes from the decomposition of the other
one, imposing an order in their execution.

We detail the analysis of the dependencies following the
example of the \lu{}.
During the analysis we use {\bf boldface} to highlight
the dependencies. 
The study commences with Task 1, 
whose output is {\small $\{L_{TL}, U_{TL}\}$}. \click{}
finds that the operands $L_{TL}$ and $U_{TL}$
are input arguments for 
Tasks 2 and 3, respectively.

\begin{enumerate}
\item $\mathbf{\{L_{TL}, U_{TL}\}} := LU(A_{TL})$ \\[-5mm]
\item $U_{TR} := \mathbf{L_{TL}}^{-1} A_{TR}$ \\[-5mm]
\item $L_{BL} := A_{BL} \mathbf{U_{TL}}^{-1}$ \\[-5mm]
\end{enumerate}
This means that two true dependencies exist: one from 
Task 1 to Task 2 and another from Task 1 to Task 3.
Next, \click{} inspects Task 2, whose output is
$U_{TR}$. $U_{TR}$ is also identified as input 
for Task 4. 
\begin{enumerate}
\setcounter{enumi}{1}
\item $\mathbf{U_{TR}} := L_{TL}^{-1} A_{TR}$ \\[-5mm]
\setcounter{enumi}{3}
\item $A_{BR} := A_{BR} - L_{BL} \mathbf{U_{TR}} $ \\[-5mm]
\end{enumerate}
Hence, a true dependency from Tasks 2 to 4 is
imposed.
A similar situation arises when inspecting
Task 3, originating a true dependency from Task 3 to Task 4.

The analysis continues with Task 4; this computes an update
of $A_{BR}$, which is then used as input by Task 5,
thus, creating one more true dependency.
\begin{enumerate}
\setcounter{enumi}{3}
\item $\mathbf{A_{BR}} := A_{BR} - L_{BL} U_{TR} $ \\[-5mm]
\item $\{L_{BR}, U_{BR}\} := LU(\mathbf{A_{BR}}) $ \\[-5mm]
\end{enumerate}
Task 5 remains to be analyzed. Since its output, $\{L_{BR}, U_{BR}\}$, 
does not appear in any of the other tasks, no new dependencies are
found.

In Figure~\ref{fig:luGraph}, the list of the dependencies for the \lu{}
is mapped onto a graph in which 
node $i$ represents Task $i$.
\begin{figure}
	\centering
	\depGraphLU{rwthblue}{rwthblue}{rwthblue}{rwthblue}{rwthblue}
	\caption{Final graph of dependencies for the \lu{}.} \label{fig:luGraph}
\end{figure}
We note that, by construction of the PME
---where equations are matched individually, not allowing
interdependencies among assignments---,
no cyclic dependencies may arise among tasks; thus,
the resulting graph is a direct acyclic graph 
(DAG).

\subsection{DAG subsets selection} \label{sec:depGraph}

Once \click{} has generated the dependency graph, it
selects all the possible subgraphs that satisfy the dependencies.
Each of them corresponds to a different loop invariant,
provided that it is feasible. A loop invariant is feasible
if it satisfies a number of constraints imposed by the FLAME 
methodology.

\begin{figure}
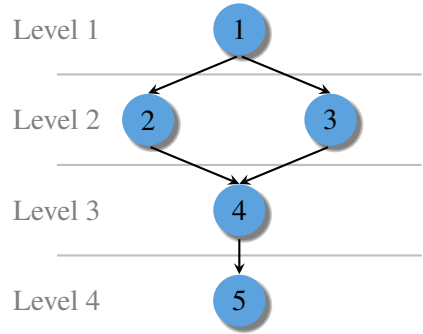

	\centering
	\LUGraphLevels{rwthblue}
	\caption{Graph of dependencies for the \lu{} with its nodes sorted by levels according to the distance from the root node.} 
	\label{fig:luGraphLevels}
\end{figure}

\click{} finds all possible subgraphs by means of Algorithm~\ref{alg:subDAG}.
The algorithm starts by sorting the nodes in the 
graph; since the graph is a DAG,
the nodes may be sorted by levels according to the longest path
from the root. For the \lu{} the sorted DAG is shown in 
Figure~\ref{fig:luGraphLevels}.
Then, the algorithm creates the list of subgraphs of the DAG incrementally,
by levels. At first it initializes the
list of subgraphs with the empty subset,
$l = [\{\}]$, which is equivalent to selecting none of 
the PME tasks. Then, at each level it extends the set
of subgraphs by adding all those resulting from appending accesible nodes
to the existing subgraphs.
A node at a given level is accesible from a
subgraph $sg$ if all the dependencies of the node are satisfied by $sg$.

\begin{center}
	\renewcommand{\lstlistingname}{Algorithm}
	\begin{minipage}{0.85\textwidth}
		\begin{lstlisting}[caption=Generation of all the subgraphs of a DAG g.,label=alg:subDAG]
		l = [{}]
		g' = sortByLevels(g)
		for each level i in g':
		  for each subgraph sg in l:
		    acc = accesibleNodesFrom(sg, g', i)
		    subsets = nonEmptySubsets(acc)
		    for each subset ss in subsets:
		      l = append(l, union(sg, ss))
		  end
		end
		\end{lstlisting}
	\end{minipage}
\end{center}

In the first iteration of the $LU$ example, 
the only accesible node from $\{\}$ at level 1 is node 1, 
hence, union($\{\}$, $\{1\}$) is added to
$l$, which becomes $[\{\}, \{1\}]$. Now, the
level is increased to 2; no node in level 2 is
accesible from $\{\}$, while both nodes 2 and 3
are accesible from $\{1\}$.
The union of $\{1\}$ with the non-empty subsets of $\{2, 3\}$ ---$\{2\}$, 
$\{3\}$ and $\{2, 3\}$--- are added to $l$, resulting in
$l = [\{\}, \{1\}, \{1, 2\}, \{1, 3\}, \{1, 2, 3\}]$.
At level 3, \click{} discovers that node 4 is accesible from 
subgraph $\{1, 2, 3\}$, thus $\{1, 2, 3, 4\}$ is added
to $l$. Finally, node 5 is accesible from $\{1, 2, 3, 4\}$.
The final list of subgraphs is:
$$[\{\}, \{1\}, \{1, 2\}, \{1, 3\}, \{1, 2, 3\}, \{1, 2, 3, 4\}, \{1, 2, 3, 4, 5\}].$$

\subsubsection{Checking the feasibility of the loop invariants}

The seven subgraphs included in the final list correspond to
predicates that are candidates
to becoming loop invariants. 
As a final step to complete the identification of loop invariants,
\click{} must check each predicate to establish its feasibility.

The FLAME methodology (see the skeleton in Figure~\ref{fig:skeleton}) imposes two 
constraints for such a predicate to be a feasible loop invariant ($\PInv$):
\begin{enumerate}
	\item[1)] There must exist a basic initialization of the operands, 
	i.e., an initial partitioning, that renders the predicate $\PInv$ true:
	\begin{center}
	\fbox{
		\begin{tabular}{ l }
		  \\[-2mm]
		  $\{ \PPre \}$ \\[1.5mm]
		  {\bf Partition} \\[1.5mm]
		  $\{ \PInv \}.$ \\[1.5mm]
		\end{tabular}
	}
	\end{center}
	\item[2)] $\PInv$ and the negation of the loop guard, $G$, must imply the 
		postcondition, $\PPost$: \\[-8mm]
	\begin{center}
	\fbox{
		$\PInv \wedge \neg G \implies \PPost$.
	}
	\end{center}
\end{enumerate}

The partitioning and the traversal of the operands play a central
role in checking the feasibility of loop invariants. 
The partitionings were already fixed in
the previous stage ({\em PME generation});
however, the traversal of the operands is not determined by their partitioning,
and thus is yet to be established.
In the LU example, all three operands, ---$L$, $U$, and $A$---
are partitioned in $2\times2$ quadrants.
In principle, each of the operands can be traversed in one of four ways:
from the top-left to the bottom-right corners (\fbox{$\searrow$}),
bottom-left to top-right (\fbox{$\nearrow$}),
bottom-right to top-left (\fbox{$\nwarrow$}), and
top-right to bottom-left (\fbox{$\swarrow$}).
\click{} determines the traversal of the operands based on the 
sorted DAG (Figure~\ref{fig:luGraphLevels}): 
The DAG indicates that computation starts
from Task 1 ($\{L_{TL}, U_{TL}\} := LU(A_{TL})$), which involves the top-left quadrants of the operands
($L_{TL}$, $U_{TL}$, and $A_{TL}$);
thus the three operands are traversed from the top-left to 
the bottom-right corners.

The system is now ready to check the feasibility of the
seven predicates.
We illustrate the process by example; we use the
the predicate corresponding to the subgraph \{1, 2\}:
\begin{equation}
	\left( 
	  \begin{array}{@{\,}c@{\,}|@{\,}c@{\,}}
	  \{ L_{TL}, U_{TL} \} = LU(A_{TL})   & U_{TR} = L_{TL}^{-1} A_{TR} \\\hline 
	     \neq                             & \neq 
	  \end{array} 
	\right).
	\label{eq:lu-linv2}
\end{equation}
The initial partitioning of the operands (statement {\bf Partition} in the skeleton)
is given by the rewrite rules in Box~\ref{box:initPart}. 
Notice that 
the top-left, top-right, and bottom-left quadrants
are, respectively, of size $0 \times 0$, $0\times m$, and
$m \times 0$, i.e., they are empty.
Thus, the initial partitioning renders the loop
invariant~\eqref{eq:lu-linv2} true. 

\begin{mybox}
{\small
$$
\begin{array}{c}
	\lowtriRuleTwoByTwo{L}{m}{m}{TL}{0}{0} \raisebox{2.8mm}{\textnormal{, }} \hspace{3mm}
	\upptriRuleTwoByTwo{U}{m}{m}{TL}{0}{0} \raisebox{2.4mm}{\textnormal{, and }} \hspace{3mm} 
	\ruleTwoByTwo{A}{m}{m}{TL}{0}{0}
\end{array}
$$
}
\caption{Initial partitioning of the operands for the \lu{}.} \label{box:initPart}
\end{mybox}

\click{} checks now the second constraint: $\PInv \wedge \neg G \implies \PPost$.
The loop guard $G$ follows from the traversal of the operands:
Initially $L_{TL}$, $U_{TL}$, and $A_{TL}$ are empty;
the loop is executed until the matrices are traversed completely.
Hence, $G$ equals $A_{TL} < A$ (and, accordingly, $L_{TL} < L$ and $U_{TL} < U$).
After the loop completes, the negation of the loop guard $\neg G$
means that the matrices $L_{TL}$, $U_{TL}$, and $A_{TL}$ equal $L$, $U$, and $A$
(and the rest of the quadrants are empty).
Predicate~\eqref{eq:lu-linv2} satisfies the second constraint:
\begin{mybox}
$$
	\begin{array}{l}
		\left( 
		  \begin{array}{@{\,}c@{\,}|@{\,}c@{\,}}
		  \{ L_{TL}, U_{TL} \} = LU(A_{TL})   & U_{TR} = L_{TL}^{-1} A_{TR} \\\hline 
			 \neq                             & \neq 
		  \end{array} 
	  \right) \; \wedge \; (A_{TL} = A \wedge L_{TL} = L \wedge U_{TL} = U) \\[8mm]
  		\qquad \qquad \Longrightarrow \quad \{L,U\} = LU(A),
	\end{array}
$$
\end{mybox}

\noindent
and therefore it represents a feasible loop invariant.

Out of the seven candidate predicates, 
the subgraph \{1, 2, 3, 4, 5\}, corresponding to the full PME, fails to satisfy the first constraint, while
the subgraph \{\}, corresponding to an empty predicate, fails to satisfy the second.
The remaining five predicates satisfy both feasibility constraints and are promoted
to valid loop invariants for the \lu{}
(Table~\ref{tab:LULoopInvs}).

\begin{table*}[!htb] \centering
\begin{tabular}{ccl} \toprule
\raisebox{2mm}{\bf \#} & \multicolumn{1}{c}{\raisebox{2mm}{\bf \footnotesize Subgraph}} & 
\multicolumn{1}{c}{\raisebox{2mm}{\bf \footnotesize Loop invariant}} \\[-2.5mm]\midrule
1 &
\raisebox{-3.2em}{\smallDepGraphLU{aicesred}{rwthblue}{rwthblue}{rwthblue}{rwthblue}} &
\renewcommand{\arraystretch}{1.4}
$
	\left( 
	  \begin{array}{@{\,}c@{\,}|@{\,}c@{\,}}
	  \{ L_{TL}, U_{TL} \} := LU(A_{TL})  & \qquad \, \neq \qquad \phantom{} \\\hline
	     \neq                             & \neq 
	  \end{array} 
	\right)
$ \\
2 &
\raisebox{-3.2em}{\smallDepGraphLU{aicesred}{aicesred}{rwthblue}{rwthblue}{rwthblue}} &
\renewcommand{\arraystretch}{1.4}
$
	\left( 
	  \begin{array}{@{\,}c@{\,}|@{\,}c@{\,}}
	  \{ L_{TL}, U_{TL} \} := LU(A_{TL})  & U_{TR} := L_{TL}^{-1} A_{TR} \\\hline 
	     \neq                             & \neq 
	  \end{array} 
	\right)
$ \\
3 &
\raisebox{-3.2em}{\smallDepGraphLU{aicesred}{rwthblue}{aicesred}{rwthblue}{rwthblue}} &
\renewcommand{\arraystretch}{1.4}
$
	\left( 
	  \begin{array}{@{\,}c@{\,}|@{\,}c@{\,}}
	  \{ L_{TL}, U_{TL} \} := LU(A_{TL})   & \qquad \, \neq \qquad \phantom{} \\\hline
	     L_{BL} := A_{BL} U_{TL}^{-1}      & \neq 
	  \end{array} 
	\right)
$ \\
4 &
\raisebox{-3.2em}{\smallDepGraphLU{aicesred}{aicesred}{aicesred}{rwthblue}{rwthblue}} &
\renewcommand{\arraystretch}{1.4}
$
	\left( 
	  \begin{array}{@{\,}c@{\,}|@{\,}c@{\,}}
	  \{ L_{TL}, U_{TL} \} := LU(A_{TL})   & U_{TR} := L_{TL}^{-1} A_{TR} \\\hline 
	     L_{BL} := A_{BL} U_{TL}^{-1}      & \neq 
	  \end{array} 
	\right)
$ \\
5 &
\raisebox{-3.2em}{\smallDepGraphLU{aicesred}{aicesred}{aicesred}{aicesred}{rwthblue}} &
\renewcommand{\arraystretch}{1.4}
$
	\left( 
	  \begin{array}{@{\,}c@{\,}|@{\,}c@{\,}}
	  \{ L_{TL}, U_{TL} \} := LU(A_{TL})   & U_{TR} := L_{TL}^{-1} A_{TR} \\\hline 
	     L_{BL} := A_{BL} U_{TL}^{-1}      & A_{BR} := A_{BR} - L_{BL} U_{TR} 
	  \end{array} 
	\right)
$ \\\bottomrule
\end{tabular}
\caption{The five loop invariants for the \lu{}.} \label{tab:LULoopInvs}
\end{table*}

We remark that the full and empty predicates for every target operation always
fail to satisfy the first and second constraints, respectively. Accordingly,
these are always discarded.

\subsection{A more complex example: the coupled Sylvester equation}
\label{sec:coupsylv}

To illustrate the potential of \click{},
we apply it to the coupled triangular Sylvester equation
(Box~\ref{box:coupOpDesc}),
an example where the complexity
of the graph of dependencies and the number of loop invariants
are such that automation becomes an indispensable tool. 

\begin{mybox}
\footnotesize
$$
\{X, Y\} := \Psi(A, B, C, D, E, F)
\equiv
\left\{
\begin{aligned}
P_{\rm pre}: \{ & \prop{Input}{A,B,C,D,E,F} \,\! \wedge \,\! \prop{Output}{X,Y} \,\, \wedge \\
			    & \prop{Matrix}{A,B,C,D,E,F,X,Y} \,\, \wedge \\
			    & \prop{LowerTriangular}{A,D} \,\, \wedge \,\, \prop{UpperTriangular}{B,E}\\
\\
P_{\rm post}: & \left\{ \begin{array}{@{}l@{}}
                          A X + Y B = C \\
                          D X + Y E = F 
						\end{array} \right.
\end{aligned}
\right.
$$
\caption{Formal description of the \cs{}.}
\label{box:coupOpDesc}
\end{mybox}

Given the description in Box~\ref{box:coupOpDesc},
\click{} finds three feasible sets of partitioning rules, 
which, in time, lead to the three
PMEs listed in Table~\ref{tab:coupsylvPMEs}.
\begin{table}[!h]
\centering
\scriptsize
\renewcommand{\arraystretch}{1.2}
\begin{tabular}{cl} \toprule
\raisebox{2mm}{\bf \#} & \multicolumn{1}{c}{\raisebox{2mm}{\bf \footnotesize Partitioned Matrix Expression}} \\[-2.5mm]\midrule
1 &
$
	\left( 
	  \begin{array}{@{\,}c@{\;}|@{\;}c@{\,}}
	    \{X_{L}, Y_{L} \} := \Psi(A, B_{TL}, C_{L}, D, E_{TL}, F_{L}) & 
	    \{X_{R}, Y_{R} \} := \Psi(A, B_{BR}, C_{R} - Y_{L} B_{TR}, D, E_{BR}, F_{R} - Y_{L} E_{TR})
	  \end{array} 
	\right)
$ \\[5mm]
2 &
$
	\left( 
	  \begin{array}{@{\,}c@{\,}}
	    \{X_{T}, Y_{T} \} := \Psi(A_{TL}, B, C_{T}, D_{TL}, E, F_{T}) \\\hline 
	    \{X_{B}, Y_{B} \} := \Psi(A_{BR}, B, C_{B} - A_{BL} X_{T}, D_{BR}, E, F_{B} - D_{BL} X_{T}) 
	  \end{array} 
	\right)
$ \\[9mm]
3 &
$
	\left( 
	  \begin{array}{@{\,}c@{\;}|@{\;}c@{\,}}
	    \{X_{TL}, Y_{TL} \} := \Psi(A_{TL}, B_{TL}, C_{TL}, D_{TL}, E_{TL}, F_{TL}) & 
		\begin{aligned}
	    \{X_{TR}, Y_{TR} \} := \Psi(& A_{TL}, B_{BR}, C_{TR} - Y_{TL} B_{TR}, \\& D_{TL}, E_{BR}, F_{TR} - Y_{TL} E_{TR})
		\end{aligned} \\\hline 
		\begin{aligned}
	    \{X_{BL}, Y_{BL} \} := \Psi(& A_{BR}, B_{TL}, C_{BL} - A_{BL} X_{TL}, \\& D_{BR}, E_{TL}, F_{BL} - D_{BL} X_{TL}) 
		\end{aligned} &
		\begin{aligned}
	    \{X_{BR}, Y_{BR} \} := \Psi(& A_{BR}, B_{BR}, C_{BR} - A_{BL} X_{TR} - Y_{BL} B_{TR}, \\& D_{BR}, E_{BR}, F_{BR} - D_{BL} X_{TR} - Y_{BL} E_{TR})
		\end{aligned}
	  \end{array}
	\right)
$ \vspace{1mm} \\
\bottomrule
\end{tabular}
\caption{The three Partitioned Matrix Expressions for the \cs{}.} \label{tab:coupsylvPMEs}
\end{table}

\noindent
We demonstrate the identification of loop invariants for the third PME.
First, \click{} traverses the PME, one quadrant at a time, to decompose
the assignments into tasks. The analysis starts from the top-left assignment; since the 
right-hand side consists of a function where all the input arguments are simple operands,
the function is yielded as a single task:
\begin{itemize}
\small
\item[1.] $\{X_{TL}, Y_{TL} \} := \Psi(A_{TL}, B_{TL}, C_{TL}, D_{TL}, E_{TL}, F_{TL})$.
\end{itemize}

Next, the top-right assignment is inspected. In this case, two of the input arguments
are not simple operands. Thus, \click{} analyzes both expressions,
{\small $C_{TR} - Y_{TL} B_{TR}$} and {\small $F_{TR} - Y_{TL} E_{TR}$},
to identify the sequence of tasks. 
Both expressions match the pattern for the matrix product ({\sc gemm}).
As a result, \click{} returns the sequence
\begin{itemize}
\small
\item[2.] $C_{TR} := C_{TR} - Y_{TL} B_{TR}$ \\[-5mm]
\item[3.] $F_{TR} := F_{TR} - Y_{TL} E_{TR}$ \\[-5mm]
\item[4.] $\{X_{TR}, Y_{TR} \} := \Psi(A_{TL}, B_{BR}, C_{TR}, D_{TL}, E_{BR}, F_{TR})$ \\[-5mm]
\end{itemize}
A similar situation occurs when studying the bottom-left assignment,
in which \click{} yields three more tasks.
\begin{itemize}
\small
\item[5.] $C_{BL} := C_{BL} - A_{BL} X_{TL}$ \\[-5mm]
\item[6.] $F_{BL} := F_{BL} - D_{BL} X_{TL}$ \\[-5mm]
\item[7.] $\{X_{BL}, Y_{BL} \} := \Psi(A_{BR}, B_{TL}, C_{BL}, D_{BR}, E_{TL}, F_{BL})$  \\[-5mm]
\end{itemize}

Only the assignment in the bottom-right quadrant 
remains to be analyzed. \click{} recognizes
that two of the input arguments to the function are expressions.
In contrast to the previous cases, the two expressions
are decomposed into more than one task. 
For instance, the expression {\small 
$C_{BR} - A_{BL} X_{TR} - Y_{BL} B_{TR}$} is
decomposed into two matrix products:
{\small $C_{BR} - A_{BL} X_{TR}$} and {\small $C_{BR} - Y_{BL} B_{TR}$}.
It is important to notice that both products are independent of one
another, i.e., they can be performed in any order.
\click{} keeps track of this fact for a correct analysis of dependencies.
In total, the analisys of the bottom-right assignment yields the following five 
tasks, two per complex input argument and the function itself:
\begin{itemize}
\small
\item[8.] $C_{BR} := C_{BR} - A_{BL} X_{TR}$ \\[-5mm]
\item[9.] $C_{BR} := C_{BR} - Y_{BL} B_{TR}$ \\[-5mm]
\item[10.] $F_{BR} := F_{BR} - D_{BL} X_{TR}$ \\[-5mm]
\item[11.] $F_{BR} := F_{BR} - Y_{BL} E_{TR}$ \\[-5mm]
\item[12.] $\{X_{BR}, Y_{BR} \} := \Psi(A_{BR}, B_{BR}, C_{BR}, D_{BR}, E_{BR}, F_{BR})$. \\[-5mm]
\end{itemize}

\click{} proceeds with the inspection of the tasks for dependencies.
The analysis commences from Task 1, whose outputs ($X_{TL}$ and $Y_{TL}$)
are inputs to Tasks 2, 3, 5 and 6. The four corresponding true dependencies
are created.
Next, Tasks 2 and 3 are inspected. Their outputs, $C_{TR}$ and $F_{TR}$,
are input to Task 4, hence enforcing two more true dependencies.
The algorithm proceeds by analyzing Task 4. One of its output operands, $X_{TR}$,
appears as an input argument of Tasks 8 and 10; two new dependencies arise.
The study of Tasks 5, 6 and 7 is analogous to that of Tasks 2, 3, and 4.
\click{} finds true dependencies from Tasks 5 and 6 to Task 7,
and from Task 7 to Tasks 9 and 11.

\begin{figure}
\centering
\includegraphics[scale=0.7]{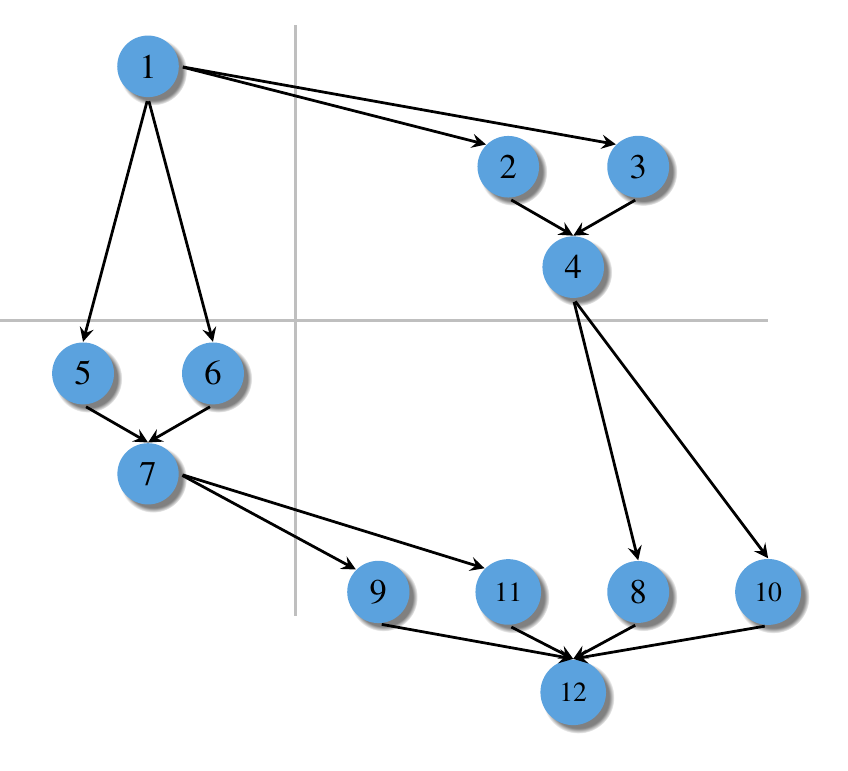}
\caption{Graph of dependencies for the \cs{}.} \label{box:coupsylvDepGraph}
\end{figure}

The analysis continues with the study of Tasks 8 and 9.
Both tasks take as input and overwrite the quantity $C_{BR}$;
however, as we pointed out earlier, 
they are independent from one another and can be computed
in any order. Therefore,
dependencies are created only from Tasks 8 and 9 to 12,
which takes $C_{BR}$ as input.
The study of Tasks 10 and 11 is led by the same
principle, originating the corresponding dependencies from
both to Task 12. 
Finally, Task 12 is analyzed. Its output, $\{X_{BR}, Y_{BR} \}$, 
does not appear in any of the other tasks, thus no new dependencies
are imposed. The final graph of dependencies
is shown in Figure~\ref{box:coupsylvDepGraph}.

Once the graph is built, \click{} executes Algorithm~\ref{alg:subDAG}
and returns a list with the predicates that are candidates to becoming loop invariants. 
Then, the predicates are checked to establish their feasibility; 
the non-feasible ones are discarded. 
In the \cs{} example, the system identifies
64 feasible loop invariants, which lead to 
64 different algorithms that solve the equation. 
In Table~\ref{tab:coupSylvLinvs}
we list a subset of the returned loop invariants. 

\begin{table}[!h] 
\scriptsize
\centering
\begin{tabular}{ccl} \toprule
\raisebox{2mm}{\bf \#} & \multicolumn{1}{c}{\raisebox{2mm}{\bf \footnotesize Subgraph}} & 
\multicolumn{1}{c}{\raisebox{2mm}{\bf \footnotesize Loop invariant}} \\[-2.5mm]\midrule
1 &
\raisebox{-8.0em}{\smallDepGraphCoupSylv{rwthblue}{rwthblue}{rwthblue}{rwthblue}{rwthblue}} &
\renewcommand{\arraystretch}{2.8}
$
	\left( 
	  \begin{array}{@{\,}c@{\;}|@{\;}c@{\,}}
	    \{X_{TL}, Y_{TL} \} = \Psi(A_{TL}, B_{TL}, C_{TL}, D_{TL}, E_{TL}, F_{TL}) & 
	    \phantom{X_{TL}, Y_{TL} } \neq \phantom{X_{TL}, Y_{TL} } \\\hline
		\neq &
		\neq
	  \end{array}
	\right)
$ \\[23mm]
2 &
\raisebox{-8.0em}{\smallDepGraphCoupSylv{aicesred}{rwthblue}{rwthblue}{rwthblue}{rwthblue}} &
\renewcommand{\arraystretch}{2.8}
$
	\left( 
	  \begin{array}{@{\,}c@{\;}|@{\;}c@{\,}}
	    \{X_{TL}, Y_{TL} \} = \Psi(A_{TL}, B_{TL}, C_{TL}, D_{TL}, E_{TL}, F_{TL}) & 
	    X_{TR} = C_{TR} - Y_{TL} B_{TR} \\\hline 
		\neq &
		\neq
	  \end{array}
	\right)
$ \\[23mm]
3 &
\raisebox{-8.0em}{\smallDepGraphCoupSylv{rwthblue}{aicesred}{rwthblue}{rwthblue}{rwthblue}} &
\renewcommand{\arraystretch}{2.8}
$
	\left( 
	  \begin{array}{@{\,}c@{\;}|@{\;}c@{\,}}
	    \{X_{TL}, Y_{TL} \} = \Psi(A_{TL}, B_{TL}, C_{TL}, D_{TL}, E_{TL}, F_{TL}) & 
	    Y_{TR} = F_{TR} - Y_{TL} E_{TR} \\\hline
		\neq &
		\neq
	  \end{array}
	\right)
$ \\[23mm]
& \LARGE \hspace{-7.7mm} $\vdots$ & \LARGE \hspace{4cm} $\vdots$ \\[4mm]
64 &
\raisebox{-8.0em}{\smallDepGraphCoupSylv{aicesred}{aicesred}{aicesred}{aicesred}{aicesred}} &
\renewcommand{\arraystretch}{2.8}
$
	\left( 
	  \begin{array}{@{\,}c@{\;}|@{\;}c@{\,}}
		\begin{aligned}
	    \{X_{TL}, Y_{TL} \} = \Psi(&A_{TL}, B_{TL}, C_{TL},\\
		                           &D_{TL}, E_{TL}, F_{TL}) 
		\end{aligned} &
		\begin{aligned}
	    \{X_{TR}, Y_{TR} \} = \Psi(&A_{TL}, B_{BR},\\
		                           &C_{TR} - Y_{TL} B_{TR},\\
								   &D_{TL}, E_{BR},\\
								   &F_{TR} - Y_{TL} E_{TR})
		\end{aligned} \\\hline 
		\begin{aligned}
	    \{X_{BL}, Y_{BL} \} = \Psi(&A_{BR}, B_{TL},\\
		                           &C_{BL} - A_{BL} X_{TL},\\
								   &D_{BR}, E_{TL},\\
								   &F_{BL} - D_{BL} X_{TL}) 
		\end{aligned} &
		\begin{aligned}
	    \{X_{BR}, Y_{BR} \} = \{&C_{BR} - A_{BL} X_{TR}\\
		                        &\phantom{C_{BR}} - Y_{BL} B_{TR},\\ 
								&F_{BR} - D_{BL} X_{TR}\\
								&\phantom{F_{BR}} - Y_{BL} E_{TR}\}
		\end{aligned}
	  \end{array}
	\right)
$\\
\bottomrule
\end{tabular}
\caption{A subset of the 64 loop invariants for the \cs{}.} \label{tab:coupSylvLinvs}
\end{table}

The large number of identified loop invariants and
the corresponding algorithms, demonstrates the necessity 
for having a system that
automates the
process.

\clearpage

\section{Algorithm construction}
\label{sec:AlgConstruction}

We discuss now the final stage in the generation of algorithms,
the Algorithm Construction. It is in this last stage where
the loop body is derived, and the algorithms are finally built.
The discussion centers around the template for a proof of correctness
introduced in Section~\ref{sec:flame}, which we reproduce here with further detail:

\begin{center}
	\begin{tabular}{|l |} \hline
		  \rowcolor[gray]{.8} \raisebox{-0.8mm}{$\{ \PPre \}$} \\[1.5mm]
		  \raisebox{-0.8mm}{{\bf Partition}} \\[1.5mm]
		  \rowcolor[gray]{.8} \raisebox{-0.8mm}{$\{ \PInv \}$} \\[1.5mm]
		  \raisebox{-0.8mm}{{\bf While} $G$ {\bf do}} \\[1.5mm]
		  \rowcolor[gray]{.8} \raisebox{-0.8mm}{\hspace*{7.5mm}$\{ \PInv \wedge G\}$} \\[1.5mm]
		  \raisebox{-0.8mm}{\hspace*{7.5mm}{\bf Repartition}} \\[1.5mm]
		  \rowcolor[gray]{.8} \raisebox{-0.8mm}{\hspace*{7.5mm}$\{ \PBefore \equiv \PInv|_{\text{\rm \raisebox{-.7mm}{Repartition}}} \}$} \\[1.7mm]
		  \raisebox{-0.8mm}{\hspace*{7.5mm} {\bf Algorithm Updates }} \\[1.5mm]
		  \rowcolor[gray]{.8} \raisebox{-0.8mm}{\hspace*{7.5mm}$\{ \PAfter \equiv \PInv|_{\text{\rm Continue with}^{-1}} \}$} \\[1.5mm]
		  \raisebox{-0.8mm}{\hspace*{7.5mm}{\bf Continue with}} \\[1.5mm]
		  \rowcolor[gray]{.8} \raisebox{-0.8mm}{\hspace*{7.5mm}$\{ \PInv \}$} \\[1.5mm]
		  \raisebox{-0.8mm}{{\bf end}} \\[1.5mm]
		  \rowcolor[gray]{.8} \raisebox{-0.8mm}{$\{ \PInv \wedge \neg G \}$} \\[2mm]
		  \rowcolor[gray]{.8} \raisebox{-0.8mm}{$\{ \PPost \}$} \\\hline
	\end{tabular}
\end{center}

\noindent
The idea is to transform the template into an algorithm
annotated with its proof of correctness by incrementally
replacing the predicates in brackets, and by filling
in the boldface labels with actual algorithm statements.

For each of the loop invariants found in the previous stage,
one such template is filled in.
Each loop invariant $\PInv$ was annotated with the traversal
of the operands, and the corresponding loop guard $G$.
With this information, together with the input equation
description ($\PPre$ and $\PPost$), the template can be partially filled in.
As an example, in Box~\ref{box:lu-template}, we provide the
partially filled in template for the third loop-invariant
of the \lu{} (Table~\ref{tab:LULoopInvs}).
We highlight in red the pieces that remain to be
derived.

\begin{table}[!ht]
\begin{center}
	\begin{tabular}{|l|}\hline
		  \rowcolor[gray]{.8}\{{\scriptsize \tt LowTri(L), UppTri(U), \ldots} \} \\
		  {{\bf Partition}} \\
		  \hspace*{0.5mm} {\scriptsize $* \quad \longrightarrow \quad\myFlaTwoByTwo{*_{TL}}{*_{TR}}{*_{BL}}{*_{BR}}$ } \\
		  \hspace*{2.5mm} {\bf where} $*_{TL}$ is $0\times0$ \\
		  \rowcolor[gray]{.8}$\{ \scriptsize \myFlaTwoByTwo{\{L_{TL}, U_{TL}\} := LU(A_{TL})}{\neq}{L_{BL} := A_{BL} U_{TL}^{-1}}{\neq}\}$ \\
		  {\bf While} $size(A_{TL}) < size(A)$ {\bf do}\\
		  \rowcolor[gray]{.8}\hspace*{7.5mm}$\{ \scriptsize \myFlaTwoByTwo{\{L_{TL}, U_{TL}\} := LU(A_{TL})}{\neq}{L_{BL} := A_{BL} U_{TL}^{-1}}{\neq} \wedge (size(A_{TL}) < size(A))\}$ \\
		  \hspace*{7.5mm}{\scriptsize $\myFlaTwoByTwo{*_{TL}}{*_{TR}}{*_{BL}}{*_{BR}} \quad 
			              \longrightarrow \quad \text{\normalsize \textcolor{red}{?}}$ } \\
		  \rowcolor[gray]{.8}\hspace*{7.5mm} \textcolor{red}{$\{ \PBefore \equiv \PInv|_{\text{\rm \raisebox{-.7mm}{Repartition}}} \}$} \\
		  \hspace*{7.5mm} \textcolor{red}{{\bf Algorithm Updates }} \\
		  \rowcolor[gray]{.8}\hspace*{7.5mm} \textcolor{red}{$\{ \PAfter \equiv \PInv|_{\text{\rm Continue with}^{-1}} \}$} \\
		  \hspace*{7.5mm}{\scriptsize $\myFlaTwoByTwo{*_{TL}}{*_{TR}}{*_{BL}}{*_{BR}} \quad 
		                  \longleftarrow \quad \text{\normalsize \textcolor{red}{?}}$ } \\
		  \rowcolor[gray]{.8}\hspace*{7.5mm}$\{ \scriptsize \myFlaTwoByTwo{\{L_{TL}, U_{TL}\} := LU(A_{TL})}{\neq}{L_{BL} := A_{BL} U_{TL}^{-1}}{\neq}\}$ \\
		  {\bf end} \\
		  \rowcolor[gray]{.8}$\{ \scriptsize \myFlaTwoByTwo{\{L_{TL}, U_{TL}\} := LU(A_{TL})}{\neq}{L_{BL} := A_{BL} U_{TL}^{-1}}{\neq} \wedge (A_{TL} = A)\}$ \\
		  \rowcolor[gray]{.8}$\{ L U = A \}$ \\\hline
	  \end{tabular}
\end{center}
\caption{Partially filled in template for $LU$'s third loop invariant (Table~\ref{tab:LULoopInvs}).}
\label{box:lu-template}
\end{table}

\begin{figure}[!t]
	\centering
	  \includegraphics[scale=1.00]{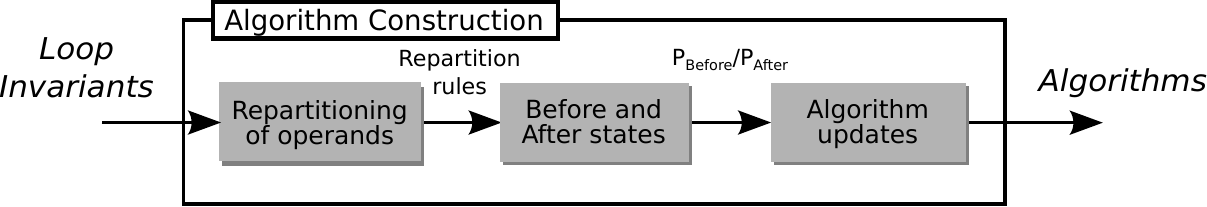}
	\caption{Steps for the construction of an algorithm from a given loop invariant.}
	\label{fig:FindingUpdates}
\end{figure}

The construction of the algorithm completes in three steps, as depicted
by Figure~\ref{fig:FindingUpdates}.
The key is to find the updates that render the $\PInv$ true at the end of the loop.
First, the traversal of the operands is formalized via the Repartition
and Continue with statements. The former exposes new parts of the matrices
to be used in the updates; the latter combines these parts to ensure
progress and termination of the loop. 
Then, the loop invariant $\PInv$ is rewritten in terms of the 
repartitioned matrices, yielding the predicates $\PBefore$ and $\PAfter$.
Finally, the algorithm updates are derived so that the computation
is taken from the state $\PBefore$ to the state $\PAfter$ and
the proof of correctness is satisfied.
In this section, we describe these steps, and detail how \click{} automates them.

\subsection{Repartitioning of the operands}

The Repartition and Continue with statements encode how the algorithm marches
through the operands, as determined by the loop invariant.
In Box~\ref{box:RepartCont}, we illustrate the statements for a generic 
matrix $A$ and a lower triangular matrix $L$; 
$A$, initially partitioned in $1 \times 2$ parts, is traversed from left to right,
while $L$, partitioned in $2 \times 2$ quadrants, is traversed from bottom-right to top-left.
The thick lines have a meaning: Initially empty, $A_{L}$ and $L_{BR}$
grow larger at each iteration; when the loop terminates, they equal the full operand.

\begin{mybox}
	\scriptsize
	\centering
	\vspace{1ex}
	\subfloat[Repartition (left) and Continue with (right) statements for a generic
                matrix $A$ traversed from left to right.] {
		$\myFlaOneByTwoI{A_{L}}{A_{R}}
		\;\rightarrow\;
		\myFlaOneByThreeRI{A_{0}}{A_{1}}{A_{2}}
		\qquad \qquad
		\myFlaOneByTwoI{A_{L}}{A_{R}}
		\;\leftarrow\;
		\myFlaOneByThreeLI{A_{0}}{A_{1}}{A_{2}}$
        \label{sbox:ExRepartContA}
    } \\
	\subfloat[Repartition (left) and Continue with (right) statements for a lower
				triangular matrix $L$ traversed from bottom-right to top-left.] {
		$\myFlaTwoByTwoI{L_{TL}}{0}{L_{BL}}{L_{BR}}
		\;\rightarrow\;
		\myFlaThreeByThreeTLI{L_{00}}{0}     {0}
							 {L_{10}}{L_{11}}{0}
							 {L_{20}}{L_{21}}{L_{22}}$
		\qquad \qquad
		$\myFlaTwoByTwoI{L_{TL}}{0}{L_{BL}}{L_{BR}}
		\;\leftarrow\;
		\myFlaThreeByThreeBRI{L_{00}}{0}     {0}
							 {L_{10}}{L_{11}}{0}
							 {L_{20}}{L_{21}}{L_{22}}$
        \label{sbox:ExRepartContL}
    }
	\caption{Two examples of Repartition and Continue with statements.}
\label{box:RepartCont}
\end{mybox}

Given the traversal for an operand, \click{} generates the corresponding
Repartition and Continue with statements.
Similarly to the inheritance of properties during the partitioning of the 
operands in the PME Generation (Section~\ref{sec:GenPME}),
the repartitioning also activates the propagation of properties.
For instance, in the example in Box~\ref{sbox:ExRepartContL},
$L_{00}$, $L_{11}$, and $L_{22}$ are lower triangular, and
$L_{01}$, $L_{02}$, and $L_{12}$ are the zero matrix.

The statements are encoded as lists of rewrite rules, which
will be used to perform
the subsequent textual substitution of the loop invariant.
For instance, for matrix $L$ in Box~\ref{sbox:ExRepartContL}, 
\click{} produces the following two lists of rules.
\bi
\item
{\em Repartition rules:}
$$ 
   \footnotesize
   \left\{ 
		   L_{TL} \;\rightarrow\; \myFlaTwoByTwo{L_{00}}{0}{L_{10}}{L_{11}} \text{, }
	       L_{BL} \;\rightarrow\; \myFlaOneByTwo{L_{20}}{L_{21}} \text{, }
	       L_{BR} \;\rightarrow\; (L_{22})
   \right\}
$$
\item
{\em Continue with rules:}
$$ 
   \footnotesize
   \left\{ 
	       L_{TL} \; \rightarrow \; (L_{00}) \text{, }
	       L_{BL} \; \rightarrow \; \myFlaTwoByOne{L_{10}}{L_{20}} \text{, }
	       L_{BR} \; \rightarrow \; \myFlaTwoByTwo{L_{11}}{0}{L_{21}}{L_{22}}
   \right\}
$$
\ei
For each of the operands in the target operation, \click{} generates such lists of rules and proceeds with
the construction of the $\PBefore$ and $\PAfter$ predicates.

\subsection{Predicates $\PBefore$ and $\PAfter$}
\label{subsec:pbef-paft}

The $\PBefore$ and $\PAfter$ predicates express the loop invariant
in terms of the repartitioned operands before and after the 
algorithm updates. These predicates are constructed in two steps:
First, the loop invariant is rewritten, replacing the partitioned 
operands by their repartitioned counterparts;
then, the resulting expressions are {\em flattened out}. 
The success of this process is dependent on \click{}'s
ability to learn the operation's PMEs.

To illustrate the generation of $\PBefore$ and $\PAfter$,
we make use of the following loop invariant for the triangular Sylvester equation:\footnote{
	This loop invariant is obtained from the third PME in Table~\ref{tab:SylvPMEs}, Section~\ref{subsec:NonUniquePME}.
}
\begin{equation}
\renewcommand{\arraystretch}{1.4}
	\left( {\begin{array}{@{}c@\;|@\;c@{}} 
		X_{TL} = C_{TL} - A_{TR} X_{BL} & 
		\neq \\\hline
		X_{BL} = \Omega(A_{BR}, B_{TL}, C_{BL}) &
		X_{BR} = \Omega(A_{BR}, B_{BR}, C_{BR} - X_{BL} B_{TR})
	\end{array}} \right).
	\label{eq:sylv-inv}
\end{equation}
All four operands ---$X$, $A$, $B$, and $C$--- are initially partitioned
in $2 \times 2$ quadrants;
$X$ and $C$ are traversed from bottom-left to top-right,
$A$ is traversed from bottom-right to top-left, and
$B$ is traversed from top-left to bottom-right. 
The rewrite rules corresponding to the Repartition and Continue statements yielded in the
previous step are provided in Boxes~\ref{box:sylv-bef-rules}~and~\ref{box:sylv-aft-rules},
respectively.

\begin{mybox}
\setlength{\arraycolsep}{0pt}
\scriptsize
\centering
	\subfloat[]{
		$\myFlaTwoByTwoI{X_{TL}}{X_{TR}}{X_{BL}}{X_{BR}}
		\; \rightarrow \;
		\myFlaTwoByTwoI{\myFlaTwoByOne{X_{00}}{X_{10}}}
					   {\myFlaTwoByTwo{X_{01}}{X_{02}}{X_{11}}{X_{12}}}
		               {(X_{20})}
					   {\myFlaOneByTwo{X_{21}}{X_{22}}}$
    } \qquad 
	\subfloat[]{
		$\myFlaTwoByTwoI{A_{TL}}{A_{TR}}{0}{A_{BR}}
		\; \rightarrow \;
		\myFlaTwoByTwoI{\myFlaTwoByTwo{A_{00}}{A_{01}}{0}{A_{11}}}
			           {\myFlaTwoByOne{A_{02}}{A_{12}}}
					   {\myFlaOneByTwo{0}{0}}
		               {(A_{22})}$
    } \\
	\subfloat[]{
		$\myFlaTwoByTwoI{B_{TL}}{B_{TR}}{0}{B_{BR}}
		\; \rightarrow \;
		\myFlaTwoByTwoI{(B_{00})}
						{\myFlaOneByTwo{B_{01}}{B_{02}}}
						{\myFlaTwoByOne{0}{0}}
						{\myFlaTwoByTwo{B_{11}}{B_{12}}{0}{B_{22}}}$
    } \qquad 
	\subfloat[]{
		$\myFlaTwoByTwoI{C_{TL}}{C_{TR}}{C_{BL}}{C_{BR}}
		\; \rightarrow \;
		\myFlaTwoByTwoI{\myFlaTwoByOne{C_{00}}{C_{10}}}
					   {\myFlaTwoByTwo{C_{01}}{C_{02}}{C_{11}}{C_{12}}}
		               {(C_{20})}
					   {\myFlaOneByTwo{C_{21}}{C_{22}}}$
    }
	\caption{{\em Repartition rules}. Repartitioning towards the generation of $\PBefore$.}
\label{box:sylv-bef-rules}
\end{mybox}

\begin{mybox}
\setlength{\arraycolsep}{0pt}
\scriptsize
\centering
	\subfloat[]{
		$\myFlaTwoByTwoI{X_{TL}}{X_{TR}}{X_{BL}}{X_{BR}}
		\; \rightarrow \;
		\myFlaTwoByTwoI{\myFlaOneByTwo{X_{00}}{X_{01}}}
		                {(X_{02})}
					    {\myFlaTwoByTwo{X_{10}}{X_{11}}{X_{20}}{X_{21}}}
		                {\myFlaTwoByOne{X_{21}}{X_{22}}}$
    } \qquad 
	\subfloat[]{
		$\myFlaTwoByTwoI{A_{TL}}{A_{TR}}{0}{A_{BR}}
		\; \rightarrow \;
		\myFlaTwoByTwoI{(A_{00})}
						{\myFlaOneByTwo{A_{01}}{A_{02}}}
						{\myFlaTwoByOne{0}{0}}
						{\myFlaTwoByTwo{A_{11}}{A_{12}}{0}{A_{22}}}$
    } \\
	\subfloat[]{
		$\myFlaTwoByTwoI{B_{TL}}{B_{TR}}{0}{B_{BR}}
		\; \rightarrow \;
		\myFlaTwoByTwoI{\myFlaTwoByTwo{B_{00}}{B_{01}}{0}{B_{11}}}
					    {\myFlaTwoByOne{B_{02}}{B_{12}}}
					    {\myFlaOneByTwo{0}{0}}
					    {(B_{22})}$
    } \qquad 
	\subfloat[]{
		$\myFlaTwoByTwoI{C_{TL}}{C_{TR}}{C_{BL}}{C_{BR}}
		\; \rightarrow \;
		\myFlaTwoByTwoI{\myFlaOneByTwo{C_{00}}{C_{01}}}
		                {(C_{02})}
					    {\myFlaTwoByTwo{C_{10}}{C_{11}}{C_{20}}{C_{21}}}
		                {\myFlaTwoByOne{C_{12}}{C_{22}}}$
    }
	\caption{{\em Continue with rules}. Repartitioning towards the generation of $\PAfter$.}
\label{box:sylv-aft-rules}
\end{mybox}

The construction of $\PBefore$ commences with the application of the 
repartition rules (Box~\ref{box:sylv-bef-rules}) to the loop invariant;
the expression is rewritten as
\begin{equation}
\setlength{\arraycolsep}{0pt}
\scriptsize
	\left( {\begin{array}{@{}c@\;|@\;c@{}} 
		\myFlaTwoByOne{X_{00}}{X_{10}} := \myFlaTwoByOne{C_{00}}{C_{10}} - \myFlaTwoByOne{A_{02}}{A_{12}} (X_{20}) &
		\myFlaTwoByTwo{X_{01}}{X_{02}}{X_{11}}{X_{12}} :=  \myFlaTwoByTwo{0}{0}{0}{0} \\\hline
		(X_{20}) := \Omega((A_{22}), (B_{00}), (C_{20})) &
		\myFlaOneByTwo{X_{21}}{X_{22}} := \Omega\left((A_{22}), \myFlaTwoByTwo{B_{11}}{B_{12}}{0}{B_{22}}, \myFlaOneByTwo{C_{21}}{C_{22}} - (X_{20}) \myFlaOneByTwo{B_{01}}{B_{02}} \right) 
	\end{array}} \right).
\label{eq:inv-bef-rules}
\end{equation}
Next, the assignments in each of the four quadrants must be simplified.
The right-hand sides consist of either explicit algebraic operations,
as in the top-left quadrant, 
or an implicit function with partitioned arguments,
as in the bottom-right quadrant.
In the first case, \click{} applies basic built-in matrix algebra knowledge.
For instance, the expression 
$$
\myFlaTwoByOne{X_{00}}{X_{10}} := \myFlaTwoByOne{C_{00}}{C_{10}} - \myFlaTwoByOne{A_{02}}{A_{12}} (X_{20})$$
is multiplied out and the assignment is distributed, resulting in
$$
\myFlaTwoByOne{X_{00} := C_{00} - A_{02} X_{20}}%
              {X_{10} := C_{10} - A_{12} X_{20}}
$$

\sloppypar
The second case, instead, requires a deeper understanding of the FLAME methodology.
In the bottom-right quadrant of Equation~\eqref{eq:inv-bef-rules}, one finds a recursive call to Sylvester
with partitioned operands. At first sight, simplifying the expression
is far from straightforward; fortunately, as \click{} generated the PMEs for Sylvester,
it learned a number of rules (Box~\ref{box:sylv-PME-rules}) to flatten out such a complicated expression.
Concretely, the rule
\begin{center}
$\begin{array}{l}
	\myFlaOneByTwo{X_L}{X_R} := \Omega \left(
	                                       (A),
										   \myFlaTwoByTwo{B_{TL}}{B_{TR}}{0}{B_{BR}},
										   \myFlaOneByTwo{C_L}{C_R}
	                                    \right)
	\longrightarrow \\[7mm]
	\hspace*{1cm} 
	\myFlaOneByTwo{X_{L} := \Omega(A, B_{TL}, C_{L})}
				  {X_{R} := \Omega(A, B_{BR}, C_{R} - X_{L} B_{TR})}
\end{array}$
\end{center}
corresponding to the first PME, enables the rewrite of 
\begin{equation}
		\myFlaOneByTwo{X_{21}}{X_{22}} := 
			\Omega \left(
				(A_{22}), 
				\myFlaTwoByTwo{B_{11}}{B_{12}}{0}{B_{22}}, 
				\myFlaOneByTwo{C_{21}}{C_{22}} - (X_{20}) \myFlaOneByTwo{B_{01}}{B_{02}} 
			\right) 
\nonumber
\end{equation}
as
$$
\myFlaOneByTwo{X_{21} := \Omega(A_{22}, B_{11}, C_{21} - X_{20} B_{01})}
              {X_{22} := \Omega(A_{22}, B_{22}, C_{22} - X_{20} B_{02} - X_{21} B_{12})}.
$$

In fact, the generation of $\PBefore$ and $\PAfter$ for the loop invariant under consideration
requires all three previously learned PMEs.
We illustrate this in Figure~\ref{fig:sylv-bef-aft}:
Above the {\em Algorithm Updates}, we provide the predicate $\PBefore$
prior and after the flattening; 
the flattening rule from PME 1 is used in the bottom-right quadrant.
Below the updates, we find the predicate $\PAfter$
also prior and after the flattening; 
rules from PMEs 2 and 3 are required to flatten the expressions in the 
bottom-right and bottom-left quadrants, respectively.

\begin{sidewaysfigure}
		\scriptsize
	\begin{tabular}{l c}
		{\normalsize $\PBefore$} &
		$
			\left( {\begin{array}{@{}c@\;|@\;c@{}} 
				\myFlaTwoByOne{X_{00}}{X_{10}} = \myFlaTwoByOne{C_{00}}{C_{10}} - \myFlaTwoByOne{A_{02}}{A_{12}} (X_{20}) &
				\myFlaTwoByTwo{X_{01}}{X_{02}}{X_{11}}{X_{12}} =  \myFlaTwoByTwo{0}{0}{0}{0} \\\hline
				(X_{20}) = \Omega(A_{22}, B_{00}, C_{20}) &
				\myFlaOneByTwo{X_{21}}{X_{22}} = 
					\Omega \left(
						(A_{22}), 
						\myFlaTwoByTwo{B_{11}}{B_{12}}{0}{B_{22}}, 
						\myFlaOneByTwo{C_{21}}{C_{22}} - (X_{20}) \myFlaOneByTwo{B_{01}}{B_{02}} 
					\right) 
			\end{array}} \right).
		$ \\[10mm]
		& 
		{\Large $\Downarrow$} Flattening
		\\ [5mm]
		&
		$
		\myFlaThreeByThree{X_{00} = C_{00} - A_{02} X_{20}}
						  {X_{01} = 0}
						  {X_{02} = 0}
						  {X_{10} = C_{10} - A_{12} X_{20}}
						  {X_{11} = 0}
						  {X_{12} = 0}
						  {X_{20} = \Omega(A_{22}, B_{00}, C_{20})}
						  {X_{21} = \Omega(A_{22}, B_{11}, C_{21} - X_{20} B_{01})}
						  {X_{22} = \Omega(A_{22}, B_{22}, C_{22} - X_{20} B_{02} - X_{21} B_{12})}
        $\\[5mm]
		& \\\hline
		& \\[5mm]
		& {\Large \em Algorithm Updates} \\[5mm]
		& \\\hline
		& \\[5mm]
		&
		$
		\scriptsize
		\myFlaThreeByThree{X_{00} = C_{00} - A_{01} X_{10} - A_{02} X_{20}}
		                  {X_{01} = C_{01} - A_{01} X_{11} - A_{02} X_{21}}
						  {X_{02} = 0}
						  {X_{10} = \Omega(A_{11}, B_{00}, C_{10} - A_{12} X_{20})}
						  {X_{11} = \Omega(A_{11}, B_{11}, C_{11} - A_{12} X_{21} - X_{10} B_{01})}
						  {X_{12} = \Omega(A_{11}, B_{22}, C_{12} - 
														   X_{10} B_{02} - 
														   X_{11} B_{12} - 
														   A_{12} X_{22})}
						  {X_{20} = \Omega(A_{22}, B_{00}, C_{20})}
						  {X_{21} = \Omega(A_{22}, B_{11}, C_{21} - X_{20} B_{01})}
						  {X_{22} = \Omega(A_{22}, B_{22}, C_{22} - X_{20} B_{02} - X_{21} B_{12})}
		$ \\[13mm]
		& 
		{\Large $\Uparrow$} Flattening
		\\ [7mm]
		{\normalsize $\PAfter$} &
		$
		\scriptsize
		\myFlaTwoByTwo%
			{\myFlaOneByTwo{X_{00}}{X_{01}} = 
					\myFlaOneByTwo{C_{00}}{C_{01}} - 
					\myFlaOneByTwo{A_{01}}{A_{02}} \myFlaTwoByTwo{X_{10}}{X_{11}}{X_{20}}{X_{21}}}
			{X_{02} = 0}
			{\myFlaTwoByTwo{X_{10}}{X_{11}}{X_{20}}{X_{21}} = 
				\Omega \left(
					\myFlaTwoByTwo{A_{11}}{A_{12}}{0}{A_{22}}, 
					\myFlaTwoByTwo{B_{00}}{B_{01}}{0}{B_{11}}, 
					\myFlaTwoByTwo{C_{10}}{C_{11}}{C_{20}}{C_{21}}
				\right)}
			{\myFlaTwoByOne{X_{12}}{X_{22}} = 
				\Omega \left(
					\myFlaTwoByTwo{A_{11}}{A_{12}}{A_{21}}{A_{22}}, 
					(B_{22}), 
					\myFlaTwoByOne{C_{12}}{C_{22}} - 
					\myFlaTwoByTwo{X_{10}}{X_{11}}{X_{20}}{X_{21}} \myFlaTwoByOne{B_{02}}{B_{12}}
				\right)}
		$\\
		\end{tabular}
		\caption{$\PBefore$ and $\PAfter$ for Sylvester's loop invariant~\eqref{eq:sylv-inv}.}
\label{fig:sylv-bef-aft}
\end{sidewaysfigure}

\subsection{Finding the updates}
\label{subsec:updates}

Finding the algorithm
updates is equivalent to finding the computation that takes the loop invariant from
the state in $\PBefore$ to the state in $\PAfter$.
Intuitively, the updates are identified by comparing the two states.
In practice, the comparison heavily relies on pattern matching and
expression rewriting.
In this section, we expose, via the Sylvester and Cholesky examples,
how \click{} derives the updates.

In our first example, we continue the Sylvester case study from the previous section. 
The comparison of the $\PBefore$ and $\PAfter$ predicates (Figure~\ref{fig:sylv-bef-aft})
is carried out quadrant by quadrant.
The quantities $X_{02}$, $X_{20}$, $X_{21}$, and $X_{22}$ hold the same value
before and after the updates; hence, for these quadrants, no computation is required.
$X_{00}$ and $X_{10}$, instead, are partially computed, i.e.,
they hold a value distinct from {\em Null} (0) at the beginning of the iteration ($\PBefore$), but
require further computation to render the loop invariant true at the end of the iteration
($\PAfter$). \click{} processes both quadrants in a similar fashion; here we discuss the update
for $X_{10}$.
The value for $X_{10}$ in $\PBefore$ is
\begin{equation}
	X_{10} := C_{10} - A_{12} X_{20}
	\label{eq:X10-bef}
\end{equation}
while the required value at the end of the loop is
\begin{equation}
	X_{10} := \Omega(A_{11}, B_{00}, C_{10} - A_{12} X_{20}).
	\label{eq:X10-aft}
\end{equation}
An inspection of both expressions quickly reveals the fact that
the quantity stored in $\PBefore$,~\eqref{eq:X10-bef},
equals the quantity required as third argument for $\Omega$ in
$\PAfter$,~\eqref{eq:X10-aft}.
Accordingly, the sought-after update is
$$X_{10} := \Omega(A_{11}, B_{00}, X_{10}),$$
where the previously computed $X_{10}$ is used as third argument.

This intuition is formalized via rewrite rules:
\click{} takes the assignment in $\PBefore$, and creates a rule of the form 
{\em right-hand side $\rightarrow$ left-hand side}.
The rule is used to rewrite the assignment in $\PAfter$:
$$
	X_{10} := \Omega(A_{11}, B_{00}, C_{10} - A_{12} X_{20}) \;\; /. \;\; C_{10} - A_{12} X_{20} \rightarrow X_{10},
$$
obtaining the update $X_{10} := \Omega(A_{11}, B_{00}, X_{10}).$

Deriving the updates in the remaining quadrants ---$X_{01}$, $X_{11}$, and $X_{12}$---
is straightforward. Since no computation is stored in $\PBefore$,
the required update equals the right-hand side of the expression in $\PAfter$.
Combining all quadrants, the final set of updates for the Sylvester example is
\begin{center}
	$\begin{aligned}
		X_{10} &:= \Omega(A_{11}, B_{00}, X_{10}) \\
		X_{11} &:= \Omega(A_{11}, B_{11}, C_{11} - A_{12} X_{21} - X_{10} B_{01}) \\
		X_{12} &:= \Omega(A_{11}, B_{22}, C_{12} - 
									   X_{10} B_{02} - 
									   X_{11} B_{12} - 
									   A_{12} X_{22}) \\
		X_{00} &:= X_{00} - A_{01} X_{10} \\
		X_{01} &:= C_{01} - A_{01} X_{11} - A_{02} X_{21}.
	\end{aligned}$
\end{center}

As a second example, we choose the following loop invariant for the Cholesky factorization:
\begin{equation}
	\myFlaTwoByTwo{ L_{TL} := \Gamma(A_{TL}) }
                { \star }
				{ L_{BL} := A_{BL} L_{TL}^{-T} }
				{ L_{BR} := A_{BR} - L_{BL} L_{BL}^T }.
	\label{eq:chol-linv3}
\end{equation}
The $\PBefore$ and $\PAfter$ predicates are given in Box~\ref{box:chol-bef-aft}.
In this case, only quadrants ``1,1'', ``2,1'', and ``2,2'' differ
between states and need to be updated.
To derive the updates for the operands $L_{11}$ and $L_{22}$, 
it suffices to build and apply rewrite rules as previously described;
the yielded updates are
\begin{center}
	$\begin{aligned}
		L_{11} &:= \Gamma(L_{11}) \\
		L_{22} &:= L_{22} - L_{21} L_{21}^T.
	\end{aligned}$
\end{center}

However, often times, a direct replacement is not sufficient to find the updates. 
This is the case, for instance, of $L_{21}$: The expression 
\begin{equation}
	L_{21} := A_{21} - L_{20} L_{10}^T
	\label{eq:L21-bef}
\end{equation}
is not directly found in the $\PAfter$ counterpart,
\begin{equation}
	L_{21} := A_{21} L_{11}^{-T} - A_{20} L_{00}^{-T} L_{10}^T L_{11}^{-T}.
	\label{eq:L21-aft}
\end{equation}
The reason why this happens is that the assignments in these two predicates
are not written in any sort of canonical form.
In such a situation, a human inspects the other quadrants 
for already computed subexpressions that may be used to rewrite
\eqref{eq:L21-bef} or \eqref{eq:L21-aft}.
For instance, the expression $A_{20} L_{00}^{-T}$ has
already been computed and stored in $L_{20}$;
therefore, $A_{20} L_{00}^{-T}$ may be
replaced in the after state \eqref{eq:L21-aft} by $L_{20}$, 
resulting in 
$$L_{21} := A_{21} L_{11}^{-T} - L_{20} L_{10}^T L_{11}^{-T}.$$
Now, the right-hand side of the before ($A_{21} - L_{20} L_{10}^T$)
is exposed in that of the after;
the simple replacement
$$L_{21} := A_{21} L_{11}^{-T} - L_{20} L_{10}^T L_{11}^{-T} \;\; /. \;\; A_{21} - L_{20} L_{10}^T \rightarrow L_{20},$$
yields the required update:
$L_{21} := L_{21} L_{11}^{-T}.$

\begin{mybox}[!ht]
	\centering
	\scriptsize
	\begin{tabular}{l c}
		{\normalsize $\PBefore$} &
		$
		\myFlaThreeByThree{L_{00} := \Gamma(A_{00})}
						  {\star}
						  {\star}
						  {L_{10} := A_{10} L_{00}^{-T}}
						  {L_{11} := A_{11} - L_{10} L_{10}^T}
						  {\star}
						  {L_{20} := A_{20} L_{00}^{-T}}
						  {L_{21} := A_{21} - L_{20} L_{10}^T}
						  {L_{22} := A_{22} - L_{20} L_{20}^T}
        $ \vspace{-2mm} \\
		& \\\hline
		& \\[-2mm]
		{\normalsize \em Updates} &
	$\begin{aligned}
      L_{11} &:= \Gamma(L_{11}) \\
	  L_{21} &:= L_{21} L_{11}^{-T} \\
	  L_{22} &:= L_{22} - L_{21} L_{21}^T.
	\end{aligned}$ \vspace{-2mm} \\
		& \\\hline
		& \\[-2mm]
		{\normalsize $\PAfter$} &
		$
		\myFlaThreeByThree{L_{00} := \Gamma(A_{00})}
						  {\star}
						  {\star}
						  {L_{10} := A_{10} L_{00}^{-T}}
						  {L_{11} := \Gamma(A_{11} - L_{10} L_{10}^T)}
						  {\star}
						  {L_{20} := A_{20} L_{00}^{-T}}
						  {L_{21} := A_{21} L_{11}^{-T} - A_{20} L_{00}^{-T} L_{10}^T L_{11}^{-T}}
						  {L_{22} := A_{22} - L_{20} L_{20}^T - L_{21} L_{21}^T}
		$\\
		\end{tabular}
	\caption{Predicates $\PBefore$ and $\PAfter$, and algorithm updates for Cholesky's loop invariant~\eqref{eq:chol-linv3}.}
	\label{box:chol-bef-aft}
\end{mybox}

\click{} makes this search systematic by first rewriting both assignments
so that redundant subexpressions are eliminated, and then
applying the direct replacement of the before in the after.
To this end, \click{} creates the two lists of rewrite rules shown
in Box~\ref{box:chol-canonical-rules};
one list per predicate, one rule per assignment.
\begin{mybox}
	\centering
	\hfill
	\subfloat[Rules created to rewrite the $\PBefore$ assignments.] 
	{
		\begin{minipage}{.35\textwidth}
			\centering
		$\begin{aligned}
			\Gamma(A_{00})           & \rightarrow  L_{00} \\
			A_{10} L_{00}^{-T}       & \rightarrow  L_{10} \\
			A_{11} - L_{10} L_{10}^T & \rightarrow  L_{11} \\
			A_{20} L_{00}^{-T}       & \rightarrow  L_{20} \\
			A_{21} - L_{20} L_{10}^T & \rightarrow  L_{21} \\
			A_{22} - L_{20} L_{20}^T & \rightarrow  L_{22} \\
		\end{aligned}$
		\end{minipage}
		\label{sbox:chol-expand-rules-bef}
    }
	\hfill
	\subfloat[Rules created to rewrite the $\PAfter$ assignments.] 
	{
		\begin{minipage}{.45\textwidth}
			\centering
		$\begin{aligned}
			\Gamma(A_{00})                                               & \rightarrow L_{00} \\
			A_{10} L_{00}^{-T}                                           & \rightarrow L_{10} \\
			\Gamma(A_{11} - L_{10} L_{10}^T)                             & \rightarrow L_{11} \\
			A_{20} L_{00}^{-T}                                           & \rightarrow L_{20} \\
			A_{21} L_{11}^{-T} - A_{20} L_{00}^{-T} L_{10}^T L_{11}^{-T} & \rightarrow L_{21} \\
			A_{22} - L_{20} L_{20}^T - L_{21} L_{21}^T                   & \rightarrow L_{22} \\
		\end{aligned}$
		\end{minipage}
		\label{sbox:chol-expand-rules-aft}
    }
	\hfill
	\phantom{.}
	\caption{Rules to rewrite the $\PBefore$ and $\PAfter$ predicates.}
	\label{box:chol-canonical-rules}
\end{mybox}
The rules are then applied to the assignments of the corresponding predicate.
While the application of the rules in Box~\ref{sbox:chol-expand-rules-bef} to the before
state does not modify $L_{21}$, the application of the rules in 
Box~\ref{sbox:chol-expand-rules-aft} to the after state, results in
\begin{equation}
	{L_{21} := A_{21} L_{11}^{-T} - L_{20} L_{10}^T L_{11}^{-T}}.
	\label{eq:L21-canonical}
\end{equation}
Now, the right-hand side of $L_{21}$'s before state may be directly
replaced in~\eqref{eq:L21-canonical}, yielding the exact same
update as derived by hand.

\subsection{The final algorithms and routines}

By repeating the process for every loop invariant obtained in the second stage 
({\em Loop Invariant Identification}), 
\click{} constructs a family of algorithms (one per loop invariant).
In Figure~\ref{fig:sylv-alg}, we present the algorithm corresponding
to Sylvester's loop invariant~\eqref{eq:sylv-inv},
the one discussed in Sections~\ref{subsec:pbef-paft}~and~\ref{subsec:updates}.
On the left side, we reproduce the blocked variant of the algorithm;
the recursive calls to Sylvester may be computed by the unblocked variant
(on the right side) which is easily obtained by setting the block size $b$ to 1.

The FLAME project offers a number of application programming interfaces
(APIs) that simplify the translation of algorithms into routines
that closely resemble the algorithmic notation.
\click{} incorporates a C code generator that makes use of the
FLAME/C API for this programming language;
the implementation generated for the blocked algorithm 
in Figure~\ref{fig:sylv-alg} (left) is displayed in Routine~\ref{code:sylv}.

\begin{figure}
	\centering
	\begin{tabular}{c c}
		\hspace*{-31mm}
		\input{Chapter4_Cl1ck/flame-algs/sylv.tex}
		\input{Chapter4_Cl1ck/flame-algs/sylv-unb.tex}
	\end{tabular}
	\caption{Blocked (left) and unblocked (right) versions of 
		     Sylvester's variant 7 (out of 16). $C$ is overwritten
			 with the solution $X$.
			 In the unblocked version, greek, lowercase and uppercase letters
			 are used, respectively, for scalars, vectors and matrices.}
	\label{fig:sylv-alg}
\end{figure}

\renewcommand{\lstlistingname}{Routine}
\noindent
\begin{minipage}{\linewidth}
\footnotesize
  \begin{lstlisting}[caption={FLAME/C code for Sylvester's blocked variant 7 as generated by \clickplain{}.},
                     label=code:sylv,
                     numbers=left,
                     basicstyle={\tt},
					 numberblanklines=true,
                     fancyvrb=true,language=Fortran,columns=fixed,basewidth=.5em,frame=b,
                      framexleftmargin=-0pt,
                      framexrightmargin=0pt,
                      xleftmargin=10pt]
void sylv_blk_var7( FLA_Obj A, FLA_Obj B, FLA_Obj X, int nb )
{
  FLA_Obj ATL, ATR, ABL, ABR, A00, A01, A02, A10, A11, A12, A20, A21, A22;
  FLA_Obj BTL, BTR, BBL, BBR, B00, B01, B02, B10, B11, B12, B20, B21, B22;
  FLA_Obj XTL, XTR, XBL, XBR, X00, X01, X02, X10, X11, X12, X20, X21, X22;

  FLA_Part_2x2( A, &ATL, &ATR, 
                   &ABL, &ABR, 0, 0, FLA_BR );
  FLA_Part_2x2( B, &BTL, &BTR, 
                   &BBL, &BBR, 0, 0, FLA_TL );
  FLA_Part_2x2( X, &XTL, &XTR, 
                   &XBL, &XBR, 0, 0, FLA_BL );
    
  while ( FLA_Obj_length( XBL ) < FLA_Obj_length( X ) ||
          FLA_Obj_width( XBL ) < FLA_Obj_width( X ) )
  {
    FLA_Repart_2x2_to_3x3( XTL, XTR, &X00, &X01, &X02,
                                     &X10, &X11, &X12,
                           XBL, XBR, &X20, &X21, &X22, nb, nb, FLA_TR );
    [...]
    
    FLA_Gemm( FLA_NO_TRANSPOSE, FLA_NO_TRANSPOSE, 
              FLA_MINUS_ONE, A02, X21, FLA_ONE, X01 );
    FLA_sylv_unb(A11, B00, X10);
    FLA_Gemm( FLA_NO_TRANSPOSE, FLA_NO_TRANSPOSE, 
              FLA_MINUS_ONE, A01, X10, FLA_ONE, X00 );
    FLA_Gemm( FLA_NO_TRANSPOSE, FLA_NO_TRANSPOSE, 
              FLA_MINUS_ONE, X10, B01, FLA_ONE, X11 );
    FLA_Gemm( FLA_NO_TRANSPOSE, FLA_NO_TRANSPOSE, 
              FLA_MINUS_ONE, A12, X21, FLA_ONE, X11 );
    FLA_sylv_unb(A11, B11, X11);
    FLA_Gemm( FLA_NO_TRANSPOSE, FLA_NO_TRANSPOSE, 
              FLA_MINUS_ONE, A01, X11, FLA_ONE, X01 );
    FLA_Gemm( FLA_NO_TRANSPOSE, FLA_NO_TRANSPOSE, 
              FLA_MINUS_ONE, X11, B12, FLA_ONE, X12 );
    FLA_Gemm( FLA_NO_TRANSPOSE, FLA_NO_TRANSPOSE, 
              FLA_MINUS_ONE, X10, B02, FLA_ONE, X12 );
    FLA_Gemm( FLA_NO_TRANSPOSE, FLA_NO_TRANSPOSE, 
              FLA_MINUS_ONE, A12, X22, FLA_ONE, X12 );
    FLA_sylv_unb(A11, B22, X12);

    [...]
    FLA_Cont_with_3x3_to_2x2( &XTL, &XTR, X00, X01, X02,
                                          X10, X11, X12,
                              &XBL, &XBR, X20, X21, X22, FLA_BL );
  }
}
  \end{lstlisting}
\end{minipage}

\section{Towards a one-click code generation}

The ultimate goal of the FLAME project in terms
of automation is the development of a system
that takes as input a high-level description
of a target operation, and returns a family
of algorithms and routines to compute the operation.
From Bientinesi's dissertation~\cite{PaulDj:PhD}:

\begin{center}
	\begin{minipage}{.7\textwidth}
		{\em ``Ultimately, one should be able to visit a website,
			 fill in a form with information about the operation
			 to be performed, choose a programming language,
			 click the `submit' button, and receive a library
			 of routines that compute the operation.''} 
	\end{minipage}
\end{center}

\begin{figure}[!h]
	\centering
	\includegraphics[scale=0.55]{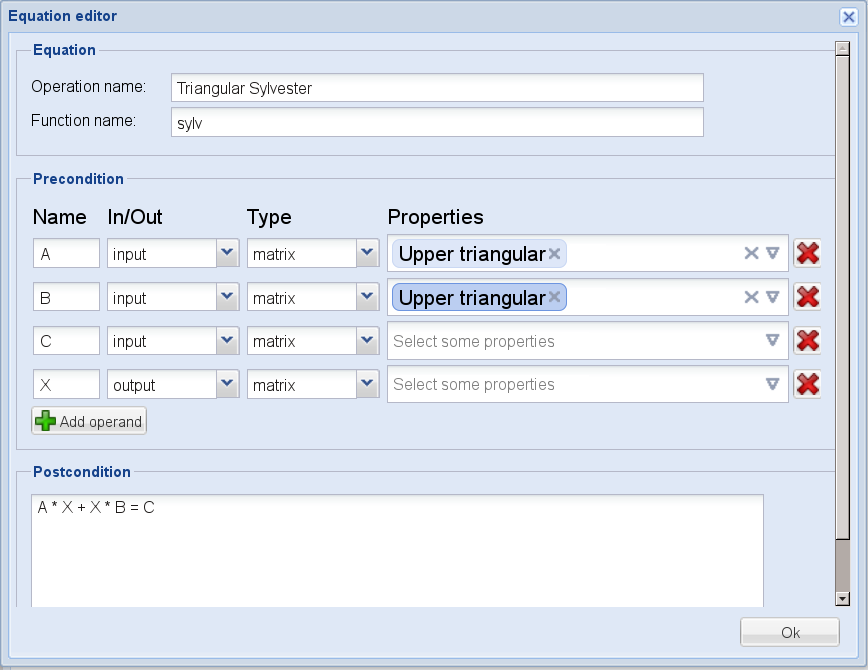}
	\caption{User-friendly web form to easily input the description of a
		target equation.
		In the example, we type the description of the Sylvester equation.}
	\label{fig:SylvInput}
\end{figure}

\begin{figure}[!h]
	\centering
	\includegraphics[scale=0.35]{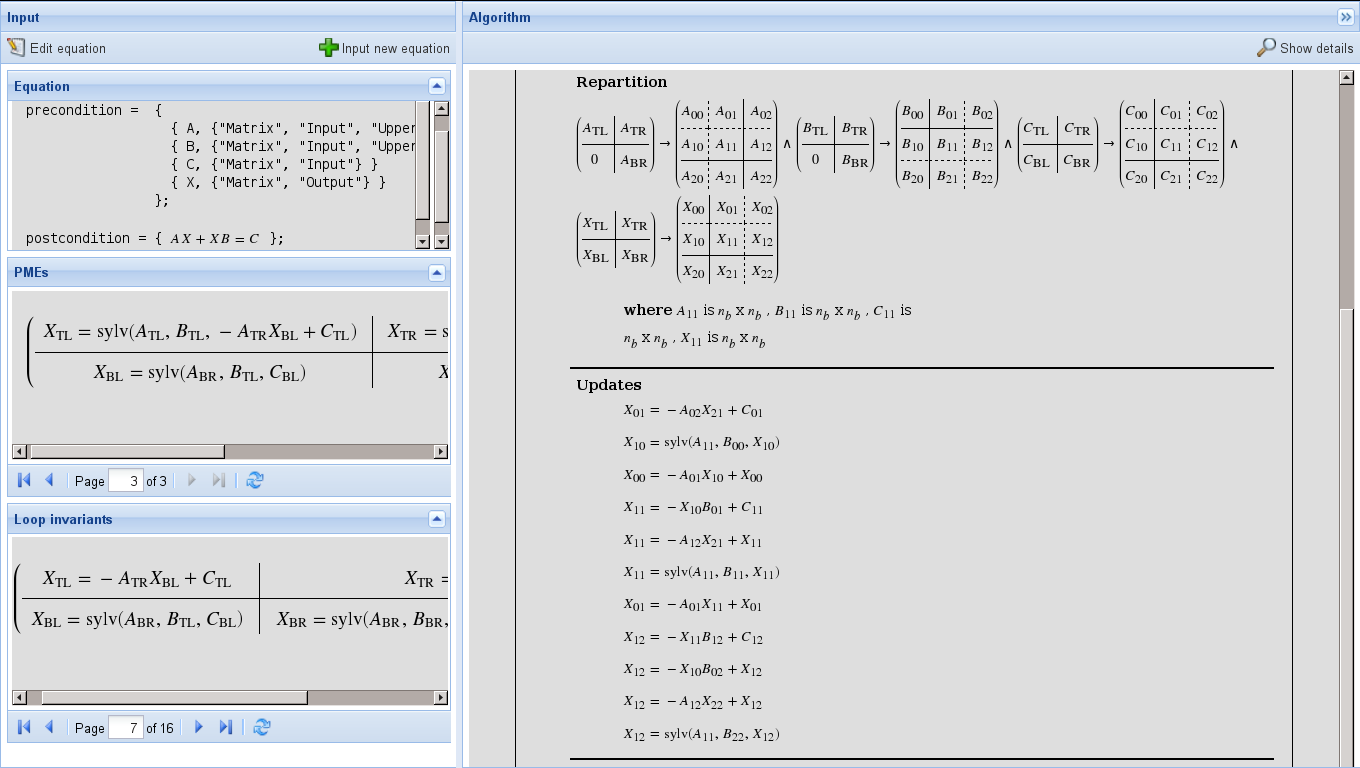}
	\caption{Output from \clickplain{}'s web interface for the input in
	         Figure~\ref{fig:SylvInput}. On the left panel, the interface
			 shows the equation itself, its PMEs, and its loop invariants;
			 on the right panel, the interface displays the algorithms.}
	\label{fig:SylvAlg}
\end{figure}

We made remarkable progress in this direction. First, 
we exposed in depth all the requirements to build such a system
and fully automated the process;
then we developed a user-friendly web interface where the
user is freed from every low level detail. 
Figures~\ref{fig:SylvInput}~and~\ref{fig:SylvAlg} contain two
screenshots of the web interface, 
corresponding to the triangular Sylvester equation used as example throughout
this chapter. 
For a comparison, we recall the formal definition of the operation:
\begin{equation} \nonumber
X := \Omega(A, B, C) \equiv
\left\{
\begin{split}
P_{\rm pre}: \{ & \prop{Input}{A} \wedge \prop{UpperTriangular}{A} \, \wedge \\
                & \prop{Input}{B} \wedge \prop{UpperTriangular}{B} \, \wedge \\
                & \prop{Input}{C} \wedge \prop{Output}{X} \} \\
\\[-2mm]
P_{\rm post}: \{ &  A X + X B = C \}.
\end{split}
\right.
\end{equation}
The first figure shows how the operation is input by the user;
as the reader can appreciate, the formal description and the input to
the interface match perfectly.

The second figure provides the output generated by the tool right after
clicking the {\em Ok} button. It corresponds to the loop invariant
number 7 from the third PME, i.e., the example used in 
Sections~\ref{subsec:pbef-paft}~and~\ref{subsec:updates}.
On the left-hand panel we find the PME and the loop invariant; 
on the right-hand panel we see the generated algorithm.
We emphasize that, in contrast to the several days that it would take by hand,
\click{} generated all 20 algorithms in only a few seconds. 

\section{Scope and limitations}

Given the mathematical definition of a target operation in terms
of the predicates Precondition and Postcondition, \click{}
produces a family of both algorithms and routines that compute it.
\click{} has been applied to a broad set of linear algebra
operations. We list a few examples:
\bi
	\item Vector-vector, matrix-vector, and matrix-matrix products (e.g., BLAS operations).
	\item Matrix factorizations, such as LU and Cholesky.
	\item Inversion of matrices.
	\item Operations arising in control theory, such as Lyapunov and Sylvester equations.
\ei
However, while it has been established that the information encoded in
the PME suffices to generate algorithms~\cite{PaulDj:PhD}, 
a precise characterization of the scope of the methodology, i.e., 
the class of operations that admit a PME, is still missing.

Beyond the scope of the methodology,
\click{} presents a number of limitations similar to those discussed 
in Section~\ref{sec:clak-scope} for \clak{}, i.e, the lack of:
1) a module to automatically select the best algorithms, 
2) code generators for multiple programming languages and programming paradigms, and
3) a mechanism to analyze the stability of the produced algorithms.
In this case, instead, promising work from Bientinesi et al.~\cite{Paolo-Stability}
proposes an extension to the FLAME methodology for the
systematic stability analysis of the generated algorithms.
While still far fetched, this extension opens up the possibility for 
the future development of a module for the automatic stability analysis of
\click{}-generated algorithms.

\section{Summary}

We presented \click{}, a prototype compiler for the automatic generation of
loop-based linear algebra algorithms. 
From the sole mathematical description of a target equation, 
\click{} is capable of generating families of algorithms that solve it.
To this end, \click{} adopts the FLAME methodology;
the application of the methodology is divided in three stages:
First, all PMEs for the target operation are generated;
then, for each PME, multiple loop invariants are identified;
finally, each loop invariant is used
to build a provably correct algorithm.
This chapter expands upon our work published in~\cite{CASC-2011-PME,ICCSA-2011-LINV}.

The list of contributions made in this chapter follows.

\bi
	\item Minimum knowledge. For a given equation, we characterize the minimum knowledge 
		required to automatically generate algorithms that solve it. 
		This is the equation itself together with the properties of its operands.
	\item Full automation. We fully automate the generation of algorithms from the sole
		mathematical description of the operation. Previous work~\cite{PaulDj:PhD} required
		a PME and a loop invariant (both manually derived) as input, and the approach to automatically find
		the algorithm updates was limited.
	\item Feasibility. Several times this project has been deemed unfeasible. 
		This chapter should serve as a precise reference on how to automate the process,
		and remove the skepticism.
\ei

For a wide class of linear algebra operations,
the developers are now relieved from tedious, often unmanageable, symbolic manipulation,
and only one \click{} separates them from the sought-after algorithms.

\chapter{\clickplain{}: High-Performance Specialized Kernels}
\chaptermark{HP Specialized Kernels}
\label{ch:flame-ad}

In the previous chapter, we demonstrated how \click{}
automates the application of the FLAME methodology
by means of multiple standard operations, such as the
LU and Cholesky factorizations.
While the application of \click{} to these operations
shows the potential of the compiler, routines to compute them
are already available from traditional libraries;
in fact, most of the algorithms
included in libFLAME~\cite{libflame} were derived using this methodology.
In this chapter, instead, we concentrate on demonstrating the
broad applicability of \click{}
by generating customized kernels for building blocks not
supported by standard numerical libraries.

When developing application libraries, it is not uncommon
to require kernels for building blocks that are closely related 
but not supported by libraries like BLAS or LAPACK.
The situation arises so often that extensions to traditional libraries 
are regularly proposed~\cite{2002:USB:567806.567807}; 
unfortunately, the inclusion of every possible kernel arising in applications is unfeasible.
While it may be possible to 
emulate the required
kernels via a mapping onto two or more available kernels,
this approach typically affects both routine's performance and developer's productivity;
alternatively, efficient customized kernels may be produced on demand.
We illustrate this issue by means of two example kernels arising in the context of 
algorithmic differentiation. 

Consider a program that solves a linear system of equations
$A X = B$, with symmetric positive definite coefficient matrix $A$,
and multiple right-hand sides $B$;
pseudocode for such a program follows.
First, $A$ is factored through a Cholesky factorization; then, two triangular linear systems
are solved to compute the unknown $X$.

$$
\begin{aligned}
	L L^T &= A        &\text{\sc (chol)} \\
	L Y   &= B \quad  &\text{\sc (trsm)} \\
	L^T X &= Y        &\text{\sc (trsm)}
\end{aligned}
$$

\noindent
When interested in the derivative of this program, e.g., for a sensitivity analysis, one must
compute the following sequence of derivative operations,
$$
\begin{aligned}
	\dv{L}{v}   \;\; L^T \; + \; L   \;\; \dv{L^T}{v} \; &= \; \dv{A}{v}  &\text{(g\sc chol)} \\
	\dv{L}{v}   \;\; Y   \; + \; L   \;\; \dv{Y}{v}   \; &= \; \dv{B}{v}  &\text{(g\sc trsm)} \\
	\dv{L^T}{v} \;\; X   \; + \; L^T \;\; \dv{X}{v}   \; &= \; \dv{Y}{v}, &\text{(g\sc trsm)} \\
\end{aligned}
$$
none of which is supported by high-performance libraries. 
For both kernels, \click{} is capable of generating high-performance algorithms
in a matter of seconds.

The aim of this chapter is two-fold:
First, we use the operation \gchol{} to provide a complete
self-contained example of the application of \click{}, 
from the description of the operation to the final algorithms.
Then, we present experimental results for both \gchol{} and \gtrsm{};
the corresponding routines attain high performance and scalability.

\section{A complete example: The derivative of the Cholesky factorization}
\label{sec:gchol}

To initiate the derivation of algorithms for the derivative of
the Cholesky factorization, \click{} requires the mathematical
description of the operation. 
Given a symmetric positive definite matrix $A$, 
the Cholesky factorization calculates a lower triangular matrix $L$ 
such that 
\begin{equation}
    \label{eq:cholesky}
    L L^T = A;
\end{equation}
its derivative is
$$\dv{L}{v} \;\; L^T \; + \;\; L \;\; \dv{L^T}{v} \;\; = \;\; \dv{A}{v},$$
where 
$L$ and $\dv{A}{v}$ are known, and
$\dv{L}{v}$ is sought after.
The quantities $L$ and $\dv{L}{v}$ are lower triangular matrices, and
$\dv{A}{v}$ is a symmetric matrix.\footnote{The rules to determine
the properties of the operands of a derivative equation were given
in Section~\ref{par:dv-inference}}
A formal description of the operation is given in Box~\ref{box:gCholOpDesc};
to simplify the notation and to avoid confusion, hereafter we replace $\dv{A}{v}$ and $\dv{L}{v}$ with $B$ and $G$, respectively.

\begin{mybox}
$$
\small
G = gChol(L, B) \equiv
\left\{
\begin{split}
	P_{\rm pre}: \{ & \prop{Output}{G} \; \wedge \; \prop{Input}{L}  \; \wedge \prop{Input}{B}  \;\; \wedge \\
	                & \prop{Matrix}{G} \; \wedge \; \prop{Matrix}{L} \; \wedge \prop{Matrix}{B} \;\; \wedge \\
	                & \prop{LowerTriangular}{G} \; \wedge \; \prop{LowerTriangular}{L} \;\; \wedge \\
                    & \prop{Symmetric}{B} \} \\
\\
P_{\rm post}: \{ &  G L^T + L G^T = B \}
\end{split}
\right.
$$
\caption{Formal description for the derivative of the Cholesky factorization.}
\label{box:gCholOpDesc}
\end{mybox}

We recall that this description is the sole input required by \click{}
to generate algorithms;
all the actions leading to the algorithms in Figure~\ref{fig:gchol-algs}
are carried out automatically.

\paragraph{Pattern Learning}

\sloppypar
Given the description of \gchol{} in Box~\ref{box:gCholOpDesc},
\click{} creates the pattern corresponding to \gchol{} (Box~\ref{box:gChol-desc}),
and incorporates it to its knowledge-base.
The system is now capable of identifying \gchol{} in the subsequent steps of the process.

\begin{mybox}
	\small
\begin{verbatim}
      
        equal[ 
          plus[ 
            times[ G_, trans[L_] ], 
            times[ L_, trans[G_] ] 
          ], 
          B_
        ] /;  isInputQ[L] && isInputQ[B] && isOutputQ[G] &&
              isMatrixQ[L] && isMatrixQ[B] && isMatrixQ[G] &&
              isLowerTriangularQ[L] && isSymmetricQ[B] && 
              isLowerTriangularQ[G]
\end{verbatim}
\caption{Mathematica pattern representing \gchol{}.}
\label{box:gChol-desc}
\end{mybox}

\subsection{Generation of the PME}

In this initial stage, \click{} 
first identifies the feasible sets of partitionings for the operands;
then, for each of these sets of partitionings, 
the system produces the corresponding partitioned postcondition, which gives raise to a number of equations;
finally, these equations are matched against known patterns, yielding the PME(s).

\subsubsection{Feasible Partitionings}

\begin{figure}
	\centering
	\includegraphics[scale=0.8]{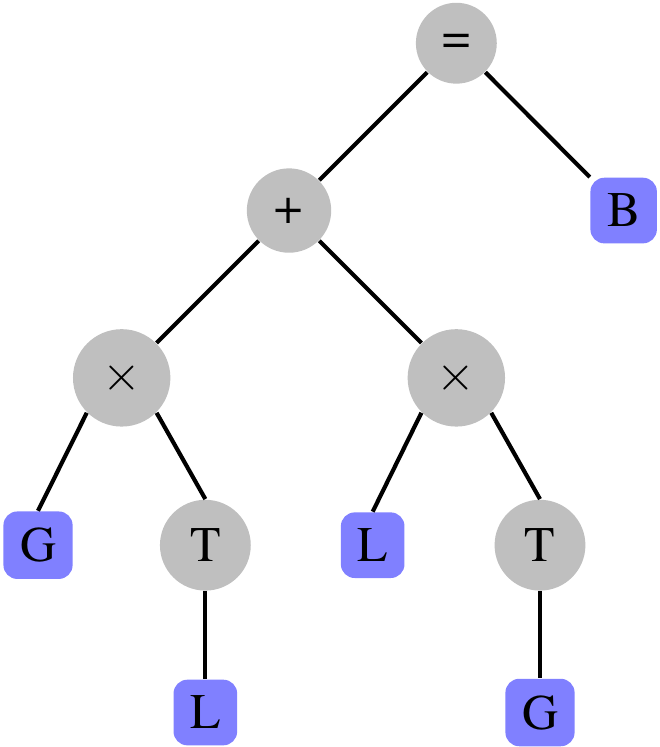}
	\caption{Tree representation of \gchol{}.}
	\label{fig:gCholtree}
\end{figure}

To find the sets of valid partitionings for the operands, 
\click{} applies the algorithm described in 
Section~\ref{subsec:automation} to the tree representation of \gchol{} (Figure~\ref{fig:gCholtree}).
The algorithm starts by creating a list of disjoint sets, 
one per dimension of the operands:
$$ [ \{ L_r \}, \{ L_c \}, \{ B_r \}, \{ B_c \}, \{ G_r \}, \{ G_c \} ]. $$
The tree is traversed in postorder.
Since $G$ is triangular, $G_c$ and $G_r$ are bound together
---the only admissible partitionings for $G$ are $1 \times 1$ and $2 \times 2$---;
similarly, the triangularity of $L$ imposes a binding between $L_r$ and $L_c$.
The corresponding sets of dimensions are merged, resulting in:
$$ [ \{ L_r, L_c \}, \{ B_r \}, \{ B_c \}, \{ G_r, G_c \} ]. $$

Next, the node for the transpose of $L$ is analyzed, not causing any binding.
Hence, \click{} continues by
studying the left-most ``$\times$'' operator. Its 
children, $G$ and $L^T$, are, respectively, of size $G_r \times G_c$ and $L_c \times L_r$;
a binding between $G_c$ and $L_c$ is thus imposed:
$$ [ \{ L_r, L_c, G_r, G_c \}, \{ B_r \}, \{ B_c \} ]. $$
A similar analysis of the subtree corresponding to $L \times G^T$, yields
no new bindings. Then, the ``+'' node is visited. The node's children
expressions are of size $G_r \times L_r$ and $L_r \times G_r$; no new bindings occur.
\click{} now proceeds with the analysis of the node corresponding to $B$; due to the symmetry of $B$, 
the sets containing $B_r$ and $B_c$ are merged:
$$ [ \{ L_r, L_c, G_r, G_c \}, \{ B_r, B_c \} ]. $$
Finally, the ``$=$'' operator, with left-hand side of size $G_r \times L_r$
and right-hand side of size $B_r \times B_c$, imposes the union of the 
remaining two sets of dimensions.
The final list of sets consists of a single group of dimensions:
$$ [ \{ L_r, L_c, G_r, G_c, B_r, B_c \} ]. $$

Since the application of the identity rule ($1 \times 1$) to all operands
does not lead to a valid partitioned postcondition, the only set of feasible
partitionings is the application of the $2 \times 2$ rule to every operand:
\begin{equation}
\renewcommand{\arraystretch}{1.4}
\label{eqn:gChol-Part}
	\begin{aligned}
      B_{m \times m}  
	  \rightarrow &
	  \left(
	    \begin{array}{c@{\;\;}|@{\;\;}c} 
		  B_{TL} & B_{BL}^T \\\hline 
		  B_{BL} & B_{BR} 
		\end{array} 
	  \right), \\
      \small \textnormal{where } & B_{TL} \textnormal{ is } k \times k
  \end{aligned}
	\begin{aligned}
      L_{m \times m}  
	  \rightarrow &
        \left( 
	      \begin{array}{c@{\;\;}|@{\;\;}c} 
	        L_{TL} & 0 \\\hline 
	        L_{BL} & L_{BR} 
	      \end{array} 
	    \right), \\
      \small \textnormal{where } & L_{TL} \textnormal{ is } k \times k
  \end{aligned}
	\begin{aligned}
      G_{m \times m}  
	  \rightarrow &
        \left( 
	      \begin{array}{c@{\;\;}|@{\;\;}c} 
	        G_{TL} & 0 \\\hline 
	        G_{BL} & G_{BR} 
	      \end{array}
	    \right). \\
      \small \textnormal{where } & G_{TL} \textnormal{ is } k \times k
  \end{aligned}
  \nonumber
\end{equation}
The newly created submatrices inherit a number of properties:
$B_{TL}$ and $B_{BR}$ are square and symmetric;
$B_{BL}$ and $B_{TR}$ are the transpose of one another;
$L_{TL}$, $L_{BR}$, $G_{TL}$, and $G_{BR}$ are lower triangular;
$L_{TR}$, and $B_{TR}$ are the zero matrix; and
$L_{BL}$, and $G_{BL}$ present no structure.
\click{} sets these properties and keeps track of them for future use.

\subsubsection{Matrix Algebra and Pattern Matching}

Next, the system replaces the operands in the postcondition by
their partitioned counterparts, producing the partitioned
postcondition

\begin{equation}
\label{eqn:gChol-PartPost}
\footnotesize
\renewcommand{\arraystretch}{1.4}
        \left( 
	      \begin{array}{c@{\;\;}|@{\;\;}c} 
	        G_{TL} & 0 \\\hline 
	        G_{BL} & G_{BR} 
	      \end{array} 
	    \right)
        \left( 
	      \begin{array}{c@{\;\;}|@{\;\;}c} 
			  L^T_{TL} & L^T_{BL} \\\hline 
	          0        & L^T_{BR} 
	      \end{array} 
	    \right)
		+
        \left( 
	      \begin{array}{c@{\;\;}|@{\;\;}c} 
	        L_{TL} & 0 \\\hline 
	        L_{BL} & L_{BR} 
	      \end{array} 
	    \right)
        \left( 
	      \begin{array}{c@{\;\;}|@{\;\;}c} 
			  G^T_{TL} & G^T_{BL}\\\hline 
	          0        & G^T_{BR} 
	      \end{array} 
	    \right)
		=
	  \left(
	    \begin{array}{c@{\;\;}|@{\;\;}c} 
		  B_{TL} & B_{BL}^T \\\hline 
		  B_{BL} & B_{BR} 
		\end{array} 
	  \right).
	  \nonumber
\end{equation}
The expression is multiplied out and the ``='' operator distributed, yielding
three equations:\footnote{The symbol $*$ means that the expression
	in the top-right quadrant is the transpose of that in the bottom-left one.}
\begin{equation}
\label{eqn:gChol-PME-Combine}
\renewcommand{\arraystretch}{1.4}
	\left( 
	  \begin{array}{c@{\;\;}|@{\;\;}c} 
		  G_{TL} L^T_{TL} + L_{TL} G^T_{TL} = B_{TL} & 
		  * \\\hline 
		  G_{BL} L^T_{TL} + L_{BL} G^T_{TL} = B_{BL} & 
		  G_{BL} L^T_{BL} + G_{BR} L^T_{BR} + L_{BL} G^T_{BL} + L_{BR} G^T_{BR} = B_{BR}
	  \end{array}
	\right).
\end{equation}

The iterative process towards the PME starts.
We recall the use of coloring to help the reader following the description:
\known{green} and \unknown{red} are used to highlight the \known{known} and \unknown{unknown}
operands, respectively.
The operands \known{$L$} and \known{$B$} are input to \gchol{}, and so are all sub-matrices resulting from 
their partitioning. \unknown{$G$} and its parts are output quantities. 
All three equations in~\eqref{eqn:gChol-PME-Combine} are in canonical form 
---on the left-hand side appear only output terms,
and on the right-hand side appear only input terms---. 
Hence, no initial algebraic manipulation is required:

\begin{equation}
\label{eqn:gChol-PME-StepOne}
\renewcommand{\arraystretch}{1.4}
	\left( 
	  \begin{array}{c@{\;\;}|@{\;\;}c} 
		  \unknown{G_{TL}} \known{L^T_{TL}} + \known{L_{TL}} \unknown{G^T_{TL}} = \known{B_{TL}} & 
		  * \\\hline 
		  \unknown{G_{BL}} \known{L^T_{TL}} + \known{L_{BL}} \unknown{G^T_{TL}} = \known{B_{BL}} & 
		  \unknown{G_{BL}} \known{L^T_{BL}} + \unknown{G_{BR}} \known{L^T_{BR}} + \known{L_{BL}} \unknown{G^T_{BL}} + \known{L_{BR}} \unknown{G^T_{BR}} = \known{B_{BR}}
	  \end{array} 
	\right).
\end{equation}

\click{} inspects~\eqref{eqn:gChol-PME-StepOne} for known patterns.
A \gchol{} operation 
is found in the top-left quadrant: 
The operation matches the pattern in Box~\ref{box:gChol-desc}, 
$L_{TL}$ and $B_{TL}$ are known matrices, $G_{TL}$ is unknown, 
$G_{TL}$ and $L_{TL}$ are lower triangular, and $B_{TL}$ is symmetric. 
The equation is rewritten as the assignment 
$G_{TL} := gChol( L_{TL}, B_{TL})$,
and the unknown quantity, $G_{TL}$, is labeled as computable and 
becomes known; this information is propagated to every appearance of the operand: 

\begin{equation}
\label{eqn:gChol-PME-StepTwo}
\renewcommand{\arraystretch}{1.4}
	\left( 
	  \begin{array}{c@{\;\;}|@{\;\;}c} 
		  \known{G_{TL}} := gChol( \known{L_{TL}}, \known{B_{TL}}) & 
		  * \\\hline 
		  \unknown{G_{BL}} \known{L^T_{TL}} + \known{L_{BL}} \known{G^T_{TL}} = \known{B_{BL}} & 
		  \unknown{G_{BL}} \known{L^T_{BL}} + \unknown{G_{BR}} \known{L^T_{BR}} + \known{L_{BL}} \unknown{G^T_{BL}} + \known{L_{BR}} \unknown{G^T_{BR}} = \known{B_{BR}}
	  \end{array} 
	\right).
\end{equation}

The bottom-left equation in~\eqref{eqn:gChol-PME-StepTwo} is not in canonical form anymore; 
a simple step of algebraic manipulation brings the equation back to canonical form:

\begin{equation}
\label{eqn:gChol-PME-StepThree}
\renewcommand{\arraystretch}{1.4}
	\left( 
	  \begin{array}{c@{\;\;}|@{\;\;}c} 
		  \known{G_{TL}} := gChol( \known{L_{TL}}, \known{B_{TL}}) & 
		  * \\\hline 
		  \unknown{G_{BL}} \known{L^T_{TL}} = \known{B_{BL}} - \known{L_{BL}} \known{G^T_{TL}} & 
		  \unknown{G_{BL}} \known{L^T_{BL}} + \unknown{G_{BR}} \known{L^T_{BR}} + \known{L_{BL}} \unknown{G^T_{BL}} + \known{L_{BR}} \unknown{G^T_{BR}} = \known{B_{BR}}
	  \end{array} 
	\right).
\end{equation}

The bottom-left equation is identified as a triangular system ({\sc trsm}). 
The output operand, $G_{BL}$, is computable, and turns green in
the bottom-right quadrant:

\begin{equation}
\label{eqn:gChol-PME-StepFour}
\renewcommand{\arraystretch}{1.4}
	\left( 
	  \begin{array}{c@{\;\;}|@{\;\;}c} 
		  \known{G_{TL}} := gChol( \known{L_{TL}}, \known{B_{TL}}) & 
		  * \\\hline 
		  \known{G_{BL}} := (\known{B_{BL}} - \known{L_{BL}} \known{G^T_{TL}}) \known{L^{-T}_{TL}} & 
		  \known{G_{BL}} \known{L^T_{BL}} + \unknown{G_{BR}} \known{L^T_{BR}} + \known{L_{BL}} \known{G^T_{BL}} + \known{L_{BR}} \unknown{G^T_{BR}} = \known{B_{BR}}
	  \end{array} 
	\right).
\end{equation}

A step of algebraic manipulation takes place to reestablish the canonical form
in the bottom-right equation, resulting in

\begin{equation}
\label{eqn:gChol-PME-StepFive}
\renewcommand{\arraystretch}{1.4}
	\left( 
	  \begin{array}{c@{\;\;}|@{\;\;}c} 
		  \known{G_{TL}} := gChol( \known{L_{TL}}, \known{B_{TL}}) & 
		  * \\\hline 
		  \known{G_{BL}} := (\known{B_{BL}} - \known{L_{BL}} \known{G^T_{TL}}) \known{L^{-T}_{TL}} & 
		  \unknown{G_{BR}} \known{L^T_{BR}} + \known{L_{BR}} \unknown{G^T_{BR}} = \known{B_{BR}} - \known{G_{BL}} \known{L^T_{BL}} - \known{L_{BL}} \known{G^T_{BL}}
	  \end{array} 
	\right).
\end{equation}

One last equation remains to be identified. Since $L_{BR}$ and $G_{BR}$ are lower triangular,
and the system can establish the symmetry of the right-hand side expression, $B_{BR} - G_{BL} L^T_{BL} - L_{BL} G^T_{BL}$,
the equation is matched as a \gchol{}. The output quantity, $G_{BR}$, becomes input.
No equation is left, the process completes, and the PME for \gchol{} (Box~\ref{box:gChol-PME})
is returned.

\begin{mybox}
\renewcommand{\arraystretch}{1.4}
	$$
		\left( 
		  \begin{array}{c@{\;\;}|@{\;\;}c} 
			  G_{TL} := gChol( L_{TL}, B_{TL}) & 
			  * \\\hline 
			  G_{BL} := (B_{BL} - L_{BL} G^T_{TL}) L^{-T}_{TL} & 
			  G_{BR} := gChol(L_{BR}, B_{BR} - G_{BL} L^T_{BL} - L_{BL} G^T_{BL})
		  \end{array} 
		\right)
	$$
	\caption{PME for the derivative of the Cholesky factorization.}
	\label{box:gChol-PME}
\end{mybox}

\subsubsection{PME Learning}

Once the PME is found, \click{} generates and stores in its knowledge-base the
rewrite rule displayed in Box~\ref{box:learn-PME},
which states how to decompose a \gchol{} problem with partitioned operands
into multiple subproblems.
We recall that this rule is essential for the flattening of the $\PBefore$ and $\PAfter$
predicates in later steps of the methodology.

\begin{mybox}
	\vspace*{4mm}
$\left( 
  \begin{array}{c@{\;\;}|@{\;\;}c} 
	G_{TL} & 0 \\\hline 
	G_{BL} & G_{BR} 
  \end{array}
\right)
$ :=
gChol $ \left(
        \left( 
	      \begin{array}{c@{\;\;}|@{\;\;}c} 
	        L_{TL} & 0 \\\hline 
	        L_{BL} & L_{BR} 
	      \end{array} 
	    \right)
	   ,
        \left( 
	      \begin{array}{c@{\;\;}|@{\;\;}c} 
			  B_{TL} & B^T_{BL} \\\hline 
	          B_{BL} & B_{BR} 
	      \end{array} 
	    \right)
	\right)
 \longrightarrow $
$$
\renewcommand{\arraystretch}{1.4}
	\left( 
	  \begin{array}{c@{\;\;}|@{\;\;}c} 
		  G_{TL} := gChol( L_{TL}, B_{TL} ) & 
		  * \\\hline 
		  G_{BL} := (B_{BL} - L_{BL} G^T_{TL}) L^{-T}_{TL} & 
		  G_{BR} := gChol(L_{BR}, B_{BR} - G_{BL} L^T_{BL} - L_{BL} G^T_{BL})
	  \end{array} 
	\right)
$$
\caption{Rewrite rule associated to \gchol{}'s PME.}
\label{box:learn-PME}
\end{mybox}

\subsection{Loop invariant identification}

From the PME, multiple loop invariants are identified 
in three successive steps: 
First, the PME is decomposed into a set of tasks;
then, a graph of dependencies among tasks is built; and
finally, the feasible subsets of the graph are returned as valid loop invariants.

\subsubsection{Decomposition into tasks}

\click{} analyzes the assignments in each quadrant of the PME, and
decomposes them into a series of tasks. 
The analysis commences from the top-left quadrant:
$ G_{TL} := gChol( L_{TL}, B_{TL} ) $. Since the right-hand
side consists of a function whose input arguments are
simple operands, no decomposition is required and
the assignment is returned as a single task.

The next inspected assignment is $G_{BL} := (B_{BL} - L_{BL} G^T_{TL}) L^{-T}_{TL}$.
No single pattern matches the right-hand side, which therefore must
be decomposed. Similarly to the decomposition undergone by \clak{}
in Chapter~\ref{ch:compiler}, \click{} first matches the expression 
$G_{BL} := B_{BL} - L_{BL} G^T_{TL}$
as a matrix-matrix product, and then identifies the remaining operation
$G_{BL} := G_{BL} L^{-T}_{TL}$ as the solution of a triangular system.
The two operations are yielded as tasks.

One last assignment remains to be studied: 
$G_{BR} := gChol(L_{BR}, B_{BR} - G_{BL} L^T_{BL} - L_{BL} G^T_{BL})$.
As in the top-left quadrant, it represents a \gchol{} function;
in this case, however, one of the input arguments is an expression.
The decomposition is carried out in two steps:
First, \click{} matches the expression 
$G_{BR} := B_{BR} - G_{BL} L^T_{BL} - L_{BL} G^T_{BL}$
with the pattern associated to the BLAS 3 operation {\sc syr2k} and returns it;
then, the function $G_{BR} := gChol(L_{BR}, G_{BR})$ is yielded.
The complete list of generated tasks is

\begin{enumerate} 
	\item $ G_{TL} := gChol( L_{TL}, B_{TL} ) $
	\item $ G_{BL} := B_{BL} - L_{BL} G^T_{TL} $
	\item $ G_{BL} := G_{BL} L^{-T}_{TL} $
	\item $ G_{BR} := B_{BR} - G_{BL} L^T_{BL} - L_{BL} G^T_{BL} $
	\item $ G_{BR} := gChol(L_{BR}, G_{BR}) $
\end{enumerate} 

\subsubsection{Graph of dependencies}

A graph of dependencies among tasks is built;
the analysis proceeds as follows 
(we recall the use of {\bf boldface} to highlight the dependencies).
The study commences with Task 1. A true dependency is found between
Tasks 1 and 2: The output operand of Task 1, $G_{TL}$,
appears as an input quantity to Task 2.
\bi
	\item[1.] $ {\bf G_{TL}} := gChol( L_{TL}, B_{TL} ) $
	\item[2.] $ G_{BL} := B_{BL} - L_{BL} {\bf G^T_{TL}}. $
\ei

\noindent
Next, Task 2 is inspected. Its output operand,
$G_{BL}$, is an input for Task 3:
\bi
	\item[2.] $ {\bf G_{BL}} := B_{BL} - L_{BL} G^T_{TL} $
	\item[3.] $ G_{BL} := {\bf G_{BL}} L^{-T}_{TL}, $
\ei
which imposes another dependency from Task 2 to 3.
Since $G_{BL}$ is also the output of Task 3,
an output dependency occurs;
the direction of the dependency is imposed during 
the decomposition of the assignment
($G_{BL} := (B_{BL} - L_{BL} G^T_{TL}) L^{-T}_{TL}$)
that originated these tasks: to ensure a correct
result, first Task 2 is computed, and then Task 3.

Two more true dependencies are found from
Task 3 to 4,
\bi
	\item[3.] $ {\bf G_{BL}} := B_{BL} L^{-T}_{TL} $
	\item[4.] $ G_{BR} := B_{BR} - {\bf G_{BL}} L^T_{BL} - L_{BL} {\bf G^T_{BL}}, $
\ei
and from Task 4 to 5
\bi
	\item[4.] $ {\bf G_{BR}} := B_{BR} - G_{BL} L^T_{BL} - L_{BL} G^T_{BL} $
	\item[5.] $ G_{BR} := gChol(L_{BR}, {\bf G_{BR}}). $
\ei
As for Tasks 2 and 3, 
an output dependency also exists between Tasks 4 and 5;
the direction is imposed
by the decomposition: First the argument to the function is
computed (Task 4), and then the function itself (Task 5).

\begin{figure}
	\centering
	\includegraphics[scale=1.0]{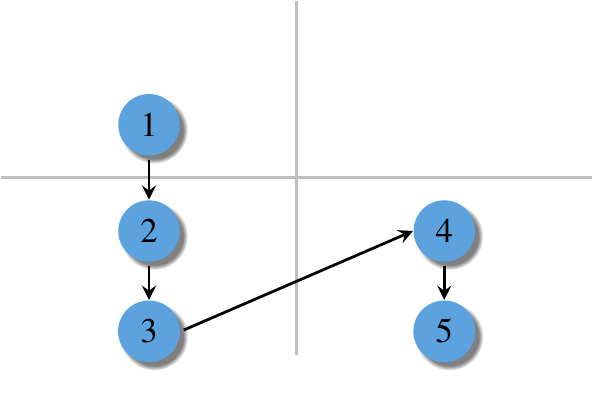}
	\caption{Task dependency graph obtained from the analysis of \gchol{}'s PME.}
	\label{fig:gChol-GraphDep}
\end{figure}

Finally, the output of Task 5, $G_{BR}$ does not appear as input
to any other task, and the analysis completes.
The resulting graph of dependencies is depicted in
Figure~\ref{fig:gChol-GraphDep}.

\subsubsection{Graph subsets selection}

Predicates candidate to be loop invariants are selected as subsets
of the graph that satisfy the dependencies.
To obtain the subsets, \click{} utilizes Algorithm~\ref{alg:subDAG}
(Section~\ref{sec:depGraph}). Since the application of the algorithm
to \gchol{}'s graph is rather straightforward, we skip the description
and give the final list of subsets:
$$[ \{\}, \{1\}, \{1, 2\}, \{1, 2, 3\}, \{1, 2, 3, 4\}, \{1, 2, 3, 4, 5\} ].$$

According to the rules stated in Chapter~\ref{ch:click}, the predicates corresponding
to the empty and full subgraphs are deemed not valid and discarded. 
The remaining four predicates lead to feasible loop invariants, which
are collected in Table~\ref{tab:gCholLoopInvs}. 
In all four loop invariants, the three operands ---$L$, $B$, and $G$---
are traversed from the top-left to the bottom-right corner.

\begin{table*}[!htb] \centering
\begin{tabular}{ccl} \toprule
\raisebox{2mm}{\bf \#} & \multicolumn{1}{c}{\raisebox{2mm}{\bf \footnotesize Subgraph}} & 
\multicolumn{1}{c}{\raisebox{2mm}{\bf \footnotesize Loop-invariant}} \\[-2.5mm]\midrule
1 &
\raisebox{-2.2em}{\smallDepGraphgChol{aicesred}{rwthblue}{rwthblue}{rwthblue}{rwthblue}} &
\renewcommand{\arraystretch}{1.4}
\scriptsize
$
	\left( 
	  \begin{array}{@{\,}c@{\;\;}|@{\;\;}c@{\,}} 
		  G_{TL} := gChol( L_{TL}, B_{TL} ) & 
		  * \\\hline 
		  \hspace{1.51cm} \neq \hspace{1.51cm} \phantom{} &
		  \qquad \, \neq \qquad \phantom{} \\
	  \end{array} 
	\right)
$ \\
2 &
\raisebox{-2.2em}{\smallDepGraphgChol{aicesred}{aicesred}{rwthblue}{rwthblue}{rwthblue}} &
\renewcommand{\arraystretch}{1.4}
\scriptsize
$
\renewcommand{\arraystretch}{1.4}
	\left( 
	  \begin{array}{c@{\;\;}|@{\;\;}c} 
		  G_{TL} := gChol( L_{TL}, B_{TL} ) & 
		  * \\\hline 
		  \hspace{3.5mm} G_{BL} := B_{BL} - L_{BL} G^T_{TL} \hspace{3.5mm} \phantom{} & 
		  \qquad \, \neq \qquad \phantom{} \\
	  \end{array} 
	\right)
$ \\
3 &
\raisebox{-2.2em}{\smallDepGraphgChol{aicesred}{aicesred}{aicesred}{rwthblue}{rwthblue}} &
\renewcommand{\arraystretch}{1.4}
\scriptsize
$
\renewcommand{\arraystretch}{1.4}
	\left( 
	  \begin{array}{c@{\;\;}|@{\;\;}c} 
		  G_{TL} := gChol( L_{TL}, B_{TL} ) & 
		  * \\\hline 
		  G_{BL} := (B_{BL} - L_{BL} G^T_{TL}) L^{-T}_{TL} & 
		  \qquad \, \neq \qquad \phantom{} \\
	  \end{array} 
	\right)
$ \\
4 &
\raisebox{-2.2em}{\smallDepGraphgChol{aicesred}{aicesred}{aicesred}{aicesred}{rwthblue}} &
\renewcommand{\arraystretch}{1.4}
\scriptsize
$
\renewcommand{\arraystretch}{1.4}
	\left( 
	  \begin{array}{c@{\;\;}|@{\;\;}c} 
		  G_{TL} := gChol( L_{TL}, B_{TL} ) & 
		  * \\\hline 
		  G_{BL} := (B_{BL} - L_{BL} G^T_{TL}) L^{-T}_{TL} & 
		  G_{BR} := B_{BR} - G_{BL} L^T_{BL} - L_{BL} G^T_{BL}
	  \end{array} 
	\right)
$ \\\bottomrule
\end{tabular}
\caption{Four loop invariants for the derivative of the Cholesky factorization.} 
\label{tab:gCholLoopInvs}
\end{table*}

\subsection{Algorithm construction}
\label{subsec:findingUpdates}

The final stage in the generation of algorithms consists in
constructing, for each loop invariant, the corresponding algorithm that computes \gchol{}.
The construction is carried out in three steps:
1) The repartitioning of the operands (Repartition and Continue with statements),
2) the rewrite of the loop invariant in terms of the repartitioned operands ($\PBefore$ and $\PAfter$ predicates), and
3) the comparison of these predicates to find the Algorithm Updates.
We continue the example by means of \gchol{}'s fourth loop invariant (Table~\ref{tab:gCholLoopInvs}).

\subsubsection{Repartitioning of the operands}

The loop invariant states that all three operands are traversed from top-left
to bottom-right. 
This traversal must be captured by the Repartition and Continue with statements;
Boxes~\ref{box:gchol-bef-rules}~and~\ref{box:gchol-aft-rules}, collect
the corresponding {\em Repartition} and {\em Continue with} rules.

\begin{mybox}
\setlength{\arraycolsep}{0pt}
\scriptsize
\centering
	\subfloat[]{
		$\myFlaTwoByTwoI{L_{TL}}{0}{L_{BL}}{L_{BR}}
		\; \rightarrow \;
		\myFlaTwoByTwoI{(L_{00})}
					   {\myFlaOneByTwo{0}{0}}
					   {\myFlaTwoByOne{L_{10}}{L_{20}}}
					   {\myFlaTwoByTwo{L_{11}}{0}{L_{21}}{L_{22}}}$
    } \qquad 
	\subfloat[]{
		$\myFlaTwoByTwoI{B_{TL}}{B_{BL}^T}{C_{BL}}{B_{BR}}
		\; \rightarrow \;
		\myFlaTwoByTwoI{(B_{00})}
					   {\myFlaOneByTwo{B_{10}^T}{B_{20}^T}}
					   {\myFlaTwoByOne{B_{10}}{B_{20}}}
					   {\myFlaTwoByTwo{B_{11}}{B_{21}^T}{B_{21}}{B_{22}}}$
    } \\
	\subfloat[]{
		$\myFlaTwoByTwoI{G_{TL}}{0}{G_{BL}}{G_{BR}}
		\; \rightarrow \;
		\myFlaTwoByTwoI{(G_{00})}
					   {\myFlaOneByTwo{0}{0}}
					   {\myFlaTwoByOne{G_{10}}{G_{20}}}
					   {\myFlaTwoByTwo{G_{11}}{0}{G_{21}}{G_{22}}}$
    }
	\caption{{\em Repartition rules} for \gchol{}'s fourth variant.}
\label{box:gchol-bef-rules}
\end{mybox}

\begin{mybox}
\setlength{\arraycolsep}{0pt}
\scriptsize
\centering
	\subfloat[]{
		$\myFlaTwoByTwoI{L_{TL}}{0}{L_{BL}}{L_{BR}}
		\; \rightarrow \;
		\myFlaTwoByTwoI{\myFlaTwoByTwo{L_{00}}{0}{L_{10}}{L_{11}}}
					   {\myFlaTwoByOne{0}{0}}
					   {\myFlaOneByTwo{L_{20}}{L_{21}}}
					   {(L_{22})}$
    } \qquad
	\subfloat[]{
		$\myFlaTwoByTwoI{B_{TL}}{B_{BL}^T}{B_{BL}}{B_{BR}}
		\; \rightarrow \;
		\myFlaTwoByTwoI{\myFlaTwoByTwo{B_{00}}{B_{10}^T}{B_{10}}{B_{11}}}
					   {\myFlaTwoByOne{B_{20}^T}{B_{21}^T}}
					   {\myFlaOneByTwo{B_{20}}{B_{21}}}
					   {(B_{22})}$
    } \\
	\subfloat[]{
		$\myFlaTwoByTwoI{G_{TL}}{0}{G_{BL}}{G_{BR}}
		\; \rightarrow \;
		\myFlaTwoByTwoI{\myFlaTwoByTwo{G_{00}}{0}{G_{10}}{G_{11}}}
					   {\myFlaTwoByOne{0}{0}}
					   {\myFlaOneByTwo{G_{20}}{G_{21}}}
					   {(G_{22})}$
    }
	\caption{{\em Continue with rules} for \gchol{}'s fourth variant.}
\label{box:gchol-aft-rules}
\end{mybox}

\subsubsection{Predicates $\PBefore$ and $\PAfter$}

To construct the $\PBefore$ and $\PAfter$ predicates, 
\click{} first rewrites the loop invariant in terms of the repartitioned operands.
To this end, \click{} applies the {\em Repartition} and {\em Continue with} rules,
producing, respectively, the top and bottom expressions in Figure~\ref{fig:gchol-bef-aft}.
Then, these expressions are flattened out using both basic matrix algebra and the 
PME rewrite rule in Box~\ref{box:learn-PME}. 
The final $\PBefore$ and $\PAfter$ predicates are also given in Figure~\ref{fig:gchol-bef-aft}.

\begin{sidewaysfigure}
\scriptsize
	\renewcommand{\arraystretch}{1.4}
	\centering
	\begin{tabular}{l c}
		$\PBefore:$ &
		$
		\myFlaTwoByTwo{(G_{00}) := gChol( (L_{00}), (B_{00}) )}
                      {*}
					  {\myFlaTwoByOne{G_{10}}{G_{20}} := \left(\myFlaTwoByOne{B_{10}}{B_{20}} - 
						                                      \myFlaTwoByOne{L_{10}}{L_{20}} (G_{00})^T\right) 
														(L_{00})^{-T}}
					  {\myFlaTwoByTwo{G_{11}}{0}{G_{21}}{G_{22}} := 
					                   \myFlaTwoByTwo{B_{11}}{B_{21}^T}{B_{21}}{B_{22}} - 
								       \myFlaTwoByOne{G_{10}}{G_{20}} \myFlaTwoByOne{L_{10}}{L_{20}}^T - 
								       \myFlaTwoByOne{L_{10}}{L_{20}} \myFlaTwoByOne{G_{10}}{G_{20}}^T}
		$ \\[9mm]
		& 
		{\Large $\Downarrow$} Flattening
		\\ [3mm]
		&
		$
		\myFlaThreeByThree{G_{00} := gChol( L_{00}, B_{00} )}
						  {0}
						  {0}
						  {G_{10} := (B_{10} - L_{10} G^T_{00}) L^{-T}_{00}}
						  {G_{11} :=  B_{11} - G_{10} L^T_{10} - L_{10} G^T_{10}}
						  {0}
						  {G_{20} := (B_{20} - L_{20} G^T_{00}) L^{-T}_{00}}
						  {G_{21} :=  B_{21} - G_{20} L^T_{10} - L_{20} G^T_{10}}
						  {G_{22} :=  B_{22} - G_{20} L^T_{20} - L_{20} G^T_{20}}
		$ \\[5mm]
        & \\\hline
        & \\[0mm]
        & {\large \em Algorithm Updates} \\[0mm]
        & \\\hline
        & \\
		$\PAfter:$ &
		$
		\myFlaThreeByThree{G_{00} := gChol( L_{00}, B_{00} )}
						  {0}
						  {0}
						  {G_{10} := (B_{10} - L_{10} G^T_{00}) L^{-T}_{00}}
						  {G_{11} := gChol( L_{11}, B_{11} - G_{10} L^T_{10} - L_{10} G^T_{10} )}
						  {0}
						  {G_{20} := (B_{20} - L_{20} G^T_{00}) L^{-T}_{00}}
						  {\begin{aligned}
							G_{21} := (B_{21} & - B_{20} L^{-T}_{00} L^T_{10} + L_{20} G^T_{00} L^{-T}_{00} L^T_{10} \\
											 & - L_{20} G^T_{10} - L_{21} G^T_{11}) L^{-T}_{11}
						  \end{aligned}}
						  {\begin{aligned}
							G_{22} :=  B_{22} & - G_{20} L^T_{20} - L_{20} G^T_{20} \\
											 & - G_{21} L^T_{21} - L_{21} G^T_{21}
						  \end{aligned}}
		$ \\[13mm]
		& 
		{\Large $\Uparrow$} Flattening
		\\ [5mm]
		&
		$
		\myFlaTwoByTwo{\myFlaTwoByTwo{G_{00}}{0}{G_{10}}{G_{11}} := 
                                      gChol\left( \myFlaTwoByTwo{L_{00}}{0}{L_{10}}{L_{11}}, 
                                                  \myFlaTwoByTwo{B_{00}}{B_{10}^T}{B_{10}}{B_{11}} 
                      \right)}
					  {*}
					  {\myFlaOneByTwo{G_{20}}{G_{21}} := 
                                     \left(\myFlaOneByTwo{B_{20}}{B_{21}} - 
                                      \myFlaOneByTwo{L_{20}}{L_{21}} \myFlaTwoByTwo{G_{00}}{0}{G_{10}}{G_{11}}^T\right) 
                                      \myFlaTwoByTwo{L_{00}}{0}{L_{10}}{L_{11}}^{-T}}
									  {\begin{aligned}
											  (G_{22}) := (B_{22}) & - \myFlaOneByTwo{G_{20}}{G_{21}} \myFlaOneByTwo{L_{20}}{L_{21}}^T \\
											                      & - \myFlaOneByTwo{L_{20}}{L_{21}} \myFlaOneByTwo{G_{20}}{G_{21}}^T
									  \end{aligned}}
		$
	\end{tabular}
	\caption{$\PBefore$ and $\PAfter$ predicates for \gchol{}'s fourth loop invariant.}
	\label{fig:gchol-bef-aft}
\end{sidewaysfigure}

\subsubsection{Finding the updates}

The final step undergone by \click{} consists in determining the
updates that take the computation from the state
in $\PBefore$ to the state in $\PAfter$.

The contents of both predicates only differ in quadrants
$G_{11}$, $G_{21}$, and $G_{22}$.
In the case of $G_{11}$ and $G_{22}$, the right-hand side of the expressions in
$\PBefore$ appear explicitly in the corresponding
quadrants of $\PAfter$.
Hence, for those quadrants, a direct replacement suffices to obtain the required
updates:
$$\begin{aligned}
	G_{11} &:= gChol( L_{11}, G_{11} ) \\
	G_{22} &:= G_{22} - G_{21} L^T_{21} - L_{21} G^T_{21}.
\end{aligned}$$

The before state for $G_{21}$, instead, is not explicitly found in the after state;
thus, both must be rewritten so that no redundant subexpressions appear. 
To avoid clutter, we anticipate that only the after state is rewritten;
\click{} uses the rule from quadrant $G_{20}$:
$$(B_{20} - L_{20} G^T_{00}) L^{-T}_{00} \rightarrow G_{20},$$
to rewrite the expression
$$
	G_{21} := (B_{21} - B_{20} L^{-T}_{00} L^T_{10} + L_{20} G^T_{00} L^{-T}_{00} L^T_{10}
					 - L_{20} G^T_{10} - L_{21} G^T_{11}) L^{-T}_{11}
$$
into
$$
{\begin{aligned}
	G_{21} &:= (B_{21} - G_{20} L^T_{10} - L_{20} G^T_{10} - L_{21} G^T_{11}) L^{-T}_{11}.
\end{aligned}}
$$

Now, the right-hand side of the before ($B_{21} - G_{20} L^T_{10} - L_{20} G^T_{10}$) is made explicit in the after,
and may be replaced, resulting in the update:

$$G_{21} := (G_{21} - L_{21} G^T_{11}) L^{-T}_{11}.$$

\noindent
The complete list of the sought-after updates is:

$$\begin{aligned}
	G_{11} &:= gChol( L_{11}, G_{11} ) \\
	G_{21} &:= (G_{21} - L_{21} G^T_{11}) L^{-T}_{11}       \\
	G_{22} &:=  G_{22} - G_{21} L^T_{21} - L_{21} G^T_{21}.
\end{aligned}$$

\subsection{The final algorithms}
\label{sec:gchol-algs}

The process described in the previous section is repeated for
each of the four loop invariants for \gchol{}.
As a result, \click{} generates the four algorithms collected 
in Figure~\ref{fig:gchol-algs}, and the corresponding routines listed in Appendix~\ref{sec:gchol-code}.
Given $B$, $L$, and $G$ of size $n \times n$,
the computational cost of the algorithms is $\frac{2}{3} n^3$.
While the derivation of these algorithms would take hours to an expert,
\click{} generates them in less than 5 seconds; it takes, literally,
more time to input the description of the operation than generating the algorithms.

\begin{figure}
\centering
	\resetsteps      





\renewcommand{\WSguard}{
	$ \text{\tt size}( B_{TL} ) < \text{\tt size}( B ) $
}


\renewcommand{\WSupdate}{
  \begin{tabular}{ c | c  } 
    \begin{minipage}[t]{5.2cm}
    \vspace*{-2pt}
    {\bf \underline{Variant 1}}\\[0.3mm]
	\vfill
	\hspace*{-2mm}
    \begin{tabular}{l l}
		\mbox{$G_{10} := B_{10} - L_{10} G^T_{00}$} & ({\sc trmm}) \\
		\mbox{$G_{10} := G_{10} L^{-T}_{00}$}       & ({\sc trsm}) \\
		\mbox{$G_{11} := B_{11} - G_{10} L^T_{10} - L_{10} G^T_{10}$} & ({\sc syr2k}) \\
		\mbox{$G_{11} := \text{gChol}(G_{11}, L_{11})$} & ({\gchol{}}) \\
	\end{tabular} 
    \end{minipage}
    &
    \begin{minipage}[t]{5.2cm}
    \vspace*{-2pt}
    {\bf \underline{Variant 2}}\\[0.3mm]
	\vfill
	\hspace*{-2mm}
    \begin{tabular}{ll}
		\mbox{$G_{10} := G_{10} L^{-T}_{00}$} & ({\sc trsm}) \\
		\mbox{$G_{11} := B_{11} - G_{10} L^T_{10} - L_{10} G^T_{10}$} & ({\sc syr2k}) \\
		\mbox{$G_{11} := \text{gChol}(G_{11}, L_{11})$} & ({\gchol{}}) \\
		\mbox{$G_{21} := B_{21} - L_{21} G^T_{11}$} & ({\sc trmm}) \\
		\mbox{$G_{21} := G_{21} - L_{20} G^T_{10}$} & ({\sc gemm}) \\
												   & \\
    \end{tabular} 
    \end{minipage}
	\\\hline
    \begin{minipage}[t]{5.2cm}
    \vspace*{-2pt}
    {\bf \underline{Variant 3}}\\[0.3mm]
	\vfill
	\hspace*{-2mm}
    \begin{tabular}{ll}
		\mbox{$G_{11} := B_{11} - G_{10} L^T_{10} - L_{10} G^T_{10}$} & ({\sc syr2k}) \\
		\mbox{$G_{11} := \text{gChol}(G_{11}, L_{11})$} & ({\gchol{}}) \\
		\mbox{$G_{21} := B_{21} - L_{21} G^T_{11}$} & ({\sc trmm}) \\
		\mbox{$G_{21} := G_{21} - L_{20} G^T_{10}$} & ({\sc gemm}) \\
		\mbox{$G_{21} := G_{21} - G_{20} L^T_{10}$} & ({\sc gemm}) \\
		\mbox{$G_{21} := G_{21} L^{-T}_{11}$} & ({\sc trsm}) \\
	\end{tabular}
    \end{minipage}
    &
    \begin{minipage}[t]{5.2cm}
    \vspace*{-2pt}
	{\bf \underline{Variant 4}}\\[0.3mm]
	\vfill
	\hspace*{-2mm}
    \begin{tabular}{ll}
		\mbox{$G_{11} := \text{gChol}(G_{11}, L_{11})$} & ({\gchol{}}) \\
		\mbox{$G_{21} := G_{21} - L_{21} G^T_{11}$} & ({\sc trmm}) \\
		\mbox{$G_{21} := G_{21} L^{-T}_{11}$} & ({\sc trsm}) \\
		\mbox{$G_{22} := G_{22} - G_{21} L^T_{21} - L_{21} G^T_{21}$} & ({\sc syr2k})\\
    \end{tabular}
    \end{minipage}\\
  \end{tabular}
}


\renewcommand{\WSpartition}{
  $
  B \rightarrow
  \FlaTwoByTwoI{B_{TL}}{\star}
               {B_{BL}}{B_{BR}},
  L \rightarrow
  \FlaTwoByTwoI{L_{TL}}{0}
               {L_{BL}}{L_{BR}},
  G \rightarrow
  \FlaTwoByTwoI{G_{TL}}{0}
               {G_{BL}}{G_{BR}}
  $
}

\renewcommand{\WSpartitionsizes}{$ 
	B_{TL}, L_{TL}, \text{ and } G_{TL} $ are $ 0 \times 0 
$}



\renewcommand{\WSrepartition}{$
  \setlength{\arraycolsep}{1pt}
  \FlaTwoByTwoI{B_{TL}}{\star}
               {B_{BL}}{B_{BR}}
  \!\!\rightarrow\!\!
  \FlaThreeByThreeBRI{B_{00}}{\star}{\star}
                     {B_{10}}{B_{11}}{\star}
                     {B_{20}}{B_{21}}{B_{22}},
  \FlaTwoByTwoI{L_{TL}}{0}
               {L_{BL}}{L_{BR}}
  \!\!\rightarrow\!\!
  \FlaThreeByThreeBRI{L_{00}}{0}{0}
                     {L_{10}}{L_{11}}{0}
                     {L_{20}}{L_{21}}{L_{22}},
  \FlaTwoByTwoI{G_{TL}}{0}
               {G_{BL}}{G_{BR}}
  \!\!\rightarrow\!\!
  \FlaThreeByThreeBRI{G_{00}}{0}{0}
                     {G_{10}}{G_{11}}{0}
                     {G_{20}}{G_{21}}{G_{22}}
$}

\renewcommand{\WSrepartitionsizes}{
	$ B_{11}, L_{11}, \text{ and } G_{11} $ are $ b \times b $
}


\renewcommand{\WSmoveboundary}{$
  \setlength{\arraycolsep}{1pt}
  \FlaTwoByTwoI{B_{TL}}{\star}
               {B_{BL}}{B_{BR}}
  \!\!\leftarrow\!\!
  \FlaThreeByThreeTLI{B_{00}}{\star}{\star}
                     {B_{10}}{B_{11}}{\star}
                     {B_{20}}{B_{21}}{B_{22}},
  \FlaTwoByTwoI{L_{TL}}{0}
               {L_{BL}}{L_{BR}}
  \!\!\leftarrow\!\!
  \FlaThreeByThreeTLI{L_{00}}{0}{0}
                     {L_{10}}{L_{11}}{0}
                     {L_{20}}{L_{21}}{L_{22}},
  \FlaTwoByTwoI{G_{TL}}{0}
               {G_{BL}}{G_{BR}}
  \!\!\leftarrow\!\!
  \FlaThreeByThreeTLI{G_{00}}{0}{0}
                     {G_{10}}{G_{11}}{0}
                     {G_{20}}{G_{21}}{G_{22}}
$}

\begin{minipage}[t]{\textwidth}
	\setlength{\arraycolsep}{2pt}
	\scriptsize 
	\myFlaAlgorithm
\end{minipage}

	\caption{The four algorithms for the derivative of the Cholesky factorization generated by {\sc Cl1ck}.}
	\label{fig:gchol-algs}
\end{figure}

\subsection{Experimental results}
\label{sec:gchol-experiments}

We turn now the attention towards the experimental results.
We show that the generated algorithms are not only of theoretical interest, 
but also of practical relevance.
Details on the computing environment for the experiments 
can be found in Section~\ref{sec:ad-clak-experiments}.

\begin{figure}[!ht]
	\centering
	\includegraphics[scale=0.82]{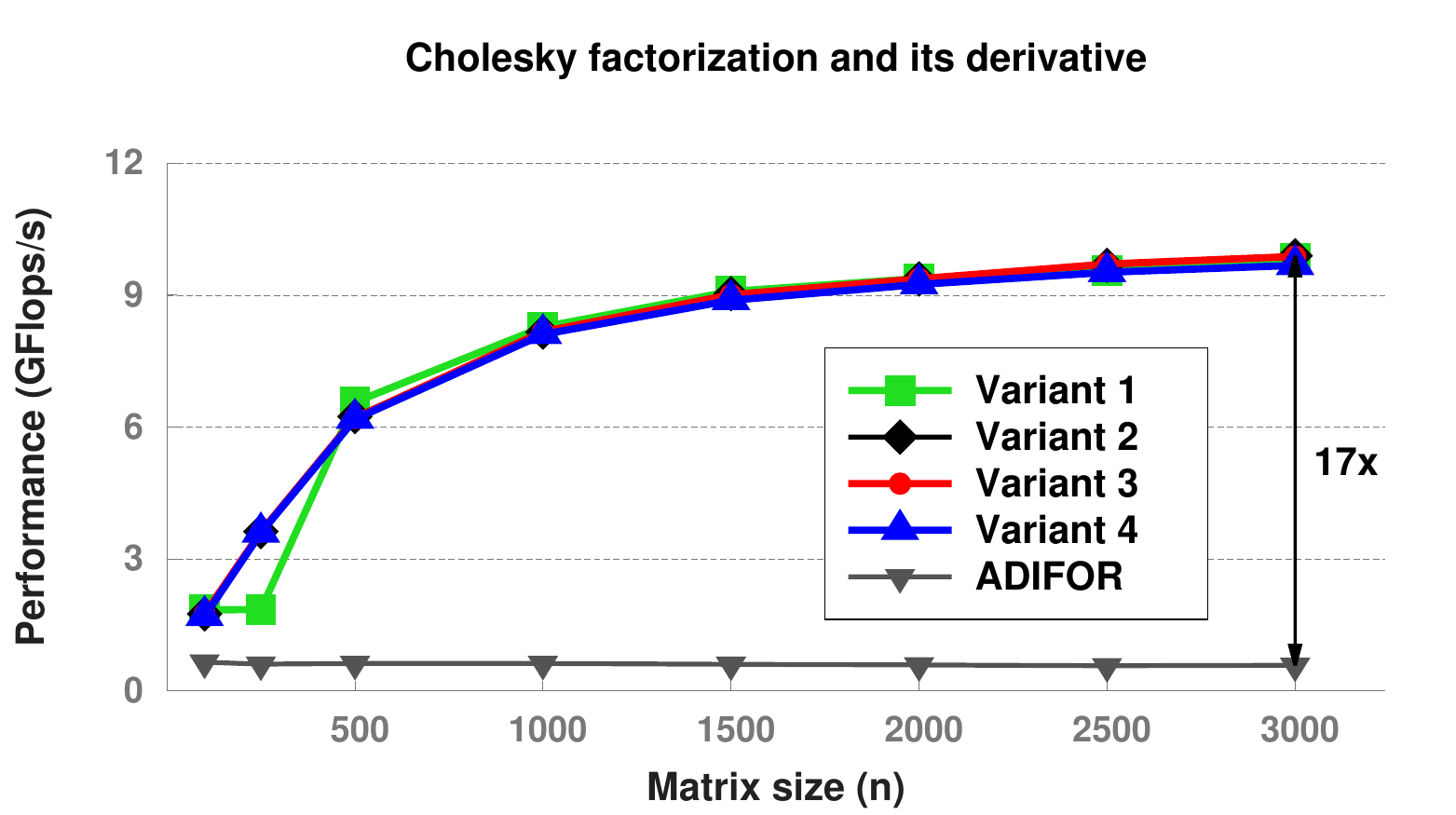}
	\caption{Comparison of the performance of \clickplain{}'s four variants for \gchol{}
		with the routine generated by ADIFOR.
	The experiments were run using a single thread.}
	\label{fig:both-single}
\end{figure}

We first compare, in Figure~\ref{fig:both-single},
the four algorithms generated by \click{}
(labeled ``Variant 1'' to ``Variant 4'')
with the routine generated by ADIFOR.
We recall that ADIFOR produces a single routine for the computation of both
the Cholesky factorization and its derivative;
therefore, to ensure a fair comparison,
we also include in the timings for our routines the execution time
of both the Cholesky factorization (via LAPACK's {\sc dpotrf})
and its derivative.
The gap in performance stands out: while ADIFOR's routine attains
about 0.6 GFlops/s,
all four \click{} variants attain between 9.5 and 10 GFlops/s.
\click{}'s fastest routine is 17 times faster than ADIFOR's.

\begin{figure}
	\centering
	\includegraphics[scale=0.82]{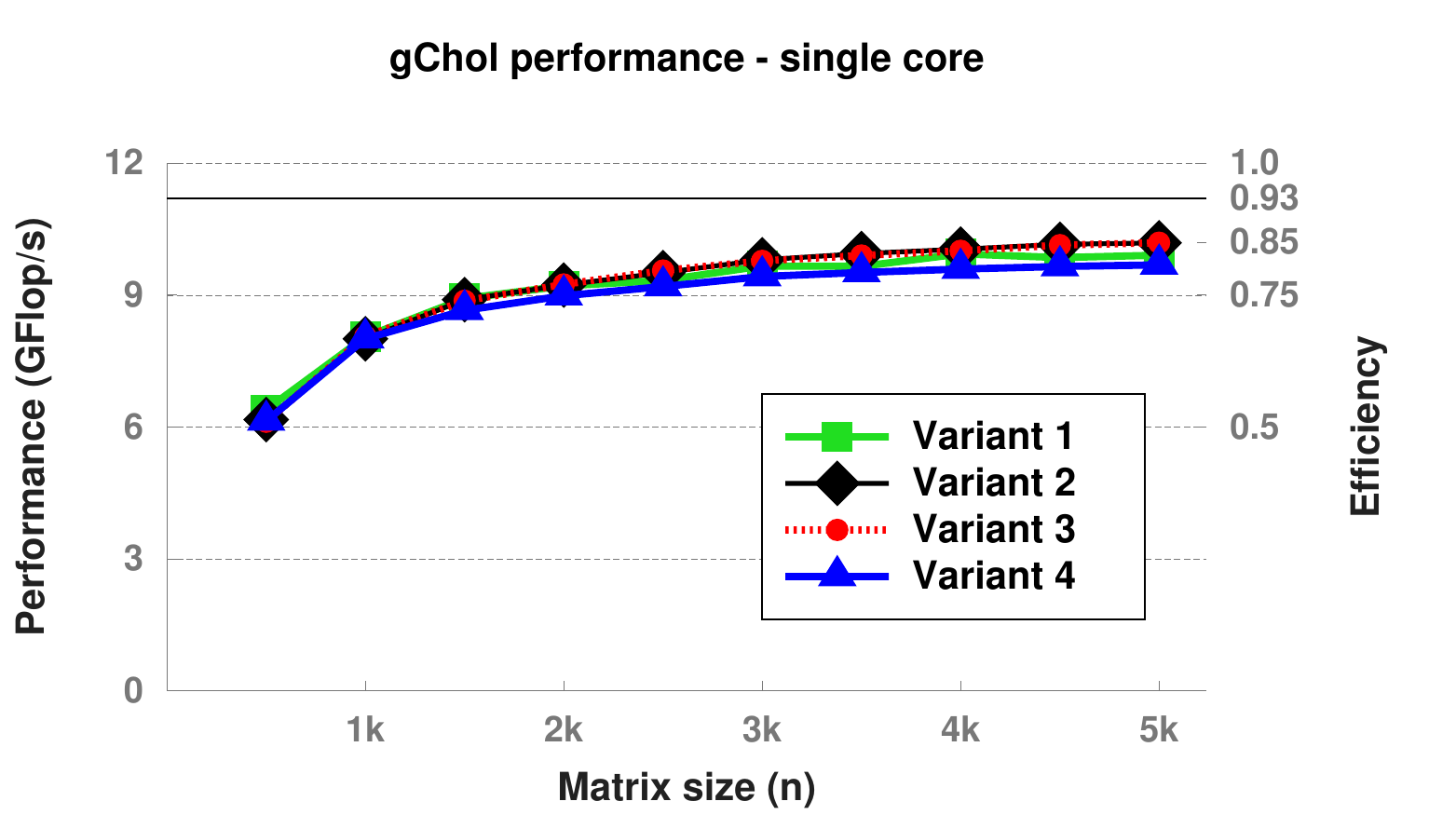}
	\caption{Comparison of the performance of \clickplain{}'s four variants for \gchol{}.
		The top border represents the architecture's theoretical
		peak performance, while the horizontal black line is the peak performance of
	{\sc gemm} (the practical peak).}
	\label{fig:gChol-single}
\end{figure}

We set now ADIFOR aside and concentrate on the performance of
\click{}'s routines for the computation of \gchol{} exclusively.
Figure~\ref{fig:gChol-single} shows results using a single core. 
The top line (12 GFlop/s) is the theoretical peak of the architecture,
and the horizontal black line represents the performance of {\sc gemm},
arguably the practical peak (93\% of the theoretical).
As the figure shows, all 4 variants are very efficient 
(over 75\% of the peak), and the best two ---Variants 2 and 3--- attain 
a performance of 10.2 GFlop/s, very close to that of {\sc gemm} (11.2 GFlop/s).

\begin{figure}
	\centering
	\includegraphics[scale=0.80]{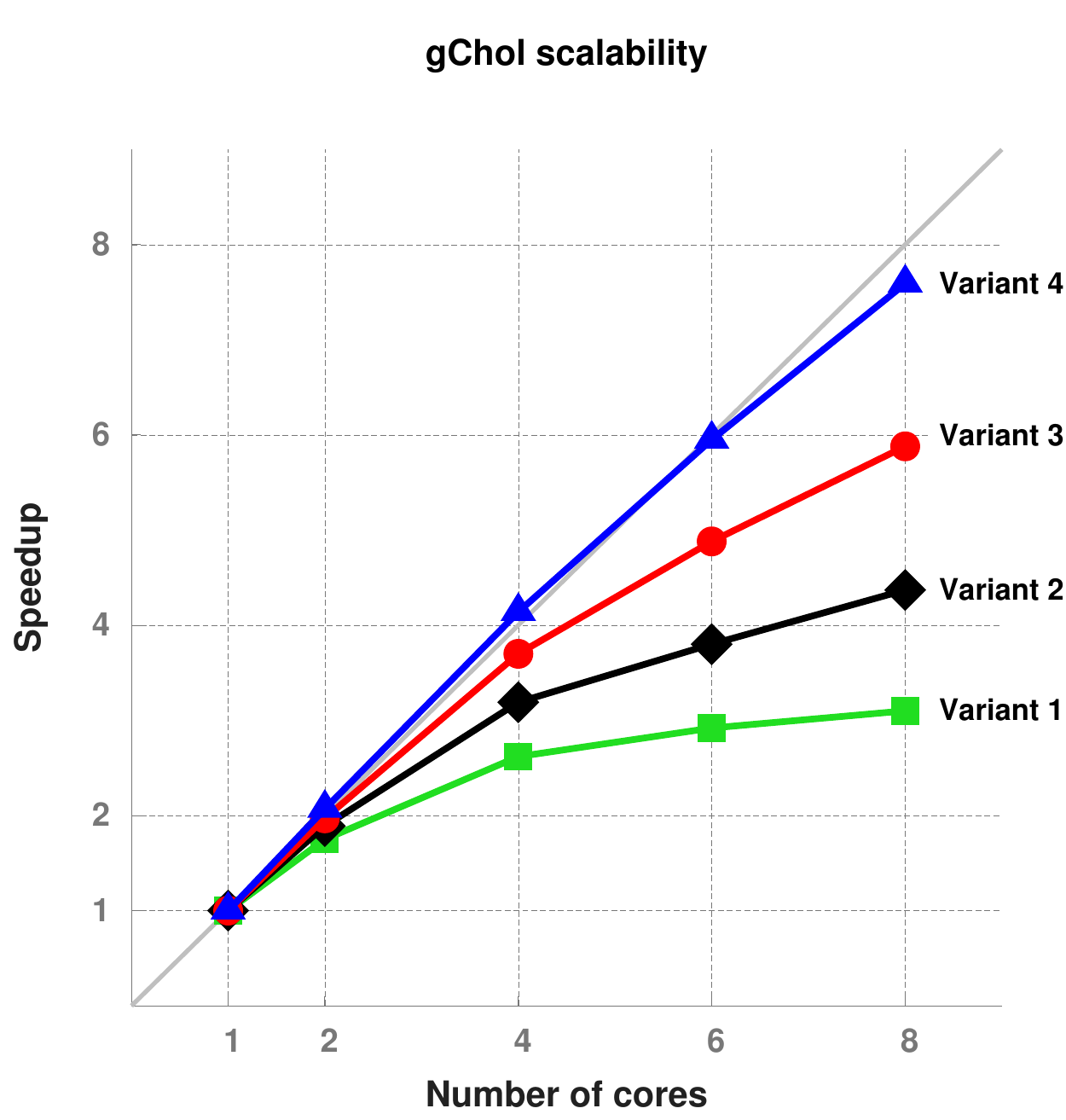}
	\caption{Scalability of the four variants for \gchol{}. 
		     The close-to-perfect scalability of variant 4 stands out, 
			 and larger speedups are expected when more cores are available.
		     The problem size is fixed to $n = 20{,}000$.}
	\label{fig:gChol-scal}
\end{figure}

We also ran experiments for parallel versions of the routines; 
parallelism is achieved via a multi-threaded version of the BLAS library. 
We first look at the scalability of the routines, and then
present performance results for their execution using 8 threads.
In Figure~\ref{fig:gChol-scal}, we show the speedup achieved for
up to 8 cores; the problem size was fixed to $n = 20{,}000$.
The diagonal gray line represents the perfect scalability 
(speedup equal to the number of cores).
The scalability of Variants 1 and 2 is rather limited
due to the type and shape of the operations
that perform the bulk of the computation:
In both cases roughly half the computation is carried out
in the {\sc trsm} operation
---$G_{10} := G_{10} L_{00}^{-T}$---,
which, in this specific shape (a small number of rows in $G_{10}$),
presents limited scalability;
in Variant 1, the other half of the computation is carried out
in the {\sc trmm} $B_{10} - L_{10} G_{00}^T$,
that due to limitations inherent to the BLAS interface,
requires memory allocation and copy, further
limiting the algorithm's scalability.
As for Variants 3 and 4, most of the computation
is performed via matrix-products (lines 4 and 5
in Variant 3, and line 4 in Variant 4).
While both scale well, the specific shape of 
the {\sc syr2k} in Variant 4 is better suited for
shared-memory parallelism and attains almost perfect
scalability; the resulting speedup is of almost 8x for 8 cores.

\begin{figure}
	\centering
	\includegraphics[scale=0.82]{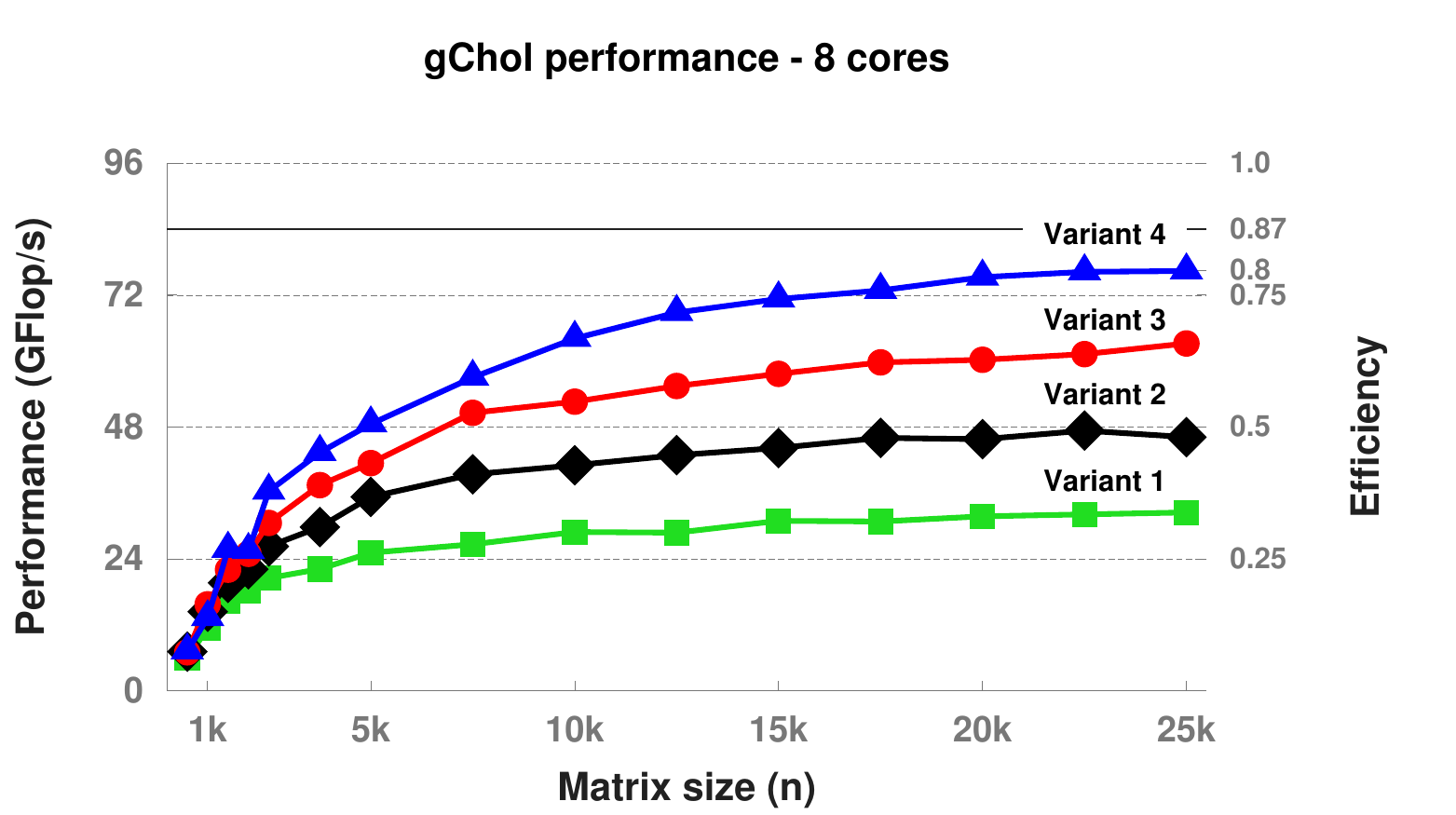}
	\caption{Comparison of the performance of \clickplain{}'s four variants for \gchol{}.
		The top border represents the architecture's theoretical
		peak performance, while the horizontal black line is the peak performance of
		{\sc gemm} (the practical peak).}
	\label{fig:gChol-eight}
\end{figure}

Finally, in Figure~\ref{fig:gChol-eight}, we collect performance results for the
four routines when using 8 cores. 
The top line (96 GFlop/s) is the theoretical peak of the architecture,
and the horizontal black line represents the performance of {\sc gemm},
arguably the practical peak (87\% of the theoretical).
The performance of the fastest routine (76.5 GFlop/s) is close to
that of {\sc gemm} (84 GFlop/s).
Interestingly, while the best variants in
the single-core case are variants 2 and 3, the best suited for multi-core architectures
is variant 4. This is one more example of why multiple variants are desired.

The bottom line of this section is that \click{} generated efficient and scalable 
routines at the effort of just one single click.

\section{The derivative of TRSM}
\label{sec:gtrsm}

To conclude the motivating example---the derivative of
the building blocks involved in the solution of an SPD linear system---,
we now focus on the solution of the \gtrsm{} equation
$$L' X + L X' = B',$$
for unknown $X'$, where
matrices $B$, $X$, $X' \in R^{m \times n}$,
and matrices $L$, $L' \in R^{n \times n}$ are lower triangular. 

We discuss two possible approaches to compute this operation efficiently:
1) Mapping it onto calls to BLAS routines, in line with \clak{}.
2) Developing a customized blocked routine.
The first solution is relatively straightforward, but an overhead is often paid either in
terms of extra computation or extra memory (operand) accesses;
the second solution, instead, is much harder to implement (by hand), but more efficient.
We show that \click{} is capable of generating automatically customized kernels that
outperform a straight mapping onto BLAS kernels.

The equation for \gtrsm{} may be rewritten as
\begin{equation}
	X' = L^{-1}(B' - L' X).
	\label{eq:gtrsm-blas}
\end{equation}
Two possible mappings of Equation~\eqref{eq:gtrsm-blas}
onto BLAS are given by Algorithms~\ref{alg:gemm-trsm}
({\sc gemm+trsm})~and~\ref{alg:trmm-trsm} ({\sc trmm+trsm}).
The difference between these two algorithms lies in the kernel
used to compute the matrix product $B' - L'X$.
Algorithm~\ref{alg:gemm-trsm} relies on the {\sc gemm} kernel,
which computes ``$\alpha A B + \beta C$'' for general matrices $A$, $B$, and $C$.
Due to the triangularity of $L'$, {\sc gemm}
performs twice the required computation;
the algorithm incurs in a 50\% redundant computation 
($3 n^2m$ flops, instead of the only $2 n^2m$ of Algorithm~\ref{alg:trmm-trsm}).
Alternatively, Algorithm~\ref{alg:trmm-trsm} exploits the triangularity of $L$
by means of the specialized BLAS kernel {\sc trmm} (line 1),
which performs the matrix product ``$\alpha A B$'' with either $A$ or $B$ triangular;
however, in this case 
the subtraction of the product $L' X$ from $B'$ is not directly supported in the call
to {\sc trmm}, and it must be calculated in a separate step (line 2).
The algorithm performs roughly $2 n^2m$, but it is penalized by the overhead due to 
the multiple sweeps through the operands, and the corresponding increase in memory traffic.

\begin{center}
\renewcommand{\lstlistingname}{Algorithm}
\begin{minipage}{0.45\linewidth}
	\begin{lstlisting}[numbers=left,caption={{\sc gemm + trsm}}, escapechar=!, label=alg:gemm-trsm]
$X' := B' - L' X$              !({\sc gemm} - $2 n^2m$)!
$X' := L^{-1} X'$              !({\sc trsm} - $  n^2m$)! !\vspace{4.3mm}!
	\end{lstlisting}
\end{minipage}
\hfill
\begin{minipage}{0.45\linewidth}
	\begin{lstlisting}[numbers=left,caption={{\sc trmm + trsm}}, escapechar=!, label=alg:trmm-trsm]
$T := L' X$              !({\sc trmm} - $n^2m$)!
$X' := B' - T$              !({\sc sub\phantom{A}} - $n^2$)!
$X' := L^{-1} X'$              !({\sc trsm} - $n^2m$)!
	\end{lstlisting}
\end{minipage}
\end{center}

The alternative is the use of customized blocked algorithms as generated by \click{}.
\gtrsm{} is seemingly simpler than \gchol{}, but applying FLAME's methodology to the operation
is far more complex due to the large number of loop invariants, and thus algorithms, found.
\click{} generates more than a hundred routines; all of them require 
$\frac{2}{3} n^2m$ flops, and perform a single sweep through the operands.
We skip the details and jump directly into the experimental results.
In Appendix~\ref{sec:gtrsm-code}, we provide the two variants used for the experiments in this section.

\subsection{Experimental results}
\label{subsec:gtrsm-experiments}

We study the performance of four routines to compute \gtrsm{}:
The first two correspond to {\sc gemm+trsm} and {\sc trmm+trsm};
the other two were generated by \click{} (variants 6 and 63).
All four routines are written in C; 
the experiments were run in the same computing environment used
for our previous tests (see Section~\ref{sec:ad-clak-experiments}).

\begin{figure}
	\centering
	\includegraphics[scale=0.82]{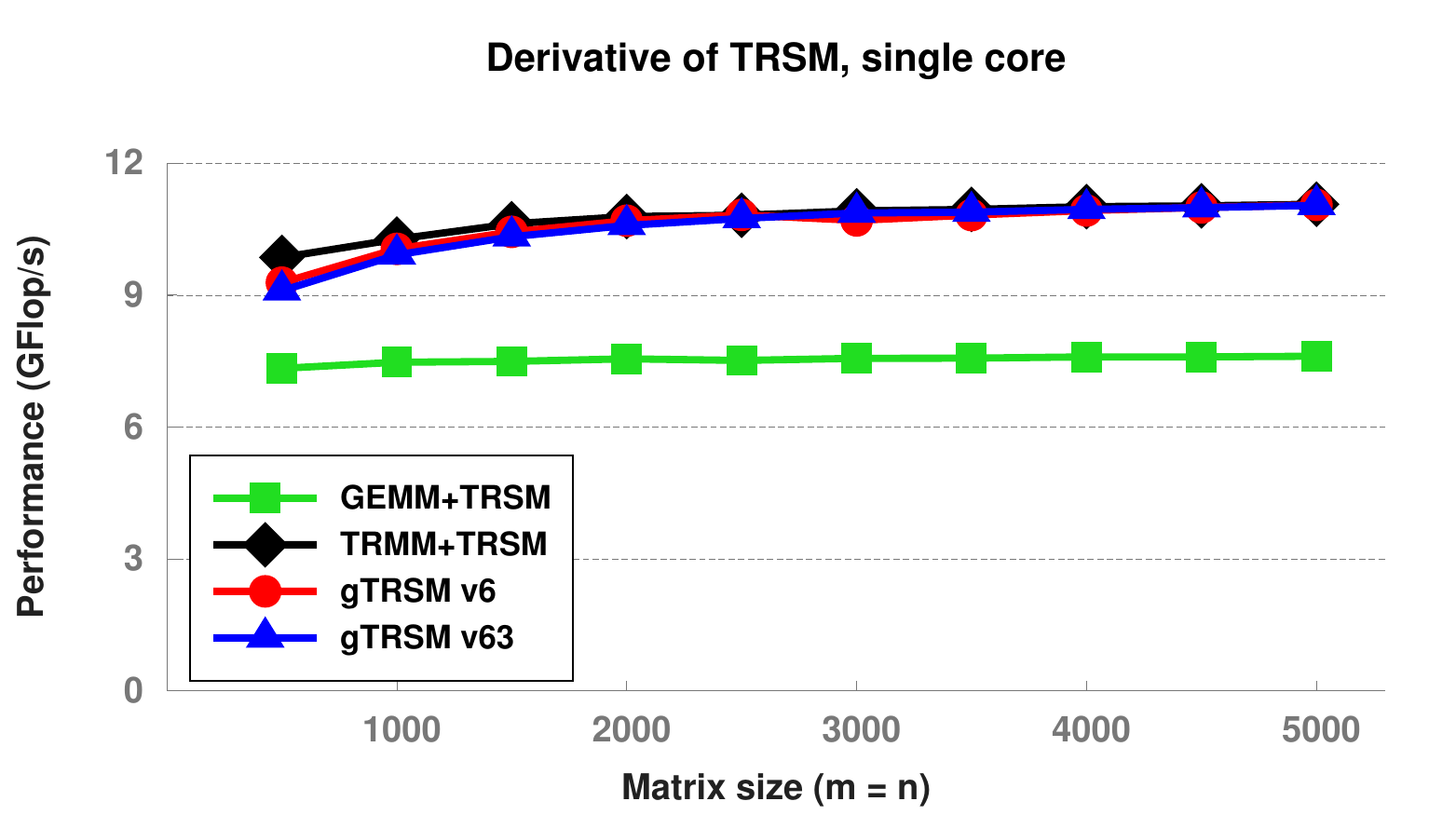}
	\caption{Performance comparison of the four routines for \gtrsm{}.
	Experiments run using a single core.}
	\label{fig:gTRSM-single}
\end{figure}

Figure~\ref{fig:gTRSM-single} contains the timings for the
single-threaded version of the routines. 
Not surprisingly, {\sc gemm+trsm} performs worst due to the
50\% extra computation. 
Despite the extra memory accesses in {\sc trmm+trsm},
the remaining three routines perform very similarly.
It is worth emphasizing that the three best routines attain
more than 90\% of the architecture's peak performance.

\begin{figure}
	\centering
	\includegraphics[scale=0.8]{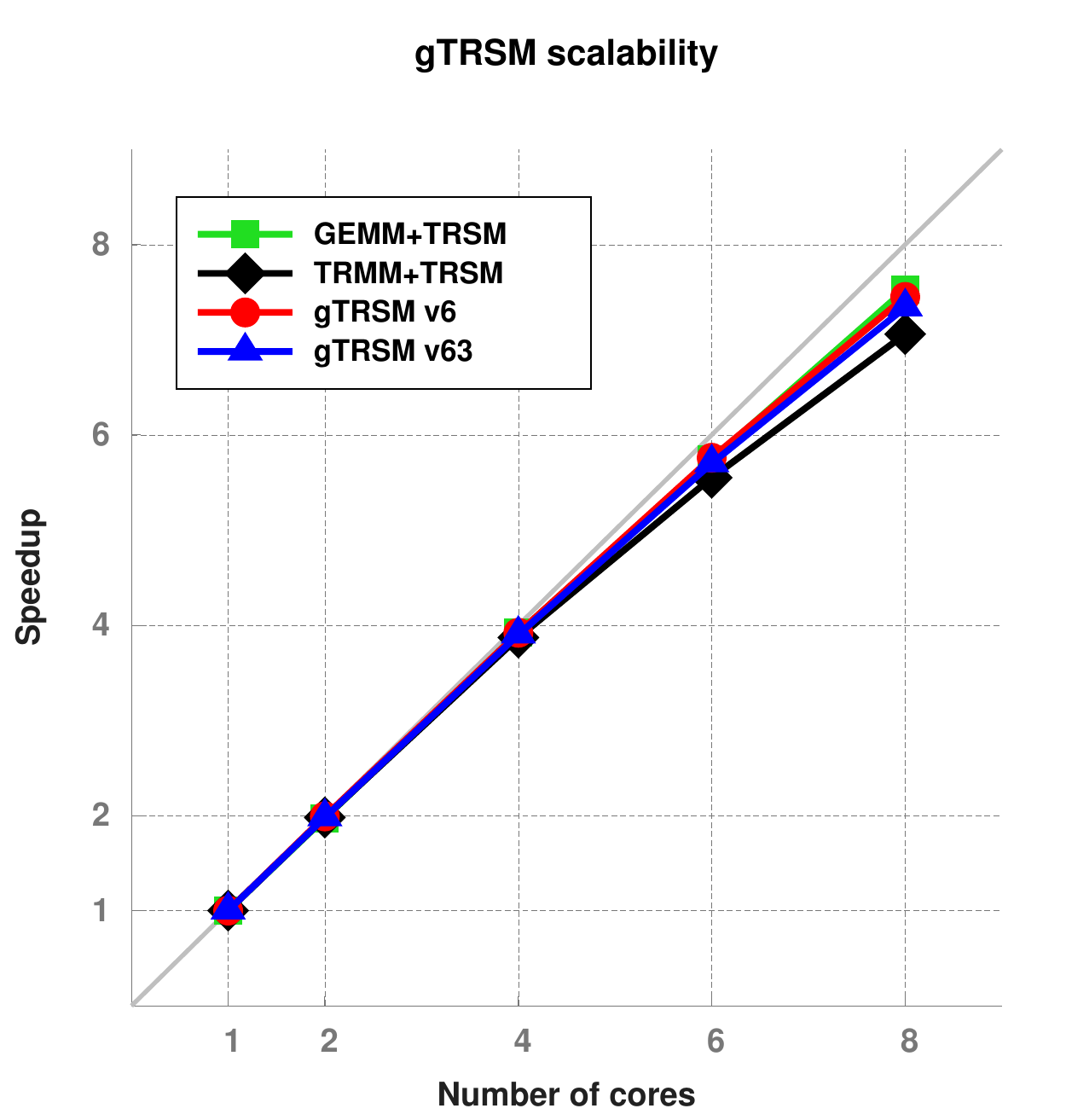}
	\caption{Scalability of the four variants to compute \gtrsm{}.
	         The gray diagonal line represents the perfect scalability.
		     Problem size: $m = n = 10{,}000$.
		     }
	\label{fig:gTRSM-scal}
\end{figure}

\begin{figure}
	\centering
	\includegraphics[scale=0.82]{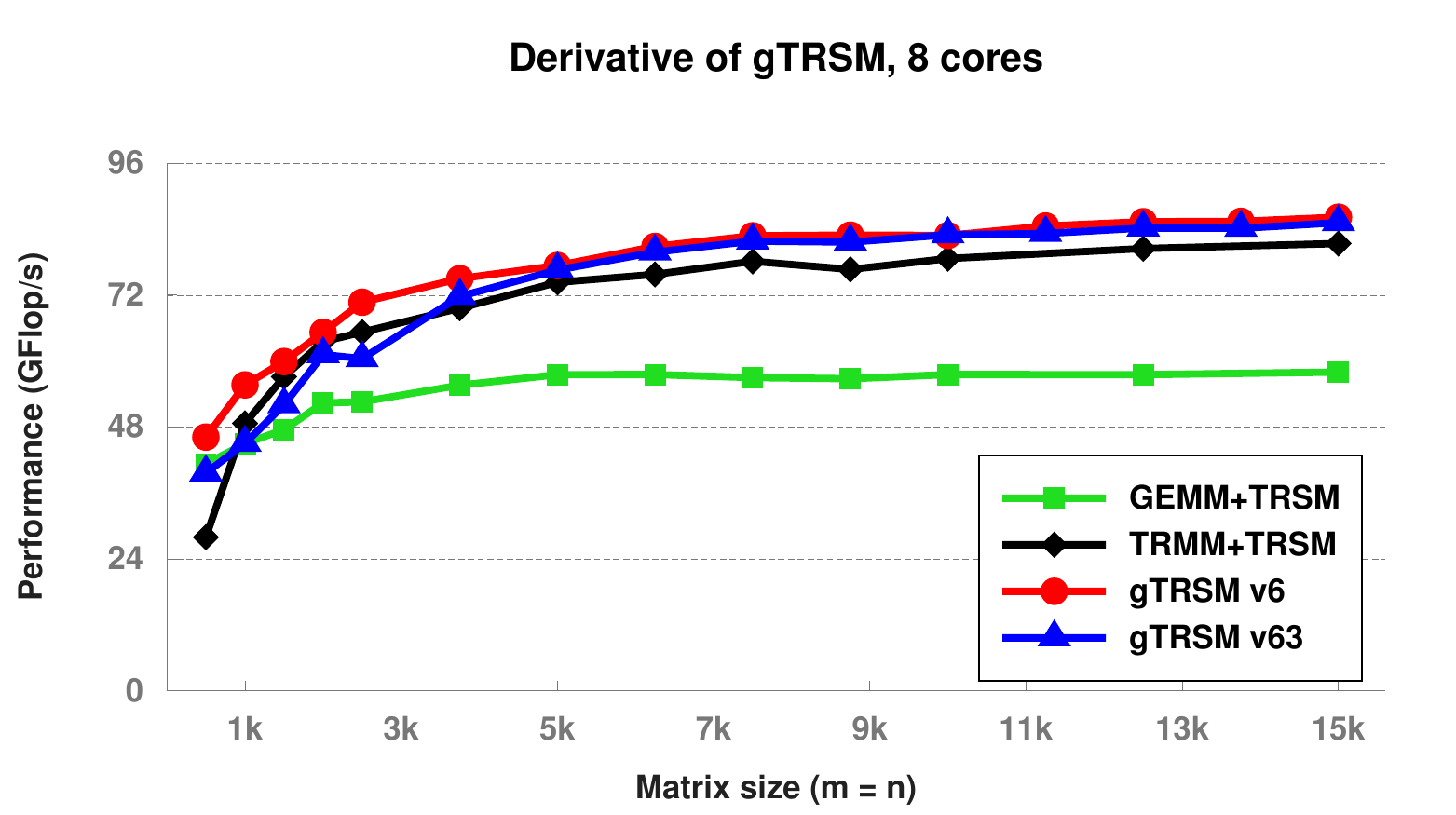}
	\caption{Performance comparison of the four routines for \gtrsm{}.
	Experiments run using the eight available cores.}
	\label{fig:gTRSM-multi}
\end{figure}

Next, we concentrate on how the routines perform in a shared-memory environment. 
Figure~\ref{fig:gTRSM-scal} provides scalability results for the four routines;
they are not only efficient, but also highly scalable: 
the speedup attained with 8 cores is of, at least, 7x.
The routine {\sc trmm+trsm} present the worst scalability
of the four due to the limited scalability
of the subtraction operation. This is reflected in the next experiment.
In Figure~\ref{fig:gTRSM-multi}, we collect the timings for the multi-threaded 
version of the routines using 8 cores. 
Due to its better scalability, \click{}'s variant 6 outperforms 
{\sc trmm+trsm} by a 6\%, and most important, 
it attains about 90\% of the peak performance.

\vspace{-2mm}
\section{Summary}
\vspace{-2mm}

In this chapter, we put the emphasis on the generation of customized
kernels for building blocks not supported by standard numerical libraries.
We used as examples two kernels arising in the derivation
of a linear system:
the derivative of the Cholesky factorization (\gchol{}) and 
the derivative of a triangular system (\gtrsm{}).

With \gchol{}, we gave a complete step-by-step example of application of \click{}; 
the example contributes a new case study
(for a non-standard linear algebra operation)
to FLAME's literature.
As a result, \click{} produced, in just a few seconds, four algorithms, 
two of which attain high performance and scalability. 
The algorithms are of interest, for instance, in the field of
statistics, when computing the variance estimation by restricted maximum likelihood (REML).
In fact, in~\cite{smith-1995} S.P. Smith discusses an unblocked version
of what we referred to as ``Variant 4'' (Figure~\ref{fig:gchol-algs}).

The \gtrsm{} operation is a characteristic example of a class of arbitrary 
building blocks that often arise in applications and for which no standard
library offers optimized kernels.
Typically, these operations may be computed as a sequence of calls
to supported kernels (in line with the approach discussed in Chapter~\ref{ch:compiler}
for \clak{}), at the expense of an overhead due to extra computation
or extra memory accesses.
We demonstrated that \click{} can generate routines based on blocked
algorithms that are competitive with or even outperform 
those based on calls to a sequence of kernels.

\chapter{Related Work}
\label{ch:relwork}

The concept of automatic program generation has been present since the appearance
of digital computers. Quoting Prywes from his 1974's survey on {\em Automatic
generation of software systems}~\cite{PrywesSurvey}:
\begin{center}
	\begin{minipage}{.8\textwidth}
	\em
		Research and development on automatic programming has been underway
		since early applications of digital computers, for nearly twenty-five
		years, and will continue for years. [...]
		Its ultimate objective is envisaged as a situation where software would
		be automatically generated for the businessman, industrialist or
		scientist, on demand. 
	\end{minipage}
\end{center}
The overarching theme in every effort towards 
automatic program generation
is the increase of productivity by 
reducing coding and maintainability cost.

\paragraph{Autotuning.}

Already in 1996, recognizing both the prominent role of the BLAS library 
in dense matrix computations and the difficulty
of providing hand-tuned optimized implementations for a broad range
of architectures, the PHiPAC project~\cite{PHiPAC} addressed the problem
of producing high-performance implementations of BLAS for
a wide range of systems with minimal effort. Instead of hand-coding the
routines, this approach consisted in a parameterized code generator
and a series of scripts to generate optimized code by varying the
parameter values; the best performing routines were selected empirically.
ATLAS~\cite{atlas-sc98} improves PHiPAC's approach in that it explores
a constrained search space, resulting in a faster optimization process.
In later stages, ATLAS combines automatically tuned code with user-contributed
hand-optimized micro-kernels.
Similarly to these two projects, FFTW~\cite{FFTW05} provides
an adaptive library for Fourier Transforms, based also on
an empirical search via actual execution and timing.

Efforts on the automatic tuning of libraries have also been made
on sparse matrix computations. For instance, OSKI~\cite{Vuduc:2004uz}
provides a collection of sparse kernels such as matrix-vector products
and the solution of triangular systems. The difference with respect to ATLAS lies
in that, since the sparsity pattern varies from matrix to matrix,
the tuning is in general deferred until run-time. While tuning affects
execution time, the cost is amortized when reused across multiple calls
to the same kernels with the same matrix, as it is the case, for instance, for iterative solvers.

The most important factor that separates both our systems, \click{} and \clak{}, 
from these projects is that, 
instead of tuning a given algorithm, we start from the mathematical description 
of the target equation and generate a family of algorithms to solve it.
The application of tuning techniques is therefore complementary to our work.

\paragraph{FLAME}

\click{} is the culmination of thorough work on the formal derivation
of loop-based linear algebra algorithms. 
In~\cite{Gunnels:2001:FFL}, Gunnels et al.~outline a series of steps for the derivation 
of correct loop-based algorithms given a loop invariant for the operation at hands.
Bientinesi extends this work in~\cite{PaulDj:PhD}, 
formalizing the entire methodology
and introducing the Partitioned Matrix Expression (PME), an object
from which loop invariants can be systematically extracted.
Bientinesi also presents evidence that an automated system is within reach. 
\click{} is the demonstration that the automation of FLAME's methodology is 
indeed feasible, and the generation of linear algebra libraries with minimal effort is possible.

FLAME also provides multiple Application Programming Interfaces (APIs) for
several programming languages and programming paradigms.
Among them, we find Elemental~\cite{elemental-toms},
a distributed-memory library that provides functionality
similar to that of FLAME and LAPACK. Elemental makes use of 
FLAME's high-level notation in its routines and may be used as
target language when code for distributed-memory architectures is desired.
In fact, 
recently, FLAME researchers prototyped DxTer~\cite{DxTer},
a system that, starting from the representation of a blocked
algorithm, like those generated by \click{}, replicates
the process carried out by domain experts to produce efficient 
distributed-memory implementations using Elemental
as target library/domain-specific language.
DxTer has been successful in generating code which is 
competitive with or even more efficient than that manually optimized by Elemental's developers.
The combination of DxTer and \click{} would enable the automatic
generation of high-performance distributed-memory implementations 
of arbitrary kernels.

\paragraph{Deduction of loop invariants.}
Software correctness is a recurrent problem in computer science.
While most software is (at most) thoroughly tested,
as Dijkstra pointed out, {\em ``Testing can only show the presence of bugs,
not their absence''}.
Proofs of correctness consist in identifying a series of predicates
assessing properties of the code at multiple points of the program.
Since developers rarely annotate their code with such predicates
or any kind of formal documentation,
a common approach to program verification is to (semi-)automatically
infer these properties.

Among the many predicates to be inferred, deduction of loop invariants
(which support the proof of correctness for loops)
have proven most complex. Automatic deduction of loop invariants dates
back to the 70's, mainly in the context of compiler technology, where asserting
them would allow code reordering and performance 
optimizations~\cite{Wegbreit:1974,Cousot:1978}.
In the last two decades, the topic gained popularity and many
techniques were developed, incorporating ideas from 
machine learning, artificial intelligence and data mining, 
among others~\cite{Ernst:2000,Flanagan:2002,Ireland:1997}.
Typically, the research efforts target general-purpose
programs, and present practical limitations for complex loops.

In sharp contrast to the above projects, FLAME's methodology
advocates a constructive approach in which, instead of
proving the correctness a posteriori, it
first identifies loop invariants  and
then builds the algorithms around them so that the invariant
is satisfied and the program is correct by construction.

\sloppypar
\paragraph{Divide-and-conquer decompositions.}
Even though the goal of the FLAME project is the derivation of 
loop-based algorithms, the underlying methodology strongly
relies on finding a recursive decomposition of the target equation in
a divide-and-conquer fashion.
This decomposition presents similarities to D. R. Smith's 
approach to formal derivation of divide-and-conquer 
algorithms~\cite{Smith:Design-of-DC}. Indeed, in~\cite{Smith:Top-Down-DC}, 
Smith demonstrates the approach
by means of sorting algorithms that may also be derived
following FLAME's techniques.
In our work, we target a class of linear algebra algorithms.

The starting point in Smith's approach for top-down
decompositions are formal specifications of the functionality
of the target problem. The functionality is expressed in terms
of input and output domains, and input and output conditions.
Such specifications are closely
related to the traditional formalism 
---precondition and postcondition predicates--- that we use
in \click{}:
the input and output domains consist of a Cartesian
product of matrix, vector, and scalar domains;
the input and output conditions are given by
the information encoded in the precondition and the
postcondition.
In both approaches, the formal specification indicates
what to solve, not how to solve it.

According to Smith~\cite{Smith:Top-Down-DC}, 
{\em ``One of the principal difficulties in top-down
design is knowing how to decompose a problem specification into
subproblem specifications''}.
In our framework, the PME represents the decomposition operator:
It states how the computation of the target problem may be
split into the computation of smaller subproblems.
The PME itself also encodes the composition operator
by indicating how the partial results, obtained from the
computation of the subproblems, are combined to assemble
the output matrices.
The base case (primitive) may be represented by the unblocked version
of the algorithms.

\paragraph{Domain-specific compilers}

After more than a decade of extensive research on domain-specific
compilers in computational science, the benefits have been shown 
in a broad variety of fields. To name a few,
the Tensor Contraction Engine (TCE)~\cite{Baumgartner05synthesisof},
the FEniCS project~\cite{Logg:2010:DAF:1731022.1731030}, and 
Spiral~\cite{Pueschel:05}, 
which target, respectively, tensor contractions,
differential equations, and linear transforms for digital signal processing.
The general approach consists in defining a high-level domain-specific
input language, and automatically generating efficient code tailored
to the target operation.

Spiral~\cite{Pueschel:05} is especially related to our work.
In its search for efficient implementations, Spiral explores
a space that  comes from the combination of breakdown rules to 
decompose the transforms in a divide-and-conquer fashion, 
and parameterized rewrite rules to incorporate knowledge of the architecture.
The PME from FLAME's methodology is closely related to the breakdown rules from Spiral.
However, Spiral targets a limited set of equations, for which 
the ``PMEs'' are taken from the literature and encoded in the system,
while in our case, the PMEs are automatically found for arbitrary input equations.
A second difference between the two projects is that we derive
loop-based algorithms that rely on computational kernels available from 
high-performance libraries, mainly BLAS, 
while Spiral derives recursive algorithms and generates its own
optimized code for the base cases.
We believe that a combination of the core ideas from both projects
has the potential to become an alternative approach for the automatic generation
of highly-optimized linear algebra libraries such as BLAS.

A more recent development is the Built-to-Order compiler (BTO)~\cite{BTO:ComposedLinAlgKernels}.
BTO addresses the generation of high-performance linear algebra kernels, with focus on
memory bound operations (e.g., BLAS 1 and BLAS 2 kernels). The main target of this compiler
are sequences of such memory bound operations to which tiling and loop fusion are applied
to reduce memory traffic. Similar results could be achieved by \click{}
provided that our compiler is extended to accept sequences of operations as input.

\paragraph{High-level languages and libraries}

High-performance computations have been traditionally associated
to low level languages such as C and Fortran.
Aiming at relieving the application developers from tedious
low level details, programming environments such as Matlab and R
act as a convenient interface to optimized libraries, at the expense
of performance.

In the last 15 years, domain-specific libraries have been developed
with the objective of extending C++ to make it more appealing to computational 
scientists and the high-performance computing community.
The introduction of Veldhuizen's Blitz++ library~\cite{BlitzArrays},
based on the so-called {\em Expression Templates} (ET)~\cite{Veldhuizen:ET}, 
opened new ways to elegant yet efficient C++ code for linear algebra computations.
Expression templates enable libraries that become linear algebra
domain-specific languages embedded within C++.
The main idea behind Blitz++ is the elimination of the overhead due to 
the creation and elimination of temporary operands, simulating, in a
sense, loop fusion.
Thus, it is specially suited for BLAS 1 and 2 operations, i.e.,
memory bound vector and matrix-vector kernels.
This same approach was later adopted by the uBLAS~\cite{uBLASOnline} library with the
objective of providing functionality similar to BLAS.

More recently, classical use of ET has been deemed
insufficient when addressing more complex operations such as matrix-matrix
products~\cite{Blaze1}. An extension, often referred to as {\em Smart}
Expression Templates (SET), is in use in modern libraries such as 
Blaze, Armadillo and Eigen~\cite{Blaze2,Armadillo,Eigen3}.
The first two address matrix-vector operations by means of classical ET
complemented with their own manually optimized code,
while for matrix-matrix products they rely on calls to optimized BLAS routines
provided by the user.
Eigen differs from Blaze and Armadillo in that it provides its own code
even for matrix-matrix products.

Beyond providing a user-friendly interface to high-performance kernels,
these libraries focus on low level optimizations for matrix and vector
products and additions.
In contrast, \clak{} targets high-level matrix equations and the discovery
of algorithms to solve them.
Also, \clak{} incorporates a number of optimizations commonly applied
by traditional general-purpose compilers~\cite{Aho:86}; these optimizations are
the logic extension to matrix operands
of techniques used by traditional compilers on vectors and scalars.
ET-based libraries can complement the decomposition performed by \clak{}
when facing memory bound computations.
Similar functionality and performance can also be achieved at compile-time by 
\click{} generated algorithms, if combined with low-level 
techniques such as the use of intrinsics for vectorization.

\paragraph{Algorithmic Differentiation}

When discussing the applicability of our compilers to algorithmic
differentiation, we compared the generated code
with that produced by ADIFOR~\cite{Bischof1996AAD}. 
While ADIFOR serves its purpose as a reference tool to which
we can compare our results, it is certainly not the only choice. 
Over the last decades, a lot of research has been carried out in the field. 
As a result, the AD landscape is populated with a
large variety of tools. Prominent examples are Tapenade~\cite{tapenade} 
and ADiMat~\cite{adimat}. The former can differentiate Fortran and C code,
while the latter targets the differentiation of Matlab code; 
both tools are based on the source code transformation approach and
can be used in forward and reverse mode.
It is worth noting that, while more modern tools like Tapenade may deliver
better performance than ADIFOR, the main results and contributions of
our work still hold.

Closely related to our work, research in progress on the 
{\em dco} AD tool~\cite{Naumann2012TAo} explores similar ideas to
those developed in this dissertation. Specifically, in~\cite{LotzNau12}
the authors study the application of AD techniques to the solution of
linear systems. By raising the level of abstraction and avoiding
a mere black-box approach, they show that the reuse of intermediate results 
leads to a reduction of the overhead incurred in computing the systems' derivatives.

\chapter{Conclusions}
\label{ch:conclusions}

In this dissertation we addressed the development
of domain-specific compilers for linear algebra
operations.
The goal was to relieve application developers from the laborious and
time consuming tasks of algorithm design and code writing,
while still matching or even surpassing the performance attained by
experts.
We presented two compilers, \clak{} and \click{};
they start from a high-level description of a target matrix operation,
together with application domain knowledge,
and return both a family of efficient algorithms that compute
the operation and the corresponding routines in the language of choice.
The main contribution of this thesis is the evidence
that linear algebra compilers, which increase experts' productivity and 
make efficiency accessible to non experts, are within reach.

In the next section, we summarize the main results of our work,
and provide references to our research publications.
We conclude with a discussion of future research directions 
to broaden and strengthen the results from this dissertation.

\section{Results}

We developed prototypes of two linear algebra compilers:
\clak{}, targeting high-level matrix equations, and
\click{}, for  building blocks.

\begin{itemize}
	\item {\bf \clak{}: Compiler for matrix equations}~\cite{CLAK-VECPAR12,CLAK-IJHPCA}.
		We presented the design of \clak{}, a domain-specific compiler for 
		linear algebra equations.
		\clak{} models the reasoning of the thought-process of a human expert,
		and extends it with the exploration power of a computer.
		The generation of algorithms centers around the decomposition of a target equation 
		into a sequence of calls to kernels provided by libraries such as BLAS and LAPACK. 
		The decomposition is not unique, and even for simple equations many alternative
		algorithms can be generated; 
		a number of heuristics, guided by both linear algebra and domain knowledge,
		are used to prune the search space while tailoring the algorithms to the application.
		In the discussion, we uncovered the modules that constitute the compiler's engine.
		The following modules were discussed: 
		1) The algebraic manipulation of expressions and knowledge management,
		2) the interface to the available building blocks,
		3) the inference of properties for the dynamic deduction of knowledge,
		4) the analysis of dependencies for the reduction of the computational cost, and
		5) the Matlab and Fortran code generation.

	\item {\bf \clickplain{}: Complete automation of FLAME's methodology}~\cite{CASC-2011-PME,ICCSA-2011-LINV}.
		The FLAME project provides a systematic methodology for the 
		derivation of correct loop-based algorithms.
		While the FLAME literature offers many examples of the manual
		application of the methodology to traditional operations,
		little evidence existed that it could be made completely mechanical.
		\click{} demonstrates that it can be, indeed, automatically 
		carried out by a computer.
		Given the sole description of an operation, \click{}
		derives families of algorithms that compute it.
		The compiler takes a three-stage approach:
		First, we detailed the generation of the PME(s),
		a recursive definition of the operation in a divide-and-conquer fashion.
		Then, we illustrated the analysis of the PME to identify a family
		of loop invariants.
		Finally, we described how each loop invariant is transformed into its
		corresponding loop-based algorithm.
		We demonstrated that the methodology applies
		not only to standard operations, but also to new
		kernels.
\end{itemize}

This dissertation also makes contributions to the fields
of algorithmic differentiation, and computational biology.

\begin{itemize}
	\item {\bf BLAS and LAPACK derivatives for algorithmic differentiation (AD).}
		The code generated by source-transformation forward-mode AD tools
		for the derivative of BLAS and LAPACK operations
		suffers from low performance.
		We illustrated how our compilers automatically generate 
		high-performance code for derivative operations.
		By raising the level of abstraction from scalars to matrices, 
		the compilers produce derivative routines that exploit
		library-provided optimized kernels;
		then, by means of a data dependency analysis, the complexity
		of the resulting code may also be reduced.
		We observed speedups with respect to ADIFOR's routines
		ranging from 5x to 80x.
		Our work contributes, first, a study of the 
		potential benefits, should high-performance differentiated 
		versions of BLAS and LAPACK be available, and
		second, a demonstration that the automatic generation
		of efficient differentiated versions 
		of these libraries is within reach.

	\item {\bf High-Performance algorithms for GWAS}~\cite{SingleGWAS,MultiGWAS,SingleTraitDistMem}.
		Genome-wide association studies carry out large-scale
		data analyses, and require performing computations ranging
		from teraflops to hundreds of petaflops. 
		While state-of-the-art libraries are satisfactory
		for short to medium problem sizes,
		they are not practical for large-scale problems.
		When applied to the GWAS equation, \clak{} yielded a family of 
		specialized algorithms that efficiently solve the equation
		and achieve a lower complexity compared to existing algorithms.
		\clak{}'s algorithms led to high-performance out-of-core 
		routines that largely outperform state-of-the-art libraries~\cite{OmicABEL}. 
		These routines have been collected in the publicly available OmicABEL
		package,\footnote{Available at \texttt{http://www.genabel.org/packages/OmicABEL}} 
		as part of the GenABEL suite for statistical genomics.
\end{itemize}

\section{Future work}
\label{sec:future-work}

The tools presented in this dissertation can be extended in 
a number of ways.
Here, we briefly discuss the most promising extensions,
which we believe are within reach.

\begin{itemize}
	\item {\bf Integration of performance analysis techniques.}
		In order to attain high-performance in a variety of scenarios, our compilers
		generate families of algorithms;
		a challenging and critical component in a compiler is
		the automatic selection of the best one. 
		So far, each produced algorithm is accompanied with its computational cost;
		however, since the mere operation count is not a reliable metric,
		we aim at incorporating advanced techniques for performance
		prediction. A promising direction relies on a sample-based approach: 
		The idea is to create performance models not for the competing algorithms, 
		but only for those routines that are used as building blocks. 
		By combining the models, it is then possible to make  
		predictions and to accurately rank the algorithms~\cite{Peise2012:50}.

	\item {\bf Support for an extended class of equations.}
		While broad, the range of supported equations in \clak{} is still rather limited.
		We aim at extending the scope of the compiler by handling more complex operations, ranging from
		explicit equations (as opposed to assignments) 
		to determinants, logarithms, and matrix functions in general.

	\item {\bf Algorithm analysis and code generation for parallel architectures.}
		Our compilers incorporate modules for the translation of the generated
		algorithms into code. However, only sequential and multi-threaded code 
		(via multi-threaded implementations of BLAS and LAPACK) is produced.
		The variety of available computing platforms 
		(e.g, multi- and many-core processors,
		clusters, and 
		co-processors such as GPGPUs)
		demands the generation of algorithms that are tailored not
		only to the application but also to the architecture.
		To this end, we envision the development of a number of modules
		responsible for the tailoring to each specific architecture and type of parallelism; 
		for instance,
		algorithms by blocks (out-of-order execution) for multi- and many-cores,
		distributed-memory for clusters, and
		the offload of computation to accelerators.

	\item {\bf Support for the reverse mode of algorithmic differentiation (AD).}
		While we only explored the use of our compilers in the forward mode of AD, 
		we believe that similar techniques may be applied to the reverse mode.
		The main extension to support the reverse mode involves the inclusion of
		transformation rules corresponding to the chain rule in the forward mode,
		and the support for additional operators like the trace of a matrix~\cite{Giles2008CMD}.
		As ongoing work in the field evidences, similar results in terms of
		reduction of complexity and increase of performance would be appreciated by
		the AD community. 

\end{itemize}

\appendix
\chapter{BLAS and LAPACK Routines}
\label{app:blas-lapack}

We list the BLAS and LAPACK operations used
across this dissertation, together with their
description.

\vspace{15mm}

\begin{table}[!ht]
\renewcommand{\arraystretch}{1.4}
\centering
  \begin{tabular}{l@{\hspace*{4mm}} l@{\hspace*{4mm}} l} \toprule
	  \multicolumn{3}{c}{\bf BLAS 1} \\ \midrule
	  {\sc scal} & Vector scaling & $y := \alpha y$ \\
	  {\sc dot}  & Dot product & $\alpha := x^T y$ \\
	  {\sc axpy} & Vector scaling and addition & $y := \alpha x + y$ \\
	  \bottomrule
	  \multicolumn{3}{c}{\bf BLAS 2} \\ \midrule
	  {\sc ger}  & Outer vector product  & $A := \alpha x y^T + A$ \\
	  {\sc gemv} & Matrix-vector product & $y := \alpha A^{\bullet} x + \beta y$ \\
	  {\sc trsv} & Triangular system with single right-hand side & $x := T^{{\bullet}^{-1}} b$ \\
	  \bottomrule
  \end{tabular}
\caption{Collection of the BLAS 1 and 2 routines used in the algorithms presented in this dissertation.
         Greek, lowercase, and uppercase letters are used for scalars, vectors, and matrices, respectively;
	 	 $T$ is a triangular matrix; $A^{\bullet}$ indicates that matrix $A$ may be used either transposed or not.}
\label{tab:blas-lapack1}
\end{table}

\begin{table}[!ht]
\renewcommand{\arraystretch}{1.4}
\centering
  \begin{tabular}{l@{\hspace*{4mm}} l@{\hspace*{4mm}} l} \toprule
	  \multicolumn{3}{c}{\bf BLAS 3} \\ \midrule
	  {\sc gemm}  & Matrix-matrix product & $C := \alpha A^{\bullet} B^{\bullet} + \beta C$ \\
	  {\sc syrk}  & Matrix-matrix product, $C$ symmetric & $C := \alpha A A^T + \beta C$ \\
	              &                                      & $C := \alpha A^T A + \beta C$ \\
	  {\sc syr2k} & Matrix-matrix product, $C$ symmetric & $C := \alpha A B^T + B A^T + \beta C$ \\
	              &                                      & $C := \alpha A^T B + B^T A + \beta C$ \\
	  {\sc trmm}  & Matrix-matrix product, $A$ triangular & $B := \alpha A^{\bullet} B$ \\
	              &                                       & $B := \alpha B A^{\bullet}$ \\
	  {\sc trsm}  & Triangular system with                           & $X := \alpha T^{\bullet^{-1}} B$ \\
	              &                        multiple right-hand sides & $X := \alpha B T^{\bullet^{-1}}$ \\
	  \bottomrule
	  \multicolumn{3}{c}{\bf LAPACK} \\ \midrule
	  {\sc posv}  &\multicolumn{2}{l}{SPD system with multiple right-hand sides} \\
	  {\sc potrf} &\multicolumn{2}{l}{Cholesky factorization} \\
	  {\sc syevr} &\multicolumn{2}{l}{Eigendecomposition of a symmetric matrix} \\
	  {\sc geqrf} &\multicolumn{2}{l}{QR factorization} \\
	  {\sc ormqr} &\multicolumn{2}{l}{Matrix-matrix product with a Q matrix as returned by {\sc geqrf}} \\
	  \bottomrule
  \end{tabular}
\caption{Collection of the BLAS 3 and LAPACK routines used in the algorithms presented in this dissertation.
         Greek, lowercase, and uppercase letters are used for scalars, vectors, and matrices, respectively;
	 	 $T$ is a triangular matrix; $A^{\bullet}$ indicates that matrix $A$ may be used either transposed or not.}
\label{tab:blas-lapack2}
\end{table}

\chapter{Code Samples}
\label{app:code}

\vspace{-4mm}
We collect a sample of the routines generated by our compilers
for the multiple experiments presented in this dissertation.
In Sections~\ref{sec:gspd-code} and~\ref{sec:gsyrk-code},
we provide examples of routines generated by \clak{}
for the derivative of the SPD linear system and the {\sc syrk} kernel,
respectively; specifically, we include the
routines for the most general derivatives.
In Sections~\ref{sec:gchol-code} and~\ref{sec:gtrsm-code},
we include, respectively, the four routines generated by \click{} 
for the derivative of Cholesky and 
two of those produced for g{\sc trsm}.


\section{g\textsc{spdsolve}}
\label{sec:gspd-code}

\vspace{-1mm}

\noindent
\renewcommand{\lstlistingname}{Routine}
\begin{minipage}{\linewidth}
\footnotesize
\begin{lstinputlisting}[caption={g\textsc{SPD}. Operands $A$ and $B$ are active.},
                     label=code:gspd,
                     numbers=left,
                     basicstyle={\tt},
					 numberblanklines=true,
                     fancyvrb=true,language=Fortran,columns=fixed,basewidth=.5em,frame=b,
                      framexleftmargin=-0pt,
                      framexrightmargin=0pt,
					  xleftmargin=10pt]{Appendix_code/gspd/gSPD_11_1.f}
\end{lstinputlisting}
\end{minipage}


\section{g\textsc{syrk}}
\label{sec:gsyrk-code}

\vfill

\noindent
\renewcommand{\lstlistingname}{Routine}
\begin{minipage}{\linewidth}
\footnotesize
\begin{lstinputlisting}[caption={g\textsc{syrk}. Operands $\alpha$, $A$, $\beta$, and $C$ are active.},
                     label=code:gsyrk,
                     numbers=left,
                     basicstyle={\tt},
					 numberblanklines=true,
                     fancyvrb=true,language=Fortran,columns=fixed,basewidth=.5em,frame=b,
                      framexleftmargin=-0pt,
                      framexrightmargin=0pt,
					  xleftmargin=10pt]{Appendix_code/gsyrk/gSYRK_1111.f}
\end{lstinputlisting}
\end{minipage}

\vfill

\section{\gchol{}}
\label{sec:gchol-code}

\noindent
\renewcommand{\lstlistingname}{Routine}
\begin{minipage}{\linewidth}
\footnotesize
\begin{lstinputlisting}[caption={\gchol{}. Variant 1.},
                     label=code:gchol-var1,
                     numbers=left,
                     basicstyle={\tt},
					 numberblanklines=true,
                     fancyvrb=true,language=Fortran,columns=fixed,basewidth=.5em,frame=b,
                      framexleftmargin=-0pt,
                      framexrightmargin=0pt,
					  xleftmargin=10pt]{Appendix_code/gchol/gChol_blk_var1.c}
\end{lstinputlisting}
\end{minipage}

\noindent
\begin{minipage}{\linewidth}
\footnotesize
\begin{lstinputlisting}[caption={\gchol{}. Variant 2.},
                     label=code:gchol-var2,
                     numbers=left,
                     basicstyle={\tt},
					 numberblanklines=true,
                     fancyvrb=true,language=Fortran,columns=fixed,basewidth=.5em,frame=b,
                      framexleftmargin=-0pt,
                      framexrightmargin=0pt,
					  xleftmargin=10pt]{Appendix_code/gchol/gChol_blk_var2.c}
\end{lstinputlisting}
\end{minipage}

\noindent
\begin{minipage}{\linewidth}
\footnotesize
\begin{lstinputlisting}[caption={\gchol{}. Variant 3.},
                     label=code:gchol-var3,
                     numbers=left,
                     basicstyle={\tt},
					 numberblanklines=true,
                     fancyvrb=true,language=Fortran,columns=fixed,basewidth=.5em,frame=b,
                      framexleftmargin=-0pt,
                      framexrightmargin=0pt,
					  xleftmargin=10pt]{Appendix_code/gchol/gChol_blk_var3.c}
\end{lstinputlisting}
\end{minipage}

\noindent
\begin{minipage}{\linewidth}
\footnotesize
\begin{lstinputlisting}[caption={\gchol{}. Variant 4.},
                     label=code:gchol-var4,
                     numbers=left,
                     basicstyle={\tt},
					 numberblanklines=true,
                     fancyvrb=true,language=Fortran,columns=fixed,basewidth=.5em,frame=b,
                      framexleftmargin=-0pt,
                      framexrightmargin=0pt,
					  xleftmargin=10pt]{Appendix_code/gchol/gChol_blk_var4.c}
\end{lstinputlisting}
\end{minipage}

\newpage

\section{\gtrsm{}}
\label{sec:gtrsm-code}

\noindent
\begin{minipage}{\linewidth}
\footnotesize
\begin{lstinputlisting}[caption={\gtrsm{}. Variant 6.},
                     label=code:gtrsm-var6,
                     numbers=left,
                     basicstyle={\tt},
					 numberblanklines=true,
                     fancyvrb=true,language=Fortran,columns=fixed,basewidth=.5em,frame=b,
                      framexleftmargin=-0pt,
                      framexrightmargin=0pt,
					  xleftmargin=10pt]{Appendix_code/gtrsm/gTRSM_blk_var6.c}
\end{lstinputlisting}
\end{minipage}

\noindent
\begin{minipage}{\linewidth}
\footnotesize
\begin{lstinputlisting}[caption={\gtrsm{}. Variant 63.},
                     label=code:gtrsm-var63,
                     numbers=left,
                     basicstyle={\tt},
					 numberblanklines=true,
                     fancyvrb=true,language=Fortran,columns=fixed,basewidth=.5em,frame=b,
                      framexleftmargin=-0pt,
                      framexrightmargin=0pt,
					  xleftmargin=10pt]{Appendix_code/gtrsm/gTRSM_blk_var63.c}
\end{lstinputlisting}
\end{minipage}


\end{document}